\documentclass[reqno,12pt,a4paper]{article}

\usepackage{amsmath,amssymb,amsthm,amsfonts,times,array,geometry,color}

\usepackage{epsfig} 
\usepackage{tikz}
\usepackage{mathrsfs}

\usetikzlibrary{calc}

\newcommand*\savesymbol[1]{%
  \expandafter\let\csname orig#1\expandafter\endcsname\csname#1\endcsname
  \expandafter\let\csname #1\endcsname\relax
}

\newcommand*\restoresymbol[2]{%
  \expandafter\global\expandafter\let\csname#1#2\expandafter\endcsname%
    \csname#2\endcsname
  \expandafter\global\expandafter\let\csname#2\expandafter\endcsname%
    \csname orig#2\endcsname
}
\savesymbol{pa}
\savesymbol{st}
\savesymbol{func}

\numberwithin{equation}{section}

\newcommand{\st}{\mspace{5mu} : \mspace{5mu}}

\newcommand{\gconn}{g}
\newcommand{\LL}{\mathsf{L}}
\newcommand{\Borelp}{\mathcal{U}_q(\mathfrak{b}^+)}
\newcommand{\Borelm}{\mathcal{U}_q(\mathfrak{b}^-)}
\newcommand{\Nilm}{\mathcal{U}_q(\mathfrak{n}^-)}
\newcommand{\Nilp}{\mathcal{U}_q(\mathfrak{n}^+)}
\newcommand{\Borelmp}{\mathcal{U}_q(\mathfrak{b}^{\mp})}
\newcommand{\Borelpm}{\mathcal{U}_q(\mathfrak{b}^{\pm})}

\newcommand{\vep}{\varepsilon}
\newcommand{\al}{\alpha}

\newcommand{\be}{\beta}
\newcommand{\ga}{\gamma}

\newcommand{\de}{\delta}
\newcommand{\De}{\Delta}
\newcommand{\ep}{\epsilon}
\newcommand{\la}{\lambda}

\newcommand{\si}{\sigma}

\newcommand{\vf}{\varphi}

\newcommand{\bw}{\bar{w}}

\newcommand{\bmu}{\bar{\mu}}

\newcommand{\CA}{{\mathcal A}}
\newcommand{\CB}{{\mathcal B}}

\newcommand{\BFE}{{\rm\mathbf E}}
\newcommand{\BFF}{{\rm\mathbf F}}

\newcommand{\CE}{{\mathcal E}}
\newcommand{\CF}{{\mathcal F}}

\newcommand{\CH}{{\mathcal H}}

\newcommand{\CJ}{{\mathcal J}}

\newcommand{\CL}{{\mathcal L}}
\newcommand{\CM}{{\mathcal M}}

\newcommand{\CO}{{\mathcal O}}
\newcommand{\CP}{{\mathcal P}}
\newcommand{\CQ}{{\mathcal Q}}
\newcommand{\CR}{{\mathscr R}}

\newcommand{\CU}{{\mathcal U}}
\newcommand{\CV}{{\mathcal V}}
\newcommand{\CW}{{\mathcal W}}

\newcommand{\CY}{{\mathcal Y}}

\newcommand{\SA}{{\mathsf A}}
\newcommand{\SB}{{\mathsf B}}
\newcommand{\SC}{{\mathsf C}}
\newcommand{\SD}{{\mathsf D}}
\newcommand{\SE}{{\mathsf E}}
\newcommand{\SF}{{\mathsf F}}

\newcommand{\SH}{{\mathsf H}}

\newcommand{\SK}{{\mathsf K}}
\newcommand{\sk}{{\mathsf k}}
\newcommand{\SL}{{\mathsf L}}

\newcommand{\SM}{{\mathsf M}}

\newcommand{\SO}{{\mathsf O}}
\newcommand{\SP}{{\mathsf P}}
\newcommand{\SQ}{{\mathsf Q}}
\newcommand{\SR}{{\mathsf R}}
\renewcommand{\SS}{{\mathsf S}}
\newcommand{\ST}{{\mathsf T}}
\newcommand{\SU}{{\mathsf U}}
\newcommand{\SV}{{\mathsf V}}

\newcommand{\SX}{{\mathsf X}}
\newcommand{\SY}{{\mathsf Y}}
\newcommand{\SZ}{{\mathsf Z}}

\newcommand{\fg}{{\mathfrak g}}

\newcommand{\sa}{{\mathsf a}}

\newcommand{\sfc}{{\mathsf c}}

\newcommand{\se}{{\mathsf e}}
\newcommand{\sg}{{\mathsf g}}
\renewcommand{\sf}{{\mathsf f}}

\newcommand{\sq}{{\mathsf q}}
\newcommand{\spp}{{\mathsf p}}
\newcommand{\sr}{{\mathsf r}}

\newcommand{\mst}{{\mathsf t}}
\newcommand{\su}{{\mathsf u}}
\newcommand{\sv}{{\mathsf v}}

\newcommand{\sw}{{\mathsf w}}

\newcommand{\sx}{{\mathsf x}}
\newcommand{\sy}{{\mathsf y}}
\newcommand{\sz}{{\mathsf z}}

\newcommand{\ot}{\otimes}
\newcommand{\fsl}{{\mathfrak s}{\mathfrak l}}
\newcommand{\fgl}{{\mathfrak g}{\mathfrak l}}
\newcommand{\pa}{\partial}
\newcommand{\ra}{\rightarrow}
\newcommand{\rf}[1]{\eqref{#1}}

\newcommand{\BR}{{\mathbb R}}

\newcommand{\BC}{{\mathbb C}}

\newcommand{\BP}{{\mathbb P}}

\newcommand{\BZ}{{\mathbb Z}}

\newcommand{\fr}[2]{{\textstyle\frac{#1}{#2}}}

\newcommand{\sst}{\scriptscriptstyle}

\newcommand{\alg}[1]{\mathfrak{#1}}

\newcommand{\func}[2]{#1 \left( #2 \right)}

\newcommand{\brac}[1]{\left( #1 \right)}
\newcommand{\sqbrac}[1]{\left[ #1 \right]}
\newcommand{\set}[1]{\left\{ #1 \right\}}

\newcommand{\ka}{\kappa}
\newcommand{\eps}{\varepsilon}

\newcommand{\wun}{\mathbf{1}}

\newcommand{\affine}[1]{\widehat{#1}}

\newcommand{\comm}[2]{\bigl[ #1 , #2 \bigr]}

\newcommand{\normord}[1]{{} : #1 : {}} 

\newcommand{\QEA}[2]{\mathcal{U}_{#1} \bigl( #2 \bigr)}

\newcommand{\qnum}[2]{\sqbrac{#1}_{#2}}

\newcommand{\eqnref}[1]{Equation~\eqref{#1}}

\DeclareMathOperator{\id}{id}
\DeclareMathOperator{\tr}{tr}

\newcommand{\qea}{quantum affine algebra}

\theoremstyle{plain}

\theoremstyle{remark}
\newtheorem{rem}{Remark}

\begin{document}
\title{Integrable light-cone lattice discretizations from the universal R-matrix}
\author{C. Meneghelli$^{1)}$, J. Teschner$^{2)}$}

\maketitle

\begin{center}
$^{1)}$ Simons Center for Geometry and Physics, \\ Stony Brook University, Stony Brook, NY 11794-3636, USA\\[1ex]
$^{2)}$ DESY Theory, Notkestr. 85, 22607 Hamburg, Germany.
\end{center}

\begin{abstract}  Our goal is to develop a more general scheme for constructing 
integrable lattice regularisations of integrable quantum field theories. Considering 
the affine Toda theories as examples, we show how to construct such lattice
regularisations using the representation theory of quantum affine algebras. 
This requires us to clarify in particular the relations between the light-cone approach to 
integrable lattice models and the representation theory of quantum affine algebras.
Both are found to be related in a very natural way, suggesting a general scheme for the
construction of generalised Baxter Q-operators.
One of the main difficulties we need to deal with is coming from the infinite-dimensionality 
of the relevant families of representations. It is handled by means of suitable renormalisation
prescriptions defining what may be called the modular double of quantum affine algebras.
This framework allows us to give a representation-theoretic proof of finite-difference equations
generalising the Baxter equation.
\end{abstract}

\setcounter{tocdepth}{2}

\tableofcontents

\section{Introduction and conclusions}

\subsection{Motivation and background}

Integrable quantum field theories offer a unique theoretical laboratory
for the exploration of several non-perturbative phenomena in quantum
field theory. Having full quantitative control about the spectrum or
even expectation values in a quantum field theory paves the way towards 
detailed 
investigations of non-perturbative effects 
like the existence of dual Lagrangian descriptions in different regions
of the parameter space.

However, up to now there are only a few examples where this has been
realised. Many two-dimensional quantum field theories of interest
are conjectured to be integrable, but this has rarely been fully 
demonstrated. Exact results have been proposed on some of these 
quantum field theories, but in most cases we do not know 
how to derive these results from first principles.
It would be desirable to have a more systematic framework for 
constructing and solving integrable quantum field theories. 

Exploiting integrability in a quantum field theoretical context is
not easy. 
One of the main problems is to regularise the
UV-divergencies in such a way that integrability is preserved. 
If this is possible, one may indeed hope that the enhanced control
provided by integrability can lead to a precise understanding of 
the dependence of physical quantities on the cut-off, and how
to remove it in the end. Lattice regularisations have been used to reach this goal 
with some success.  Prominent examples are the 
massive Thirring / Sine-Gordon models for which some exact results have 
been obtained by using the XXZ or the XYZ 
spin chains as a lattice regularisation.

Up to now there does not seem to exist a systematic procedure for 
constructing integrable lattice regularisations for a given Lagrangian field theory.
A proposal in this direction was made in \cite{RT}. This proposal
was inspired by the well-known relations between integrable lattice models and 
the representation theory of quantum groups.  Possible hopes that relations of this 
type may hold even in a quantum field theoretical context are 
supported in particular by the works \cite{BLZ3,BHK} where beautiful
relations between the integrable structure of conformal field theory and 
quantum group representation theory were found.
Starting from a Lagrangian description of the field theory of 
interest it was proposed in \cite{RT} to 
\begin{itemize}
\item identify the relevant quantum group using the algebra of interaction terms in the 
light-cone formulation of the dynamics,
\item and construct the main ingredients of integrable lattice regularisations like Lax-matrices and 
R-matrices from the representation theory of this quantum group.
\end{itemize}
The feasibility of such a program was illustrated by constructing integrable
lattice regularisations of some Lagrangian field theories on the kinematical level.
Taking into account the form of the Lax matrices expressing integrability on the classical 
level leads to almost unique answers for R- and Lax-matrices defining the integrable lattice
regularisation quantum-mechanically. A more general approach to identifying the quantum algebraic
structures behind integrable perturbations of conformal field theories was proposed in \cite{BR}.

Our goal in this paper is to illustrate how the crucial next steps in this program can be performed: the 
definition of an integrable time-evolution and the construction of Baxter Q-operators. 

\subsection{Approach}

To reach our goals we will use the light-cone approach to integrable lattice models introduced in 
\cite{FV92}, and further developed in \cite{BBR}, see in particular \cite{BS15} for 
recent developments of this approach. It has been pointed out in \cite{RT} that this 
approach is particularly well-suited for using quantum group representation theory 
to construct integrable lattice regularisations of more general Lagrangian field theories. 
A new feature
introduced in \cite{RT} is the possibility to have a natural relation between light-cone directions
and Borel sub-algebras of the relevant quantum groups. 
Previous versions of the light-cone lattice formalism
used a slightly different formulation in which this is not manifest\footnote{See Remark \ref{sl2rem} in 
Section \ref{sec:alternating} for a comparison}.
This feature is important for the further development of the formalism as 
it leads in particular to a very 
natural relation between the lattice time-evolution operators and the universal R-matrix.

For simplicity we will focus on the affine Toda theories where the relevant quantum groups 
are the quantum affine algebras $\CU_q(\widehat{\fsl}_M)$, but we expect the resulting 
scheme to be of much wider applicability. The integrable field theories related to quantum affine
super-algebras discussed in \cite{RT}, for example, should be within reach.

For the cases of our interest we will
explain how to construct time-evolution and Baxter Q-operators from the universal R-matrix
of the relevant quantum groups. Our main tool will be the product formula for the 
universal R-matrix found in \cite{KT}.
The main difficulties in constructing
time-evolution and Baxter Q-operators from the universal R-matrix are due to the fact that
we need to evaluate the R-matrix in infinite-dimensional representations. 
This feature appears to be inevitable if one wants to have tailor-made lattice discretisations 
of field theories having non-compact target space.
The product formula 
represents the R-matrix as an infinite product over factors which are infinite
sums over powers of the generators of the quantum affine algebra. It is therefore not obvious 
how to produce well-defined operators from the product formula for the universal
R-matrix if the representations of interest are infinite-dimensional.  

Our approach to handle the resulting difficulties is based on two main elements:
\begin{itemize}
\item
We will observe that the representations needed to get 
light-cone Lax matrices and evolution operators 
from the universal R-matrix have a remarkable property: The infinite 
products resulting from the product formulae for the universal R-matrix 
truncate automatically to finite products. The use of the light-cone lattice approach therefore
allows us to solve one of the two problems coming from the infinite-dimensionality of the 
relevant representations.

\item 
The infinite-dimensional representations that we need for our goals
have the useful feature that the generators of the quantum affine algebras 
are represented by positive self-adjoint operators. This feature will allow us to replace the infinite
sums over powers of the generators appearing in the product formula 
by well-defined operator-functions. We will demonstrate that 
this replacement preserves the validity of all relevant relations satisfied by
the universal R-matrix in the representations of our interest. 
 \end{itemize}

Our choice of representations 
is motivated by the fact that the positive self-adjoint
operators representing the quantum group generators correspond to positive 
quantities in the affine Toda theories.

\subsection{Conclusions}

The main conclusions we'd like to draw from our results are the following: Combining the
light-cone lattice approach with the representation theory of quantum affine algebras
gives us a systematic way to construct integrable lattice discretisations of the affine 
Toda theories. Non-compactness of the space in which the fields take values motivates us 
to consider infinite-dimensional representations of the relevant quantum affine algebras.
However, we only need to consider the simplest nontrivial representations of this type.
Infinite-dimensionality can be handled by expressing the main objects (time evolution- and
Q-operators) in terms of the non-compact quantum dilogarithm function. One thereby gets 
a natural renormalisation of the formal expressions obtained from the universal R-matrix,
leading to fairly simple explicit formulae for the time evolution- and Q-operators.
The relevant properties (commutativity, functional relations) 
all boil down to known properties of the 
non-compact quantum dilogarithm. Verifying this in some detail accounts for a fair amount
of the work that went into this paper, but once this is understood in these cases it should be
possible to generalise our approach to wider classes of theories without excessive efforts.

\subsection{Summary of main results}

As our paper is quite long, we will now offer more detailed
overviews over the main results.


As indicated above, one of our main goals is to clarify the relation between the universal 
R-matrix of $\CU_q(\widehat{\fsl}_M)$ and the Baxter Q-operators from which the 
evolution operators are recovered by specialising the spectral parameter. It  will be
obtained by a variant of the scheme proposed in \cite{BLZ3}. The necessary modifications
are two-fold. The place of the infinite-dimensional representations 
of the Borel sub-algebras $\Borelpm$ of $\CU_q(\widehat{\fsl}_M)$
of q-oscillator type employed in \cite{BLZ3} in auxiliary space 
will be taken by representations which are neither of 
highest nor lowest weight type. This appears to be inevitable in order to get
operators with favourable analytic properties.  In quantum space
we will use representations of $\CU_q(\widehat{\fsl}_M)$ that can be represented as 
tensor products of the same type of representations as used in auxiliary space. 
The tensor products display a staggered structure reflecting a factorisation of the 
monodromy matrix into factors associated to light-like segments.


Our main results include a derivation of generalised
Baxter T-Q-relations. The Baxter equations are found to follow from the reducibility 
of certain tensor products of representation at particular values of their 
parameters, in this respect resembling previous derivations of functional equations
for transfer matrices from the representation theory 
of quantum affine algebras given in \cite{BLZ3,AF}.
Two features of our derivation appear to be new. Our derivation on the one hand 
uses an interesting finite-dimensional representation constructed from 
fermionic oscillators. This allows us to leads to simplify algebraic 
aspects of the derivation. We furthermore need to handle
the additional issues originating from the fact that our representations do not 
have extremal weight vectors.

We furthermore find fairly simple explicit formulae for the kernels
representing the Baxter Q-operators.  The formulae are simplest when a
variant of the quantum affine algebra $\CU_q(\widehat{\fsl}_M)$ is used for the
construction of integrable lattice models that differs from the standard one by a
Drinfeld twist. The resulting expressions resemble the formulae found in \cite{BKMS,DJMM} 
for the transfer-matrices of generalised Chiral Potts Models.
Having explicit formulae for the kernels of the Q-operators should allow us to 
determine the analytic properties of these operators by generalising the 
results of \cite{ByT}. Our results thereby lay the foundations for future 
analytic studies of the spectrum of the affine Toda field theories.

\subsection{Mathematical aspects}

As indicated above, one of our main tasks is to give a sense to the formal 
expressions obtained by evaluating the product formula for the universal R-matrix 
in the infinite-dimensional representations of our interest. 
These representations are in some respects similar to the representations 
of q-oscillator type employed in \cite{BLZ3,BHK}. The terminology
pre-fundamental representations was introduced in \cite{HJ} for a family of
representations of the Borel sub-algebras of quantum affine algebras
generalising the representations of q-oscillator type considered in \cite{BLZ3,BHK}.
As opposed to \cite{BLZ3,BHK,HJ} we will here
be interested in representations of the q-oscillator algebra that 
have no extremal weight. This being understood we will adopt the terminology 
"pre-fundamental" for the simple representations of the Borel sub-algebras that 
will be used as building blocks for the class of representations of our interest.

What will allow us to regain mathematical 
control in the absence of extremal weights is the fact that the generators are 
represented in terms of 
positive self-adjoint operators. This implies that our representations behave in some 
respects similar to the 
representations of the modular double of $\CU_q(\fsl_2)$ introduced in \cite{PT99,F99}.
The terminology modular double refers to the fact that these representations are 
simultaneously 
representations of the algebra obtained by replacing the deformation parameter
$q=e^{-\pi i b^2}$ by the parameter $\tilde{q}=e^{-\pi i /b^2}$. 
Taking tensor products of pre-fundamental representations
will generate various other representations including 
evaluation representations of modular double type.

We will observe that the special features of pre-fundamental representations of 
modular double type allow us to define a canonical renormalisation of the 
formal expressions obtained by evaluating the universal R-matrix in such representations.
The infinite products representing the universal R-matrix 
get automatically truncated to a finite product when evaluated on 
pre-fundamental
representations. Most of the remaining factors 
are expressed in terms of the 
quantum exponential function. Replacing this function by the non-compact quantum dilogarithm
preserves the relevant algebraic properties and 
produces expressions that are well-defined in representations of modular-double type.
The most delicate aspect is to find renormalised versions of the contributions of the 
imaginary roots in the product formula.  This is crucial 
in particular for giving representation-theoretic proofs of generalised Baxter equations.
We will show that there is an essentially canonical renormalisation
for these contributions as well. In order to see this, it will be necessary to study some aspects
of the behaviour of the product formula under the action of the co-product 
that do not seem to be discussed in the literature. 

\subsection{Relations to previous work}

The affine Toda theories have been extensively studied already. 
A lot is known about the affine Toda theories in infinite volume including factorised S-matrices
\cite{AFZ,BCDS,CM1,CM2} and form-factors \cite{L97,OL}\footnote{To keep the length of the list of references within reasonable bounds we only quote literature studying affine Toda theories of higher rank ($M>2$) which are
the main objects of interest in our paper.}. This can be used to 
predict the ground-state energy in the finite volume via the thermodynamic 
Bethe ansatz \cite{FKS}. 

The full finite-volume spectrum is not easily accessible in this way, motivating the use
of lattice regularisations. Lattice Lax-matrices and an 
integrable lattice dynamics
have been proposed in \cite{KaR}. A Lie-theoretic framework for constructing discrete versions
of the Toda flow on the classical level was presented in \cite{HKKR}.

The connection to the quantum affine algebra $\CU_q(\widehat{\fsl}_M)$ implies relations
to spin chains of XXZ-type on the algebraic level. Operators that are similar to
the Q-operators constructed in our paper 
have been introduced in the study of generalised  chiral Potts model 
in \cite{BKMS,DJMM}. The Q-operators to be studied in our paper may be
seen as non-compact analogs of those from \cite{BKMS,DJMM}.

\subsection{Perspectives}

It should be possible to generalise the approach described in this paper to 
the models related to quantum affine 
super-algebras studied in \cite{RT}. A product formula for the universal R-matrices
of these quantum groups is known \cite{Ya}. We may furthermore note that the representations 
defined in \cite{RT} are of a similar type as the prefundamental representations
 studied in 
this paper. Renormalised versions of the universal R-matrix have been 
studied for representations of modular double type of
the quantum super-algebra $\CU_{q}({\mathfrak o\mathfrak s \mathfrak p}(1|2))$ in \cite{IZ}.
This work gives a first hint that the 
renormalisation of the universal R-matrices can be carried out for quantum affine 
super-algebras in a similar way as done in this paper.
This gives us hope that evolution and Q-operators can be constructed
for the lattice models defined in \cite{RT} by using a generalisation of the
techniques developed here.

We have found reasonably simple formulae for the kernel of the Baxter Q-operator
which are natural generalisations of the formulae found in \cite{ByT}. This should allow us  to 
deduce the analytic properties of the Q-operators by generalising the 
arguments from \cite{ByT}. The information on the analytic properties of the Baxter Q-operator
defines the space of all solutions to the generalised Baxter equation which can correspond to
eigenvalues of this operator. Baxter equation and analytic properties represent the pieces 
of information that completely characterise the spectrum. It should be possible
to translate this description of the spectrum into equivalent formulations described either in terms
of non-linear integral equations or using partial differential equations, generalising the results known
for the Sinh-Gordon model \cite{Z00,L00,ByT,LZ}.

Our results finally suggest that the representation theory of quantum affine algebras may have
a mathematically rich and interesting extension to certain categories of infinite-dimensional 
representations. In the finite-dimensional case it was observed in \cite{ByT02} that the 
R-operator of the modular double of  $\mathcal{U}_q(sl(2,\BR))$ \cite{F99} may be seen as a "more universal R-matrix" in 
the following sense. The representations of the modular double of $\mathcal{U}_q(sl(2,\BR))$ 
considered in \cite{PT99,F99,ByT02} have dual representation that are realised on 
certain spaces of distributions. The dual representations 
contain highest weight representations as sub-representations. 
It was verified in \cite{ByT02} that the action of the R-matrix defined in \cite{F99} on tensor products of 
the dual representations restricts  to the action of the usual universal R-matrix on tensor products
of highest weight representations. The R-operator of the modular double is therefore "more universal" than the
universal R-matrix in the sense that it unifies the R-matrices defined on finite- and certain infinite-dimensional 
representations. It would be interesting to make this point of view more 
precise, and to extend it to the case of quantum affine algebras.

\subsection{Guide to the paper}

The paper is quite long. However, there are some important parts of our story that
can be understood without having digested all of our paper. To help the reader finding the
parts of most immediate interest we will here offer a brief overview over the sections. 
The introduction of each section 
contains a slightly more detailed description of its contents.

Section 2 reviews some basic background on the classical theory and possible 
approaches to the quantisation of the affine Toda theories. 

The following Section 3 develops the light-cone lattice approach introduced in the pioneering papers
\cite{FV92,FV94,BBR}. In order to have manifest locality, we are working with a slightly redundant parameterisation
of the degrees of freedom. A gauge symmetry is introduced 
allowing us to identify the physical degrees of freedom as gauge-invariant combinations of the basic variables.

Section 4 offers a review of the basic background on quantum affine algebras together with a short
summary of the available hints indicating that the integrability of the affine Toda theories can be
understood using the representation theory of quantum affine algebras.

Section 5 describes first steps towards the definition and calculation of Lax-matrices and R-operators
based on the universal R-matrix of quantum affine algebras. The main tool for this purpose are the 
formulae representing the universal R-matrix as an infinite product going back to 
Khoroshkin and Tolstoy.  We start explaining how to renormalise the formal expressions obtained by
evaluating the product formula in the infinite-dimensional representations of our interest in 
the case of $\CU_q(\widehat{\fsl}_2)$.

This analysis is generalised in the next Section 6 for the case of $\CU_q(\widehat{\fsl}_M)$. We describe how 
to obtain the fundamental R-operators for the lattice affine Toda models
from the representation theory of $\CU_q(\widehat{\fsl}_M)$. Different types of explicit representations
for the fundamental R-operators are derived. For a twisted version of the quantum affine algebras we 
find a particularly convenient representation, leading to useful representations
for the generalised Baxter Q-operators constructed from the fundamental R-operators as integral operators.

For the derivation of functional relations satisfied by the Q-operators like generalised Baxter equations
it is crucial to analyse the contributions coming from  the factors in the
product formula involving imaginary root generators.  Such an analysis is carried out in Section 7
for the case of of $\CU_q(\widehat{\fsl}_2)$. A uniform prescription is found for renormalising the 
contributions associated to  imaginary roots for a large family of representations including the representations
relevant for the lattice Sinh-Gordon model. We verify  the consistency of this prescription with 
taking co-products, and use all this to give a derivation of the Baxter equation valid for the infinite-dimensional
representations of our interest. 

The generalisation of this analysis to the case of $\CU_q(\widehat{\fsl}_M)$ is presented in 
Section 8. We begin by describing a fairly simple representation-theoretic proof of generalised Baxter
equations which is valid provided the renormalisation prescription preserves the relevant properties of 
the R-operators under the co-product. The fact that it does is verified afterwards, studying the fairly 
intricate mixing between real and imaginary roots under the co-product. Our results also allow us 
to derive functional relations of quantum Wronskian type. Together with the 
analytic properties of the kernel of the Q-operators we have thereby obtained all the information
necessary to study the spectrum of the lattice affine Toda theories 
generalising the case of the Sinh-Gordon
model studied in \cite{ByT}.

Various more technical details are deferred to appendices. Appendix \ref{ShGapp} in particular contains
a detailed comparison with previously known results on the Sinh-Gordon model and to the 
Faddeev-Volkov model.

\subsection{Acknowledgements}

We would like to thank V. Bazhanov, D. B\"ucher, I. Runkel and F. Smirnov for discussions and interest in this project.
Support from the DFG in the framework of the SFB 676 {\it Particles Strings, and the Early 
Universe} is gratefully acknowledged.


\section{Background}

Our main example in this paper will be the affine $\fsl_M$-Toda theories, 
which are classically defined
in the Hamiltonian formalism by introducing field $\phi_i(x,t)$,  canonical conjugate 
momenta $\Pi_i(x,t)$ and Poisson brackets 
\begin{equation}
\{\,\Pi_i(x,t)\,,\,\phi_j(x',t)\,\}\,=\,\pi\,\de_{ij}\de(x-x')\,,\qquad
\begin{aligned}
&\{\,\phi_i(x,t)\,,\,\phi_j(x',t)\,\}\,=\,0\,\\
&\{\,\Pi_i(x,t)\,,\,\Pi_j(x',t)\,\}\,=\,0\,.
\end{aligned}
\end{equation}
The  dynamics is generated by the Hamiltonian
\begin{equation}
H\,=\,\int d^2z\;\sum_{i=1}^{M}\bigg(\frac{1}{2\pi}\big(\Pi^2_i+(\pa_x\phi_i)^2\big)+\mu 
e^{2b(\phi_i-\phi_{i+1})}
\bigg)\,.
\end{equation}
The resulting equations of motion for $\vf_i:=\phi_i-\phi_{i+1}$ can be represented in the form
\begin{equation}\label{EOM}
(\pa_t^2-\pa_x^2)\,\vf_i\,=\,-2{\pi}b\mu\big(2e^{2b\vf_i}-e^{2b\vf_{i+1}}
-e^{2b\vf_{i-1}}\big)\,.
\end{equation}
As the motion of $\bar\phi(x,t)=\sum_{i=1}^M\phi_i(x,t)$ decouples, $(\pa_t^2-\pa_x^2)\bar\phi=0$,
it is possible to impose the condition that $\bar\phi(x,t)=0$.

\subsection{Classical integrability}

The starting point is a zero curvature representation of the classical dynamics, 
taken to be of the form
\begin{equation}\label{ZCC}
[\,\pa_x-A_x(\la)\,,\,\pa_t-A_t(\la)\,]\,=\,0\,.
\end{equation}
We may here take $A_x(\la)=A_+(\la)-A_-(\la),$
$A_t(\la)\,=\,A_+(\la)-A_-(\la)$,
where
\begin{equation}\label{classLax}
\begin{aligned}
&A_+(\la)\,=\,
\sum_{i=1}^M\left(-{b}(\pa_+\phi_i)\SE_{ii}+
me^{b(\phi_i-\phi_{i+1})}\SE_{i,i+1}\right)\,,\\
&A_-(\la)\,=\,
\sum_{i=1}^M\left(+{b}(\pa_-\phi_i)\SE_{ii}-
me^{b(\phi_i-\phi_{i+1})}\SE_{i+1,i}\right)\,,
\end{aligned}
\end{equation}
using the notations $\pa_\pm=\frac{1}{2}(\pa_t\pm\pa_x)$.
The zero curvature condition \rf{ZCC} will reproduce the equation of
motion \rf{EOM} provided that $m^2=\pi \mu b^2$.

Integrability of the classical dynamics is closely related to the existence
of infinitely many conserved quantities which can be constructed from 
the monodromy matrix
\begin{equation}
M(\la)\,=\,\CP\exp\left(\int_{0}^{R}dx\;A_x(\la)\right)\,.
\end{equation}
as the trace
\begin{equation}\label{TransferClass}
T(\la)\,=\,{\rm Tr}(M(\la))\,.
\end{equation}
The Poisson structure of the
field theory implies Poisson bracket relations of the form
\begin{equation}\label{clMMr}
\{\,M(\la)\,\overset{\otimes}{,}\,M(\mu)\,\}\,=\,[\,r(\la/\mu)\,,\,M(\la)\ot M(\mu)\,]\,,
\end{equation}
with $r(\la)$ being a certain numerical matrix. These relations imply $\{T(\la),T(\mu)\}=0$.
As the Hamiltonian $H$ appears in the asymptotic expansion of $M(\la)$ at infinity it follows
that $T(\la)$ is conserved for all values of $\la\in\BC$.

\subsection{Light-cone representation}

It is also possible to take the values of the basic field restricted 
to the light-like segments as Cauchy-data. Let us
define 
the ``saw-blade'' contours 
${\mathcal{C}}_N=\bigcup_{k=1}^N{\mathcal{C}}_k^+\cup
{\mathcal{C}}_k^-$, where ${\mathcal{C}}_k^\pm$ are the light-like segments
\begin{equation}
\begin{aligned}
{\mathcal{C}}_k^+&=
\big\{(k\De+u,t+u)\st 0 \leqslant u \leqslant \De/2\big\},\\
{\mathcal{C}}_k^- &=
\big\{(k\De+v,t+\De-v)\st \De/2 \leqslant v \leqslant \De\big\}
\end{aligned}
\qquad\text{($\De:=R/N$).}
\end{equation}


In the light-cone picture for the classical dynamics, one takes
the values of the field $\phi$ on the two light-like segments of
${\mathcal C}_1$,
\begin{equation}
\phi_i^+(2u)=\phi_i(u,u) \quad \text{and} \quad
\phi_i^-(2v)=\phi_i(\tfrac{R}{2}-v,\tfrac{R}{2}+v), 
\qquad  0 \leqslant u,v \leqslant \tfrac{R}{2},
\end{equation}
as initial values
for the time-evolution from which $\phi_i(x,t)$
can be found for all $x$ and $t$
by solving the equations of motion. 
 The dynamics may still be represented in the Hamiltonian form by using the Poisson structure
\begin{equation}\label{PBlc}
\{\, \phi_i^+(u) \,,\, \phi_j^+(u') \,\} \,=\, \frac{\pi}{4}\,\de_{ij}\,
{\rm sgn}_R(u-u') ,\quad
\{\, \phi_i^-(v) \,,\, \phi_j^-(v') \,\} \,= \,\frac{\pi}{4}\,
\de_{ij}\,{\rm sgn}_R(v-v')
\end{equation}
on the light-cone data $\phi_i^+$ and $\phi_i^-$ defined on segments ${\mathcal C}^+_k$ and 
${\mathcal C}^-_k$, respectively.  
The evolution of  $\pa_+\phi^+(x_+)$ in the  $x_-$-direction 
can now be represented in the Hamiltonian form as
\begin{equation}
\pa_-^{}(\pa_+^{}\phi_i^+)\, = \,
\big\{\, H_-^{} \,,\, \pa_+^{}\phi_i^+ \,\big\}\,,
\end{equation}
where 
\begin{equation}\label{H_-def}
H_- = 
{\mu}\int_0^R{dx_+}\,\sum_{i=1}^M \,e^{2b(\phi_i^+-\phi_{i+1}^+)} .
\end{equation}
A very similar equation of motion obtained by exchanging the roles
of $\phi_i^+$ and $\phi_i^-$ governs the evolution of $\pa_-\phi^-(x_-)$ in the  $x_+$-direction. 

Vanishing of the curvature of the Lax-connection
allows us to deform the contour in definition of the monodromy
matrix, leading to a representation of $M(\la)$ 
as an integral over light-like segments.
The zero curvature condition \rf{ZCC} implies that
\begin{equation}\label{Tmod}
M(\la)  =  \CP\exp\left(\int_0^{R}dx\;A_x(x,t;\lambda)\right)
 = \CP\exp\left(\int_{\mathcal{C}} ds\;
\frac{dx^{\al}}{ds} A_{\al}(\la)\right) ,
\end{equation}
for any contour $\mathcal{C}$ that can be deformed into
${\mathcal{C}}_0=\{(x,t) : 0 \leqslant x \leqslant R \}$, preserving
 the start and end points. This allows us to rewrite $M(\la)$ as
\begin{equation}\label{Tmodmod}
M(\la)  =  L_N^{-}(\la)L_N^+(\la)\cdots
L_1^{-}(\la)L_1^+(\la) ,
\end{equation}
where
\begin{equation}\label{lctransp}
L_k^{+}(\la) :=
\CP\exp\bigg(\int_{{\mathcal{C}}_k^{+}} dx_+\;A_{+}(\la)\bigg) ,\quad
L_k^{-}(\la) :=
\CP\exp\bigg(\int_{{\mathcal{C}}_k^{-}} dx_-\;A_{-}(\la)\bigg).
\end{equation}
When $\De\ra 0$, $N\ra\infty$ with 
$R=N\De$ finite one expects to be able to approximate the fields by piecewise  constant values 
along ${\mathcal C}_k^\pm$. 
The representation \rf{Tmodmod} of $M(\la)$ 
suggests a natural lattice discretisation resembling a staggered spin chain. 

\subsection{Continuum approaches}

A very useful approach to the quantisation of such an integrable system
is provided by the quantum inverse scattering method (QISM). A central 
object in this approach is the so-called quantum monodromy matrix
$\SM(\la)$, the matrix formed from the operators that are obtained by
quantising the matrix elements of the classical monodromy matrix $M(\la)$.
If it is possible to construct a matrix $\CM(\la)$ out of the 
quantised degrees of freedom of the field theory of interest in such a way
that the Poisson bracket relations \rf{clMMr} get quantised into  
quadratic commutation relations of the form
\begin{equation}\label{RMM}
R(\la/\mu)\,(\CM(\la)\ot{\rm I})\,({\rm I}\ot \CM(\mu)) 
\,=\,({\rm I}\ot \CM(\mu)) \,(\CM(\la)\ot{\rm I})\, R(\la/\mu)    \,,
\end{equation}
one would get the conserved quantities of the 
quantized field theory from
\begin{equation}
\ST(\la)\,=\,{\rm Tr}(\SM(\la))\,.
\end{equation}
However, this dream is hard to realise in practise. 
In canonical quantisation it is by no means straightforward to construct an operator-valued matrix 
$\SM(\la)$ out of the quantised local fields that would satisfy nice quadratic
relations of the form \rf{RMM}. It is furthermore not clear which 
numerical matrices $R(\la)$ could appear in the relations \rf{RMM}. 
Even though $R(\la)$ is severely restricted by the 
Yang-Baxter relation 
\begin{equation}\label{YBESEC2}
R_{12}(\la/\mu)\,R_{13}(\la/\nu)\,R_{23}(\mu/\nu)\,=\,
R_{23}(\mu/\nu)\,R_{13}(\la/\nu)\,R_{12}(\la/\mu)
\,,
\end{equation}
following from the consistency of \rf{RMM} with the associativity of operator
products, one still has a large supply of possible choices for 
$R(\la)$ to consider.


The situation appears to be slightly better in the light-cone representation.
Following \cite{BLZ4} let us 
note that the Poisson brackets \rf{PBlc} are those a massless free field.
The quantization is therefore standard. Let us write the expansion
of $\phi_i^{\pm}(x_{\pm})$ into Fourier modes in the form
\begin{equation}\label{modes}
\phi_i^\pm(x_{\pm}) = \sq_i + \frac{2\pi}{R}\spp_i x_{\pm} +
\phi^{\pm}_{i,<}(x_{\pm})+\phi^{\pm}_{i,>}(x_{\pm}) ,
\end{equation}
where
\begin{equation}
\phi^{\pm}_{i,<}(x_{\pm}) = \sum_{n<0} \frac{{\mathrm i}}{n}
\sa_{i,n}^{\pm}
e^{-2\pi {\mathrm i}nx_{\pm} / R} ,\quad
\phi^{\pm}_{i,>}(x_{\pm}) =\sum_{n>0} \frac{{\mathrm i}}{n}
\sa_{i,n}^{\pm}
e^{-2\pi {\mathrm i}nx_{\pm} / R} .
\end{equation}
The modes $\sa_{i,n}^\eps$ ($\eps = \pm$), $\sq_i$ and $\spp_i$ 
are required to satisfy the canonical
commutation relations
\begin{equation}\label{CCR}
\comm{\sq_i}{\spp_j} = \frac{i}{4}\de_{ij}, \qquad
\comm{\sa_{i,m}^{\ep}}{\sa_{i,n}^{\ep'}} = \frac{1}{4} \,m \,\de_{m+n,0} \,\de_{ij}
\,\de_{\ep\ep'}.
\end{equation}
Quantum analogs of the exponential functions $e^{2\al_i\phi_i^\pm}$ are then constructed by normal ordering:
\begin{equation}\label{normord}
\normord{e^{2\al_i\phi_i^\pm(x_\pm)}} \equiv
\exp\big(2\al_i\phi_{i,<}^\pm(x_\pm)\big)\,
e^{2\al_i (\sq_i+2\pi\spp_i x_\pm/R)}\, \exp\big(2\al_i\phi_{i,>}^\pm(x_\pm)\big) .
\end{equation}
The quantum Hamiltonians $\SH_+$ and $\SH_-$
corresponding to $H_+$ and $H_-$, respectively, will similarly be defined
by normal ordering.The quantum equation of motion for an
observable $\SO_\pm$ built from $\pa_\pm\phi^{\pm}(x_\pm)$ can then be represented in the form
\begin{equation}\label{q-EOM}
-{\mathrm i}\,\pa_\mp\SO \,=\, 
\big[ \,\SH_\mp \,,\, \SO_\pm \,] 
\,,\qquad \SH_\mp=\mu\sum_{i=1}^{M}\SQ_i^\pm\,,
\end{equation}
where the so-called screening charge operators $\SQ_i^\ep$ are defined as
\begin{equation}
\SQ_i^\ep = \int_0^R dx \; \SV_i^\ep(x), \qquad
\SV_i^\ep(x) = \normord{e^{2b(\phi^\ep_i(x)-\phi^\ep_{i+1}(x))}}\,,\qquad
\ep=\pm\,.
\end{equation}

If the parameter 
$b=i\be$ is purely imaginary, it should be possible to 
define a natural candidate for the quantum 
monodromy matrix $\SM(\la)$ by following the approach
of Bazhanov, Lukyanov and Zamolodchikov.  
For $\be$ in a certain range of values 
it would allow us to define quantum monodromy matrices 
associated to the segments ${\mathcal C}_k^\pm$ of the saw-blade contour 
as series of ordered 
integrals over products of 
normal ordered exponentials of the free fields. 

Such an approach 
has not been developed  in full detail yet. Even if it were, it could not easily be generalised to 
the case $b\in\BR$ of our main interest. The UV-problems are more 
delicate for $b\in\BR$, causing serious problems for the definition 
of the quantum monodromy matrices
along the lines of \cite{BLZ1,BLZ3,BLZ4}.

\subsection{Lattice regularization}

Another method to treat these problems is the lattice discretization.
The initial values $\phi^i(x)\equiv \phi^i(x,t)|_{t=0}^{}$, 
$\Pi^i(x)\equiv \Pi^i(x,t)|_{t=0}^{}$
of the fields at time $t=0$ are replaced by 
variables $\phi_n^i$, $\Pi_n^i$ 
defined on a one-dimensional lattice which has $N$ sites labelled by the
index $n$. The variables  $\phi_n^i$, $\Pi_n^i$ may be thought of as 
averages of the initial values, 
\begin{equation}\label{discrete}
\phi_n^i = \frac{1}{\Delta}\int_{n\Delta}^{(n+1)\Delta}dx\;\phi^i(x) ,\qquad
\Pi_n^i = \frac{1}{4\pi}\int_{n\Delta}^{(n+1)\Delta}{dx}\;\Pi^i(x) .
\end{equation}
The quantization of these variables
will yield operators which satisfy the commutation relations
\begin{equation}
\big[ \phi_n^i , \Pi_m^j \big] = \frac{i}{2} \de^{ij}\de_{n,m} .
\end{equation}
The space of states of the regularized model may therefore be identified with
as $L^2(\BR^{MN})$.

A regularized version $\SM_{N}(\la)$ of the 
monodromy matrix $\SM(\la)$ may be constructed 
as a product of local Lax matrices
\begin{equation}\label{SMdefSEC2}
\CM_N(\la) = \CL_{N}(\la)\CL_{N-1}(\la)\cdots
\CL_{1}(\la)  \,,
\end{equation}
where the lattice Lax matrices $\CL_{n}(\la)$ are to be 
constructed from the discretized variables $(\vf_n^i,\Pi_n^i)$.
It will be shown that the
matrices $\CM_N(\la)$ can be constructed in such a way 
that they satisfy the algebra 
\begin{equation}\label{reg-RMM}
R(\la/\mu)\,(\CM_N(\la)\ot{\rm I})\,({\rm I}\ot \CM_N(\mu)) 
\,=\,({\rm I}\ot \CM_N(\mu)) \,(\CM_N(\la)\ot{\rm I})\, R(\la/\mu)    \,,
\end{equation}
with coefficients $R(\la/\mu)$ that are independent of $N$ and $\Delta$.
If the continuum limit $N\ra\infty$ of $\CM_{N}(\la)$ exists in 
a suitable sense, the relations \rf{reg-RMM} will ensure that
the monodromy matrix $\CM(\la)$ defined by that limit satisfies the
crucial algebraic relations \rf{RMM}.

In the case of the Sinh-Gordon model corresponding to $M=2$ 
it was shown in 
\cite{ByT,ByT2} that the lattice discretisation leads to exact results
for the energy spectrum. The excellent agreement with results 
from the thermodynamic Bethe
ansatz and from the existing relations with Liouville theory \cite{ByT2} 
indicates that the lattice approach is indeed suitable for the
construction and solution of the affine Toda theories.




\section{Integrable light-cone lattice models -- algebraic framework}
\label{scheme}

The use of the lattice light-cone approach is inspired by previous works
 \cite{FR,FV92,FV94,BBR,KaR} on the lattice light-cone discretisation
of the Sine- and Sinh-Gordon models. In order to maintain manifest locality it will be
useful to parameterise the degrees of freedom in a somewhat redundant way.
The physical degrees of freedom can be identified using a gauge-symmetry. We describe
how to define a natural time-evolution for gauge-invariant quantities.

\subsection{Overview on the light-cone lattice approach}
\label{overviewSEC_timeevo}

It turns out to be very useful to preserve 
a certain democracy in the treatment of spacial
and time-like directions by working on a rhombic space-time lattice
\begin{equation}
\Gamma\,=\,\{\,(\si,\tau)\,|\,\si\in\BZ/N\BZ\,,\,\tau\in\BZ\,,\si+\tau\;\;{\rm even}\,\}\,.
\end{equation}
This lattice is generated  by the vectors $v_+=(1,1)$ and  $v_-=(-1,1)$
 which connect nearest neighbor sites, see Figure \ref{lightconelattice}.
%

%
\begin{center}
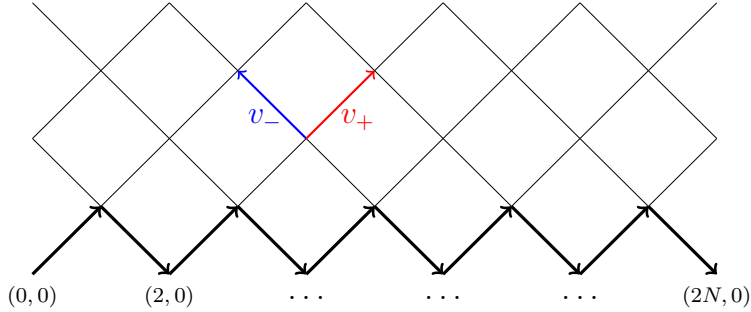
\begin{figure}[h]
\label{lightconelattice}
\begin{center}
\begin{tikzpicture}[scale=1.8]
\draw[->,very thick] node[anchor=north]{\scriptsize{$(0,0)$}}(0,0)  --(0.5,0.5);
\draw[->,very thick] (0.5,0.5)  -- (1,0);
\draw[->,very thick] (1,0)node[anchor=north]{\scriptsize{$(2,0)$}}  -- (1.5,0.5);
\draw[->,very thick] (1.5,0.5)  -- (2,0);
\draw[->,very thick] (2,0) node[anchor=north ]{$\stackrel{}{\dots}$} -- (2.5,0.5);
\draw[->,very thick] (2.5,0.5)  -- (3,0);
\draw[->,very thick] (3,0)  node[anchor=north ]{$\stackrel{}{\dots}$}  -- (3.5,0.5) ;
\draw[->,very thick] (3.5,0.5)  -- (4,0);
\draw[->,very thick] (4,0) node[anchor=north ]{$\stackrel{}{\dots}$}  -- (4.5,0.5);
\draw[->,very thick] (4.5,0.5)  -- (5,0) node[anchor=north]{\scriptsize{$(2N,0)$}} ;
\draw[ultra thin] (0,1)  -- (0.5,0.5);
\draw[ultra thin] (0.5,0.5)  -- (1,1);
\draw[ultra thin] (1,1)  -- (1.5,0.5);
\draw[ultra thin] (1.5,0.5)  -- (2,1);
\draw[ultra thin] (2,1)  -- (2.5,0.5);
\draw[ultra thin] (2.5,0.5)  -- (3,1);
\draw[ultra thin] (3,1)  -- (3.5,0.5);
\draw[ultra thin] (3.5,0.5)  -- (4,1);
\draw[ultra thin] (4,1)  -- (4.5,0.5);
\draw[ultra thin] (4.5,0.5)  -- (5,1);
\draw[ultra thin] (0,1)  -- (0.5,1.5);
\draw[ultra thin] (0.5,1.5)  -- (1,1);
\draw[ultra thin] (1,1)  -- (1.5,1.5);
\draw[<-][blue,thick] (1.5,1.5)  -- (2,1) node[pos=0.4,below] {$v_-$};
\draw[->][red,thick] (2,1)  -- (2.5,1.5) node[pos=0.6,below] {\,\,\,\,$v_+$};
\draw[ultra thin] (2.5,1.5)  -- (3,1);
\draw[ultra thin] (3,1)  -- (3.5,1.5);
\draw[ultra thin] (3.5,1.5)  -- (4,1);
\draw[ultra thin] (4,1)  -- (4.5,1.5);
\draw[ultra thin] (4.5,1.5)  -- (5,1);
\draw[ultra thin] (0,2)  -- (0.5,1.5);
\draw[ultra thin] (0.5,1.5)  -- (1,2);
\draw[ultra thin] (1,2)  -- (1.5,1.5);
\draw[ultra thin] (1.5,1.5)  -- (2,2);
\draw[ultra thin] (2,2)  -- (2.5,1.5);
\draw[ultra thin] (2.5,1.5)  -- (3,2);
\draw[ultra thin] (3,2)  -- (3.5,1.5);
\draw[ultra thin] (3.5,1.5)  -- (4,2);
\draw[ultra thin] (4,2)  -- (4.5,1.5);
\draw[ultra thin] (4.5,1.5)  -- (5,2);
\end{tikzpicture}
\end{center}
\caption{Light-cone lattice $\Gamma$.}
\end{figure}
\end{center}
A collection of elements $\{\chi^i_{\sigma,\tau}\}_{i=1,\dots,M}$ 
of the quantum algebra of observables $\mathcal{A}_{M,N}$ to be defined below
is attached to  each vertex $(\sigma, \tau)$ of the dual lattice 
 $\Gamma^{\vee} $
defined by the condition  $\sigma+\tau$ odd. 
For each vertex of $\Gamma$  a relation between the variables $\chi^i_{\sigma,\tau}$  
associated to the neighbouring faces is required to hold.
 Such relations are called 
\emph{quantum  discrete equations of motion} as
they reduce to the equations of motion \eqref{EOM} in the classical continuum limit.

Let us describe the dynamics more explicitly. The algebra of observables $\mathcal{A}_{M,N}$  
will be
generated by invertible elements 
\begin{equation}
 \chi_{i,m}^{}\,,\qquad i\,\in\,\mathbb{Z}/M\mathbb{Z}\,,\,\,\,\,\,
m\,\in\,\mathbb{Z}/2N\mathbb{Z}\,.
\end{equation}
satisfying certain relations.
The only non-trivial commutation relations are
\begin{equation}\label{chialg}
\chi_{i,2a-1}^{}\,\chi_{j,2a}^{}\,=\,q^{+2\,c_{ij}}\,\chi_{j,2a}^{}\, \chi_{i,2a-1}^{}\,,
\qquad
\chi_{i,2a}^{}\, \chi_{j,2a+1}^{}\,=\,q^{-2\,c_{ij}}\,\chi_{j,2a+1}^{}\,\chi_{i,2a}^{}\,,
\end{equation} 
where $c_{ij}=-\left(\delta_{ij}-\delta_{i j+1}\right)$.
In this paper we are mostly interested in the case   $|q|=1$. 
In this case the generators $\chi_{i,m}$ will be realized as positive self-adjoint operators.

We will introduce two automorphisms $\tau_{\pm}$ of the algebra $\mathcal{A}_{M,N}$ such that upon defining
\begin{equation}\label{taupm}
  \chi^i_{\sigma\pm 1,\tau+1}\,:=\,
\tau_{\pm}\left( \chi^{i}_{\sigma,\tau}\right)\,,
\end{equation}
with initial conditions  $\chi^i_{2a-1,0}\,:=\,\chi_{i,2a-1}$,
$\chi^i_{2a,1}\,:=\,\chi_{i,2a}$, 
the following quantum equations of motion are satisfied
\begin{equation}\label{equationofmotiondiscrete}
 q^{2} \, \chi^{i}_{\si,\tau}\,\chi^{i}_{\si,\tau+1}=
   \chi^{i}_{\si-1,\tau}\,\chi^{i}_{\si+1,\tau}\,
\frac{1+q^{+1}\ka^2\,\chi^{i+1}_{\si-1,\tau}}
{1+q^{-1}\ka^2\,\chi^{i}_{\si-1,\tau}}\,
\frac{1+q^{+1}\ka^2\,\chi^{i-1}_{\si+1,\tau}}
{1+q^{-1}\ka^2\,\chi^{i}_{\si+1,\tau}}\,.
\end{equation}
These equations allow  to define the values of the variables $\chi^{i}_{\si,\tau}$ 
on the entire lattice from the initial values associated to the faces nearest to the bold saw in Figure \ref{lightconelattice}. 
It is easy to check that the evolution 
equation \rf{equationofmotiondiscrete} reproduces the equation of motion 
\rf{EOM} if one identifies $\chi^i_{\si,\tau}$ with $e^{2b\vf_i(\De\si,\De\tau)}$ and takes the limit 
$q=e^{-i\pi b^2}\ra 1$ and $\Delta \ra 0$ with $\kappa=m\Delta$ and $\chi_i$  fixed.
 
The equations of motion above will be shown to follow from the zero curvature condition
 \begin{equation}\label{zerocurv}
 \gconn^-_{\sigma+1,\tau+1}(\lambda)\,\, \gconn^+_{\sigma+1,\tau}(\lambda)
\,\,=\,\,
\gconn^+_{\sigma,\tau+1}(\lambda)\,\, \gconn^-_{\sigma,\tau}(\lambda)\,
\qquad \qquad
\sigma+\tau\,\,\text{even}
\end{equation}
for certain operator valued matrixes attached to the edges of the lattice $\Gamma$.
This is a quantum discrete analogue of \eqref{ZCC} encoding quantum integrability of the time evolution 
defined above.
The relation \eqref{zerocurv} corresponding to each face in the  the lattice $\Gamma$,
 see Figure \ref{lightconelattice}, can be depicted as follows 
\begin{center}
\begin{tikzpicture}
\draw[->,very thick] (0,0)
 node[anchor=north]{\scriptsize{$(\sigma,\tau)$}} -- (1,1) node[anchor = south west]{\scriptsize{$(\sigma\!+\!1,\tau\!+\!1)$}} ;
\draw[->,very thick] (1,1)  -- (0,2) node[anchor=south]{\scriptsize{$(\sigma,\tau\!+\!2)$}} ;
\draw[ultra thin] (0,2)  -- (-1,1) node[anchor=south east]{\scriptsize{$(\sigma\!-\!1,\tau\!+\!1)$}};
\draw[ultra thin] (-1,1)  -- (0,0)  ;
\node (hh)  at (4,1) {\LARGE{$=$}};
\end{tikzpicture}
\qquad
\begin{tikzpicture}
\draw[ultra thin] (0,0)
 node[anchor=north]{\scriptsize{$(\sigma,\tau)$}} -- (1,1) node[anchor=south west ]{\scriptsize{$(\sigma\!+\!1,\tau\!+\!1)$}} ;
\draw[ultra thin] (1,1)  -- (0,2) node[anchor=south]{\scriptsize{$(\sigma,\tau\!+\!2)$}};
\draw[<-,very thick] (0,2)  -- (-1,1) node[anchor=south east]{\scriptsize{$(\sigma\!-\!1,\tau\!+\!1)$}};
\draw[<-,very thick] (-1,1)  -- (0,0)  ;
\end{tikzpicture}
\end{center}
Notice that the  matrices $\gconn^+_{\sigma,\tau}(\lambda)$ and
 $\gconn^-_{\sigma,\tau}(\lambda)$ represent parallel transport on the lattice from 
$(\sigma-1,\tau)$ to $(\sigma,\tau+1)$ and $(\sigma,\tau)$ to $(\sigma-1,\tau+1)$ respectively.
 It follows that  
\begin{equation}\begin{aligned}
&\gconn^+_{\sigma,\tau}(\lambda)\qquad \text{is defined for $\sigma+\tau$  odd,  }\\
&\gconn^-_{\sigma,\tau}(\lambda) \qquad \text{is defined for $\sigma+\tau$  even.}
\end{aligned}
\end{equation}
The rule to associate an operator valued matrix to a path on the lattice follows 
from the basic property of the path ordered exponential
 $\Omega_{\gamma}$, i.~e.~$\Omega_{\gamma_1+\gamma_2}=\Omega_{\gamma_2}\Omega_{\gamma_1}$
 when the final point of the path $\gamma_1$ coincides with the initial point of $\gamma_2$.
 
The explicit form of the Lax operators of discretized affine $\mathfrak{gl}_M$-Toda theory will be
 \begin{equation}\label{Laxdef}\begin{aligned}
 &\gconn^+(\lambda)\,=\,
\sum_{i=1}^M\,\left(\su_i^{+1}\,\SE_{ii}\,+\,q^{-\frac{1}{2}}\,\kappa\,\lambda^{-1}\,\sv_i\,\SE_{ii+1} \right)\,,\\
 &\gconn^-(\lambda)\,=\,
\sum_{i=1}^M\,\left(\su_{i}^{-1}\,\SE_{ii}\,-\,q^{-\frac{1}{2}}\,\kappa\,\lambda^{+1}\,\sv_i\,\SE_{i+1i} \right)\,,
\end{aligned}
\end{equation}
where $\SE_{ij}$ are the matrices having a non-vanishing matrix element equal to one only
in the $i$-th row and $j$-th column,  $\kappa=m\Delta$ 
and we suppressed the explicit dependence of $\su_i, \sv_i$ on $\sigma$ and $\tau$.
This choice of quantum Lax operators is motivated by 
the form of the classical flat connection in light-cone coordinates, compare to \eqref{classLax}. 
We will later see that the matrices $\gconn^\pm(\lambda)$
satisfy quadratic relations of  the form \rf{RMM} with $\SM(\la)$ replaced by 
$\gconn^\pm(\lambda)$
relations iff the commutation relations of $\su_i$, $\sv_i$ are
\begin{equation}\label{algebraWdef}
 \su_i\,\su_j\,=\, \su_j\,\su_i \qquad \sv_i\,\sv_j\,=\, \sv_j\,\sv_i\qquad \su_i\,\sv_j\,=\,q^{c_{ij}}\, \sv_j\,\su_i\,,
\end{equation}
where $c_{ij}=-(\delta_{ij}-\delta_{ij+1})$.
We further impose $\prod_i\su_i\,=\,\prod_i\sv_i=1$ as they are central.
We call $\mathcal{W}_M$  the algebra generated by $\su_i$, $\sv_i$ and their inverses.


In this description, the quantum algebra of observables $\mathcal{A}_{M,N}$ emerges as
 a quotient of the enlarged algebra 
 $\mathcal{A}'_{M,N}=\left(\mathcal{W}_M\right)^{\otimes 2 N}$, associated to the saw-blade contour in
 Figure \ref{lightconelattice},
 by certain gauge transformations. One may get rid of gauge redundancies at the price of giving up ultralocality,
 which is the requirement that at fixed $\tau$ the matrix entries of 
$\gconn^{\epsilon_1}_{\sigma_1,\tau}$ commute with the matrix entries of $\gconn^{\epsilon_2}_{\sigma_2,\tau}$ 
when $\sigma_1\neq\sigma_2$.



\subsection{The monodromy matrices}

\subsubsection{An alternating spin-chain}
\label{sec:alternating}

The  monodromy matrix $\CM(\la)$ of the lattice model is constructed 
as a product of local Lax matrices as
\begin{equation}\label{SMdefSEC3}
\CM(\la) = \CL_{N}(\la)\CL_{N-1}(\la)\cdots
\CL_{1}(\la)\,.
\end{equation}
In the lightcone representation  $\CL_{a}^{}(\la)$ takes the factorized form
\begin{equation}\label{L+-def}
\CL_{a}^{}(\la) =\bar{\mathsf{L}}^{-}_{2a}(q^{\frac{1}{2}}\kappa^{+1}\lambda)\, 
\mathsf{L}^{+}_{2a-1}(q^{\frac{1}{2}}\kappa^{-1}\lambda)\,,
\end{equation}
where 
\begin{equation}\label{LpbarLminusfromLparenthesis}
 \mathsf{L}^{+}(q^{\frac{1}{2}}\kappa^{-1}\lambda)\,:=\,\gconn^+(\lambda)\,,
\qquad
 \bar{\mathsf{L}}^{-}(q^{\frac{1}{2}}\kappa^{+1}\lambda)\,:=\,
\left(1-q^{-1}\lambda^{M}\right) 
\left(\gconn^-(\lambda)\right)^{-1}\,,
\end{equation}
with $\gconn^{\pm}(\lambda)$ given  in \eqref{Laxdef}. 
The scalar factor multiplying $\left(\gconn^-(\lambda)\right)^{-1}$
 in \rf{LpbarLminusfromLparenthesis} can be identified with the
 quantum determinant $\text{q-det}(\gconn^-(\lambda))$ as defined in Appendix \ref{app:minors}. 
The definition \eqref{LpbarLminusfromLparenthesis} may be written more explicitly as
\begin{align}\label{moreexpl1}
&\bar{\mathsf{L}}^{-}(\lambda)\,=\left(1-q^{-1}\lambda^{M}\right)\,
\Bigg{[}\sum_{i=1}^M\,\left(\su^{-1}_i\,\SE_{ii}\,-\,q^{-1}\,\lambda\,\sv_i\,\SE_{i+1i} \right)\Bigg{]}^{-1}\,,\\
\label{moreexpl2}
&\mathsf{L}^{+}(\lambda)\,=\,
\sum_{i=1}^M\,\left(\su_i^{+1}\,\SE_{ii}\,+\,\lambda^{-1}\,\sv_i\,\SE_{ii+1} \right)\,.
\end{align}
The monodromy matrix \eqref{SMdefSEC3} is the operator-valued matrix  associated to the bold path
 in Figure \ref{lightconelattice} upon setting 
$\gconn^{+}_{2a-1}:=\gconn^{+}_{2a-1,0}$ and $\gconn^{-}_{2a}:=\gconn^{-}_{2a,0}$. 
The index $m$ on $\gconn^{\pm}_m$ denotes the embedding of $\mathcal{W}_M$
 in the $m$-th thensor factor of $\left(\mathcal{W}_M\right)^{\otimes 2 N}$.
It is thus clear that the matrix entries of quantum Lax operators 
associated to different sites of the chain commute.

The algebra $\mathcal{W}_M$ admits a simple realization in $L^2(\mathbb{R}^M)$ given as follows
\begin{equation}\label{fromuvtopq}
 \su_i\,=\,e^{-2\pi b\,\mathsf{p}_i}\,,
\qquad
\sv_i\,=\,e^{\pi b\,(\mathsf{q}_i-\mathsf{q}_{i+1})}\,,
\qquad
[\mathsf{p}_i,\mathsf{q}_j]\,=\,(2\pi i)^{-1}\delta_{ij}
\end{equation}
with $q=e^{-i \pi b^2}$.
The quantum space on which the matrix entries of the monodromy 
matrix act may be taken to be $\mathcal{H}'_{M,N}\,:=\,L^2(\mathbb{R}^{NM})$.
Alternatively one may impose the constraint 
$\sum_{i=1}^M\,\mathsf{p}_i=0$ for each spin-chain site, leading to a representation 
of $\CW$ in a subspace $\mathcal{H}_{M,N}$ of $\mathcal{H}'_{M,N}$ isomorphic to $L^2(\mathbb{R}^{N(M-1)})$.

Both Lax matrices $\mathsf{L}^+(\la)$ and $\bar{\mathsf{L}}^-(\la)$ 
satisfy relations of the form 
\begin{equation}\label{RLLLLR}
\mathsf{R}(\la,\mu) \bigl(\mathscr{L}(\la)\otimes 1\bigr)
\bigl(1\otimes \mathscr{L}(\mu)\bigr)
 = \bigl(1\otimes \mathscr{L}(\mu)\bigr)
\bigl(\mathscr{L}(\la)\otimes 1\bigr)\mathsf{R}(\la,\mu)\,,
\end{equation}
with the same auxilliary R-matrix $\SR(\lambda,\mu)$ given as 
\begin{align}\label{RfundfundTEXT}
& {\mathsf R}(\la,\mu) \,=\,\sum_{i=1}^M\,{\mathsf E}_{ii}\otimes{\mathsf E}_{ii}\,+\,
\nu\,\sum_{i\neq j}\,{\mathsf E}_{ii}\otimes{\mathsf E}_{jj}\,+\,
\sum_{i\neq j}\,\kappa_{(i-j)_M}\,{\mathsf E}_{ij}\otimes{\mathsf E}_{ji}\,, \\
& \nu\,=\,\frac{\mu^M-\la^M}{q^{-1}\,\mu^M-\,q^{+1}\,\la^M}\,,\qquad
\kappa_{\ell}\,=\,\frac{q^{-1}-q^{+1}}{{q^{-1}\,\mu^M-\,q^{+1}\,\la^M}}\,\mu^{M-\ell}\,\la^{\ell}\,,
\notag\end{align}
where $(i-j)_M$ denotes $(i-j)$ modulo $M$.
The monodromy matrix constructed in \rf{SMdefSEC3} therefore satisfies
the relations \rf{reg-RMM}, as desired. 
This implies in particular 
that the one-parameter family of
operators $\ST(\la)$
\begin{equation}\label{Ttransferdef-1}
\ST(\la)\,=\,{\rm Tr}_{\mathbb{C}^M}(\mathcal{M}(\la))\,,
\end{equation}
is mutually commutative
\begin{equation}
[\,\ST(\la)\,,\ST(\mu)\,]\,=\,0\,.
\end{equation}
The family of operators $\ST(\la)$ will represent conserved quantities for the time-evolution defined 
above.

\begin{rem}
\label{sl2rem} In the case  of $\mathfrak{sl}_2$ one has $\su^{}_1=\su^{-1}_2\,=\su$,
$\sv^{}_1=\sv^{-1}_2\,=\sv$ and 
the definition \eqref{L+-def} reads
\begin{equation}
\mathsf{L}^{+}(\la)\,=\,\bigg(\,\begin{matrix}
\su & {\la}^{-1}\,\sv\\
 {\la}^{-1}\,\sv^{-1} &\su^{-1}
\end{matrix}\,\bigg)\,,
\qquad \,\,\,\,\,\,
\bar{\mathsf{L}}^{-}(\la)\,=\,
\bigg(\,\begin{matrix}
\su & {\la}^{+1}\,\sv^{-1}\\
 {\la}^{+1}\,\sv^{} &\su^{-1}
\end{matrix}\,\bigg)\,.
\end{equation}
Our formulation of the  light-cone lattice approach is in this case similar to the one described in \cite{FV92,FV94,BBR}.
An important difference is due to the fact that 
$\bar{\mathsf{L}}^{-}(\la)$ is taken to be equal to $\mathsf{L}^{+}(\la)$ in \cite{FV92,FV94,BBR}.  
The two formulations are equivalent for even $N$, as will be discussed in  Section \ref{reltoXXZ} below.
The relations with the representation theory of 
quantum affine algebras appear to be more natural in our formulation.
 \end{rem}

 \begin{rem}
 The inverse of the matrix $\bar{\mathsf{L}}^{-}(\la)$ given in \eqref{moreexpl1} can be written more explicitly using the following observation:
 For any matrix of the form
$\SF(a)\,:=\,1-\,\sum_{i=1}^M\,a_i\,\SE_{i+1i}$,
one has
\begin{equation}
 (1-a_M\cdots\,a_2a_1)\,
 \left(\SF(a)\right)^{-1}\,=\,
\left(1\,+\,
\sum_{i\neq j}\,\left(a_{i-1}a_{i-2}\,\cdots\,a_j\right)\,\SE_{i,j} \right)\,.
\label{InverseF}
\end{equation}
Norice that in order to derive \eqref{InverseF} no commutation relation between $a_s$ 
have been used.

 \end{rem}

\subsubsection{Relation to XXZ-type spin-chains}\label{reltoXXZ}

It will be useful to note that there is a closely related Lax-matrix which is defined as 
\begin{equation}\label{Tdefsec3}
\CL_{a}^{\rm\sst XXZ}(\la)\,=\,\CL_{a}^{}(\la)
 \,\ST\,,\qquad\ST=\sum_{i=1}^M\SE_{i,i+1}\,.
\end{equation}
The Lax matrix $\CL_{a}^{\rm\sst XXZ}(\la)$
satisfies the same equation \rf{RLLLLR} that is satisfied by $\CL_{a}^{}(\la)$, as follows from the fact that
$\SR(\la,\mu)$ commutes with $\ST\ot\ST$.
It furthermore has a dependence on the spectral parameter $\la$ of the form
\begin{equation}\label{XXZform}
\CL_{a}^{\rm\sst XXZ}(\la)=\sum_{i=1}^M\SE_{ii}(\CE_{a,ii}^{}+\la^M\CE_{a,ii}')+\sum_{i<j}(\la^{M+i-j}\SE_{ij}\,\CE_{a,ji}
+\la^{j-i}\SE_{ji}\,\CE_{a,ij})\,.
\end{equation}
It follows from \rf{RLLLLR} together with the form \rf{XXZform} that the matrix elements $\CE_{ij}$ generate a
representation of the quantum group $\CU_{q}(\fsl_M)$,  as will be further discussed 
in Section \ref{evalextension} below. 

Note furthermore that 
\begin{equation}
\ST\,\CL_{a}^{\rm\sst XXZ}(\la)\,\ST^{-1}\,=\,\Omega^{-1}\cdot\CL_{a}^{\rm\sst XXZ}(\la)\cdot\Omega\,,
\end{equation}
where $\Omega$ is the automorphism of the algebra of generated by the matrix elements
of $\CL_{a}^{\rm\sst XXZ}(\la)$ defined as
\begin{equation}
\Omega^{-1}_a\cdot\CE_{a,i,j}\cdot\Omega_a\,=\,\CE_{a,i-1,j-1}\,.
\end{equation}
The automorphism $\Omega_a$ allows one to relate the monodromy matrix $\CM(\la)$ to the monodromy 
matrix $\ST^{-N}\CM^{\rm\sst XXZ}(\la)$, 
\begin{equation}
\CM^{\rm\sst XXZ}(\la) = \CL_{N}^{\rm\sst XXZ}(\la)\CL_{N-1}^{\rm\sst XXZ}(\la)\cdots
\CL_{1}^{\rm\sst XXZ}(\la)\,.
\end{equation}
The automorphism $\Omega_a$ has order $M$, $(\Omega_a)^M={\rm id}$. 
If $N$ is divisible by $M$, the spectral problem for $\ST(\la)$ therefore becomes
equivalent to the spectral 
problem for $\ST^{\rm\sst XXZ}(\la)={\rm Tr}_{\mathbb{C}^M}(\mathcal{M}^{\rm\sst XXZ}(\la))$.

The close relation between spin chains of XXZ-type and lattice regularisations of the affine Toda theories will make
it natural and often useful to discuss both of them in parallel. 

\subsection{Light-cone time-evolution}
\label{lightconeEVOderivation}

We will now derive the quantum equations of motion 
\eqref{equationofmotiondiscrete}. 
The derivation will  be based on an explicit construction of the light-cone
 evolution operators $\mathsf{U}_{\kappa}^{\pm}$, see
\eqref{taupm}. The latter will later be shown to belong to a large family 
of commuting operators contructed as transfer matrices in Section \ref{FunR/Q}.

Before proceeding to the derivation an important remark is in order.
The zero curvature condition \eqref{zerocurv} can not specify by itself
 a unique time evolution for the variables $\su^{i}_{\si,\tau}$, $\sv^{i}_{\si,\tau}$.
The reason is that if $\gconn^{\pm}_{\si,\tau}$ satisfy the zero curvature condition, then also
\begin{equation}\label{gaugetransf}
 \left(\gconn^{+}_{\si,\tau}\right)^\prime\,=\,
\mathsf{D}_{\si,\tau+1}\,\gconn^{+}_{\si,\tau}\,\mathsf{D}^{-1}_{\si-1,\tau}\,,
\qquad
 \left(\gconn^{-}_{\si,\tau}\right)^\prime\,=\,
\mathsf{D}_{\si-1,\tau+1}\,\gconn^{-}_{\si,\tau}\,\mathsf{D}^{-1}_{\si,\tau}\,,
\end{equation}
do. In  \eqref{gaugetransf} $\mathsf{D}_{\si,\tau}$ are taken to be diagonal 
matrices in  order to preserve the form of $\gconn^{\pm}_{\si,\tau}$ given in 
 \eqref{Laxdef}. 
We refer to the transformations \eqref{gaugetransf} as gauge transformations.
The transformations \eqref{gaugetransf} reflect
 the transformation properties of the path order exponential 
$\Omega_{\gamma}\mapsto \mathsf{D}^{}_B\,\Omega_{\gamma}\,\mathsf{D}^{-1}_A$,
 where $\gamma$ is a path connecting
the point  $A$ to the point $B$. It will be shown that
the zero curvature condition specifies  a unique time evolution for the gauge invariant sub-algebra 
of $\left(\mathcal{W}_M\right)^{\otimes 2N}$. 


\subsubsection{Identification of physical observables}\label{Phys-Obs}

We first want to 
clarify how the quantum algebra of observables $\mathcal{A}_{M,N}$ 
emerges form the enlarged algebra $\left(\mathcal{W}_M\right)^{\otimes 2N}$
generated by the operators $\su_{i,r}$, $\sv_{i,r}$,
$i=1,\dots,M$, $r=1,\dots,2N$.

Consider the products $\bar{\mathsf{L}}^{-}_{2a}(\mu)\mathsf{L}^+_{2a-1}(\nu)$
and $\mathsf{L}^+_{2a+1}(\nu)\bar{\mathsf{L}}^{-}_{2a}(\mu)$, which may be represented as
\begin{subequations}\label{Lpmsecond}
\begin{align}
\bar{\mathsf{L}}^{-}_{2a}(\mu)\mathsf{L}^+_{2a-1}(\nu)&=(1-q^{-1}\mu^M)
\bigg(1-\mu\sum_{i=1}^M\SY_{i,2a}^-\,\SE_{i+1,i}^{} \bigg)^{-1}\\[-1ex]
&\qquad\qquad\qquad\qquad\cdot\Lambda(\su_{2a}\su_{2a-1})
\bigg(1+\frac{1}{\nu}\sum_{i=1}^M
\mathsf{Y}_{i,2a-1}^+\,
\SE_{i,i+1}^{} \bigg)\notag\\
  \mathsf{L}^+_{2a+1}(\nu)\bar{\mathsf{L}}^{-}_{2a}(\mu)&=
 \bigg(1+\frac{1}{\nu}\sum_{i=1}^M\tilde{\mathsf{Y}}_{i,2a+1}^+\,\SE_{i,i+1}^{} \bigg)\Lambda(\su_{2a+1}\su_{2a})\\[-1ex]
&\qquad\qquad\qquad\qquad\cdot (1-q^{-1}\mu^M)
\bigg(1-\mu\sum_{i=1}^M\tilde{\SY}_{i,2a}^-\,\SE_{i+1,i}^{} \bigg)^{-1}\!\!
\notag\end{align}
\end{subequations}
where ${\mathsf \Lambda}(x):=\sum_{i=1}^M\,x_i\,\SE_{ii}$ and
\begin{equation}
\begin{aligned}\label{generatorW2}
&{{\mathsf Y}}^-_{i,r}\,:=\,\su_{i+1,r}\sv_{i,r}\,,\\
&\tilde{{\mathsf Y}}^-_{i,r}\,:=\,\sv_{i,r}\su_{i,r}\,,
\end{aligned}
\qquad 
\begin{aligned}
&{\mathsf Y}^+_{i,r}
\,:=\,\su^{-1}_{i,r}\sv^{}_{i,r}\,,\\
&\tilde{\mathsf Y}^+_{i,r}
\,:=\,\sv^{}_{i,r}\su^{-1}_{i+1,r}\,.
\end{aligned}
\end{equation}
The group of gauge-transformations on a  time slice is generated from the transformations 
\begin{equation}
G_{2a-1}^{}:\left\{\begin{aligned}
\mathsf{L}^+_{2a-1}(\lambda)&\ra D_{2a-1}^{-1}\,\mathsf{L}^+_{2a-1}(\lambda)\\
\bar{\mathsf{L}}^-_{2a}(\lambda)&\ra \bar{\mathsf{L}}^-_{2a}(\lambda)\,D_{2a-1}^{}\end{aligned}
\right\}\quad
G_{2a}^{}:\left\{
\begin{aligned}
\mathsf{L}^+_{2a-1}(\lambda)&\ra \mathsf{L}^+_{2a-1}(\lambda)\,D_{2a}^{}\\
\bar{\mathsf{L}}^-_{2a}(\lambda)&\ra D_{2a}^{-1}\,\bar{\mathsf{L}}^-_{2a}(\lambda)
\end{aligned}\right\}
\end{equation}
Using the factorised expressions \rf{Lpmsecond} it is easy to see that ${\mathsf Y}^+_{i,r}$ and 
${\mathsf Y}^-_{i,r}$ are invariant under $G_{2a-1}^{}$, while
$\tilde{\mathsf Y}^+_{i,r}$ and 
$\tilde{\mathsf Y}^-_{i,r}$ are invariant under $G_{2a}^{}$. 
Note furthermore that the combinations $\su_{i,2a}^{}\su_{i,2a-1}^{-1}$ which are 
{\it not} invariant under $G_{2a-1}^{}$ do {\it not} appear in the 
product $\mathsf{L}^-_{2a}(\lambda)\mathsf{L}^+_{2a-1}(\lambda)$. 
A similar statement holds for the combinations $\su_{i,2a+1}^{}\su_{i,2a}^{-1}$ which are 
{\it not} invariant under $G_{2a}^{}$. 

The next step will be to identify operators that implement the gauge transformations
$G_{2a-1}^{}$ and $G_{2a}^{}$ within the chosen Hilbert space representation
of $\CA'_{M,N}$. To this aim let us introduce the operators  
\begin{equation}\label{center}
\begin{aligned}
&\mathsf{c}^{}_{i,2a-1}
=\,
\big(\su^{}_{i,2a}\sv_{i,2a}^{-1}\sv^{}_{i,2a-1}\su^{}_{i+1,2a-1}\big)^{\frac{1}{2}}\,,\\
&\mathsf{c}^{}_{i,2a}=
\big(\su^{}_{i+1,2a}\sv^{-1}_{i,2a}\sv^{}_{i,2a+1}\su^{}_{i,2a+1}\big)^{\frac{1}{2}}\,.
\end{aligned}\end{equation}
It is easy to see that $\mathsf{c}^{}_{i,2a-1}$ commutes with ${\mathsf Y}^+_{i,r}$ and 
${\mathsf Y}^-_{i,r}$, but it does not commute with $\su_{i,2a}^{}\su_{i,2a-1}^{-1}$. 
This allows us to identify $\log\mathsf{c}^{}_{i,2a-1}$ as an infinitesimal generator for 
$G_{2a-1}^{}$. 
By very similar reasoning one may identify $\log\mathsf{c}^{}_{i,2a}$ as an infinitesimal generator for 
${G}_{2a}^{}$. Having related $\mathsf{c}^{}_{i,r}$ with the generators of the 
gauge symmetry motivates us to define the algebra $\CA_{M,N}^{}$ of "physical" observables to be
the sub-algebra of $\CA_{M,N}'$ generated by the operators commuting with all $\mathsf{c}^{}_{i,r}$,
more precisely
\begin{equation}\begin{aligned}
\CA_{M,N}^{}:=\,\big\{ \,\SO\in\CA_{M,N}';\,& \;(\mathsf{c}^{}_{i,r})^{is}\cdot\SO\cdot(\mathsf{c}^{}_{i,r})^{-is}\,=\,\SO\\
&\quad \forall\; i=1,\dots,M, \; \forall \;r=1,\dots,2N,\,\;\forall\;s\in\BR\,\big\}\,.
\end{aligned}\end{equation}
It is easy to find an explicit set of generators for $\CA_{M,N}^{}$: It is given by the
operators 
\begin{equation}\label{chidef}\begin{aligned}
&\chi_{i,2a-1}^{}\,:=\,
\su^{}_{i,2a}\sv^{}_{i,2a}\sv^{}_{i,2a-1}\su^{-1}_{i+1,2a-1}\,\\
&\chi^{}_{i,2a}:=
 \sv^{}_{i,2a}\su^{}_{i+1,2a}\su_{i,2a+1}^{-1}\sv^{}_{i,2a+1}\,.
 \end{aligned}
\end{equation}
One may easily check that the operators $\chi_{i,r}^{}$ defined in \rf{chidef}
commute with $\mathsf{c}^{}_{i,r}$ for all 
allowed values of $i$ and $r$, and satisfy the commutation relations \eqref{chialg}.

One may note that the operators $\chi_{i,r}^{}$ and $\mathsf{c}^{}_{i,r}$ with $r=2a-\ep$
either odd ($\ep=1$) or even ($\ep=0$) 
generate commutative subalgebras of $\CA_{M,N}'$. They can therefore be simultaneously 
diagonalised, leading to representations where states are represented by wave-functions 
$\psi'(x,c)$, with $x$ and $c$ being vectors with components
$x_{i,a}$ and $c_{i,a}$ for $i=1,\dots,M$ and $a=1,\dots,N$,  respectively. The representations 
are defined such that
\begin{equation}
\chi_{i,2a-\ep}^{}\psi'(x,c)\,=\,x_{i,a}\psi'(x,c)\,,
\qquad\mathsf{c}_{i,2a-\ep}^{}\psi'(x,c)\,=\,c_{i,a}\psi'(x,c)\,.
\end{equation}
Whenever a physical operators $\SO'$ can be represented as an integral operator, one may 
assume that this representation takes the form 
\begin{equation}\label{intop}
(\SO'\psi')(x,c)\,=\,\int dx'\;K_{\SO'}(x,x';\ga_c)\psi'(x,c).
\end{equation}
The kernel $K_{\SO'}(x,x';\ga_c)$ may depend on the values $\ga_c$ of the central elements that
the algebra generated by the $\mathsf{c}_{i,r}^{}$ has.

One may then define a natural projection sending $\psi'(x,c)$ to 
$\psi(x)=\psi'(x,\mathbf{1})$, where $\mathbf{1}$ has components $c_{i,a}=1$
for $i=1,\dots,M$ and $a=1,\dots,N$.
Physical operators are projected to the operators 
\begin{equation}
(\SO\psi)(x)\,=\,\int dx'\;K_{\SO}(x,x')\psi(x)\,,\qquad
K_{\SO}(x,x')\equiv K_{\SO'}(x,x';\ga_{\mathbf{1}})\,.
\end{equation}

\subsubsection{Hamiltonian formalism}

In a Hamiltonian framework one may describe the time evolution of 
arbitrary observables $\CO_{\si,\tau}$ by means of operators 
$\SU_{\ka}^\pm$, see \eqref{taupm}, which generate the light-cone evolution by one time step in the 
following sense:
\begin{equation}\label{discevol}
\CO_{r+1,\tau+1}^{}:=(\SU_{\ka}^+)^{-1}\cdot\CO_{r,\tau}^{}\cdot\SU_{\ka}^+\,,\qquad
\CO_{r-1,\tau+1}^{}:=(\SU_{\ka}^-)^{-1}\cdot\CO_{r,\tau}^{}\cdot\SU_{\ka}^-\,.
\end{equation}
The corresponding discrete time evolution operator $\SU_{\kappa}$ is given as
\begin{equation}
\SU_{\kappa}\,=\,\SU_{\kappa}^-\cdot\SU_{\kappa}^+\,=\,\SU_{\kappa}^+\cdot\SU_{\kappa}^-\,.
\end{equation}
Notice that this operator shifts the time variable $\tau$ by two units.
The main ingredient to construct the light-cone evolution
operators will be an operator $\sr^{\dot{-}+}_{m,n}(\la,\nu)$
that satisfies
\begin{equation}\label{rminusplusequation}
(\sr_{m,n}^{\dot{-}+}(\la,\nu))^{-1}\cdot
\bar{\mathsf{L}}_{m}^-(\la)\,\mathsf{L}_{n}^+(\nu)\cdot\sr^{\dot{-}+}_{m,n}(\la,\nu)\,=\,
\mathsf{L}_{n}^+(\mu)\,\bar{\mathsf{L}}_{m}^-(\la)\,.
\end{equation}
The motivation for introducing the notation $\dot{-}$ will become clear in the following. 
Having such an operator we may construct 
 $\SU_\kappa^\pm$ in the following form
\begin{equation}\label{SUdef}
\SU_{\kappa}^+\,=\,
\Bigg[\,\prod_{a=1}^N \sr_{2a,2a-1}^{\dot{-}+}(\bar{\mu},\mu)\,\Bigg]
\cdot{\mathsf C}_{\rm\sst odd}^{}\,,
\qquad
\SU_{\kappa}^-\,=\,
\Bigg[\,\prod_{a=1}^N \sr_{2a,2a-1}^{\dot{-}+}(\bar{\mu},\mu)\,\Bigg]
\cdot{\mathsf C}_{\rm\sst even}^{-1}\,.
\end{equation}
where $\kappa^2=\mu^{-1}\bar{\mu}$.
The operators ${\mathsf C}_{\rm\sst odd}$ and 
${\mathsf C}_{\rm\sst even}$
are defined such that
\begin{equation}\label{shiftdef}
\begin{aligned}
&{\mathsf C}_{\rm\sst odd}\cdot\SO_{2a+1}\,=\,\SO_{2a-1}\cdot
{\mathsf C}_{\rm\sst odd}\,,\qquad
{\mathsf C}_{\rm\sst odd}\cdot\SO_{2a}\,=\,\SO_{2a}\cdot
{\mathsf C}_{\rm\sst odd}\,,\\
&{\mathsf C}_{\rm\sst even}\cdot\SO_{2a-1}\,=\,\SO_{2a-1}\cdot
{\mathsf C}_{\rm\sst even}\,,\qquad
{\mathsf C}_{\rm\sst even}\cdot\SO_{2a}\,=\,\SO_{2a-2}\cdot
{\mathsf C}_{\rm\sst even}\,,
\end{aligned}
\end{equation}
for all operators $\SO_m$ which act nontrivially only on the tensor
factor with label $m$ in  $\left(\mathcal{W}_M\right)^{\otimes 2N}$.
It follows that $(\SU_{\kappa}^-)^{-1}\SU_{\kappa}^+$ generates space-shift of two lattice units, as it should.
It is then easy to show that the zero curvature condition 
\eqref{zerocurv} will be satisfied in the time evolution generated by
$\SU_\kappa^\pm$:
\begin{align*}
\bar{\mathsf{L}}_{2a+1,\tau+1}^-(\bar\mu)\,&\mathsf{L}_{2a,\tau+1}^+(\mu)\,=\\
&\overset{\rf{discevol}}{=}(\SU_\kappa^+)^{-1}\cdot \bar{\mathsf{L}}_{2a,\tau}^-(\bar\mu)\,
\mathsf{L}_{2a-1,\tau}^+(\mu)\cdot\SU_\kappa^+\\
&\overset{\rf{SUdef}}{=}\SC_{\rm\sst odd}^{-1}\cdot(\sr^{\dot{-}+}_{2a,2a-1}(\bar{\mu},\mu))^{-1}\cdot
\bar{\mathsf{L}}_{2a,\tau}^-(\bar\mu)\,\mathsf{L}_{2a-1,\tau}^+(\mu)
\cdot
\sr_{2a,2a-1}^{\dot{-}+}(\bar{\mu},\mu)\cdot\SC^{}_{\rm\sst odd}\\
&\overset{\rf{rminusplusequation}}{=}\,\SC^{-1}_{\rm\sst odd}
\cdot
 \mathsf{L}_{2a-1,\tau}^+(\mu)\,\bar{\mathsf{L}}_{2a,\tau}^-(\bar\mu)\cdot\SC^{}_{\rm\sst odd}\\
&\overset{\rf{shiftdef}}{=}\mathsf{L}_{2a+1,\tau}^+(\mu)\,
\bar{\mathsf{L}}_{2a,\tau}^-(\bar\mu)\,.
\end{align*}
The fact that $\ST(\la)$ give in \eqref{Ttransferdef-1} generates
 quantities conserved in this time-evolution,
\begin{equation}
(\SU_{\ka}^\pm)^{-1}_{}\cdot\ST(\la)\cdot\SU_{\ka}^{\pm}\,=\,\ST(\la)\,,
\end{equation}
may now be checked directly using \rf{rminusplusequation},
\rf{shiftdef} and the
cyclicity ot the trace.

\subsubsection{Evolution of physical degrees of freedom}
\label{sec:evolutionHamiltonian}

We will now derive the evolution equations  \eqref{equationofmotiondiscrete}
from the Hamiltonian point of view. 
To do so we will use an explicit solution of \rf{rminusplusequation}:
\begin{equation}\label{sr-expl}
\sr^{\dot{-}+}_{2a,2a-1}(\bar{\mu},\mu)\,=\,
\left[\prod_{i=1}^M\,\mathcal{J}_{\ka}(\chi_{i,2a-1})\right]\,q^{\mst_{2a,2a-1}}\,,
\end{equation}
where $q^{\mst_{2a,2a-1}}$ is the operator
\begin{equation}
\mst_{2a,2a-1}\,=\,\frac{1}{(\pi b^2)^2}\sum_{i=1}^M\log(\su_{i,2a})\log(\su_{i,2a-1})\,,
\end{equation}
while $\mathcal{J}_{\ka}(x)$ is a special function satisfying the 
functional relation
\begin{equation}\label{J-funrel}
 \frac{\mathcal{J}_{\ka}(q^{-1}x)}{\mathcal{J}_{\ka}(q^{+1}x)}\,=\,1+\kappa^2 x\,.
\end{equation}
Note that $q^{\mst_{2a,2a-1}}$ satisfies
\begin{equation}\label{qtadef}\begin{aligned}
& q^{\mst_{2a,2a-1}}\,\sv^{}_{i,2a-1}\,q^{-\mst_{2a,2a-1}}\,=\,\su^{}_{i,2a}\,\su_{i+1,2a}^{-1}\,\sv^{}_{i,2a-1}\,,\\
&q^{\mst_{2a,2a-1}}\,\sv^{}_{i,2a}\,q^{-\mst_{2a,2a-1}}\,=\,\su^{-1}_{i+1,2a-1}\,\su^{}_{i,2a-1}\,\sv^{}_{i,2a}\,,
\end{aligned}
\end{equation}
and commutes with $\su_{i,2a}$, $\su_{i,2a-1}$.
The fact that the operator defined in \rf{sr-expl} satisfies \rf{rminusplusequation} can be
verified by straightforward calculations.
As we will see in Section \ref{canren}
 the functional relation \eqref{J-funrel} supplemented by the requirement
that the time evolution is unitary will determine a solution $\mathcal{J}_{\ka}(x)$ of \rf{J-funrel} almost 
uniquely.

From the explicit form of $\sr^{\dot{-}+}(\bar{\mu},\mu)$ given in  \eqref{sr-expl}
it is easy to derive the quantum discrete equations of motion.
Let $\tau_{\pm}(\mathsf{z}):=
\left( \SU_{\ka}^{\pm}\right)^{-1}\!\cdot\mathsf{z}\cdot\SU_{\ka}^{\pm}.
$
Using
the definitions \rf{sr-expl}, \rf{qtadef}, the algebra \rf{chialg} and the
functional relation \rf{J-funrel}  one obtains 
\begin{subequations}
\label{basictimeev}\begin{align}
 \tau_{+}\left(\chi_{i,2a-1}\right)\,&=\,
 \tau_{-}\left(\chi_{i,2a+1}\right)\,=\,\chi_{i,2a}\,,\\
\label{secondtauplus}
 \tau_{+}\left(\chi_{i,2a}\right)\,&=\,\tau_{-}\left(\chi_{i,2a+2}\right)\,= \\[-1ex] 
 &=\,
\chi_{i,2a}\,\chi_{i,2a+1}^{-1}\,\chi_{i,2a+2}\,\frac{1+q^{-c}\ka^2\chi_{i+1,2a}}{1+q^{+c}\ka^2\,\chi_{i,2a}}\,
\frac{1+q^{-c}\ka^2\chi_{i-1,2a+2}}{1+q^{+c}\ka^2\,\chi_{i,2a+2}}\,,
\notag\end{align}
\end{subequations}
which implies the discrete time-evolution \rf{equationofmotiondiscrete}. Note furthermore 
that 
\begin{align}
 \tau_{+}\left(\sfc_{i,2a-1}\right)=
 \tau_{-}\left(\sfc_{i,2a+1}\right)=\sfc_{i,2a}\,,\quad
 \tau_{+}\left(\sfc_{i,2a}\right)=\tau_{-}\left(\sfc_{i,2a+2}\right)=
\sfc_{i,2a}^{}\,\sfc_{i,2a+1}^{-1}\,\sfc_{i,2a+2}^{}
\end{align}
This means that the evolution of the unphysical degrees of freedom represented by the operators
$\sfc_{i,r}$ decouples completely from the evolution of the physical observables $\chi_{i,r}$.

One may 
notice that the equation \eqref{rminusplusequation} does not specify $\sr^{-+}$ uniquely, 
see Section \ref{Fundsect6} for more details. This is related to the fact that the 
zero curvature condition does not specify a unique time evolution for the enlarged algebra
 $\left(\mathcal{W}_M\right)^{\otimes 2N}$. However, the ambiguity left by equation \eqref{rminusplusequation}
 does not affect the time-evolution \rf{equationofmotiondiscrete} of the physical degrees of freedom.


 

\subsection{Fundamental R-matrices and Q-operators}\label{FunR/Q}

One of the simplest possible ways to make integrability manifest is
realised if the operators $\SU^{\pm}$ for the light-cone 
evolution are obtained from 
a family of commuting operators $\mathcal{Q}_{\pm}(\la)$, 
by specializing the parameter $\la$
to a certain value, $\SU^{\pm}=\left[\mathcal{Q}_{\pm}(\la^{\pm}_\ast)\right]^{\mp 1}$
for a certain $\la^{\pm}_\ast\in\BC$.
This is achieved naturally when the model is
 defined by an alternating spin chain 
as the one introduced in 
Section \ref{sec:alternating}, see \cite{FR,FV92}.

We will later see that the  operators $\mathcal{Q}_{\pm}(\la)$ 
are natural generalizations of the Baxter Q-operators, as the
notation anticipates.

\subsubsection{Fundamental R-matrices}

A standard tool for the construction of local lattice Hamiltonians
are the so-called fundamental\footnote{The name fundamental refers to 
the fact that they play a  fundamental role in the integrability of the model.
It should not be confuses with the adjective fundamental used to 
 attributed to the fundamental 
representation. } R-matrices which
are defined by the commutation relations
\begin{equation}\label{SRLL}
\big(\mathcal{R}_{AB}^{}(\bar\mu,\mu;\bar\nu,\nu)\big)^{-1}\,
\CL_A^{}(\bar\mu,\mu)\CL_{B}^{}(\bar\nu,\nu)\,
\mathcal{R}_{AB}^{}(\bar\mu,\mu;\bar\nu,\nu)\,=\,
\CL_B^{}(\bar\nu,\nu)\CL_A^{}(\bar\mu,\mu)\,
\,.
\end{equation}
In our case we are dealing with lattice Lax matrices that factorize as
\begin{equation}\label{L+-gendef}
\CL_{A}^{}(\bar\mu,\mu)
= \bar{\mathsf{L}}^{-}_{\bar{a}}(\bar\mu) \mathsf{L}_{a}^{+}(\mu)\,,
\end{equation}
where
$\mu=q^{\frac{1}{2}}\kappa^{-1}\la$ 
and  $\bar{\mu}=q^{\frac{1}{2}}\kappa^{+1}\la$.
This factorized form implies 
in particular that the fundamental R-matrices can be constructed 
as
\begin{equation}\label{fundRfact}
\mathcal{R}_{AB}(\bar\mu,\mu;\bar\nu,\nu)\,=\, \sr_{a,\bar{b}}^{+\dot{-}}(\mu,\bar\nu)\,
\sr_{a,b}^{++}(\mu,\nu)\,\sr_{\bar{a},\bar{b}}^{\dot{-}\dot{-}}(\bar\mu,\bar\nu)\,
\sr_{\bar{a},b}^{\dot{-}+}(\bmu,\nu)\,,
\end{equation}
provided that be operators $\sr_{m,n}^{\epsilon_1,\epsilon_2}(\mu,\nu)$ 
satisfy the relations
\begin{subequations}\label{srll}
\begin{align}
\mathsf{L}^{+}_{m}(\mu)
\,\bar{\mathsf{L}}^{-}_{n}(\nu)\,
\mathsf{r}^{+\dot{-}}_{m,n}(\mu,\nu)\,=\,&
\mathsf{r}^{+\dot{-}}_{m,n}(\mu,\nu)\,
\bar{\mathsf{L}}^{-}_{n}(\nu)\,
\mathsf{L}^{+}_{m}(\mu)\,,\\
\mathsf{L}^{+}_{m}(\mu)
\,\mathsf{L}^{+}_{n}(\nu)\,
\mathsf{r}^{++}_{m,n}(\mu,\nu)\,=\,&
\mathsf{r}^{++}_{m,n}(\mu,\nu)\,
\mathsf{L}^{+}_{n}(\nu)\,
\mathsf{L}^{+}_{m}(\mu)\,,\\
\bar{\mathsf{L}}^{-}_{m}(\mu)
\,\bar{\mathsf{L}}^{-}_{n}(\nu)\,
\mathsf{r}^{\dot{-}\dot{-}}_{m,n}(\mu,\nu)\,=\,&
\mathsf{r}^{\dot{-}\dot{-}}_{m,n}(\mu,\nu)\,
\bar{\mathsf{L}}^{-}_{n}(\nu)\,
\bar{\mathsf{L}}^{-}_{m}(\mu)\,,\\
\bar{\mathsf{L}}^{-}_{m}(\mu)
\,\mathsf{L}^{+}_{n}(\nu)\,
\mathsf{r}^{\dot{-}+}_{m,n}(\mu,\nu)\,=\,&
\mathsf{r}^{\dot{-}+}_{m,n}(\mu,\nu)\,
\mathsf{L}^{+}_{n}(\nu)\,
\bar{\mathsf{L}}^{-}_{m}(\mu)\,.
\end{align}
\end{subequations}
The regularity property for the fundamental R-operator,
 i.e.~ $\mathcal{R}_{AB}(\bar\mu,\mu;\bar\mu,\mu)=\mathbb{P}_{AB}$,
 which is often used to construct \emph{local} 
 conserved charges from the fundamental transfer matrix, will hold if the conditions
\begin{equation}\label{redid}
\sr_{r,s}^{++}(\mu,\mu)\,=\,\mathbb{P}_{rs}\,,
\qquad
\sr_{r,s}^{\dot{-}\dot{-}}(\bar\mu,\bar\mu)\,=\,\mathbb{P}_{rs}\,,
\qquad
\sr_{r,s}^{+\dot{-}}(\mu,\bar\mu)\,\sr_{s,r}^{\dot{-}+}(\bar\mu,\mu)\,=\,1\,,
\end{equation}
are satisfied,
where $\mathbb{P}_{ij}$ is the operator of permutation of the tensor factors
with labels $i$ and $j$.

We will later discuss how operators
$\sr_{rs}^{\ep\ep'}(\mu,\nu)$ satisfying \rf{srll} and \rf{redid}
can be constructed using the representation  theory of quantum affine algebras.
It will turn out that the dependence on the spectral parameters is of the form 
\begin{equation}\label{differenceproperties}
\sr_{r,s}^{\epsilon_1\,\epsilon_2}(\mu,\nu)\,=\,
\sr_{r,s}^{\epsilon_1\,\epsilon_2}(\mu^{-1}\,\nu)\,.
\end{equation}

In Section \ref{reltoXXZ} we had introduced the Lax-matrices $\CL_{a}^{\rm\sst XXZ}(\la)$. It is easy to
see that the fundamental R-operators 
\begin{equation}\label{RfunfromRXXZ}
\mathcal{R}_{AB}^{\rm\sst XXZ}(\bar\mu,\mu;\bar\nu,\nu):=\Omega_A^{}\cdot\mathcal{R}_{AB}^{}(\bar\mu,\mu;\bar\nu,\nu)
\cdot\Omega_B^{-1}\,,
\end{equation}
will satisfy the commutation relations \rf{SRLL} with $\CL$ replaced by
$\CL^{\rm\sst XXZ}$.

Our next goal is to show that the operators 
$\sr_{rs}^{\ep\ep'}(\mu,\nu)$ allow us to construct generalized  commuting transfer
matrices which are 
conserved in the time evolution.

\subsubsection{$\mathcal{Q}$-operators}

We may then use the fundamental R-matrices to define generalised transfer matrices as
\begin{equation}\label{fundtransfer}
\mathcal{T}(\bar\mu,\mu;\bar\nu,\nu)=
{\rm Tr}_{\CH_0^-\ot\CH_0^+}^{}\big(\,
\mathcal{R}_{0N}^{}(\bar\mu,\mu;\bar\nu,\nu)
\,\dots\,\mathcal{R}_{01}^{}(\bar\mu,\mu;\bar\nu,\nu)\,\big)\,.
\end{equation}
It follows from \eqref{fundRfact}  that \eqref{fundtransfer} factorizes into 
the product of two more fundamental transfer matrices as
\begin{equation}\label{Tfactor-G}
\mathcal{T}(\bar\mu,\mu;\bar\nu,\nu)\,=\,\CQ_{+}(\mu;\bar\nu,\nu)\cdot
\CQ_-(\bar\mu;\bar\nu,\nu)\,,
\end{equation}
where 
\begin{align}\label{QdefSECTION3}
& \CQ_-(\mu;\bar\nu,\nu)={\rm Tr}_{\CH_0^{\dot{-}}}^{}\big(\,\sr^{\dot{-}\,\dot{-}}_{0,2N}(\mu,\bar\nu)
\sr^{\dot{-}\,+}_{0,2N-1}(\mu,\nu)\;\dots\;\sr^{\dot{-}\,\dot{-}}_{0,2}(\mu,\bar\nu)
\sr^{\dot{-}\,+}_{0,1}(\mu,\nu)\,\big)\,.\\
& \CQ_+(\mu;\bar\nu,\nu)={\rm Tr}_{\CH_0^+}^{}\big(\,\sr^{+\,\dot{-}}_{0,2N}(\mu,\bar\nu)
\sr^{+\,+}_{0,2N-1}(\mu,\nu)\;\dots\;\sr^{+\,\dot{-}}_{0,2}(\mu,\bar\nu)
\sr^{+\,+}_{0,1}(\mu,\nu)\,\big)\,.
\end{align}

Each of the operators 
 $\CQ_\ep(\la;\bar\mu,\mu)$, $\ep=\pm$ will generate a 
mutually commutative family 
\begin{equation}\label{Q-comm}
\CQ_{\ep_1}(\la_1;\bar\mu,\mu)\cdot
\CQ_{\ep_2}(\la_2;\bar\mu,\mu)=\CQ_{\ep_2}(\la_2;\bar\mu,\mu)\cdot
\CQ_{\ep_1}(\la_1;\bar\mu,\mu)\,,
\end{equation}
of operators provided that the constituent R-operators $\sr^{\epsilon_1\,\epsilon_2}$
satisfy the Yang-Baxter equations
\begin{equation}\label{rrr}
\sr^{\epsilon_1\,\epsilon_2}_{m,\,n}(\la\,\mu^{-1})\,
\sr^{\epsilon_1\,\epsilon_3}_{m,\,p}(\la\,\nu^{-1})\,
\sr^{\epsilon_2\,\epsilon_3}_{n,\,p}(\mu\,\nu^{-1})
\,=\,
\sr^{\epsilon_2\,\epsilon_3}_{n,\,p}(\mu\,\nu^{-1})\,
\sr^{\epsilon_1\,\epsilon_3}_{m,\,p}(\la\,\nu^{-1})\,
\sr^{\epsilon_1\,\epsilon_2}_{m,\,n}(\la\,\mu^{-1})
\end{equation}
where we have used the so-called difference property
\eqref{differenceproperties}.

Recall that  $\bar{\mu}\,\mu^{-1}=\ka^2$
is a fixed parameter of the model. It follow from the explicit definition 
\eqref{QdefSECTION3} and from the properties \eqref{redid},
that the  transfer matrices $\CQ_\ep(\la;\bar\mu,\mu)$ for special value of the spectral
parameter $\la$ satisfy
\begin{equation}\label{Qspecial}
\mathcal{Q}_+(\mu;\bar\mu,\mu)\,=\,
\left(\mathsf{U}^+_{\ka}\right)^{-1}\,,
\qquad
\mathcal{Q}_-(\bar\mu;\bar\mu,\mu)\,=\,
\mathsf{U}^-_{\ka}\,,
\end{equation}
where $\mathsf{U}_{\ka}^{\pm}$ are given in \eqref{SUdef}.
It follows 
from \rf{Q-comm} that 
$\mathcal{Q}_\ep(\nu;\bar\mu,\mu)$ commute with $\SU_\ka^\pm$, and therefore represent
conserved quantities for the evolution  generated by them.

We will later see that the operators $\mathcal{Q}_{\pm}(\la)\equiv\CQ_{\pm}(\la;\bar\mu,\mu)$,
defined in \rf{QdefSECTION3}
satisfy finite difference equations constraining the $\la$-dependence which
generalise the Baxter equations. 
This  motivates us to call these operators (generalised) Baxter 
Q-operators.

It is useful to note, however, that multiplying the family of operators $\mathcal{Q}_{\pm}(\la)$ 
by an operator that is not $\la$-dependent will yield another solution of the generalised Baxter
equations. It may, for example, be useful to consider 
\begin{equation}
\mathbf{Q}_{\bar\mu,\bar\nu,\nu}(\la)\,:=\,\mathcal{T}(\bar\mu,\la;\bar\nu,\nu)\,=\,\CQ_{+}(\la;\bar\nu,\nu)\cdot
\CQ_-(\bar\mu;\bar\nu,\nu)\,,
\end{equation}
as an alternative definition of (generalised) Baxter 
Q-operators. 
The operators $\mathbf{Q}_{\bar\mu,\bar\nu,\nu}(\la)$
represent another useful family of conserved quantities. 
Somewhat surprisingly we will find kernels representing the operators $\mathbf{Q}_{\bar\mu,\bar\nu,\nu}(\la)$
that are simpler than those we could for $\mathcal{Q}_{\pm}(\la)$.

\section{Background on quantum affine algebras}

This section first reviews the basic background on quantum affine algebras used in this paper.
We then summarise the available hints that this algebraic structure is the one underlying the integrability
of the affine Toda theories. 

\subsection{Quantum affine algebras} \label{QuaAffAlg}

To begin with, let us briefly review the necessary background on quantum affine Lie-algebras.

Let 
$\widehat{\alg{g}}$
 be the (untwisted) affine Kac-Moody algebra associated to the
 simple Lie algebra $\alg{g}$.  We let $r$ denote the rank of $\alg{g}$ and assume, 
for simplicity, that all the real roots of $\widehat{\alg{g}}$ have the same length 
(this is the only case that will concern us).  The \qea{} $\QEA{q}{\widehat{\alg{g}}}$ 
may then be defined \cite{Dr1,J} as the Hopf algebra generated by the elements 
$\wun$ (the unit), $e_i$, $f_i$, $k_i = q^{H_i}$ ($i = 0 , 1 , \ldots , r$), and $q^D$, 
subject to the following relations:
\begin{subequations} \label{eqnQEARels}
\begin{gather}
k_i e_j = q^{A_{ij}} e_j k_i, \qquad k_i f_j = q^{-A_{ij}} f_j k_i, 
\qquad e_i f_j - f_j e_i = \delta_{ij} \frac{k_i - k_i^{-1}}{q - q^{-1}}, \label{eqnQEARels1} \\
q^D e_i = q^{\delta_{i0}} e_i q^D, \qquad k_i k_j = k_j k_i,
 \qquad q^D k_i = k_i q^D, \qquad q^D f_i = q^{-\delta_{i0}} f_i q^D, \label{eqnQEARels2} \\
\sum_{n=0}^{1-A_{ij}} \brac{-1}^n \genfrac{[}{]}{0pt}{}{1-A_{ij}}{n}_q e_i^n e_j e_i^{1-A_{ij}-n} =
 \sum_{n=0}^{1-A_{ij}} \brac{-1}^n \genfrac{[}{]}{0pt}{}{1-A_{ij}}{n}_q f_i^n f_j f_i^{1-A_{ij}-n} = 0.
 \label{eqnQEASerre}\end{gather}
\end{subequations}
Here, $A$ is the Cartan matrix of $\widehat{\alg{g}}$ and we use the standard $q$-number notation
\begin{equation}
\genfrac{[}{]}{0pt}{}{m}{n}_q = \frac{\qnum{m}{q}!}{\qnum{n}{q}! \qnum{m-n}{q}!}, \qquad \qnum{n}{q}! =
 \qnum{n}{q} \qnum{n-1}{q} \cdots \qnum{1}{q}, \qquad \qnum{n}{q} = \frac{q^n - q^{-n}}{q - q^{-1}}.
\end{equation}
\eqnref{eqnQEASerre} is known as the Serre relations.  This is supplemented by a coproduct $\Delta$ given by
\begin{subequations} \label{eqnQEACoproduct}
\begin{align}
\func{\Delta}{e_i} &= e_i \otimes k_i + \wun \otimes e_i, & \func{\Delta}{k_i} &= k_i \otimes k_i, \\
\func{\Delta}{f_i} &= f_i \otimes \wun + k_i^{-1} \otimes f_i, & \func{\Delta}{q^D} &= q^D \otimes q^D.
\end{align}
\end{subequations}
There is also a counit and antipode, though their explicit forms are not important for us, 
except in noting that there exist Hopf
subalgebras $\Borelp$ and $\Borelm$ generated by the $e_i$, $k_i$, $q^D$ and the
$f_i$, $k_i$, $q^D$, respectively. These are the analogs of Borel subalgebras and 
we will refer to them as such.  The subalgebras $\Nilp$ and $\Nilm$ generated 
by the $e_i$ and the $f_i$, respectively, will be called the nilpotent subalgebras.  
They are not Hopf subalgebras.

As in the classical case ($q = 1$) above, we will generally be interested in level $0$ representations.
  Because of this, we will often denote a quantum affine algebra by $\QEA{q}{\affine{\alg{g}}_0}$,
 understanding that the linear combination of Cartan generators giving the level has been set to $0$.
  As the level is dual to the derivation $D$ under the (extended) Killing form, 
it is therefore often also permissible to ignore $D$ in our computations.

The quantum affine algebra $\mathcal{U}_q(\widehat{\mathfrak{sl}}_M)$, which will be the main focus of this paper,
 is defined as above
upon taking the  Cartan matrix  to be
 $A_{ij}=2\delta_{i,j}-\delta_{i+1,j}-\delta_{i,j+1}$, 
where indices are identified modulo $M$. 
The finite group $\mathbb{Z}_M$ is realized as  automorphisms of the Dynkin diagram of $\widehat{\mathfrak{sl}}_M$. 
We denote by $\Omega$ the corresponding generator.  
Due to their central role in the following analysis we report the form of the Serre relations in this special case ($M>2$)
\begin{equation}
 e_i^2\,e^{\phantom{2}}_{i\pm 1}
-(q+q^{-1})\,e^{\phantom{2}}_i\,e^{\phantom{2}}_{i\pm 1}\,e^{\phantom{2}}_i\,
+\,e^{\phantom{2}}_{i\pm 1}\,e_i^2=0\,,
\label{Serreeasy1}
\end{equation}
\begin{equation}
e_{i}\,e_j\,=\,e_{j}\,e_{i}\,,
\qquad \text{if $i\neq j\pm 1$}\,,
\label{Serreeasy2}
\end{equation}
and similarly for $f_i$. Notice that the Serre relations are unchanged under $q\rightarrow q^{-1}$.
 The quantum affine algebra $\mathcal{U}_q(\widehat{\mathfrak{gl}}_M)$ can be defined introducing 
the generators $\{q^{\bar{\epsilon}_i}\}_{i=1,\dots,M}$  related to $k_i$ in  \eqref{eqnQEARels} as
\begin{equation}
 k_i\,=\,q^{H_i}\,=\,q^{\bar{\epsilon}_i-\bar{\epsilon}_{i+1}}\,.
\label{introduceebar}
\end{equation}
The generator $\bar{\epsilon}:=\sum_{i=1}^M\,\bar{\epsilon}_i$ is central. 
If it is set to zero we recover $\mathcal{U}_q(\widehat{\mathfrak{sl}}_M)$. 
Notice that the simple roots of $\widehat{\mathfrak{gl}}_M$, see Appendix \ref{sec:convexorderSLM}, 
satisfy $H_i(\alpha_j)=A_{ij}$. This follows form $\bar{\epsilon}_i(\epsilon_j)\,=\,\delta_{ij}$ 
and justifies the notation. 

Finally, we remark that the automorphism $\Omega$ of the Dynkin diagram of 
$\widehat{\mathfrak{sl}}_M$ induces an 
 automorphism of $\mathcal{U}_q(\widehat{\mathfrak{sl}}_M)$
\begin{equation}
 \Omega\,\circ\,\left(e_i,\,f_i,\,k_i\right)\,:=\,
\left(e_{\Omega(i)},\,f_{\Omega(i)},\,k_{\Omega(i)}\right)\,,
\label{OmegaonUq}
\end{equation}
where $\Omega(i)=i+1$.

\subsection{Universal R-matrix}
\label{SSunivR}

The physical relevance of quantum
affine algebras stems from the existence \cite{D} of the so-called universal R-matrix $\CR$.
This is a formally invertible infinite sum of tensor products of algebra elements
\begin{equation} \label{eqnUnivRExp}
\CR  = \sum_i a_i \otimes b_i, \qquad a_i, b_i \in \QEA{q}{\affine{\alg{g}}},
\end{equation}
which must satisfy three properties:
\begin{subequations} \label{eqnUnivRAxioms}
\begin{gather}
\CR \func{\Delta}{x} = \func{\Delta^{\textup{op}}}{x} \CR \qquad 
\text{for all $x \in \QEA{q}{\affine{\alg{g}}}$,} \label{eqnUnivRIntertwiner} \\
\func{\brac{\Delta \otimes \id}}{\CR} = \CR_{13} \CR_{23}
\qquad \text{and} \qquad
\func{\brac{\id \otimes \Delta}}{\CR} = \CR_{13} \CR_{12}. \label{eqnUnivROther}
\end{gather}
\end{subequations}
Here, $\func{\Delta^{\textup{op}}}{x}$ denotes the ``opposite'' coproduct of $\QEA{q}{\affine{\alg{g}}}$, 
 defined as $\De^{\rm op}(x)=\si(\De(x))$,
where the permutation $\si$ acts as
\begin{equation}
\si(x\ot y) = y\ot x.
\end{equation}
We have also used the standard shorthand $\CR_{12} = \sum_i a_i \otimes b_i \otimes \wun$, $\CR_{13} 
= \sum_i a_i \otimes \wun \otimes b_i$ and $\CR_{23} = \sum_i \wun \otimes a_i \otimes b_i$.

Quantum affine algebras have an abstract realisation in terms of a so-called 
quantum double \cite{D} which proves the existence of their universal R-matrices.  
This realisation moreover shows that these R-matrices can be factored 
so as to isolate the contribution from the Cartan generators:
\begin{equation}\label{redRdef}
\CR = q^t \bar{\CR} , \qquad t = \sum_{i,j} 
\bigl( \affine{A}^{-1} \bigr)_{ij}  H_i \otimes H_j.
\end{equation}
Here, $\affine{A}$ denotes the non-degenerate extension of the Cartan matrix
 to the entire Cartan subalgebra (including $D$).  
This is achieved by identifying this matrix with that of the (appropriately normalised)
 standard invariant bilinear form on the Cartan subalgebra.
The so-called \emph{reduced R-matrix} $\bar{\CR}$ is
a formal linear combination of monomials of the form
$\BFE_I\ot \BFF_J:=e_{i_1}\cdots e_{i_k}\ot f_{j_1}\cdots
f_{j_{\ell}}$ ($I = \set{i_1 , \ldots , i_k}$, $J =
\set{j_1 , \ldots j_{\ell}}$).
It is worth noting \cite{KT} that $\CR$
is already uniquely defined up to a scalar multiple
by \eqref{eqnUnivRIntertwiner}
and \eqref{redRdef}.

We note that a second solution to the defining properties \eqref{eqnUnivRAxioms}
 is given by \cite{D}
\begin{equation} \label{DefR-}
\CR^- = \brac{\func{\si}{\CR}}^{-1}.
\end{equation}
This alternative universal R-matrix $\CR^-$ is then of the form
\begin{equation}\label{R-factor}
\CR^- = \bar\CR^- q^{-t},
\end{equation}
in which $\bar\CR^-$ is a formal series in monomials of the form $\mathbf{F}_I \ot \mathbf{E}_J$.
 In order to emphasise the symmetry between the two universal R-matrices
we shall also use the notation $\CR^+:=\CR$.
$\CR^+$ and $\CR^-$ may also be
related by the anti-automorphism $\zeta$ given by
\begin{equation} \label{DefZeta}
\zeta(e_i) = f_i, \qquad \zeta(f_i) = e_i, \qquad \zeta(H_i) = H_i, \qquad \zeta(D) = D, \qquad \zeta(q) = q^{-1}.
\end{equation}
This action can be continued to tensor products via $\zeta(x\ot y)=\zeta(x)\ot\zeta(y)$.
In terms of $\zeta$, we can represent $\CR^-$ as
$\CR^-=\zeta(\CR^+).$

In order to get an idea how property \eqref{eqnUnivRIntertwiner} determines the universal R-matrix
let us first note that $q^t$ satisfies the equations
\begin{align}
&q^t\,\left(f_i\otimes 1\right)\,=\,\left(f_i\otimes k_i^{-1}\right)\,q^t\,,
\qquad
q^t\,\left(1\otimes f_i\right)\,=\,\left(k_i^{-1}\otimes f_i\right)\,q^t\,,\\
&q^t\,\left(e_i\otimes 1\right)\,=\,\left(e_i\otimes k_i^{+1}\right)\,q^t\,,
\qquad
q^t\,\left(1\otimes e_i\right)\,=\,\left(k_i^{+1}\otimes e_i\right)\,q^t\,,
\end{align}
The intertwining property \eqref{eqnUnivRAxioms} 
 implies the following relations for the reduced R-matrices $\bar{\mathscr{R}}^{\pm}$
\begin{align}
& \left[\bar{\mathscr{R}}^+,f_i\otimes 1 \right]\,=\,
\left(k_i\otimes f_i\right)\,\bar{\mathscr{R}}^+-
\bar{\mathscr{R}}^+\,\left(k_i^{-1}\otimes f_i\right)\,,
\label{Rp1}\\
& \left[\bar{\mathscr{R}}^-,e_i\otimes 1 \right]\,=\,
\left(k_i\otimes e_i\right)\,\bar{\mathscr{R}}^- -
\bar{\mathscr{R}}^-\,\left(k_i^{-1}\otimes e_i\right)\,.
\label{Rm1}
\end{align}  
These equations can be solved recursively in the order of the monomials
$\mathbf{E}_I \ot \mathbf{F}_J$ or
$\mathbf{F}_I \ot \mathbf{E}_J$, the first few terms for $\bar{\mathscr{R}}^-$ being
\footnote{
We obtained this expansion for  $\mathcal{U}_q(\mathfrak{g}(A))$, 
where $\mathfrak{g}(A)$ is the Kac-Moody algebra associated to the  (symmetrizable)  generalized Cartan matrix $A$. 
In this case the relation \eqref{eqnQEARels1}
generalizes to 
\begin{equation}
k_i e_j = q^{(\alpha_i,\alpha_j)} e_j k_i, \qquad k_i f_j = q^{-(\alpha_i,\alpha_j)} f_j k_i, 
\qquad e_i f_j - f_j e_i = \delta_{ij} \frac{k_i - k_i^{-1}}{q_i - q_i^{-1}}\,,
\end{equation}
where $(\alpha_i,\alpha_j)=(\alpha_j,\alpha_i)$  and 
the Serre relations take the same form as in  \eqref{eqnQEASerre}  with  
the 
 Cartan Matrix given by
$A_{ij}\,=\,2\frac{(\alpha_i,\alpha_j)}{(\alpha_i,\alpha_i)}$.
}
\begin{align}
\bar{\mathscr{R}}^-\,=\,1&+\sum_{i=0}^r\,\left(q_i^{-1}-q_i\right)\,\left(f_i\otimes e_i\right)+
\sum_{i=0}^r\,\frac{(q_i^2-1)^2}{q_i^2+1}\,e_i^2\,\otimes\,f_i^2\\ &+\,
\sum_{i\,\neq\,j}\,\frac{(q_i-q_i^{-1})(q_j-q_j^{-1})}{q^{(\alpha_i,\alpha_j)}-q^{-(\alpha_i,\alpha_j)}}\,
\left(q^{(\alpha_i,\alpha_j)}\,e_i\,e_j-e_{j}\,e_i\right)\,\otimes\,f_i\,f_j+\dots\,.
\notag\end{align}
Notice that the quadratic Serre relation $e_i\,e_j=e_{j}\,e_i$ for  $(\alpha_i,\alpha_j)=0$
 follows as a necessary condition  for the existence of the universal R-matrix.

For the case of $\mathcal{U}_q(\widehat{\mathfrak{gl}}_M)$ of our main interest we may note that
introducing the Cartan generators $\bar{\epsilon}_i$ 
simplifies the expression for  $t$ entering the universal R-matrix  as
\begin{equation}\label{qtdefSLM}
t\,=\, \sum_{i,j=1}^{M-1}\,\left(A^{-1}\right)_{ij}\,\left(H_i\otimes H_j\right)\,=\,
\sum_{i=1}^M\,\bar{\epsilon}_i\otimes \bar{\epsilon}_i\,-\,
\frac{1}{M}\,\bar{\epsilon}\otimes \bar{\epsilon}\,.
\end{equation}
Note furthermore  that in the case of $\mathcal{U}_q(\widehat{\mathfrak{gl}}_M)$ 
the universal R-matrices $\mathscr{R}^{\pm}$ are $\BZ_M$-symmetric,
\begin{equation}
 \left( \Omega\otimes  \Omega\right)\,\circ\,\mathscr{R}^{\pm}\,=\,
\mathscr{R}^{\pm}\,.
\label{OmegaonR}
\end{equation}
as follows from the uniqueness\footnote{The automorphism  $\Omega$
 does not alter the ansatz for the universal R-matrix that enters the uniqueness theorem in \cite{KT92}.} 
of the universal R-matrix.

It finally follows from the defining properties \eqref{eqnUnivRAxioms} that $\CR^+$ and $\CR^{-}$
 satisfy the abstract Yang-Baxter equations \begin{subequations}
 \label{YBE}\begin{align}
&\hspace{3cm}\CR_{12}^+ \,\CR_{13}^+\, \CR_{23}^+ \,=\, \CR_{23}^+ \,\CR_{13}^+\, \CR_{12}^+\,,\label{YBE++}\\
&\CR_{12}^+ \,\CR_{13}^-\, \CR_{23}^- \,=\,
\CR_{23}^- \,\CR_{13}^-\, \CR_{12}^+\,,\qquad
\CR_{23}^+ \,\CR_{13}^-\, \CR_{12}^- \,=\,
\CR_{12}^- \,\CR_{13}^-\, \CR_{23}^+\,,\label{YBE+-}\\
&\CR_{12}^- \,\CR_{13}^+\, \CR_{23}^+ \,=\,
\CR_{23}^+ \,\CR_{13}^+\, \CR_{12}^-\,,\qquad
\CR_{23}^- \,\CR_{13}^+\, \CR_{12}^+ \,=\,
\CR_{12}^+ \,\CR_{13}^+\, \CR_{23}^-\,,
\label{YBE-+}\\
&\hspace{3cm}\CR_{12}^- \,\CR_{13}^-\, \CR_{23}^- \,=\, \CR_{23}^- \,\CR_{13}^-\, \CR_{12}^-\,.\label{YBE--}
\end{align}\end{subequations}
The equations \rf{YBE} will imply in particular the crucial relations \rf{RLLLLR} when evaluated in 
suitable representations.



\subsection{Drinfeld twist}\label{Twists}

One may modify the defining data of the quantum affine algebras by means of Drinfeld twists,
represented by invertible elements $\mathscr{F}\in\QEA{q}{\widehat{\alg{g}}}\ot\QEA{q}{\widehat{\alg{g}}}$
\begin{equation}
\tilde{\De}(x)\,=\,\mathscr{F}^{-1}\,\De(x)\,\mathscr{F}\,,\quad\forall\; x\in\CA\,,\qquad
\tilde{\CR}\,=\,\si(\mathscr{F}^{-1})\,\CR\,\mathscr{F}\,.
\end{equation}
We will only consider elements $\mathscr{F}$ preserving co-associativity of the co-product (cocycles). 
For a very particular choice of $\mathscr{F}$ we will later find useful simplifications in the expressions
for the fundamental R-operators. This choice is $\mathscr{F}=\si(q^{-f})$, where
\begin{equation}\label{f-defn}
f\,=\,-\frac{1}{2}X_{ij}\bar{\ep}_i\ot\bar\ep_j\,,\qquad X_{ij}\,=\,\frac{2}{M}(i-j)_{{\rm mod}\;M}^{}\,.
\end{equation}
Useful properties of the coefficients $X_{ij}$ are
\begin{equation}\label{Xprop}
\begin{aligned}
&X_{i+1,j}-X_{i,j}\,=\,+\frac{2}{M}-2\de_{i+1,j}\\
&X_{i,j+1}-X_{i,j}\,=\,-\frac{2}{M}+2\de_{i,j}
\end{aligned}\qquad
X_{i,j}+X_{j,i}\,=\,\frac{2}{M}-2\de_{i,j}\,.
\end{equation}
We may furthermore note that \rf{Xprop} implies that
\begin{equation}
q^f\,\si(q^f)\,=\,q^t\,.
\end{equation}
This identity allows us to write $\tilde{\CR}^+$ and $\tilde{\CR}^-$ in the forms 
\begin{align}
\label{twistedRpm}
&\tilde{\CR}^+\,=\,q^{2f}\,\big[\si(q^f)\,\bar\CR^+\,\si(q^{-f})\big]\,,
\\
&\tilde{\CR}^-\,=\, \big[q^f\,\bar{\CR}^-\,q^{-f}\big]\,\si(q^{-2f})\,.
\end{align}
These formulae, together with 
\begin{align}
\si(q^f)\,(e_i\ot f_i)\,\si(q^{-f})&=\,e_i\,q^{\bar\ep_{i+1}-\frac{\bar\ep}{M}}\ot q^{\bar\ep_i-\frac{\bar\ep}{M}}f_i\,,\\
q^f\,(f_i\ot e_i)\,q^{-f}&=\,f_i\,q^{\bar\ep_i-\frac{\bar\ep}{M}}\ot q^{\bar\ep_{i+1}-\frac{\bar\ep}{M}}\,e_i\,,
\end{align}
are useful for computing the Lax- and R-matrices from the twisted universal R-matrices.

\begin{rem}
Parts of the literature use conventions where $\CR^+$ is factorised as 
$\CR^+=\bar{\CR}^+{}'\,q^t$, compare to \eqref{redRdef}.
 The factor $\bar{\CR}^+{}'$ is constructed from the generators $e_i'=e_ik_i^{-1}$ and 
$f_i'=k_if_i$. We have
\begin{align}
&\tilde\De(e_i')\,=\,e_i'\ot q^{\bar\ep_i-\frac{\bar\ep}{M}}+q^{\frac{\bar\ep}{M}-\ep_i}\ot e_i'\,,\\
&\tilde\De(f_i')\,=\,f_i'\ot q^{\frac{\bar\ep}{M}-\ep_{i+1}} +q^{\bar\ep_{i+1}-\frac{\bar\ep}{M}}\ot f_i'\,,
\end{align}
indicating that our choice $\mathscr{F}=\si(q^{-f})$ is indeed a particularly natural one to consider.
\end{rem}

\subsection{Relevance for affine Toda theories}

Before we continue with more formal developments let us pause to review 
some important hints indicating that the representation theory of quantum affine 
algebras will be the proper framework for establishing and exploiting the quantum integrability
of the affine Toda theories.

\subsubsection{Continuum approaches}

One of the key observations \cite{BMP} pointing in this direction is the fact that the screening charges
$\SQ_i^\pm$ generate representations of the the nilpotent sub-algebras 
$\Nilm$, $\Nilp$,
\begin{equation}\label{scrrepdef}
\pi_{\rm\sst FF}^+(f_i):=\frac{q}{q^2-1}\,\SQ_i^+\,, \qquad
\pi_{\rm\sst FF}^-(e_i):=\frac{q}{q^2-1}\,\SQ_i^-\,.
\end{equation}
Indeed, it can be verified by direct calculations that the Serre-relations are satisfied \cite{BMP,BLZ3}.
This observation relates the 
interaction terms in the light-cone Hamiltonians to the representation theory of 
the quantum affine algebra 
$\CU_q(\widehat{\fsl}_M)$. It can be used to construct the local conserved charges of 
the affine Toda theories in the light-cone representation \cite{FF1,FF2}.

The representations $\pi_{\rm\sst FF}^+$
can be extended to representations of the Borel sub-algebras 
$\Borelm$ $\Borelp$ by setting 
\begin{equation}
\pi_{\rm\sst FF}^+(h_i):=\frac{2{\mathrm i}}{b}(\spp_i-\spp_{i+1})\,, \qquad
\pi_{\rm\sst FF}^-(h_i):=-\frac{2{\mathrm i}}{b}(\spp_i-\spp_{i+1})\,.
\end{equation}
A beautiful observation was made in \cite{BLZ3} and \cite{BHK} in the cases $M=2$ and $M=3$,
respectively: It is indeed 
possible to evaluate the
universal R-matrix in the tensor product of 
representations $\pi^{\rm f}_\la\ot \pi_{\rm\sst FF}^+$, where $\pi_{\rm\sst FF}^+$ is the 
free-field representation defined above, and
$\pi^{\rm f}_\la$ is defined as
\begin{equation}
 \pi^{\rm{f}}_{\lambda}(e_i)\,=\,\lambda^{-1}\,\SE_{i,i+1}\,,\qquad
 \pi^{\rm{f}}_{\lambda}(f_i)\,=\,\lambda\,\SE_{i+1,i}\,,\qquad
 \pi^{\rm{f}}_{\lambda}(h_i)\,=\,\SE_{i,i}-\SE_{i+1,i+1}\,;
\label{fundrepTEXT1}
\end{equation}
the matrices $\SE_{ij}$ are the matrix units 
$\SE_{ij}\,\SE_{kl}\,=\,\delta_{jk}\,\SE_{il}$. For a certain range of imaginary values of the parameter
$b=i\be$, the matrix elements of 
\begin{equation}\label{BLZ}
\SM^+(\la):=(\pi^{\rm f}_\la\ot \pi_{\rm\sst FF}^+)(\CR)\,,
\end{equation}
represent 
well-defined operators on the Fock space underlying the representation $\pi_{\rm\sst FF}^+$. 
The matrices $\SM^+(\la)$ represent quantum versions of the 
monodromy matrices representing the
integrable structure of the massless limit of the affine Toda theories. These results were later generalised to 
$M>3$ in \cite{Ko}.

The massless limit decouples left- and right-moving degrees of freedom. 
By a careful analysis of the massless limit it was shown in \cite{RT} that
the monodromy
matrices $\SM^+(\la)$  and $\SM^-(\la):=(\pi^{\rm f}_\la\ot \pi_{\rm\sst FF}^-)(\CR^-)$  
describe the decoupled integrable structures of the right- and left-moving 
degrees of freedom, respectively.
This means that there is a correspondence between light-cone 
directions and Borel sub-algebras. This observation
will be very useful for us.

For the cases $b=i\be$ it might be possible to define monodromy cases for the massive
theories by considering
\begin{equation}
\SM(\la)\,=\,\SM^-(\la)\SM^+(\la)\,,
\end{equation}
as suggested by the representation \rf{Tmodmod} of the classical monodromy matrix for 
$N=1$. 
Unfortunately it is not straightforward to generalise \rf{BLZ}
to the cases of our interest, $b\in\BR$. The short-distance singularities are 
more severe in these cases. 
It may nevertheless be possible to  define monodromy
matrices $\SM(\la)$ by using a renormalised version of the right hand side of 
\rf{BLZ}. They key observation that \rf{scrrepdef} defines 
representations of $\Nilp$, $\Nilm$ remains valid, after all. 
However, this approach has not been developed yet.

\subsubsection{Lattice discretisation}

In order to gain full control, we will instead employ 
a lattice regularisation.
As will be discussed in more detail below, it is then possible to obtain the lattice Lax matrices
from the universal R-matrix in a way that is quite similar to \rf{BLZ},
\begin{equation}\label{lattice-BLZ}
\begin{aligned}
&\SL^{+}(\la):=\frac{1}{\theta^+(\la\mu^{-1})}[(\pi^{\rm f}_\la\ot \pi^{+}_\mu)(\CR_\pm)]_{\rm ren}^{}\,,\\
&\bar\SL^{-}(\la):=\frac{1}{\bar\theta^{{-}}(\la\mu^{-1})}[(\pi^{\rm f}_\la\ot \bar{\pi}^{-}_\mu)(\CR_\pm)]_{\rm ren}^{}\,,
\end{aligned}
\end{equation}
where the 
representations $\pi^{+}$ and $\bar{\pi}^{-}$ are  defined as 
\begin{align}
& \pi_{\lambda}^+(f_i)\,=\,\frac{\lambda}{q-q^{-1}}\,\su_i^{-1}\,\sv_i\,,\qquad
 \pi_{\lambda}^+(k_i)\,=\,\su_i\,\su^{-1}_{i+1}\,,
\label{repplus-0}\\
&  \bar{\pi}_{\lambda}^-(e_i)\,=\,\frac{\lambda^{-1}}{q^{-1}-q}\,\sv_i\,\su_{i+1}\,,\qquad
  \bar{\pi}_{\lambda}^-(k_i)\,=\,\su^{-1}_i\,\su_{i+1}\,.
\label{repminusbar-0}
\end{align}
$\{\sv_i,\su_{i}\}_{i=1,\dots,M}$ generate the algebra $\mathcal{W}$, see \eqref{algebraWdef}. 
It is easy to verify that  \eqref{repplus-0} and \eqref{repminusbar-0} satisfy, respectively,  the defining relations  of $\mathcal{U}_q(\mathfrak{b}^+)$ and $\mathcal{U}_q(\mathfrak{b}^-)$.
The notation $[\dots]_{\rm ren}^{}$ indicates the application of a 
certain renormalisation procedure, which will be necessary to get
well-defined results in the cases where the representations 
$\pi^{\pm}$ are infinite-dimensional. The normalisation factors $(\theta^+(\la\mu^{-1}))^{-1}$ and 
$(\bar\theta^{{-}}(\la\mu^{-1}))^{-1}$ in \rf{lattice-BLZ} are proportional to the identity operator and will be fixed later.

We get another strong hint that the representation theory of quantum affine algebras
is well-suited for our purpose by observing that it gives us a 
very natural way to obtain the light-cone evolution operator from the universal R-matrix.
We had observed above 
in order to build an evolution operator we need to find an operator $\sr^{+\dot{-}}(\mu/\la)$
satisfying
\begin{equation}\label{rLL-rep-0}
(\sr^{+\dot{-}}(\mu/\la))^{-1}\cdot \LL^+(\la)\,\bar\LL^-(\mu)\cdot\sr^{+\dot{-}}(\mu/\la)\,=\,\bar\LL^-(\mu)\,\LL^+(\la)\,.
\end{equation}
A solution to this equation in the sense of formal power series in the parameters $\mu$, $\la$
 can be obtained from the universal R-matrix,
\begin{equation}
\sr^{+\dot{-}}(\mu/\la)\,=\,
(\pi^+_{\la^{-1}}\ot\bar\pi^{-}_{\mu^{-1}})(\CR^-)\,,
\end{equation}
as follows by 
applying $\pi^{\rm f}_{1}\ot\pi^+_{\la^{-1}}\ot\bar\pi^-_{\mu^{-1}}$ to the 
Yang-Baxter equation \rf{YBE+-}.
We will later discuss the renormalisation of $(\pi^+_{\la^{-1}}\ot\bar\pi^{-}_{\mu^{-1}})(\CR^-)$
needed to turn $\sr^{+\dot{-}}(\mu/\la)$ into a well-defined operator. 
The definition \rf {lattice-BLZ} realises the
link between light-cone directions and
Borel sub-algebras of $\CU_q(\widehat{\fsl}_M)$ observed in \cite{RT}
within the lattice discretisation.  It is crucial for making
the relation between the evolution operator and the universal R-matrix as direct as possible.


\section{R-operators from the universal R-matrix - case of $\CU_q(\widehat{\fsl}_2)$}

\subsection{Overview}

We had observed in Section \ref{scheme} that basic building blocks 
of the QISM are the operators $\sr_{rs}^{\ep\ep'}(\mu,\nu)$ which are
required to be solutions to the RLL-relations 
\begin{align}\label{srll-a}
\big(\sr_{rs}^{\ep\ep'}(\mu,\nu)\big)^{-1}\,{\mathsf{L}}_r^{\ep}(\mu)\mathsf{L}_{s}^{\ep'}(\nu)\,\sr_{rs}^{\ep\ep'}(\mu,\nu)\,=\,
\mathsf{L}_{s}^{\ep'}(\nu)\mathsf{L}_r^{\ep}(\mu)\,,
\quad\ep,\ep'=\pm\,.
\end{align}
The operators $\sr_{rs}^{\dot{-}+}(\la,\mu)$ are in particular needed
for the construction of an integrable time-evolution.

The framework of quantum affine algebras will allow us to 
systematically obtain solutions of the 
equations \rf{srll-a} from the universal 
R-matrix of the quantum affine algebraa 
$\CU_q(\widehat{\fsl}_M)$. This fact is known in the case of spin chains
of XXZ-type, where it is sufficient to evaluate the 
universal R-matrices in finite-dimensional or infinite-dimensional 
representations of highest or lowest weight type. The main issue to 
be addressed in our case originates from the fact that some of the 
relevant representations will not have a highest or a lowest weight.
On first sight this causes very serious problems: Evaluating the 
universal R-matrices in infinite-dimensional representations 
will generically produce infinite series in monomials of the 
operators representing the generators of $\CU_q(\widehat{\fsl}_M)$.
These series turn out not to be convergent in the cases of our interest. 

It will nevertheless be found that there exists an essentially canonical
renormalisation of the universal R-matrices. The main tool for establishing this claim will be the 
product formulae for $\CR^\pm$ found by Khoroshkin and Tolstoy.
The product formulae  are particularly
well-suited for our task: They disentangle  
the infinity from the infinite extension of the root system from the
infinite summations over powers of the root generators. 
We will identify simple representations 
 such that only finitely many real root generators
will be represented nontrivially.
More general representation of our interest
can be constructed by taking tensor 
products of the simple representations, curing the first type of problem.
The second type of divergence can be dealt with for representations
in which the root generators are represented by positive self-adjoint 
operators. Replacing the quantum exponential functions appearing in the
product formulae by 
a special function related to the non-compact quantum dilogarithm
produces well-defined operators which will satisfy all relevant 
properties one would naively expect to get from the evaluation of the 
universal R-matrices.

 
 A review of the product formulae will be given in Subsection \ref{sec:prodformula} below.
We then start discussing how to renormalise the expressions obtained by
evaluating the the product formulae in the representations of our interest.
In order to disentangle difficulties of algebraic nature from analytic
issues we will begin discussing  the necessary renormalisation for the 
case of $\CU_q(\widehat{\fsl}_2)$.
The cases $\CU_q(\widehat{\fsl}_M)$ will be discussed in the next section.
Some of the factors obtained by evaluating the product formulae will be proportional 
to the identity operator. These contributions, associated to what are called the imaginary roots,
will be discussed later in Sections \ref{Imrenorm-2} and \ref{sec:ImrootsSLM} below.

\subsection{The product formula for the universal R-matrix}
\label{sec:prodformula}


In this section we begin by reviewing the explicit formula 
for the universal R-matrix  obtained by  Khoroshkin and Tolstoy. 
We will follow the conventions in  \cite{KT2}. A guide to the original literature
can be found in Section \ref{sec:originofprod} below.

\subsubsection{Construction of root generators}
\label{sec:rootgen}

Recall that $\Delta_+(\widehat{\mathfrak{g}})=\Delta^{\text{re}}_+
(\widehat{\mathfrak{g}})\cup \Delta^{\text{im}}_+(\widehat{\mathfrak{g}})$ where
\begin{equation}\label{affinerootRE}
\Delta^{\text{re}}_+(\widehat{\mathfrak{g}})\,=\,
\{\gamma+k\delta\,|\gamma \in \Delta_+(\mathfrak{g}),\,k\in \mathbb{Z}_{\geq 0}\}\,\cup 
\{\left( \delta-\gamma\right)+k\delta\,|\gamma \in \Delta_+(\mathfrak{g}),\,k\in \mathbb{Z}_{\geq 0}\}
\end{equation}
\begin{equation}\label{affinerootIM}
\Delta^{\text{im}}_+(\widehat{\mathfrak{g}})\,=\,\{k\delta\,|\,k\in \mathbb{Z}_{> 0}\}
\end{equation}

The first step of the procedure is to choose a special ordering in $\Delta_+(\widehat{\mathfrak{g}})$.
We say that an order $\prec$ on $\Delta_+\left(\widehat{\mathfrak{g}}\right)$ 
is normal (or convex) if it satisfy the following condition:
\begin{equation}\label{convexorddef}
 (\alpha,\beta)\in \left(\Delta_+\times \Delta_+\right)/\left(\Delta_+^{\text{im}}\times\Delta_+^{\text{im}}\right)\,,
\,\,\, \alpha\prec\beta\,,\,\,\,
\alpha+\beta\in \Delta_+\,\,\,
\Rightarrow\,\,\,\,\alpha\prec\alpha+\beta\prec\beta\,
\end{equation}
This definition can be applied to any Kac-Moody Lie algebra.
For finite dimensional Lie algebras there is a one to one correspondence between normal orders
 and reduced expressions for the longest element of the Weyl group, see e.g.~\cite{CP}.
For untwisted  affine Lie algebras a convex order splits the positive real roots in two parts:
 those that are greater than $\delta$ and those that are smaller than $\delta$,
 see \cite{Ito} and appendix \ref{sec:convexorderSLM}.
 Without loss of generality, roughly up to the action of the Weyl group of $\mathfrak{g}$, we further impose
\begin{equation}
 \gamma+\mathbb{Z}_{\geq 0}\,\delta\,\prec\,
\mathbb{Z}_{> 0}\,\delta
\,\prec\,
\left(\delta-\gamma\right)+\mathbb{Z}_{\geq 0}\,\delta\,,\qquad
\qquad \gamma\in \Delta_{+}(\mathfrak{g})\,.
\label{precsucc}
\end{equation}
In applications we will as well use the opposite ordering compared to \eqref{precsucc}.
 From the definition it is clear that given 
a convex ordering the opposite ordering is convex as well.
This ordering reflects a triangular decomposition of 
$\Nilp\simeq \mathcal{U}^+_q(\prec) \otimes \mathcal{U}^+_q(\sim) \otimes \mathcal{U}^+_q(\succ)$ 
(see e.g.~\cite{Lu} 40.2.1), 
and is manifest in the the structure of the product formula  for the universal R-matrix given below.

The second step of the procedure is to construct the generators corresponding to the positive
roots of  $\widehat{\fg}$, where imaginary roots are counted with multiplicities, from the generators
corresponding to the simple positive roots
$e_{\alpha_0} = e_{\delta - \theta}$ and $e_{\alpha_i}$.
The procedure goes as follows
\begin{itemize}
 \item[1.]
Let $\alpha,\beta,\gamma \in \Delta^{re}(\widehat{\mathfrak{g}})$ with 
$\gamma=\alpha+\beta$ and  $\alpha\prec \gamma\prec\beta$  be a  minimal sequence,
 i.e.~there are no other positive roots $\alpha'$ and $\beta'$ between $\alpha$ and $\beta$
such that $\gamma=\alpha'+\beta'$,  then we set
\begin{equation}\label{step1}
 e_{\gamma}\,:=\,\left[e_{\alpha},e_{\beta}\right]_{q^{-1}}\,
:=\,e_{\alpha}\,e_{\beta}\,-\,q^{-(\alpha,\beta)}\,e_{\beta}\,e_{\alpha}\,.
\end{equation}
Notice that when, for a fixed normal order, the minimal sequence is not unique,
 the root vector does not depend on the choice of minimal sequence. 
This is ensured by the Serre relations.
In this way one construct all root vectors $e_{\gamma}$, $e_{\delta-\gamma}$,  for $\gamma\in\Delta_+(\mathfrak{g})$.
\item[2.] Next, set
\begin{subequations}
\label{iterativeIMrootsandmore}
\begin{align}
  e^{(i)}_{\delta}\,&:=\,\left[e_{\alpha_i},e_{\delta-\alpha_i}\right]_{q^{-1}}\,,\qquad i=1,\dots,\text{rank}(\mathfrak{g})\,,\\
 e_{\alpha_i+k\,\delta}\,&:=\,[(\alpha_i,\alpha_i)]_{q}^{-k}\,\big(-\text{Adj}\,e^{(i)}_{\delta} \big)^k
\,\cdot\,e_{\alpha_i}\,,\\
 e_{(\delta-\alpha_i)+k\,\delta}\,&:=\,[(\alpha_i,\alpha_i)]_{q}^{-k}\,\big(\text{Adj}\,e^{(i)}_{\delta} \big)^k
\,\cdot\,e_{\delta-\alpha_i}\,, \\
 e'^{(i)}_{k\delta}\,&:=\,\left[e_{\alpha_i+(k-1)\delta},e_{\delta-\alpha_i}\right]_{q^{-1}}\,.
\end{align}\end{subequations}
In the case in which the Cartan matrix is symmetric one has $(\alpha_i,\alpha_i):=\,a^{sym}_{ij}=2$.
\item[3.] Construct the remaining real root vectors $e_{\gamma+k\delta}$ and $e_{(\delta-\gamma)+k\delta}$
 for all $\gamma\in\Delta_+(\mathfrak{g})$, $k\geq 1$
using the same procedure as step one. 
\item[4.] Define the imaginary root vectors $e^{(i)}_{k\delta}$ from $e'^{(i)}_{k\delta}$ as follows:
\begin{equation}
 E_i(z)\,=\,\ln\left(1\,+\,E'_i(z)\right)\,,
\end{equation}
where
\begin{equation}\label{Imrootscurrents}
 E_i(z)\,=\,\left(q^{-1}-q^{+1} \right)\,\sum_{k=1}^{\infty}\,e^{(i)}_{k\delta}\,z^{-k}\,,
\qquad 
 E'_i(z)\,=\,\left(q^{-1}-q^{+1} \right)\,\sum_{k=1}^{\infty}\,e'^{(i)}_{k\delta}\,z^{-k}\,.
\end{equation}
\end{itemize}
The root vectors corresponding to the negative roots are obtained 
with the help of Cartan anti-involution  \eqref{DefZeta}.
Notice that once we fix the normal ordering as in \eqref{precsucc} the root vectors $e^{(i)}_{\delta}$,
$e_{\alpha_i+k\,\delta}$,
$e_{(\delta-\alpha_i)+k\,\delta}$
are independent on the specific choice of root ordering, see \cite{Dam2}.

The constructed  root vectors satisfy a number of remarkable properties. 
Among others, the following property explains the attribute \emph{convex} associated
to the constructed basis. 
For $\alpha\prec \beta$, $\alpha,\beta\in \Delta_{+}(\widehat{\mathfrak{g}})$ one has
\begin{equation}
 e_{\alpha}\,e_{\beta}\,-q^{-(\alpha,\beta)}\,e_{\beta}\,e_{\alpha}\,=\,
\sum_{\alpha\prec\gamma_1\prec\dots\prec\gamma_{\ell}\prec \beta}c_{\gamma}(k)\, 
\left(e_{\gamma_1}\right)^{k_1}\,\dots\,\left(e_{\gamma_{\ell}}\right)^{k_{\ell}}\,
\label{eeCONVEX}
\end{equation}
where $c_{\gamma}(k)$ are rational function of $q$ non vanishing only for 
$\alpha+\beta=\sum_{i=1}^{\ell}\,k_{i}\,\gamma_{i}$.

An other important property of the imaginary root generators, see \cite{Dam2}, is the following
\begin{equation}
 \Delta(E_i(z))-E_i(z)\otimes 1 -1\otimes E_i(z)\,
\in\,
\mathcal{U}^+(\prec)\otimes \mathcal{U}^0\,\mathcal{U}^+(\succ)\,.
\end{equation}
We will discuss the coproduct of imaginary roots in greater details in 
Section
\ref{sec:coprIM_root_slM}.

For the case of $\mathcal{U}_q(\widehat{\mathfrak{gl}}_M)$
  a distinguished normal order and the explicit  definition of some relevant
 root vector are presented in Appendix \ref{sec:rootSLM}.

\subsubsection{Statement of the product formula}

The expression for the universal $R$-matrix 
has the form
\begin{equation}\label{KT}
\mathscr{R}^- \,=\,\bar{\mathscr{R}}^-\,q^{-t}\,=\,
\mathscr{R}^-_{\prec \delta} \, \mathscr{R}^- _{\sim \delta} \, \mathscr{R}^-_{\succ \delta}\,q^{-t} \,.
\end{equation}
The quantity $\bar{\mathscr{R}}^-$ is an infinite ordered product
 over the positive roots $\Delta_+(\widehat{\mathfrak{g}})$. 
The order of factors is the same 
as the convex order used in the definition of root vectors. 
The infinite product decomposes into three parts as follows form \eqref{precsucc}
and enphasized in \rf{KT} by the notation $\prec \delta,\,\sim \delta,\, \succ \delta$. 
To each real positive root we associate the factor
\begin{equation}
\mathscr{R}^-_{\gamma} \,=\,\exp_{q^{(\gamma, \gamma)}} \left( (q^{-1} - q)
\, s^{-1}_{\gamma} \, f_{\gamma}\otimes e_{\gamma} \right)\,
\qquad \gamma\in\Delta^{\text{re}}_+(\widehat{\mathfrak{g}})\,,
\label{rgm}
\end{equation}
with $\exp_q(x)$ the quantum exponential 
 \begin{equation}\label{q-exp}
\exp_q(x)\,=\,\sum_{n=0}^{\infty}\frac{1}{(n)_q!}x^n\,,\qquad
(k)_q=\frac{q^{k}-1}{q-1}\,,
\qquad
(n)_q!=(1)_q\,(2)_q\dots(n)_q\,.
\end{equation}
The quantities $s_{\gamma}$ in \rf{rgm}  are determined by the relation
\[
[e_{\gamma}, \, f_{\gamma}] = s_{\gamma} \, \frac{q^{h_{\gamma}} - q^{-h_{\gamma}}}{q - q^{-1}},
\]
where $h_{\gamma} = \sum_i k_i h_i$ if $\gamma = \sum_i k_i \alpha_i$. 
In the case $\mathfrak{g}=\mathfrak{sl}_M$ we simply have $s_{\gamma}=1$.

The contribution of positive imaginary roots is given by
\begin{equation}\label{rpd}
\mathscr{R}^-_{\sim \delta} = \exp \bigg( (q^{-1} - q^{+1}) \sum_{m \in \BZ_+} \sum_{i, j
= 1}^r u_{m, \, ij} \, f^{(i)}_{m \delta} \otimes e^{(j)}_{m \delta} \bigg),
\end{equation}
where $r$ is the rank of the Lie algebra $\mathfrak{g}$,
and the quantities $u_{m, \, ij}$, $m\in\BZ_+$,
are the elements of the matrix
$u_m$ inverse to the matrix $t_m$ with elements
\begin{equation}\label{t-matrix}
t_{m, \, ij} = (-1)^{m(1 - \delta_{ij})} m^{-1} [m a_{ij}]_q,
\end{equation}
entering the commutation relations
\begin{equation}\label{eecomm_gives_t}
[e_{\alpha_i + m \delta}, \, e^{(j)}_{n \delta}\,] = t_{n, i j} \, e_{\alpha_i + (m + n)\delta}.
\end{equation}
In the case $\mathfrak{g}=\mathfrak{sl}_M$, the coefficients $u_{m,ij}$ appearing in 
  \eqref{rpd}  can be represented  explicitly as
\begin{equation}\label{umij}
u_{m,ij}\,=\,\frac{m}{[M\,m]_q}\,[M-\text{max}(i,j)]_{q^m}\,[\text{min}(i,j)]_{q^m}\,(-1)^{m(i-j)} \,,
\end{equation}
where $\text{min}(i,j)$, $\text{max}(i,j)$ denotes the minimum and maximum value among $i$ and $j$.

While the root generators and their algebra depend on the choice of convex order,
the universal R-matrix is independent of this choice. 
This is a non-trivial fact that follows from the uniqueness (under certain assumptions) 
of the universal R-matrix, see e.g. \cite{KT92}.




\subsubsection{A guide to the literature on the product formula}
\label{sec:originofprod}

In the following we collect  some  references that should help the interested
 reader in understanding the  origin of the product formula for the universal R-matrix. 

An explicit formula for the universal R-matrix of $\mathcal{U}_q(\mathfrak{sl}_2)$ 
was presented by Drinfeld in \cite{Dr86}. 
Shortly after it was given for  $\mathcal{U}_q(\mathfrak{sl}_M)$ in  \cite{Ro},
 for any finite dimensional simple Lie algebra in \cite{KiR}, \cite{LS}
and for finite dimensional Lie super-algebras in  \cite{KT91b} and \cite{Y}
. In  the affine case, both twisted and untwisted, an explicit expression for
 the universal R-matrix has first been  given by  Khoroshkin and Tolstoy
 in \cite{KT91a}, \cite{KT92} and later in \cite{LSS} (for $\mathcal{U}_q(\widehat{\mathfrak{sl}}_2)$)
 and \cite{Dam2}, \cite{Dam3} using different techniques.  Product formulae for quantum affine
 super-algebras were presented in \cite{Ya}.

Using Drinfeld double construction \cite{Dr86}, the problem of finding 
explicit expressions for the universal R-matrix reduces to
 the determination of  basis of $\Borelmp$ which are orthonormal with respect to the standard pairing
between $\Borelp$ and $\Borelm$.
The key idea is to find a convenient basis, with simple properties 
under product and coproduct, that simplifies the calculation of the pairing.
In parallel to the $q=1$ case, one  construct so called (convex) 
basis of Poincar\'e-Birkhoff-Witt (PBW) type as ordered product of root vectors.
Thus, one must first define analogues of root vectors associated to non-simple roots of $\mathfrak{g}$.
There is an elegant construction of such root vectors. 
If $\mathfrak{g}$ is finite dimensional all roots are in the trajectory 
under the Weyl group of a simple root.
As the  Weyl group  can be be extended  to  a braid group action  
on $\mathcal{U}_q(\widehat{\mathfrak{g}})$ \cite{Lu} one can construct non-simple root vectors from simple ones
following this observation, see \cite{CP}.
In the affine case the situation is more involved as imaginary roots, 
by definition \cite{Kac}, are not in the orbit of simple ones under the Weyl group. 
The construction of imaginary root vectors in this case has been
 carried over in \cite{Dam1}, \cite{LSS}, \cite{Beck1}, \cite{Beck2}, \cite{Dam2}.
While explicit  proofs in the literature concerning properties
 of PBW basis use techniques connected to the braid group action, in the following 
we will use a different construction.

Convex bases in the affine case have also been constructed in 
 \cite{Tol}, \cite{KT91a}, \cite{KT93a} \cite{KT93b}.
In these references the braid group action is not used and  explicit proofs are mostly omitted.
 The construction of root vectors, referred to as Cartan-Weyl basis,
is guided by the authors experience with so called extremal projectors, see \cite{Tol2}. 
This construction of root vectors is convenient when dealing
 with representations and will be used in the following.
We  remark that the root vectors constructed by this procedure 
are closely related to the quantum current type generators appearing
in the Drinfeld's second realization of $\mathcal{U}_q(\widehat{\mathfrak{g}})$  \cite{Dr87}, 
see \cite{KT93a} and \cite{Beck1}.


\subsection{Simple representations of $\CU_{q}(\widehat{\fsl}_2)$}
\label{sec:simplerep}


\subsubsection{Evaluation representations}
\label{evHOMOsl2}

To begin with, let us recall that there is a well-known way to 
get representations of the loop algebras $\CU_{q}(\widehat{\fsl}_2)_0$ from 
representations of the quantum group $\CU_q(\fsl_2)$. It is based
on the following homomorphism of algebras:
Let  $\CU_q(\fsl_2)$ be the algebra generated by 
$\SE$, $\SF$ and $\SK^{\pm 1}$ with  relations
\begin{equation}\label{Uqsl2}
\begin{aligned}
\SK\SE=q^{+1}\SE\SK\,,\\
\SK\SF=q^{-1}\SF\SK\,,
\end{aligned}\qquad
[\,\SE\,,\,\SF\,]\,=\,\frac{\SK^{2}-\SK^{-2}}{q-q^{-1}}\,,
\end{equation}
then 
\begin{equation}\begin{aligned}
&\text{ev}_{\la}(e_1)\,=\,\la^{-1}\,q^{\frac{1}{2}}\,
\SK^{-{1}}\SE\,,\qquad
\text{ev}_{\la}(e_0)\,=\,\la^{-1}\,q^{\frac{1}{2}}\,
\SK^{+1}\,\SF\,,\qquad
\text{ev}_{\la}(k_1)\,=\,
\SK^{+2}\,,\\
& \text{ev}_{\la}(f_1)\,=\,\la^{+1}\,q^{\frac{1}{2}}\,
\SK^{+1}\,\SF\,,\qquad
\text{ev}_{\la}(f_0)\,=\,\la^{+1}\,q^{\frac{1}{2}}\,
\SK^{-{1}}\SE\,,\qquad
\text{ev}_{\la}(k_0)
\,=\,\SK^{-2}\,.
\end{aligned}\end{equation}
satisfy the defining relations of $\CU_{q}(\widehat{\fsl}_2)_0$. This claim 
can be verified by a straightforward calculation.
The center of $\CU_q(\fsl_2)$ is generated by the Casimir $\SC$
defined as
\begin{equation}\label{sl2casimir}
\SC\,:=\,
\SF\SE+\frac{q\,\SK^2+q^{-1}\,\SK^{-2}-2}{(q-q^{-1})^2}\,=\,
\frac{q^{2x}+q^{-2x}-2}{(q-q^{-1})^2}\,.
%
\end{equation}
The last equality in  this equation is a convenient parametrization of the Casimir $\SC$.

There are two types of representations of $\CU_q(\fsl_2)$ that will be
relevant for us: The usual finite-dimensional representations labelled by 
 $j\in\frac{1}{2}\BZ_{\geq 0}$ and
 certain infinite-dimensional representations  
for which $\SE$, $\SF$ and 
$\SK$ are realized by positive self-adjoint operators.
Let us discuss them in more details.

\paragraph{Finite-dimensional evaluation representations.}
We denote the $(2j+1)$-dimensional representation of 
$\CU_q(\fsl_2)$  by $\pi^{\rm f.d.}_{j}$ where
 $j\in\frac{1}{2}\,\BZ_{\geq 0}$. 
 In this case  $\SK$ has spectrum $\{q^{-j},q^{-j+1},\dots,q^{j-1},q^{j}\}$
 and the parameter $j$ is related to the Casimir $\SC$ defined in \eqref{sl2casimir} as
\begin{equation}
\pi^{\rm f.d.}_{j}\left(\SC\right)
\,=\,\frac{q^{2j+1}+q^{-2j-1}-2}{(q-q^{-1})^2}\,.
\end{equation}
We further define 
 $\pi^{\rm f.d.}_{\la,j}= \pi^{\rm f.d.}_{j}\circ \text{ev}_{\la}$.
Of particular importance will be the fundamental 
representation $\pi^{\rm f}_{\la}$ corresponding to $\pi^{\rm ev}_{\la,j}$, $j=1/2$, where we 
may take
\begin{equation}
\pi^{\rm f.d.}_{1/2}\left(\SE\right)
\,=\,\bigg(\,\begin{matrix} 0 & 1 \\ 0 & 0 \end{matrix}\,\bigg)\,,\qquad
\pi^{\rm f.d.}_{1/2}\left(\SF\right)
\,=\,\bigg(\,\begin{matrix} 0 & 0 \\ 1 & 0 \end{matrix}\,\bigg)\,,\qquad
\pi^{\rm f.d.}_{1/2}\left(\SK\right)
\,=\,\bigg(\,\begin{matrix} q^{+\frac{1}{2}} & 0 \\ 0 & q^{-\frac{1}{2}} \end{matrix}\,\bigg)\,.
\end{equation}
The representations $\pi^{\rm f.d.}_{\la,j}$ for $j>1/2$ 
can be generated from $\pi^{\rm f}_{\la}$ by taking tensor 
products and quotients.

\paragraph{Evaluation representations of modular double type.}
We will also be interested in
infinite-dimensional evaluation
representations $\pi^{\rm m.d.}_{\la,s}$, $s\in\BR$, 
of modular double type where $\SE$, $\SF$ and 
$\SK$ are realized by positive self-adjoint operators. 
A representation $\CP_s$ of $\CU_q(\fsl_2)$
can be constructed using self-adjoint operators $\spp$ and $\sq$
satisfying $[\spp,\sq]=(2\pi i)^{-1}$ as follows,
\begin{equation}\label{uqsl-rep}
\begin{aligned}
&
\pi^{\rm m.d.}_{s}\left(\SE\right)\,=\,
\mathsf{e}_s\,:=\,e^{+\pi b \sq}\frac{\cosh\pi b(\spp-s)}{\sin\pi b^2}e^{+\pi b \sq}\,,\\
&\pi^{\rm m.d.}_{s}\left(\SF\right)\,=\,
\mathsf{f}_s\,:=\,e^{-\pi b \sq}\frac{\cosh\pi b(\spp+s)}{\sin\pi b^2}e^{-\pi b \sq}\,,
\end{aligned}
\qquad \,\,
\pi^{\rm m.d.}_{s}\left(\SK\right)\,=\,
\mathsf{k}_s\,:=\,e^{\pi b \spp}\,.
\end{equation}
These operators satisfy the relations \rf{Uqsl2} with $q=e^{-\pi i b^2}$.
The operators $\mathsf{e}_s$, $\mathsf{f}_s$ and 
$\mathsf{k}_s$ are unbounded. There is a canonical subspace $\CP_s$ of $L^2(\BR)$ 
representing a maximal domain of definition for $\CU_q(\fsl_2)$. The 
terminology modular double type refers to the fact that positivity
of the operators $\mathsf{e}_s$, $\mathsf{f}_s$ and 
$\mathsf{k}_s$ allows us to construct operators
$\tilde{\mathsf{e}}_s$, $\tilde{\mathsf{f}}_s$ and 
$\tilde{\mathsf{k}}_s$
from 
$\mathsf{e}_s$, $\mathsf{f}_s$ and 
$\mathsf{k}_s$ which generate a representation of $\CU_{\tilde{q}}(\fsl_2)$ 
with $\tilde{q}=e^{-\pi i /b^2}$, see also Remark \ref{moddoubrem} below.

The Casimir $\SC$ of $\CU_q(\fsl_2)$ 
defined in \eqref{sl2casimir} 
 is now represented as 
\begin{equation}
\label{mdCASIMIR}
\pi^{\rm m.d.}_{s}\left(\SC\right)\,=\,
\frac{
e^{+2\pi b (s+\frac{i}{2}b^{-1})}
+e^{-2\pi b (s+\frac{i}{2}b^{-1})}
-2}{(q-q^{-1})^2}\,=\,
\left(\frac{\cosh(\pi b s)}{\sin(\pi b^2)}\right)^2\,.
\end{equation}
The middle equation makes it manifest that
for this representation
$q^{\pm2x}\mapsto -e^{\pm2\pi b s}$.
Notice that
$\pi^{\rm m.d.}_{\la,s}= \pi^{\rm m.d.}_{s}\circ \text{ev}_{\la}$
corresponds to 
positive self-adjoint operators for $\la\,\in\,\mathbb{R}_{\geq 0}$.

\subsubsection{Prefundamental representations} \label{prefunddef}

For our physical application we introduce representations $\pi^{\pm}_\mu$
 of  the Borel-subalgebras 
$\Borelmp$ of $\CU_q(\widehat{\fsl}_2)$ such that
\begin{equation}\label{LpmfromR}
\LL^{\pm}(\la)\,=\,\bigg(\,\begin{matrix}
\su & {\la}^{\mp 1}\,\sv^{\pm 1}\\
 {\la}^{\mp 1}\,\sv^{\mp 1} &\su^{-1}
\end{matrix}\,\bigg)\,=\,\frac{1}{\theta^\pm(\la)}[(\pi^{\rm f}_{\la}\ot\pi^{\pm}_{1})(\CR^\pm)]_{\rm ren}^{}\,,
\end{equation}
where  $\su$, $\sv$  are operators satisfying 
$\su\sv=q^{-1}\sv\su$, and ${\rho^\pm(\la)}$ is proportional 
to the identity operator.
The notation $[\dots]_{\rm ren}^{}$ indicates 
that the formal expressions following from the universal R-matrix 
will require a certain renormalization. 

It is easy to see that we need to have
\begin{align}
& \pi_{\la}^+(f_1)=\frac{\la}{q-q^{-1}}\,\su^{-1}\,\sv\,,\quad
\pi_{\la}^+(f_0)=\frac{\la}{q-q^{-1}}\,\su\,\sv^{-1}\,,\quad
 \pi_{\la}^+(k_1)=\su^2=\pi_{\la}^+(k_0^{-1})\,,
\label{piplus}\\
&  {\pi}_{\la}^-(e_1)=\frac{\la^{-1}}{q^{-1}-q}\,
\sv\,\su^{-1}\,,\quad
{\pi}_{\la}^-(e_0)=\frac{\la^{-1}}{q^{-1}-q}\,
\sv^{-1}\,\su\,,\quad
  {\pi}_{\la}^-(k_i)=\su^{-2}=\pi_{\la}^+(k_0^{-1})\,.
\label{piminus}
\end{align}
In order to see that these definitions are indeed necessary to get a relation of the form \rf{LpmfromR}, let
us first consider $\LL^-(\la)$ and remind ourselves that 
$\bar{\mathscr{R}}^-=1+\sum_{j}\left(q^{-1}-q\right)\left(f_i\otimes e_i\right)+\dots$ up 
to higher order terms, which implies that 
\begin{equation}
(\pi^{\rm f}_{\la}\ot\pi^{-}_1)(\CR^-)\,=\,
\bigg(\begin{matrix}
1  & \la\,\sv^{-1}\su \\ \la\,\sv\,\su^{-1} & 1
\end{matrix}\bigg)\bigg(\,\begin{matrix} \su & 0 \\ 0  & \su^{-1}\end{matrix}\,\bigg)+\CO(\la^2)\,.
\end{equation}
The case of $\LL^+(\la)$ is very similar.

The representations $\pi^{\pm}_{\la}$ will play a fundamental role for us. They 
are the simplest examples  of what is called a prefundamental 
representation in \cite{HJ}. 
To motivate this terminology let us anticipate that all representations of our interest will
be found within the tensor products of such representations. We may therefore regard the
representations $\pi^{\pm}_{\la}$ as elementary building blocks for the category of representations
we are interested in.

One of the most basic and fundamental observations is that the operators $\sf_i:=\pi^+_\la(f_i)$, $i=0,1$ satisfy
the relations of a q-oscillator algebra,
\begin{equation}\label{q-oscrel}
\sf_0\,\sf_1-q^{-2}\,\sf_1\,\sf_0\,=\,\frac{\la^2}{q-q^{-1}}\,.
\end{equation}
This implies that the operator representing the imaginary root element $f^{(1)}_{\de}$ is proportional to the 
identity operator. It follows immediately from the iterative definition \eqref{iterativeIMrootsandmore}, 
that the operators representing the  higher real root generators $f_{\al_i+k\de}$ vanish identically. 
This observation will later be very useful.

\begin{rem}\label{moddoubrem}
For $|q|<1$ 
one may consider representations of highest or lowest weight type, as done in \cite{BLZ3}.
In this paper we will mainly be interested in infinite-dimensional 
representations where $\su$ and $\sv$ are realized by positive-selfadjoint
operators, for example
\begin{equation}
\su\,=\,e^{\pi b \sx}\,,\qquad \sv\,=\,e^{2\pi b \sy}\,,\qquad
[\sx,\sy]=\frac{i}{2\pi}\,.
\end{equation}
The positive-selfadjointness  of the operators $\su$ and $\sv$ implies a remarkable
duality phenomenon: Using the operators $\tilde\su:=\su^{\frac{1}{b^2}}$, and 
 $\tilde\sv:=\sv^{\frac{1}{b^2}}$, and replacing $q=e^{-\pi i b^2}$ by 
 $\tilde{q}=e^{-\pi i b^{-2}}$ one may use the formulae above
 to realise representations of the Borel subalgebras $\tilde{\CB}_\pm$ 
 of $\CU_{\tilde{q}}(\widehat{\fsl}_2)$ on the same space on which $\CB_\pm$ are
 realised. This has profound consequences, as was first observed in \cite{PT99,F99} for the 
 similar case of $\CU_q(\fsl_2)$.  Representations exhibiting this duality phenomenon
 will generally be referred to as representations of modular double type.
\end{rem}

\subsection{Evolution operators from the universal R-matrix}\label{canren}

In order to build an evolution operator we need to find an operator $\sr^{+-}(\mu/\la)$
satisfying
\begin{equation}\label{rLL-rep}
(\sr^{+-}(\mu/\la))^{-1}\cdot \LL^+(\la)\,\LL^-(\mu)\cdot\sr^{+-}(\mu/\la)\,=\,\LL^-(\mu)\,\LL^+(\la)\,.
\end{equation}
A formal solution to this equation is given
 by
 $(\pi^+_{\la^{-1}}\ot\pi^{-}_{\mu^{-1}})(\CR^-)$.
Indeed, formally
applying $\pi^{\rm f}_{1}\ot\pi^+_{\la^{-1}}\ot\pi^-_{\mu^{-1}}$ to the 
Yang-Baxter equation \rf{YBE+-} seems to indicate that 
$(\pi^+_{\la^{-1}}\ot\pi^{-}_{\mu^{-1}})(\CR^-)$ solves \rf{rLL-rep}. However, 
it is far from clear how to make sense out of 
$(\pi^+_{\la^{-1}}\ot\pi^{-}_{\mu^{-1}})(\CR^-)$ due to 
the infinite summations over monomials of generators defining the universal R-matrix.
Our main goal in this paper will be to generalise the definition of the universal R-matrix 
in such a way that evaluations like $(\pi^+_{\la^{-1}}\ot\pi^{-}_{\mu^{-1}})(\CR^-)$
become well-defined and satisfy  all
the relevant properties. The product formula will 
be very useful for this aim. In this subsection we will describe a first step
in this direction.

We had observed in Section \ref{prefunddef} that $\pi^+_{\mu}(f_{\al_i+k\de})=0$ for 
$i=0,1$, $k>0$. This implies immediately that the 
infinite products representing\footnote{To simplify the following formulae we rewrite
$(\pi^+_{\la^{-1}}\ot\pi^{-}_{\mu^{-1}})(\CR^-)=
(\pi^+_{\mu}\ot\pi^{-}_{\la})(\CR^-)$.}
$(\pi^+_\mu\ot \pi^-_\la)(\CR^-_{\prec \de})$ and 
$(\pi^+_\mu\ot \pi^-_\la)(\CR^-_{\succ \de})$
 truncate to a single factor.
 We furthermore observed after equation \rf{q-oscrel}
that the imaginary roots are represented by central elements
in the representations $\pi^+_\mu$ and $\pi^-_\la$. 
We conclude that the product formula  \eqref{KT}
yields a well-defined {\it formal} series in powers of $\mu/\la$ of the form 
\begin{equation}\label{R-factor0}
\sr^{+-}_{\rm\sst formal}(\mu/\la)\,=\,\rho(\mu/\la)\,
\vep_{q}(-\tau_q^2\,\sf_1\otimes\se_1)\,
\vep_q(-\tau_q^2\,\sf_0\otimes\se_0)\, q^{-\mst}\,,
\end{equation}
where $\sf_i:=\pi^+_{\mu}(f_i)$ and $\se_i:=\pi^-_{\la}(e_i)$ for $i=0,1$, 
 $q^{-\mst}:=(\pi^+_\la\ot \pi^-_\mu)(q^{-t})$, and $\tau_q=q-q^{-1}$.
 The function
$\vep_q(w)$ is related to the quantum exponential as $\vep_q(w):=\exp_{q^2}((q-q^{-1})^{-1}w)$ 
introduced in \eqref{q-exp}.
It can be written as  
\begin{equation}\label{thetadef}
\vep_q(w):=\exp(\theta_q(w))\,,
\qquad
\theta_q(w):=\sum_{k=1}^{\infty}\,\frac{(-1)^{k+1}}{k}\,\frac{w^k}{q^{k}-q^{-k}}\,.
\end{equation}
 The factor
$\rho(\mu/\la)$ in \rf{R-factor0} is a central element collecting the contributions coming from the imaginary roots,
\begin{equation}
\rho(\mu/\la):=\,(\pi^+_{\mu}\ot \pi^-_{\la})(\CR^-_{\sim\de})\,.
\end{equation}
By means of a straightforward calculation one may check that the expression \rf{R-factor0}
will satisfy \rf{rLL-rep} in the sense of formal power series 
thanks to the fact that $\vep_q(w)$ satisfies the functional relation 
\begin{equation}\label{g-funrel}
 \frac{\varepsilon_q(qw)}{\varepsilon_q(q^{-1}w)}\,=\,{1+w}\,.
\end{equation}

Our ultimate goal, however, is to construct an {\it operator} representing $(\pi^+_{\mu}\ot\pi^{-}_\la)(\CR^-)$ on the 
vector space carrying the representation $\pi^+_{\mu}\ot\pi^{-}_\la$ of $\Borelm \ot\Borelp$.
One of the main ingredients in the definition of the product formula
is the 
function $\varepsilon_q(x)$ which is well-defined
for 
 $|q| \neq 1$.
We are here interested in the case $q=e^{-\pi i b^2}$, $b\in\BR$.
The function $\varepsilon_q(x)$ can not be used in this case: The series \eqref{thetadef}
defining $\varepsilon_q(x)$ is clearly singular for all rational values of $b^2$,
and has bad convergence properties otherwise. However,
in order to preserve the most important properties of the universal R-matrix after 
renormalisation it will be sufficient to replace the
function $\varepsilon_q(x)$
by a new special function which is 
well-defined for   $q=e^{-\pi i b^2}$, $b\in\BR$, and 
which has all the relevant properties $\varepsilon_q(x)$ has.


\subsubsection{Canonical solution}

We had seen above that 
the functional equation \rf{g-funrel} plays a key role
for ensuring that the product formula satisfies the
defining properties of the universal R-matrix. 
We therefore need to find a 
function $\CE_{b^2}(w)$ that is well-defined for $|q|=1$ and satisfies the 
functional equation \rf{g-funrel}.
The physical application we have in mind forces us to impose another important
requirement: We want that the operator $\sr^{+-}(\mu/\la)$ is unitary for real $\mu/\la$,
which is necessary to get a unitary time evolution operator. Unitarity will hold if the 
function $\CE_{b^2}(w)$ replacing $\vep_q(w)$ in \rf{R-factor0} satisfies $|\CE_{b^2}(w)|=1$ for 
real positive $w$.
We are now going to explain that unitarity fixes a unique  solution to the functional
relation \rf{g-funrel} when $|q|=1$. 

It is by now pretty well-known how to find 
such a function $\CE_{b^2}(w)$: A good replacement for $\vep_{q}(w)$ will be the 
function $\CE_{b^2}(w)$ defined as
\begin{equation}\label{Eb2-def}
\CE_{b^2}(w)\,=\,\exp\big(\Theta_{b^2}\big(\fr{1}{2\pi b}\log(w)\big)\big)\,, 
\quad
\Theta_{b^2}(x):=\int_{\BR+i0}\frac{dt}{4t}\,\frac{e^{-2itx}}{\sinh({bt)\sinh(t/b)}}\,.
\end{equation}
The function $\CE_{b^2}(w)$ defined in \rf{Eb2-def} is easily seen to 
fulfil the requirements formulated above. It is closely related to the function ${\mathbf e}(x):=\CE_{b^2}(e^{2\pi bx})$
called non-compact quantum dilogarithm
in \cite{F99}. References containing useful lists of properties and further references
include \cite{FKV,ByT,Vo}.
The functional relation \rf{g-funrel} is equivalent to the following
finite difference equation for $\Theta_{b^2}(x)$,
\begin{equation}
{\mathrm D}_b\Theta_{b^2}(x):=
\Theta_{b^2}(x+ib/2)-\Theta_{b^2}(x-ib/2)\,=\,-\log(1+e^{2\pi b x})\,,
\end{equation}
which has a canonical solution 
obtained by Fourier-transformation
\begin{align}
\Theta_{b^2}(x)=& 
-{\mathrm D}_b^{-1}
\log(1+e^{2\pi b x})\,=\,\\
&=\,{\mathrm D}_b^{-1} \int_{\BR+i0}\frac{dt}{2t}\,\frac{e^{-2itx}}{\sinh(t/b)}=
\int_{\BR+i0}\frac{dt}{4t}\,\frac{e^{-2itx}}{\sinh({bt)\sinh(t/b)}}\,.
\notag\end{align} 
The second equality  in the last equation can be  verified by summing over residues.

We will now argue that replacing
$\vep_q(w)$ by  $\CE_{b^2}(w)$ is the essentially unique choice that
not only solves the functional relation \rf{g-funrel}, but is  also unitary.

Note that  \rf{g-funrel} is formally equivalent to 
$(\vep_{q}(\su))^{-1}\cdot\sv^2\cdot \vep_{q}(\su)=\sv^2+\sv\su\sv$
for any operators $\su$, $\sv$ satisfying the Weyl-algebra
$\su\sv=q^{-1}\sv\su$. 
We are going to argue that $\CE_{b^2}(w)$ is essentially the unique 
function of $w$ which satisfies 
$|\CE_{b^2}(w)|=1$ for $w\in\BR^+$ and
\begin{equation}\label{SUdiag}
(\CE_{b^2}(\su))^{-1}\cdot\sv^2\cdot \CE_{b^2}(\su)\,=\,\sv^2+\sv\su\sv\,,
\end{equation}
for {\it positive self-adjoint operators} $\su$, $\sv$ satisfying the 
Weyl-algebra $\su\sv=q^{-1}\sv\su$. 
As the function $\CE_{b^2}(\su)$ defined in \rf{Eb-def} is unitary, 
it follows from \rf{SUdiag} that
$\sv^2+\sv\su\sv$ is self-adjoint. Working in a representation
in which $\sv$ is diagonal, one may use $\CE_{b^2}(\su)$ 
to map $\sv^2+\sv\su\sv$ to diagonal form. Uniqueness of the spectral
decomposition of the self-adjoint operator $\sv^2+\sv\su\sv$
implies that the most general operator 
which satisfies \rf{SUdiag}
is the form $\SD=g(\sv)\CE_{b^2}(\su)$. The only operators depending 
only on $\su$ which do this job are scalar multiples 
of $\CE_{b^2}(\su)$.

\subsubsection{Minimality of the renormalization}

To round off the discussion, we are going to argue that 
replacing  $\vep_q(w)$ by $\CE_{b^2}(w)$ is in a precise sense
the {\it minimal subtraction} of the divergencies
$\vep_q(w)$ has when $q$ approaches the unit circle.

Let us note that $\CE_{b^2}(w)$ can be analytically
continued to complex values of $b$, allowing us to define 
it in the case where $|q|<1$. We may then compare
$\CE_{b^2}(w)$ to $\vep_q(w)$ in this regime. The integral defining 
$\Theta_{b^2}(w)$ may be evaluated as a sum over 
residues in this case, giving
\begin{equation}\label{S-factor}
\CE_{b^2}(w)\,=\,\varepsilon_q(w)\varepsilon_{\tilde{q}}(\tilde{w})\,,\qquad
\tilde{q}:=e^{-\pi i /b^2}\,,\quad\tilde{w}:=w^{\frac{1}{b^2}}\,.
\end{equation}
This means that $\CE_{b^2}(w)$ and $\varepsilon_q(w)$ differ by 
{\it quasi-constants}, functions $f(w)$
of $w$ which satisfy $f(q^2w)=f(w)$. Such quasi-constants 
represent an ambiguity
in the solution of the difference equation \rf{g-funrel} 
that needs to be fixed by additional requirements,
in general.

The particular choice of quasi-constants appearing in 
\rf{S-factor} can be seen as the {\it minimal} modification of the 
function $\vep_q(w)$ which is needed to get a function 
well-defined for all $q$ on the unit circle $|q|=1$.
 In order to see this,
let us consider the function $\theta_q(w)$ introduced in \eqref{thetadef}
as function of the complex parameter $q$. 
We will be interested in the behaviour
of $\theta_q(w)$ when
$q=e^{-\pi i b^2}$, $b^2=k/l+i\ep$. The terms with $n=rl$ in the sum defining
$\theta_q(w)$ will be singular for $\ep\ra 0$. They behave as
\begin{equation}\label{div-1}
-\frac{(-1)^{r(l+k)}}{\pi (rl)^2\ep}\,w^{rl}\,.
\end{equation}
The terms with $n=rk$ in the series defining  $\theta_{\tilde{q}}(\tilde{w})$ will similarly
behave for $\tilde{q}=e^{\pi i \tilde{b}^2}$, $\tilde{b}^2=l/k+i\tilde{\ep}$ as
\begin{equation}\label{div-2}
-\frac{(-1)^{r(l+k)}}{\pi (rk)^2\tilde{\ep}}\,\tilde{w}^{rk}\,.
\end{equation}
The divergent pieces in  \rf{div-1} and \rf{div-2}
will exactly cancel each other if $\tilde{w}=w^{\frac{1}{b^2}}$ and 
$\tilde{\ep}=-\ep\,l^2/k^2$, as is necessary to have
$\tilde{b}^2=b^{-2}+\CO(\ep^2)$. We thereby recognise 
the factor $\varepsilon_{\tilde{q}}(\tilde{w})$ in \rf{S-factor} 
as a minimal choice of a quasi-constant that cancels all the divergences that
$\varepsilon_q(w)$ develops 
when $q$ approaches the unit circle.

Taken together, the observations above motivate
us to call  $\CE_{b^2}(w)$ the 
{\it canonical renormalisation} of the function 
$\vep_q(w)$ 
which is defined for $|q|=1$. 
The considerations above motivate us to regard the operator
\begin{equation}\label{R-factor-ren}
\hat{\sr}^{+-}_{rs}(\la)\,=\,\rho_{\rm\sst ren}(\la)\sr^{+-}(\la)\,,\qquad\sr^{+-}(\la):=
\CE_{b^2}(\la \sf_{rs}^{+})\,
\CE_{b^2}(\la /\sf_{rs}^+)\,
e^{\frac{2i}{\pi  b^2}\log\su_r\log\su_s}
\,,
\end{equation}
where $\sf_{rs}^+:=\su^{-1}_r\sv_r^{}\,\sv_s^{}\su^{-1}_s$, 
as a {\it renormalised} version of the formal expression 
$\sr^{+-}_{\rm\sst formal}(\la/\mu)$. The definition of the 
scalar factor $\rho_{\rm\sst ren}(\la)$ will be discussed later.
And it is indeed straightforward to check 
that the evolution operator constructed from $\sr^{+-}_{rs}(\la)$ reduces to 
the one constructed previously in Section \ref{sec:evolutionHamiltonian}.

\subsection{Building R-operators}

We've seen that the renormalization of the universal R-matrix provides us
with $\sr^{+-}(\la/\mu)$, the main ingredient for the construction 
of the time-evolution operator. In order to build the Q-operators we
need a second ingredient, the operator $\sr^{++}(\la/\mu)$. There is 
a fairly easy way to get $\sr^{++}(\la/\mu)$ from 
$\sr^{+-}(\la/\mu)$. Note that 
\begin{equation}\label{Lpmrel}
L^{-}_r(\mu)\,=\,\CF^{-1}_r\cdot\mu^{-1}\,L^{+}_r(\mu)\,
\si_1\cdot\CF_{r}^{} \,,
\end{equation}
where $\si_1=\big(\begin{smallmatrix} 0 & 1 \\ 1 & 0 
\end{smallmatrix}\big)$, 
and $\CF_r$ is the operator of Fourier-transformation
which maps
\begin{equation}
\CF^{-1}_r\cdot\su^{}_r\cdot\CF^{}_r\,=\,\sv^{-1}_r\,,\qquad
\CF^{-1}_r\cdot\sv^{}_r\cdot\CF^{}_r\,=\,\su^{}_r\,.
\end{equation}
Observing that $\si_1^{}L_{r}^{+}(\mu)\si_1^{}=\CF_{r}^2\cdot
L_{r}^{+}(\mu)\cdot \CF_r^{-2}$ one may easily check that
\begin{equation}\label{r++fromr+-}
\tilde{\sr}^{++}_{rs}(\la/\mu):=\CF_s^{}\cdot\sr^{+-}_{rs}(\la/\mu)\,
\CF_r^{-2}\cdot\CF_s^{-1}\,,
\end{equation}
will satisfy the defining relations \rf{srll}. 

It is furthermore not hard to show that the most general operator 
satisfying \rf{srll} can be written in the form
$\tilde{\sr}^{++}_{rs}(\la/\mu)H(\sz_{rs}^+)$, where
$\sz_{rs}^+:=\su^{}_r\sv_r^{-1}\,\sv_s^{}\su^{}_s$. 
The choice of the function $H(z)$ will turn out to be irrelevant for our 
applications to the lattice Sinh-Gordon model, and may therefore be fixed by the
convenient normalisation condition
${\sr}^{++}_{rs}(1)=P_{rs}$.

In order to get useful explicit formulae for $\sr^{+-}(\la/\mu)$ and
$\sr^{++}(\la/\mu)$ we may start from 
\rf{R-factor-ren}. 
An alternative representation will be particularly
useful:
\begin{equation}\label{r+-factor1}
\sr^{+-}_{rs}(\la)\,=\,\SP_{rs}^{}\cdot
{\rho_{\la}^{}(\sf_{rs}^{+},\sg_{rs}^-)}
\cdot
\CF_r\CF_s\,,
\end{equation}
using the notations 
$\sg_{rs}^-:=\su_r^{}\sv_r^{}\,\sv_s^{-1}\su_s^{-1}$,
and 
\begin{equation}
\rho_{\la}^{}(w,z):=\,\frac{\rho_{\la}^{}(w)}{\rho_0^{}(z)}\,,\qquad
\begin{aligned}
&\rho_\la(w):=
e^{-\frac{i}{4\pi  b^2}(\log w)^2}\,
\CE_{b^2}^{}(\la w)\CE_{b^2}^{}(\la w^{-1})\\
&\rho_0(z):=
e^{-\frac{i}{4\pi  b^2}(\log z)^2}
\,.
\end{aligned}
\end{equation}
In order to derive \rf{r+-factor1} one may use the identity
\begin{equation}\label{identity}
e^{\frac{2i}{\pi  b^2}\log\su_r\log\su_s}\,=\,
e^{\frac{i}{4\pi  b^2}((\log \sg^-_{rs})^2-(\log \sf^+_{rs})^2)}
\cdot\SP_{rs}\cdot\CF_r\CF_s\,,
\end{equation}
which can be verified by computing the matrix elements.

By combining \rf{r++fromr+-} and \rf{R-factor1} one finds immediately that
$\tilde\sr^{++}_{rs}(\la)=
\SP_{rs}^{}\,\rho_{\la}(\sf_{\check{r}s}^{+},\sz_{{r}s}^+),
$
where $\sf_{\check{r}s}^+:=\CF_r\cdot\sf_{rs}^+\cdot\CF_r^{-1}=\su_r^{}\sv_r^{}\,\su_s^{}\sv_s^{-1}$,
noting that 
$\sz_{{r}s}^+:=\CF_r\cdot\sg_{rs}^{-}\cdot\CF_r^{-1}$. We may now conclude that 
\begin{equation}\label{R-factor2}
\sr^{++}_{rs}(\la):=\,
\SP_{rs}^{}\,\rho_{\la}^{}(\sf_{\check{r}s}^{+})\,,
\end{equation}
is the unique solution of \rf{srll} which 
satisfies the normalisation condition $\sr^{++}_{rs}(1)=\SP_{rs}$.
For the case of $\CU_q(\widehat{\fsl}_2)$ we have thereby completed the calculation of the 
main ingredients needed to construct fundamental R-operators and the corresponding 
transfer matrices. The development of the theory in this case is
continued in Appendix \ref{ShGapp} where it is shown how to reproduce the Q-operators 
for the lattice Sinh-Gordon model previously constructed in 
\cite{ByT} by other methods 
from our formulae for $\sr^{+-}$ and $\sr^{++}$ found above. In the main text we shall
continue with the generalisation of these results for the case of $\CU_q(\widehat{\fsl}_M)$.

\section{R-operators from the universal R-matrix - case of $\CU_q(\widehat{\fsl}_M)$}\label{Fundsect6}


We now generalise the discussion of the renormalisation of the real root contributions to the 
cases of $\CU_q(\widehat{\fsl}_M)$. 
To begin with, we explain how to obtain the evolution operator from the universal R-matrix.
One of the new issues that arises for $M>2$ is due to the fact that
we will need to consider a slightly larger family of representations.
Instead of the representations $\pi^\pm$ we will need to consider 
pairs of mutually conjugate representations $(\pi^+,\pi^{\dot{+}})$ and $(\pi^-,\pi^{\dot{-}})$. 
In Subsection \ref{FunR-repth} we will explain how the 
factorised representations for the  fundamental R-operators like \rf{fundRfact} follow from the representation 
theory of $\CU_q(\widehat{\fsl}_M)$.

In the rest of this section we derive useful explicit representations
for the resulting R-operators, including a representation as an 
integral operator with
an explicit kernel. The kernel becomes simplest  when we consider a variant of the lattice model
obtained from the twisted universal R-matrices $\tilde{\mathscr{R}}^\pm$ introduced in 
Section \ref{Twists}. The fundamental transfer matrix $\mathcal{T}$ is shown to be 
a physical observable in the sense defined in Section \ref{Phys-Obs}, 
and the projection to the physical degrees of freedom is described precisely
as an integral operator with explicit kernel.

\subsection{Representations in quantum space}

The connection between the integrable model defined
in Section \ref{scheme} and the representation theory of $\mathcal{U}_q(\widehat{\mathfrak{gl}}_M)$
 is encoded in the following relations
\begin{equation}\label{LpmanduniR}
\begin{aligned}
\mathsf{L}^{+}(\lambda\mu^{-1})\,&=\,
\frac{1}{\theta^{+}(\lambda\mu^{-1})}\,
\left[\left(\pi^{\rm{f}}_{\la}\otimes \pi_{\mu}^+\right)\left(\mathscr{R}^{+}\right)\right]_{\text{ren}} \\
{\mathsf{L}}^{\dot{-}}(\lambda\mu^{-1})\,&=\,
\frac{1}{\theta^{\dot{-}}(\lambda\mu^{-1})}\,
\big[\big(\pi^{\rm{f}}_{\la}\otimes {\pi}_{\mu}^{\dot{-}}\big)\left(\mathscr{R}^{-}\right)\big]_{\text{ren}} 
\end{aligned}
\end{equation}
where $\mathsf{L}^{+}(\lambda)$, ${\mathsf{L}}^{\dot{-}}(\lambda)\equiv \bar{\mathsf{L}}^{-}(\lambda)$  
were defined in Section \ref{sec:alternating}, $\mathscr{R}^{\pm}$ are two  universal R-matrices given in
Section \ref{sec:prodformula} and $\theta^{+}(x)$,  $\theta^{\dot{-}}(x)$ are certain scalar factors.
The relevant representations entering \eqref{LpmanduniR} are defined as follows
\begin{equation}
 \pi^{\rm{f}}_{\lambda}(e_i)\,=\,\lambda^{-1}\,\SE_{i,i+1}\,,\qquad
 \pi^{\rm{f}}_{\lambda}(f_i)\,=\,\lambda\,\SE_{i+1,i}\,,\qquad
 \pi^{\rm{f}}_{\lambda}(h_i)\,=\,\SE_{i,i}-\SE_{i+1,i+1}\,,
\label{fundrepTEXT}
\end{equation}
where $\SE_{ij}$ are the matrix units 
$\SE_{ij}\,\SE_{kl}\,=\,\delta_{jk}\,\SE_{il}$ and
\begin{equation}
 \pi_{\lambda}^+(f_i)\,=\,\frac{\lambda}{q-q^{-1}}\,\su_i^{-1}\,\sv_i\,,\qquad
 \pi_{\lambda}^+(k_i)\,=\,\su_i\,\su^{-1}_{i+1}\,,
\label{repplusTEXT}
\end{equation}
\begin{equation}
  {\pi}_{\lambda}^{\dot{-}}(e_i)\,=\,\frac{\lambda^{-1}}{q^{-1}-q}\,\sv_i\,\su_{i+1}\,,\qquad
  {\pi}_{\lambda}^{\dot{-}}(k_i)\,=\,\su^{-1}_i\,\su_{i+1}\,.
\label{repminusbarTEXT}
\end{equation}
$\{\sv_i,\su_{i}\}_{i=1,\dots,M}$ generate the algebra $\mathcal{W}$, see \eqref{algebraWdef}. 
It is easy to verify that  \eqref{repplusTEXT}, \eqref{repminusbarTEXT} satisfy, respectively,  the defining relations  of
$\Borelm$, $\Borelp$, see \eqref{eqnQEARels}.
In particular the Serre relations \eqref{Serreeasy1}, \eqref{Serreeasy2} follow from the exchange relations
\begin{equation}\label{exchangerelTEXT}
\pi_{\mu}^+\left(f_i\,f_j \right)\,=\,q^{+(\delta_{i+1,j}-\delta_{j+1,i})}\,\pi_\mu^+\left(f_j\,f_i \right)\,,\qquad
{\pi}_{\mu}^{\dot{-}}\left(e_i\,e_j \right)\,=\,q^{-(\delta_{i+1,j}-\delta_{j+1,i})}\,{\pi}_{\mu}^{\dot{-}}\left(e_j\,e_i \right)\,.
\end{equation}

We postpone to Section \ref{sec:examplerenIMroot}
 the derivation of  \eqref{LpmanduniR}
 by applying the relevant representations to the infinite product formula for 
the universal R-matrix given in Section \ref{sec:prodformula}.
A simple way to verify the identification \eqref{LpmanduniR} is to notice that 
\begin{equation}
\left(\pi^{\rm{f}}_{\lambda}\otimes \pi_{\mu}^+\right)\left(q^{+t}\right)\,=\,
\sum_{i=1}^M\,\SE_{ii}\,\su_i\,=\,
\left(\pi^{\rm{f}}_{\lambda}\otimes \bar{\pi}_{\mu}^-\right)\left(q^{-t}\right)\,
\end{equation}
and  check that the image of the reduced R-matrix satisfies the relations \eqref{Rp1}, \eqref{Rm1}. 
As opposed to the direct evaluation of the product formula for the universal R-matrix, 
this procedure does not allow to determine the scalar factors $\theta^{+}(x)$, $\theta^{\dot{-}}(x)$.

The relations \eqref{RLLLLR} follow from the universal Yang-Baxter equation \rf{YBE}
 upon noticing that the matrix $\mathsf{R}(x,y)$ is obtained from the universal 
R-matrix as explained in Appendix \ref{App:Rfund}.

\subsection{Light-cone evolution operator from the universal R-matrix}

After we have identified the relevant representations in quantum space, 
see equations \eqref{repplusTEXT}, \eqref{repminusbarTEXT},  
we will show how to obtain the operators $\sr^{\epsilon_1\epsilon_2}$
 from  the product formula for the universal R-matrix.   
In order to clarify certain features of such expression for the infinite
 dimensional representations we are considering, we will focus our attention on
\begin{equation}\label{RpmfromuniR}
 \sr^{+\dot{-}}(\lambda\,\mu^{-1})\,=\,\frac{1}{\rho^{+\dot{-}}(\lambda\mu^{-1})}
 \,\left[\left(\pi^{+}_{\lambda}\otimes \pi^{\dot{-}}_{\mu}\right)\left(\mathscr{R}^{-}\right)\right]_{\text{ren}}\,. 
\end{equation}
As in the previous section, the notation $[\dots]_{\rm ren}$ indicates the use of a certain 
prescription for defining the infinite sums in the definition 
of $\mathscr{R}^{\pm}$. 
The operator $ \sr^{\dot{-}+}$ can be obtained in a similar way, 
or just using the relation $ \sr^{\dot{-}+}\sigma\big(\sr^{+\dot{-}}\big)=1$, 
where $\sigma$ exchanges the tensor factors.
The case $\epsilon_1=\epsilon_2$ requires further considerations 
as both tensor factors correspond to the same Borel half. 
This case is considered in some details in Section \ref{sec:rppfromuniR}.

The expression \eqref{RpmfromuniR} provides us  with a formal solution to the
relations \eqref{srll} which characterize the building block for the light-cone evolution operator. 
The key relation \rf{srll} follows directly from the special case 
\[
\mathscr{R}_{12}^+ \,\mathscr{R}_{13}^-\, \mathscr{R}_{23}^-\,=\,
\mathscr{R}_{23}^- \,\mathscr{R}_{13}^-\, \mathscr{R}_{12}^+\,
\]
of the universal Yang-Baxter equations
\rf{YBE} by applying $\pi^{\rm f}\ot\pi_{\la^{-1}}^+\ot \pi_{\mu^{-1}}^{\dot{-}}$ to this relation.
In order to obtain the generators of the
discrete time-evolution for the lattice models from \eqref{RpmfromuniR},
it is crucial that $\mathsf{L}^{+}$ and $\bar{\mathsf{L}}^{-}$ are obtained via \eqref{LpmanduniR}
from $\mathscr{R}^-$ and $\mathscr{R}^+$, respectively. This fact
reflects the respective orientations in the integration
over light-like segments defining the classical counterparts of
$\mathsf{L}^{+}$ and $\bar{\mathsf{L}}^{-}$, as was observed in \cite{RT}.

As summarized in Section \ref{sec:prodformula}, the evaluation
 of the universal R-matrix consists of the following three steps: 
fix a convex order in $\Delta_+(\widehat{\mathfrak{sl}}_M)$, 
evaluate root vectors and finally substitute the root vectors
 in the product formula  \rf{KT}.
This procedure is straightforward upon following the instructions
 in Section \ref{sec:rootgen} and Appendix \ref{sec:rootSLM}.
 We proceed performing the first step.

\subsubsection{The image of root vectors under $\pi^+$ and $\bar{\pi}^-$}
\label{sec:Imagerootvectors}

A key observation is that for the representations $\pi^{+}$, $\bar{\pi}^{-}$ most of the root vectors, 
for a specific choice of convex order of positive roots, evaluate to zero.
In the case of  $\pi^+$, using the  root ordering specified in Appendix 
\ref{sec:convexorderSLM}, the only non-vanishing root vectors are given by
\begin{align}\label{eipiplus}
&\pi^{+}(f_{\epsilon_{i}-\epsilon_{i+1}})\,=:\,{\mathsf f}_i\,,
\\ \label{eisuccpiplus}
&\pi^{+}(f_{\delta-\left(\epsilon_{i}-\epsilon_M\right)})\,=
\,(q-q^{-1})^{i-1}\,{\mathsf f}_{i-1}\,\dots\,{\mathsf f}_1\,{\mathsf f}_{0}\,,
\\ 
\label{piplusimaginary}
&\pi^{+}(f^{(M-1)}_{k\delta})\,=\,
\frac{(-1)^{k+1}}{k}\,\frac{1}{q-q^{-1}}\,\mathsf{c}_+^k\,.
\end{align}
where $\mathsf{c}_+:=q^{-1}(q-q^{-1})^M\,{\mathsf f}_{M}\,\dots\,{\mathsf f}_1=\lambda^M$ is central.
For $\bar{\pi}^-$, using the shorthand notation $\bar{\mathsf{e}}_i=\bar{\pi}^-(e_i)$, 
one obtains that the non-vanishing root vectors are given by
\begin{align}\label{eipibarminus}
&\bar{\pi}^-(e_{\epsilon_i-\epsilon_j})\,=\,(q^{-1}-q)^{j-i-1}\,\bar{\mathsf{e}}_{j-1}\dots\bar{\mathsf{e}}_i\,,
\qquad i<j\,,
\\ \label{eisuccbarpiminus}
& \bar{\pi}^-(e_{\delta-(\epsilon_1-\epsilon_j)})\,=\,(q^{-1}-q)^{M-j}\,\bar{\mathsf{e}}_{M}\dots\bar{\mathsf{e}}_j\,,
\\ \label{eiimbarminus} 
&
 \bar{\pi}^-(e^{(1)}_{k\delta})\,=\,\frac{(-1)^{k+1}}{k}\frac{1}{q^{-1}-q}\,\bar{\mathsf{c}}_-^k\,.
\end{align}
where $\bar{\mathsf{c}}_-:=q(q^{-1}-q)^M\,\bar{{\mathsf e}}_{M}\,\dots\,\bar{{\mathsf e}}_1=\lambda^{-M}$ 
is central.
Notice that we suppressed the dependence on the spectral parameter
 from the notation $\pi^{+}$, $\bar{\pi}^-$.  

The equations above  immediately follow from the exchange relations \eqref{exchangerelTEXT} 
and the definition of root vectors given in Section \ref{sec:rootgen} and Appendix \ref{sec:rootSLM}.

\subsubsection{The image of the universal R-matrix  under $\pi^+\otimes\bar{\pi}^-$}
\label{Sec:uniimage1}

The representations $\pi^+$, $\bar{\pi}^-$ posses the remarkable 
feature that the imaginary root vectors are central.
As a consequence, the contribution from the positive imaginary root
 to the universal R-matrix is a scalar factor given by 
\begin{equation}
\rho^{+\dot{-}}(\lambda\mu^{-1})\,:=\,
\left[\left(\pi^{+}_{\lambda}\otimes \pi^{\dot{-}}_{\mu}\right)
\left(\mathscr{R}^{-}_{\sim \delta}\right)\right]_{\text{ren}} \,. 
\end{equation}
We postpone the discussion about the renormalisation of this expression
 for $q$ on the unit circle, which is the case of interest to this paper,  to Section \ref{sec:ImrootsSLM}.

Concerning the contribution of real positive roots
 $\gamma\,\in\,\Delta_+^{\text{re}}(\widehat{\mathfrak{gl}}_M)$, 
compare to \eqref{rgm}, the results of Section  \ref{sec:Imagerootvectors}
 immediately imply
\begin{equation}
 \left(\pi^{+}_{\lambda}\otimes \bar{\pi}^{-}_{\mu}\right)(f_{\gamma}\otimes e_{\gamma})\,=\,
\begin{cases}
 \mathsf{f}_i\otimes \bar{\mathsf{e}}_i\, & \text{if}\,\,\gamma=\alpha_{i\,\in\,\{1,\dots,M\}}\\
 0 & \text{otherwise}
\end{cases}
\end{equation}
Moreover, these operators commute among themselves 
\begin{equation}
\check{\chi}_i\,\check{\chi}_j=\check{\chi}_j\,\check{\chi}_i\,,
\qquad \qquad 
\check{\chi}_i:=-(q-q^{-1})^2\,\mathsf{f}_i\otimes \bar{\mathsf{e}_i}\,.
\end{equation}
It follows that
 \begin{equation} \label{pibarmpip}
\left[\left(\pi^{+}_{\lambda}\otimes \bar{\pi}^{-}_{\mu}\right)\,\left(\mathscr{R}^-\right)\right]_{\text{ren}} \,=\,
 \rho^{+\dot{-}}(\lambda\mu^{-1})\,(F(\check\chi))^{-1}\,
 q^{-T},
 \end{equation}
 where
 \begin{equation}(F(\check\chi))^{-1}\,=\,
 \prod_{i=1}^M\,\mathcal{E}_{b^2}(\check{\chi}_i)\,,\qquad
 q^{T}:=
e^{\log\su {\otimes} \log\su /{i \pi b^2}}\,,
 \end{equation}
where as in Section \ref{canren} we took $\mathcal{E}_{b^2}(\omega)=\left[\varepsilon_q(\omega)\right]_{\text{ren}}$,
 with $\mathcal{E}_{b^2}(\omega)$
  given in \eqref{Eb2-def}  and  $\varepsilon_q(\omega)=\exp_{q^{2}}((q-q^{-1})^{-1}\omega)$.
Let us compare this expression with \eqref{sr-expl}. 
Using the definition \eqref{RpmfromuniR} and the result \eqref{pibarmpip} one finds
\begin{equation}\label{r-+fromr+-viaUNI}
 \mathsf{r}^{-+}(\mu\lambda^{-1})\,=\,
\left[\sigma\left(r^{+-}(\lambda\mu^{-1})\right)\right]^{-1}\,=\,F(\check{\chi}')
\, q^{T}\,,
\end{equation}
where
\begin{equation}
 \check{\chi}'_i\,:=\,\sigma\left(q^{T}\,\check{\chi}_i\,q^{-T}\right)\,=\,
\lambda\mu^{-1}\,\left(\su_i\sv_i\otimes \sv_i\su_{i+1}^{-1}\right)\,.
\end{equation}
The expression \eqref{r-+fromr+-viaUNI} coincides with \eqref{sr-expl} 
upon taking $\mathcal{J}_{\kappa}(x)\,=\,\left[\mathcal{E}_{b^2}(\kappa^2x)\right]^{-1}$.
The fact that the renormalized expression \eqref{pibarmpip} satisfies 
the intertwining relation \eqref{srll} follows from the fact that 
 \eqref{sr-expl} does.
Notice that $ \check{\chi}'_i$ is a positive self-adjoint operator 
 for $\lambda\mu^{-1}$ positive.
%

\subsection{Fundamental R-operator from the universal R-matrix}\label{FunR-repth}

After having constructed the evolution operator, the next step is to construct the fundamental R-operators.
Our goal in this section will be to elaborate on the representation-theoretic meaning of the 
factorised form \rf{fundRfact} for the fundamental R-operators observed in Section \ref{FunR/Q}.
It will be useful to first consider $\CR^{\rm\sst XXZ}$ which turns out to have the most direct
relation to the universal R-matrix.
The fundamental R-operator $\CR^{\rm \sst AT}$ 
can then be obtained simply via \rf{RfunfromRXXZ}.

\subsubsection{More Lax-matrices}

First, let us note that $\CL^{\rm\sst XXZ}(\la)$ admits a factorisation similar to \rf{L+-def}. 
We shall represent the matrix $ \mathsf{L}^{+}(\lambda)\,\ST$ appearing as a factor
in $\CL^{\rm\sst XXZ}(\la)$ in the form
\begin{equation}\label{L+fromL-}
 \mathsf{L}^{+}(\lambda)\,\ST\,=\,
\lambda^{-1}\,\mathcal{F}\,\mathsf{L}^-(\lambda)\, \mathcal{F}^{-1}\,,
\end{equation}
where ${\mathsf T}=\sum_i \SE_{i+1,i}$,  the automorphism $\CF$ is defined via
$\mathcal{F}^{-1}\,(\su_i,\,\sv_{i})\,\mathcal{F}\,=\,(\sv_{i-1},\,\su_i)$,
and 
\begin{equation}
\label{Lminsaux}
 \mathsf{L}^-(\lambda)\,:=\,
\sum_{i=1}^M\,\left(\su_i\,\SE_{ii}\,+\,\lambda\,\sv_i\,\SE_{i+1i} \right)\,.
\end{equation}
We note that the matrix ${\mathsf T}$ is the generator 
of the automorphism $\mathbb{Z}_M$ in the fundamental representation.
$\Omega_{\mathcal{W}}:=\mathcal{F}^2$ represents  
the same generator of $\mathbb{Z}_M$ on $\mathcal{W}$.

Let us consider, a bit more
generally
\begin{equation}\label{LXXZfactor}
{\mathcal{L}}_A^{\ep_1,\ep_2}(\bar\mu,\mu)\,=\, 
{\mathsf{L}}^{\ep_1}_{\bar{a}}(\bar{\mu})\,
 \mathsf{L}^{\ep_2}_{a}(\mu)
 \,\,\,\in\,\,\,
\text{End}(\mathbb{C}^M)\ot
\mathcal{W}\otimes\mathcal{W}\,,\quad\ep_1,\ep_2\in\{+,\dot{+},-,\dot{-}\}\,.
\end{equation}
where   $\SL^+(\la)$ and ${\mathsf{L}}^{\dot{-}}(\la)\equiv\bar{\SL}^-(\la)$
were defined in  \rf{moreexpl2} and \rf{moreexpl1}, respectively, while and $\mathsf{L}^-(\la)$ and
${\mathsf{L}}^{\dot{+}}(\lambda)$ are introduced as
\begin{align}\label{barL+}
&{\mathsf{L}}^{\dot{+}}(\lambda)\,=\,\left(1-q\,\la^{-M}\right)\,
\bigg{[}\sum_{i=1}^M\,\left(\su^{-1}_i\,\SE_{ii}\,-\,q^{}\,\lambda^{-1}\,\sv_i\,\SE_{i,i+1} \right)\bigg{]}^{-1}\,. 
\end{align}
The Lax operators $\mathsf{L}^-(\lambda)$ and 
${\mathsf{L}}^{\dot{+}}(\lambda)$ can be obtained
 from the universal R-matrix as
\begin{align}
&{\mathsf{L}}^{\dot{+}}(\lambda\mu^{-1})\,=\,\frac{1}{\theta^{\dot{+}}(\lambda\mu^{-1})}
\,\big[\big(\pi^{\rm f}_{\la}\otimes {\pi}_{\mu}^{\dot{+}}\big)(\mathscr{R}^{+})\big]_{\text{ren}}\,,\\
&\label{LminusUNIR}
{\mathsf{L}}^{-}(\lambda\mu^{-1})\,=\,\frac{1}{\theta^{-}(\lambda\mu^{-1})}
\,\big[\big(\pi^{\rm f}_{\la}\otimes \pi_{\mu}^-\big)(\mathscr{R}^{-})\big]_{\text{ren}}\,,
\end{align}
where
\begin{align}
& 
\pi_{\lambda}^{\dot{+}}(f_i)\,=\,\frac{\lambda}{q-q^{-1}}\,\su_{i+1}^{}\sv_i^{}\,,\qquad
\pi_{\lambda}^{\dot{+}}(k_i)\,=\,\su^{}_i\,\su_{i+1}^{-1}\,,\\
&
\label{repminus}
 \pi_{\lambda}^-(e_i)\,=\,\frac{\lambda^{-1}}{q^{-1}-q}\,\sv_i\,\su_i^{-1}\,,\qquad
 \pi_{\lambda}^-(k_i)\,=\,\su^{-1}_i\,\su_{i+1}\,.
\end{align}
In order to find a relation between $\CL_A^{\dot{+}+}$ and $\CL_A^{\dot{-}-}$ let us note that
\begin{align}
&\ST^{-1}\SL^{-}(\la)\,=\,\la\,\CF\cdot\SL^+(\la)\cdot\CF^{-1}\,,\\
&\SL^{\dot{-}}(\la)\ST\,=\,-q\la^{-1}\,\dot\CF\cdot\SL^{\dot{+}}(\la)\cdot\dot\CF^{-1}\,,
\end{align}
where $\CF$ and $\dot\CF$ are the automorphisms of $\CW$ defined as
\begin{equation}
\begin{aligned}
&\CF\cdot\su_i\cdot\CF^{-1}=\sv_i\,,\\
&\CF\cdot\sv_i\cdot\CF^{-1}=\su_{i+1}\,,
\end{aligned}
\qquad
\begin{aligned}
&\dot\CF\cdot\su_i^{}\cdot\dot\CF^{-1}=\sv_i^{-1}\,,\\
&\dot\CF\cdot\sv_i^{}\cdot\dot\CF^{-1}=\su_{i+1}^{-1}\,.
\end{aligned}
\end{equation}
It follows that ${\mathcal{L}}_A^{\dot{+}+}(\bar\mu,\mu)$ and 
${\mathcal{L}}_A^{\dot{-}-}(\bar\mu,\mu)$ are related by a similarity transformation,
\begin{equation}
{\mathcal{L}}_A^{\dot{-}-}(\bar\mu,\mu)
\,=\,-q\,\mu\bar\mu^{-1}\,\dot\CF_{\bar a}\CF_a\cdot
{\mathcal{L}}_A^{\dot{+}+}(\bar\mu,\mu)
\cdot(\dot\CF_{\bar a}\CF_a)^{-1}\,.
\end{equation}
This implies that $\mathcal{R}^{\rm\sst XXZ}_{AB}$ can be obtained from  an operator 
$\mathcal{R}'{}_{AB}^{\rm\sst XXZ}$
satisfying
\begin{equation}\label{SRLL-mod}
\CL_A^{\dot{+}+}(\bar\mu,\mu)\CL_{B}^{\dot{-}-}(\bar\nu,\nu)\,
\mathcal{R}'{}_{AB}^{\rm\sst XXZ}(\bar\mu,\mu;\bar\nu,\nu)=
\mathcal{R}'{}_{AB}^{\rm\sst XXZ}(\bar\mu,\mu;\bar\nu,\nu)\,
\CL_B^{\dot{-}-}(\bar\nu,\nu)\CL_A^{\dot{+}+}(\bar\mu,\mu)\,.
\end{equation}
simply by applying the similarity transformation $\CF_A:=\dot\CF_{\bar a}\CF_a$.

\subsubsection{Factorisation}

As observed previously we may get operators $\mathcal{R}'{}^{\rm\sst XXZ}$ satisfying \rf{SRLL-mod} 
from the universal 
R-matrix in the following form
\begin{equation}
\mathcal{R}'{}^{\rm\sst XXZ}(\bar\mu,\mu;\bar\nu,\nu)\,:=\,
\big[\rho_{\rm\sst XXZ}(\bar\mu,\mu;\bar\nu,\nu)\big]^{-1}\,
\big[\,(\pi^+_{\mu^{-1}}\otimes {\pi}^{\dot{+}}_{\bar\mu^{-1}})\otimes (\pi^-_{\nu^{-1}}\otimes {\pi}^{\dot{-}}_{\bar\nu^{-1}})
\left(\mathscr{R}^{-}\right)\,\big]_{\text{ren}}\,.
\end{equation}
The product of Lax-matrices appearing in the factorisation \rf{LXXZfactor} represents the 
tensor product of representations $\pi^-_\mu\ot\pi^{\dot{-}}_{\bar\mu}$ 
of the Borel-subalgebra $\Borelp$.
It then follows from $\func{\brac{\Delta \otimes \id}}{\CR} = \CR_{13} \CR_{23}$
and
$\func{\brac{\id \otimes \Delta}}{\CR} = \CR_{13} \CR_{12}$
that the operator $\mathcal{R}'{}^{\rm\sst XXZ}$ can be factorised as 
\begin{equation}\label{RXXZfactorrep}
\mathcal{R}'{}_{AB}^{\rm\sst XXZ}(\bar\mu,\mu;\bar\nu,\nu)=
\sr_{a,\bar{b}}^{+\dot{-}}(\mu,\bar\nu)\,
\sr_{a,b}^{+-}(\mu,\nu)\,\sr_{\bar{a},\bar{b}}^{\dot{+}\dot{-}}(\bar\mu,\bar\nu)\,
\sr_{\bar{a},b}^{\dot{+}-}(\bmu,\nu)\,,
\end{equation}
with factors $\sr^{\ep_1,\ep_2}(\mu,\nu)$ all obtained from  the universal R-matrix by evaluation
in suitable representations as
\begin{equation}\label{srSEC6_all}
\sr^{\ep_1,\ep_2}(\mu,\nu):=
\frac{1}{\rho^{\ep_1\ep_2}(\mu^{-1}\nu)}\,
\left[\left(\pi^{\ep_1}_{\nu}\otimes \pi^{\ep_2}_{\mu}\right)
\left(\mathscr{R}^{-}\right)\right]_{\text{ren}},\;\;
\ep_1\in\{+,\dot{+}\}\,,\quad\ep_2\in\{-,\dot{-}\}\,.
\end{equation}
The factorised representation \rf{RXXZfactorrep} for 
$\mathcal{R}'{}_{AB}^{\rm\sst XXZ}$ implies similar representations for
$\mathcal{R}_{AB}^{\rm\sst XXZ}$ and ${\mathcal{R}}_{AB}^{}$,
as anticipated in \rf{fundRfact}.

\begin{rem}
The representations $\bar{\pi}^-$ and $\pi^-$ can the considered the  conjugate of each other in the following sense:
\begin{equation}\label{antipodestar}
\bar{\pi}^{\pm}_{-q^{-1}\la}(a)\,=\,\left(\pi^{\pm}_{\la}(\mathscr{S}(a)) \right)^*\,,
\end{equation}
where $\mathscr{S}$ is the antipode
\begin{equation}
\mathscr{S}(e_i)\,=\,-e_i\,k_i^{-1}\,,\qquad
\mathscr{S}(f_i)\,=\,-k_i^{+1}\,f_i\,,\qquad
\mathscr{S}(k_i)\,=\,k_i^{-1}\,.
\end{equation}
and the involution $*$ is the anti-automorphism of the algebra $\mathcal{W}$
defined by $(\su_i,\sv_i)^*\,=\,(\su_i^{-1},\,\sv_i)$.
Notice that in \eqref{antipodestar} we introduced the representation $\bar{\pi}^{+}$ relevant for the following sections.
In the case of  $\mathcal{U}_q(\widehat{\mathfrak{sl}}_2)$ one has $\bar{\pi}^-_{\lambda}=\pi^-_{\lambda}$,
$\bar{\pi}^+_{\lambda}=\pi^+_{\lambda}$. 
One may further  notice that
$\pi_{\lambda}^+(f_i)\,=\,\pi_{-q^{-1}\lambda^{-1}}^-(e_i)$,
$\pi^+(k_i)\,=\,\pi^{-}(k_i^{-1})$.
\end{rem}

\subsubsection{Decoupling}
\label{sec:decoupling}

The representation $\pi^-_\mu\ot\pi^{\dot{-}}_{\bar\mu}$ is reducible, as can be formally expressed as
\begin{equation}\label{factortriv}
  \big(\pi^-_{\mu^{-1}}\otimes {\pi}^{\dot{-}}_{\bar\mu^{-1}}\big)\Delta(a)\,=\,
\left(\pi^{\text{triv}}_{}\ot\pi^{\text{min}}_{\la,\ka}\right)\Delta(a)\,,
\qquad a\in\Borelp\,,
\end{equation}
where $\pi^{\text{min}}_{s,t}$ and $\pi^{\text{triv}}_{}$ are defined via
\begin{align}
&\pi^{\text{min}}_{\la,\ka}(e_i)\,=\,\frac{\la^{-1}}{q^{-1}-q}
\,{\mathsf{V}}_{i}^{\frac{1}{2}}\,
\big(\ka^{}\,{\mathsf{U}}^{}_{i}+\ka^{-1}{\mathsf{U}}_i^{-1}\big)\,
{\mathsf{V}}_{i}^{\frac{1}{2}}\,,
\qquad
\pi^{\text{min}}_{\la,\ka}(q^{\bar{\epsilon}_i})\,=\,
{\mathsf{U}}_i^{-1}\,,\\
& \pi^{\text{triv}}_{}(e_i)\,=\,0\,,\qquad
\pi^{\text{triv}}_{}(q^{\bar{\epsilon}_i})\,=\,{\mathsf{C}}_{i}^{-1}\,,
\end{align}
provided that we define
\begin{align}
& {\mathsf{U}}^2_i\,=\,\su_i^{}\sv^{-1}_i\ot\su_i^{}\,\sv_i^{}\,,
\qquad
{\mathsf{V}}_{i}^{}\,=\,\sv_i\ot\su_{i+1}\cdot\SC_i^{-1}\,,
\qquad
{\mathsf{C}}^2_i\,=\,\su_i^{}\sv^{}_i\ot\su_i^{}\,\sv_i^{-1}\,,
\\
%
 \label{sim2}
& \la:=\,q^{\frac{1}{2}}\left({\mu\bar\mu}\right)^{-\frac{1}{2}}
\qquad
\ka:=\,(\bar\mu\mu^{-1})^{\frac{1}{2}}\,.
\end{align}
Note that the
operators $\SU_i$, $\SV_i$ satisfy the commutation relations 
of the algebra $\mathcal{W}$, wile 
${\mathsf{C}}^2_i$ are central in the algebra generated by 
${\mathsf{U}}^2_i$, ${\mathsf{V}}^2_i$ and ${\mathsf{C}}^2_i$.
The relation \rf{factortriv} can be easily verified on  the generators $e_i$, $q^{\bar{\epsilon}_i}$
 using the definition of the coproduct and the representations $\pi^{\dot{-}}$, $\pi^-$
 given in \eqref{repminusbarTEXT}, \eqref{repminus}.

The factorisation \rf{factortriv} can alternatively  be shown  in the language of Lax matrices as follows.
 It is straightforward  to see that 
${\mathcal{L}}_A^{\dot{-}-}(\bar\mu,\mu)$ can be factorised as
 \begin{equation}\label{factortriv-matrix}
{\mathcal{L}}_A^{\dot{-}-}(\bar\mu,\mu)\,=\,\SL^{\text{min}}(\bar\mu,\mu)\,\Lambda(\SC)\,,
  \end{equation}
  where 
  \begin{equation}
  \SL^{\text{min}}(\bar\mu,\mu)=
  \left(1-q^{-1}\bar{\mu}^M\right)
 \bigg[\sum_i(\SE_{ii}-\bar\mu\,\SU_i\SV_i\,\SE_{i+1,i})\bigg]^{-1}\sum_i(\SE_{ii}\SU_i-\mu\,\SV_i\,\SE_{i+1,i})
\,,
 \end{equation}
 using the definitions above. The fact that there exists ${\rho}^{\text{min}}_\ka(\lambda)$ such that
 \begin{equation}
  \SL^{\text{min}}(\bar\mu,\mu)\,=\,
 \frac{1}{{\rho}^{\text{min}}_\ka(\lambda)}
\,\big[\big(\pi^{\rm f}_{1}\otimes {\pi}_{\la,\ka}^{\text{min}}\big)(\mathscr{R}^{-})\big]_{\text{ren}}
\end{equation}
can either be verified directly, or follows more elegantly from the observations made below in 
Subsection \ref{evalextension}. 
Keeping in mind that  matrix multiplication represents the action of the co-product one may deduce 
\rf{factortriv} from \rf{factortriv-matrix}.

The factorisation \rf{factortriv} will be the representation theoretic root of the decoupling of
"unphysical" degrees of freedom observed in Section \ref{sec:physdofandFACT}. It will imply that the 
operator $\mathcal{R}'{}_{AB}^{\rm\sst XXZ}$ constructed in the factorised form \rf{RXXZfactorrep} 
admits a similar factorisation into two factors acting nontrivially only in $\pi^{\text{min}}_{\la,\ka}$
and $\pi^{\text{triv}}_{}$, respectively, as will be verified by explicit calculation below.

\subsubsection{Relation to evaluation representation}\label{evalextension}

It is useful to notice that the representation $\pi^{\text{min}}$
can be extended to a representation of the full affine algebra $\CU_q(\widehat{\fgl}_M)$, as can be 
seen in the following way. Note that the spectral parameter dependence of the matrix entries of 
$\SL^{\text{min}}(\bar\mu,\mu)$ takes the following form 
\begin{equation}\begin{aligned}
\SL^{\text{min}}(\bar\mu,\mu)\,=\,
&
\sum_{i=1}^M\,\SE_{ii}\left(\SL_{ii}+\mu\bar\mu^{M-1}\bar{\SL}_{ii}\right)\\
&+
\bar{\mu}^{\frac{M}{2}}
\left(\frac{\mu}{\bar\mu}\right)^{\frac{1}{2}}\sum_{i> j}\big(\bar{\mu}^{-\frac{M}{2}}\bar\mu^{i-j}\,\SE_{ij}\,\SL_{ji}+
\bar{\mu}^{\frac{M}{2}}\bar\mu^{j-i}\,\SE_{ji}\,\SL_{ij}\big) \,,
\label{specpardep}
\end{aligned}
\end{equation}
where $\SL_{ii}$, $\bar{\SL}_{ii}$ and $\SL_{ij}$ are operators independent of the spectral parameter
and $\SL_{ii}\bar{\SL}_{ii}$ is central.
The fact that the representation $\pi^{\text{min}}_{s,t}$ 
extends from $\Borelp$ to a representation of $\CU_q(\widehat{\fgl}_M)$
follows from the known fact 
that Lax matrices satisfying \rf{RLLLLR} which have
the form \rf{specpardep} with central $q^{\ga}:=\SL_{ii}\bar{\SL}_{ii}$ one 
may get a representation of $\CU_q(\widehat{\fgl}_M)$ by setting
\begin{align}
&\pi^{\text{ev}}_\la(e_i)\,=\,\frac{\la^{-1}}{q^{-1}-q}\, \SL_{i,i+1}^{}\,\SL_{ii}^{-1}\,,
\qquad
\pi^{\text{ev}}_\la(f_i)\,=\,\frac{\la}{q-q^{-1}}\, \bar{\SL}_{ii}^{-1}\,\SL_{i+1,i}\,\,,
\label{evdef1}\\
& \pi^{\text{ev}}_\la\big(q^{\bar{\epsilon}_i}\big)\,=\,
\SL_{ii}^{-1}\,.
\label{evdef2}
\end{align}
The extension of the representation $\pi^{\text{min}}_{s,t}$ to all
of $\CU_q(\widehat{\fgl}_M)$ is thereby recognised as a particular representation $\pi^{\text{ev}}_\la$
of evaluation type.

\begin{rem}\label{highertensor}
By a similar analysis as the one presented in Section \ref{sec:decoupling} it is easy to argue that
\begin{equation}\label{evaluationgeneric}
\left(\pi^+_{\la_1}\,\otimes \dots\otimes \pi^+_{\la_M}\right)\Delta^{(M)}\,
\,=\,\big(\pi^{\text{triv}}\,\ot\,
\pi^{\{\la\}}_{\text{m.~d.~}}\circ\text{ev}\big)\Delta\,,
\end{equation}
where $\pi^{\text{triv}}(f_i)=0$ and $\pi^{\{\la\}}_{\text{m.~d.~}}$ denotes a representation of 
$\mathcal{U}_q(\mathfrak{sl}_M)$ on $L^{2}(\mathbb{R}^{\frac{M(M-1)}{2}})$.
If the parameters $\{\la_s\}$ are generic, this is an irreducible representation.
\end{rem}



\subsection{ $ \sr^{++}$ from the universal R-matrix}
\label{sec:rppfromuniR}

In order to construct  fundamental R-matrices and Q-operators,
 see Section \ref{FunR/Q}, we need to determine the other building 
blocks. We shall  start with 
\begin{equation}
 \sr^{+-}(\la\,\mu^{-1})\,=\,
\frac{1}{\rho^{+-}(\la\mu^{-1})}
\left[(\pi^{+}_{\la}\otimes \pi_{\mu}^-)(\mathscr{R}^-)\right]_{\text{ren}}\,,
\end{equation}
appearing in the expression \rf{RXXZfactorrep} for the fundamental R-operator
$\mathcal{R}'{}_{AB}^{\rm\sst XXZ}(\bar\mu,\mu;\bar\nu,\nu)$

In the following we obtain a regularized version of $(\pi^{+}\otimes \pi^-)\mathscr{R}^-$ 
from the product formula for the universal R-matrix and 
show explicitly that  $\sr^{+-}(\la)$ satisfies the intertwining property \eqref{srll}.


In the case of  $\pi^-$, using the root ordering specified in Appendix 
\ref{sec:convexorderSLM}, the only non-vanishing root vectors are given by
\begin{subequations} \label{sec:Imagerootvectors2}
\begin{align}\label{eipiminus}
&\pi_{\la}^{-}(e_{\epsilon_{i}-\epsilon_{i+1}})\,=:\,{\mathsf e}_i\,=\,
\frac{q^{-\frac{1}{M}}\lambda^{-1}}{q^{-1}-q}\,\mathsf{y}^{-1}_{i}\mathsf{y}_{i+1}^{}\,,
\\ \label{eisuccpiminus}
&\pi^{-}_{\la}(e_{\delta-\left(\epsilon_{i}-\epsilon_M\right)})\,=
\,(q^{-1}-q)^{i-1}\,{\mathsf e}_0\,{\mathsf e}_1\dots {\mathsf e}_{i-1}\,=\,
\frac{q^{-\frac{i}{M}}\,\lambda^{-i}}{q^{-1}-q}\,\mathsf{y}^{-1}_{M}\,\mathsf{y}_{i}^{}\,\,,
\\ 
 \label{eiIMpiminus}
&\pi^{-}_{\la}(e^{(M-1)}_{k\delta})\,=\,
\frac{(-1)^{k+1}}{k}\frac{1}{q^{-1}-q}\,\mathsf{c}_-^k\,.
\end{align}
\end{subequations}
where
$\mathsf{c}_-\,=\,
q\,(q^{-1}-q)^{M}\,{\mathsf e}_1\dots {\mathsf e}_{M}=\lambda^{-M}$ 
is central. 


As in Section \ref{Sec:uniimage1} the contribution form the positive  imaginary root 
to the universal R-matrix is a scalar factor given by 
\begin{equation}
\rho^{+-}(\lambda\mu^{-1})\,:=\,
\left[\left(\pi^{+}_{\lambda}\otimes \pi^{-}_{\mu}\right)\left(\mathscr{R}^{-}_{\sim \delta}\right)\right]_{\text{ren}}\,. 
\end{equation}
The renormalization prescription and its explicit form are discussed in Section  \ref{sec:ImrootsSLM}.

Concerning the contribution of real positive roots
 $\gamma\,\in\,\Delta_+^{\text{re}}(\widehat{\mathfrak{gl}}_M)$ 
the result of Section  \ref{sec:Imagerootvectors} together with \rf{sec:Imagerootvectors2}
 immediately implies
\begin{equation}
 \left(\pi^{+}_{\lambda}\otimes \pi^{-}_{\mu}\right)(f_{\gamma}\otimes e_{\gamma})\,=\,
 -\tau_q^{-2}
\begin{cases}
 \check{\mathsf{w}}_i\, & \text{if}\,\,\gamma=\alpha_{i}\,,\,\,\,\,\,i\in\{1,\dots,M-1\}\\
q^{1-i}\check{\mathsf{w}}_0\dots\,\check{\mathsf{w}}_{i-1} & \text{if}\,\,\gamma=
\delta-(\epsilon_i-\epsilon_M)\,,\,i\in\{1,\dots,M-1\}\\
 0 & \text{otherwise}
\end{cases}
\end{equation}
Where $\tau_q=q-q^{-1}$ and 
$\check{\mathsf{w}}_i:=-\tau_q^{2}\,\,\mathsf{f}_i\otimes \mathsf{e}_i$
satisfy the abelian current algebra on the lattice, see e.g.~\cite{FV93}
\begin{equation}
\label{wcheckcurrent}
 \check{\mathsf{w}}_i\,\check{\mathsf{w}}_j\,=\,
q^{2\left(\delta_{i+1,j}-\delta_{i,j+1} \right)}\,\check{\mathsf{w}}_j\,\check{\mathsf{w}}_i
\end{equation}
It follows that 
\begin{equation}\label{pimpip}
\begin{aligned}
 \left[\left({\pi^+_{\lambda}}\otimes {\pi^-_{\mu}} \right)(\mathscr{R}^-)\right]_{\text{ren}}\!\!=
 \rho^{+-}(\lambda\mu^{-1})&
\overbrace{\mathcal{E}_{b^2}(\check{\mathsf{w}}_1)\dots \mathcal{E}_{b^2}(\check{\mathsf{w}}_{M-1})}^{M-1\,\,\text{factors}}\times \\
\times&
\underbrace{\mathcal{E}_{b^2}(q^{2-M}\check{\mathsf{w}}_{0} \dots\check{\mathsf{w}}_{M\!-\!2})\dots \mathcal{E}_{b^2}(\check{\mathsf{w}}_{0})}_{M-1\,\,\text{factors}}
\,q^{-T}
\end{aligned} \end{equation}
where $q^T$ is given in \eqref{pibarmpip}
and the renormalization prescription for the quantum 
exponential is  the same as in \eqref{pibarmpip}.
 Notice that $\check{\mathsf{w}}_i=\lambda\mu^{-1}\,\su_i^{-1}\sv_i^{}\ot \sv^{}_i\su^{-1}_i$ are positive self
 adjoint operator for $\lambda\mu^{-1}$ real and positive.

\subsection{Intertwining properties and useful expressions for  $\sr^{++}$}

We now consider the operator $ \sr^{++}(\la)$ appearing in 
the factorised representation  \rf{fundRfact} for ${\mathcal{R}}_{AB}^{}$. The case of $ \sr^{--}(\la)$
is very similar.
We first introduce an operator
 $ \sr'{}^{++}(\la)$  related to the operator $ \sr^{+-}(\la)$ constructed in the previous subsection 
as
\begin{equation}\label{rppfromuni}
 \sr'{}^{++}(\la)\,=\,
\left(1\otimes \mathcal{F}\right)\cdot \sr^{+-}(\la)\cdot
(\Omega\otimes \mathcal{F})^{-1}\,.
\end{equation}
Our next goal will be to  verify that our renormalisation prescription for the universal R-matrix
guarantees that  the intertwining relations \eqref{srll} are satisfied.
 To this aim we shall identify conditions that ensure that an operator 
 $\mathsf{r}^{++}_{\,a\,b}(\lambda_a,\lambda_b)$ represents
a solution  of 
\begin{equation}\label{LLRpluplus}
\big[\,\mathsf{r}^{++}_{\,a\,b}(\lambda_a,\lambda_b)\,\big]^{-1}  
\mathsf{L}_a^{+}(\lambda_a)\,\,\mathsf{L}_b^{+}(\lambda_b)\,\,\mathsf{r}^{++}_{\,a\,b}(\lambda_a,\lambda_b)\,\,=\,\,\,\,\mathsf{L}_b^{+}(\lambda_b)\,\, \mathsf{L}_a^{+}(\lambda_a)\,,
\end{equation}
where $\mathsf{L}^+(\lambda)$ is defined in \eqref{moreexpl2}.
It will then be  easy to verify 
that the operator $\sr^{++}$ given by \eqref{rppfromuni}, \eqref{pimpip} 
satisfies these conditions.
It will be convenient to introduce
\begin{equation}\label{Rcheck}
\check{\mathsf{r}}^{++}_{\,a\,b}\,=\,\mathbb{P}_{a\,b}^{}\,\mathsf{r}^{++}_{\,a\,b}\,.
\end{equation}
We temporarily  suppress the dependence on the spectral parameters $\lambda_{a,b}$ in our notations.
The intertwining relation  \eqref{LLRpluplus} implies the following commutation relations
\begin{equation}\label{centr}
 \left[ \,\check{\mathsf{r}}^{++}_{\,a\,b},\,\sv_{i+1,a}\sv_{i,b}\,\right]\,=\,0\,,\qquad
 \left[\, \check{\mathsf{r}}^{++}_{\,a\,b},\,\su_{i,a}\su_{i,b}\,\right]\,=\,0\,,
\end{equation}
\begin{equation}\label{R++step} 
  \left(\lambda^{-1}_a\,\sv_{i,b}\su_{i+1,a}\,+\,
  \lambda^{-1}_b\,\su_{i,b}\sv_{i,a} \right)\,\check{\mathsf{r}}^{++}_{\,a\,b}
\,=\,
\check{\mathsf{r}}^{++}_{\,a\,b}\,
 \left(\lambda_b^{-1}\,\sv_{i,b}\su_{i+1,a}\,+\,\lambda_a^{-1}\,\su_{i,b}\sv_{i,a} \right)\,.
\end{equation}
In order to solve \eqref{centr} we define
\begin{align}\label{wdef}
&{\mathsf w}_i\,:=\,\sv_{i,b}^{-1}\,\sv_{i,a}\,\su_{i,b}\,\su^{-1}_{i+1,a}\,,\\
\label{etadef}
&{\mathsf \eta}_i\,:=\,\sv_{i,b}\,\sv_{i,a}^{-1}\,\su_{i,b}\,\su^{-1}_{i+1,a}\,.
\end{align}
One may  put extra $a\,b$ indices  on ${\mathsf w}, {\mathsf \eta}$,
 this will not be done here as no ambiguity arises. 
These variables generate the subalgebra of $\mathcal{W}\otimes \mathcal{W}$
that commutes with  $\sv_{i+1,a}\sv_{i,b}$ and $\su_{i,a}\su_{i,b}$, compare to \eqref{centr}.
They give rise to  two  mutually commutative copies of the
 $U(1)$ current algebra on the lattice, compare to \eqref{wcheckcurrent}, 
with opposite charge, 
\begin{align}\label{wwexch}
& {\mathsf w}_i\,{\mathsf w}_j\,=\,q^{-2\left(\delta_{i,j+1}-\delta_{i+1,j}\right)}\,{\mathsf w}_j\,{\mathsf w}_i\,,\\
\label{etaetaexch}
& {\mathsf \eta}_i\,{\mathsf \eta}_j\,=\,q^{+2\left(\delta_{i,j+1}-\delta_{i+1,j}\right)}\,
{\mathsf \eta}_j\,{\mathsf \eta}_i\,,\\
\label{wetaexch}
& {\mathsf \eta}_i\,{\mathsf w}_j\,=\,{\mathsf w}_j\, {\mathsf \eta}_i\,.
\end{align}
Any function of the operators $\sw_i$, ${\mathsf \eta}_i$ will satisfy \eqref{centr}. Turning our attention to the 
conditions 
\eqref{R++step}, let us note that these equations can be rewritten as
\begin{equation}\label{eqforR++}
\left( \sv_{i,b}\,\su_{i+1,a}\right)\,\check{\mathsf{r}}^{++}_{\,a\,b}\,\left( \sv_{i,b}\,\su_{i+1,a}\right)^{-1}\,=\,
\left( z\,+\,q\,{\mathsf w}_i\right)^{-1}\,\check{\mathsf{r}}^{++}_{\,a\,b}\,\left( 1\,+z\,\,q\,{\mathsf w}_i\right)\,,
\end{equation}
where $z:=\lambda^{}_b\lambda_a^{-1}$. 
Noting that $\left( \sv_{i,b}\,\su_{i+1,a}\right)\,\sw_j^{}\left( \sv_{i,b}\,\su_{i+1,a}\right)^{-1}
=q^{2(\de_{i,j}-\de_{i+1,j})}\sw_j^{}$ one recognises \rf{eqforR++} as a difference equation 
restricting the dependence of $\check{\mathsf{r}}^{++}_{\,a\,b}$ on the operators $\sw_i$.


\subsubsection{First formula for $\check{\mathsf{r}}^{++}$}
\label{sec:R++first}

In this section we will show that any expression of the form 
\begin{equation}\label{R++sol}
\begin{aligned}
 \check{\mathsf{r}}^{++}_{\,a\,b}\,=\,&
\frac{h(z\,{\mathsf w}_1^{-1})}{\theta({\mathsf w}_1)}\,
\frac{h(z\,{\mathsf w}_{1\_\,2}^{-1})}{\theta({\mathsf w}_{1\_\,2})}\, 
\dots
\frac{h(z\,{\mathsf w}_{1\_\,(M-1)}^{-1})}{\theta({\mathsf w}_{1\_\,(M-1)})}\,\\
&\qquad\times h(z^{M-1}\,{\mathsf w}_{1\_\,(M-1)})
\,\dots\,
h(z^2\,{\mathsf w}_{1\_\,2})\,
h(z\,{\mathsf w}_1)\,,
\end{aligned}\end{equation}
will represent a solution of \eqref{eqforR++}  
provided that the function $h(x)$ satisfies the relations
\begin{equation}\label{specialfunchtheta}
 h(q^{+1}x)\,=\,h(q^{-1}x)\,\left(1+x \right)\,,\qquad \,\theta(x):=\,h(x)h(x^{-1})\,.
\end{equation}
In \rf{R++sol} we have been using the notations
${\mathsf w}_{i\_\,j}:=q^{i-j}\,{\mathsf w}_i\,{\mathsf w}_{i+1}\,\dots\,{\mathsf w}_j$.
Notice that the $q$ power in front of this expression is such 
that ${\mathsf w}_{i\_\,j}\,=\,e^{\sum_{k=i}^j\,\log \mathsf{w}_k}$.
One furthermore has $\theta(q^{+1}x)\,=\,x\,\theta(q^{-1}x)$.
It is manifest  that for $z=1$ one has $\check{\mathsf{r}}^{++}_{\,a\,b}=1$.
The fact that \eqref{R++sol} solves \eqref{eqforR++} for $i=1$ is
 immediately verified by using the properties
\eqref{specialfunchtheta}  and observing that the products
 ${\mathsf w}_{1\_\,k}$ are invariant under the 
conjugation
 in the left hand side of \eqref{eqforR++} for $2\leq k\leq M$. In order to complete the proof that 
\eqref{R++sol} provides the desired solution it is enough to show that it is
 cyclic invariant, i.e.~it does not
change upon substituting ${\mathsf w}_j$ with ${\mathsf w}_{j+1}$.
 In order to do so  we find it convenient to rewrite
 \begin{equation}\label{R++sol2}
 \check{\mathsf{r}}^{++}_{\,a\,b}\,=\,
\Upsilon_{\mathsf w}\,h(z\,{\mathsf w}_2)\,
h(z\,{\mathsf w}_3)\, 
\dots
h(z\,{\mathsf w}_{M})\,
h(z^{M-1}\,{\mathsf w}_{1\_\,(M-1)})
\,\dots\,
h(z^2\,{\mathsf w}_{1\_\,2})\,
h(z\,{\mathsf w}_1)\,,
\end{equation}
where 
\begin{equation}\label{upsdef}
 \Upsilon_{\mathsf w}\,:=\,
\frac{1}{\theta({\mathsf w}_1)}\,
\frac{1}{\theta({\mathsf w}_{1\_\,2})}\, 
\dots
\frac{1}{\theta({\mathsf w}_{1\_\,(M-1)})}
\,=\,
\frac{1}{\theta({\mathsf w}_{M-1})}\,\dots\,
\frac{1}{\theta({\mathsf w}_2)}\,
\frac{1}{\theta({\mathsf w}_1)}\,.
\end{equation}
The quantity $\Upsilon_{\mathsf w}$ is cyclic invariant by itself.
 This can be shown moving the the last factor on the right of \eqref{upsdef}
to the left and using basic properties of the function $\theta(x)$.
 In order to show that the remaining factor in \eqref{R++sol2}
 is cyclic one uses the
 pentagon relation
\begin{equation}\label{pentagonh}
 h({\mathsf y})\, h({\mathsf x})\,=\, h({\mathsf x})\, h(q^{+1}{\mathsf x}\,{\mathsf y})\, h({\mathsf y})\,,\qquad 
 {\mathsf x}\,{\mathsf y}\,=\,q^{-2}\, {\mathsf y}\,{\mathsf x}\,.
\end{equation}
Details are left to Appendix \ref{ciclRpp}.
We have thus singled out \eqref{specialfunchtheta} and \eqref{pentagonh} as the
 properties of the special function $h(x)$ necessary in order for \eqref{R++sol} to solve 
 \eqref{eqforR++}.
 These properties are satisfied by $\mathcal{E}_{b^2}(x)$ \cite{F99,FKV,Vo}, so we will set 
$h(x)=\mathcal{E}_{b^2}(x)$. The function $\mathbf{e}_b(x)=\mathcal{E}_{b^2}(e^{2\pi b x})$
satisfies the inversion relation 
\begin{equation}
\mathbf{e}_b(x)\mathbf{e}_b(-x)=
\zeta_b
e^{i \pi x^2}\,,
\qquad\zeta_b=e^{\frac{i\pi }{12}(b^2+b^{-2})}\,,
\end{equation}
which  implies that
$\theta(e^{2\pi b z})=\zeta_b\,e^{i\pi z^2}$.

\subsubsection{$\mathsf{r}^{++}$ via the Universal R-matrix: comparison}

One should note, however that equation \eqref{LLRpluplus} can not determine ${\mathsf{r}}^{++}_{\,a\,b}$
uniquely. Recall that the  variables $\eta$ 
commute with the variables $\mathsf{w}$ and satisfy \eqref{centr}. They 
furthermore commute with $\sv_{i,b}\,\su_{i+1,a}$. Multiplying a given solution $\mathsf{r}^{++}_{\,a\,b}$
of \eqref{LLRpluplus} by any function 
of the operators $\eta_i$ will therefore give us
another solution of \eqref{LLRpluplus}.

 In Section \ref{sec:R++first} we demonstrated that the operator $\mathsf{r}^{++}_{\,a\,b}(\la,\mu)$
 defined using \eqref{R++sol} in \rf{Rcheck} is a solution to \eqref{LLRpluplus}. We expect
 that the operator  ${\mathsf{r}}'{}^{++}(\mu\la^{-1})$ defined 
 by a suitable renormalisation of the formal
 expression following from the universal R-matrix in Section \ref{sec:rppfromuniR} 
 represents another solution to \eqref{LLRpluplus}.
 We shall now clarify the relation between the two operators.
 It is expressed by the following formula
 \begin{equation}\label{r++andR++}
 \sr'{}^{++}_{a\,b}(\mu\la^{-1})\,=\,
\mathsf{r}^{++}_{\,a\,b}(\la,\mu)\,\Upsilon_{\eta}\,,
 \end{equation}
 where
 $\Upsilon_{\mathsf{\eta}}$ takes the same form as 
$\Upsilon_{\mathsf{w}}$, defined in \eqref{upsdef},
 with the function $\theta(x)$
 replaced by its inverse $\theta^{-1}(x)$.
 It follows immediately from relation \eqref{r++andR++}
that the operator $\sr'{}^{++}$ indeed 
solves the intertwining relation \eqref{LLRpluplus}, as expected.
 The presence of the factor $\Upsilon_{\mathsf{\eta}}$ reflects the ambiguity 
 in the solution of  \eqref{LLRpluplus} noted above.


\noindent{\emph{Proof of \eqref{r++andR++}}:} It follows from \eqref{rppfromuni} and  \eqref{pimpip} that 
 \begin{equation}\label{RppviaUniR}
\mathbb{P}^{\phantom{(++)}}_{a\,b}\!\!\!\!\!\sr'{}^{++}_{a\,b}(\mu\la^{-1})\,=\,
\mathcal{N}^{}_{ab}\,\,
\mathcal{S}^{}_{ab}\,
 F(\hat{\mathsf{w}})
\mathcal{S}^{-1}_{ab}\,,
\end{equation}
where
 \begin{equation}
 F(\hat{\mathsf{w}})=
\mathcal{E}_{b^2}(\check{\mathsf{w}}_1)\dots \mathcal{E}_{b^2}(\check{\mathsf{w}}_{M-1})
\,
\mathcal{E}_{b^2}(q^{2-M}\check{\mathsf{w}}_{0} \dots\check{\mathsf{w}}_{M\!-\!2})\dots \mathcal{E}_{b^2}(\check{\mathsf{w}}_{0})
\end{equation}
\begin{equation}
\mathcal{N}_{ab}\,:=\,\mathbb{P}_{a\,b}\,\mathcal{F}_b\,q^{-T_{ab}}\,\mathcal{F}_b^{-1}\,\Omega_{a}^{-1}\,,
\qquad 
\mathcal{S}_{ab}\,:=\,\Omega_a\,\mathcal{F}_b\,q^{+T_{ab}}\,,
\end{equation}
and $\check{\mathsf{w}}_i=\mu\la^{-1}\,\su_i^{-1}\sv_i^{}\ot \sv^{}_i\su^{-1}_i$.
In order to derive  \eqref{r++andR++} from \eqref{RppviaUniR} we need to take two simple steps:
(i) Study the action of the similarity transform $\mathcal{S}_{ab}$, 
and (ii) derive  a useful expression for  the operator  $\mathcal{N}_{ab}$.
 Concerning point (i),  it is easy to show that 
\begin{equation}
\label{similarity}
 \mathcal{S}^{}_{ab}\,\,
\check{\mathsf{w}}_i\,\,
\mathcal{S}^{-1}_{ab}\,=\,z\,\mathsf{w}_{i+1}\,,
\end{equation}
where $\mathsf{w}_i$ is defined in  
\eqref{wdef},  while 
 $z=\mu^{}\,\la^{-1}$.
For taking the second step (ii), it is useful to note the following identity
\begin{equation}
\label{NUps}
 \mathcal{N}_{ab}\,=\,\Upsilon_{\mathsf{w}}\,\Upsilon_{\eta}\,,
\end{equation}
where $\Upsilon_{\mathsf{w}}$ is defined in \eqref{upsdef} and
 $\Upsilon_{\mathsf{\eta}}$ takes the same form as 
$\Upsilon_{\mathsf{w}}$ with the function $\theta(x)$
 replaced by its inverse $\theta^{-1}(x)$.
This identity can be shown by computing the matrix elements of both sides.
One may further notice  that $\Upsilon_{\eta, \mathsf{w}}$
satisfy the relations
\begin{equation}
 \Upsilon_{\mathsf w}\,{\mathsf w}_i\,\Upsilon^{-1}_{\mathsf w}\,=\,{\mathsf w}_{i-1}^{-1}\,,
\qquad
 \Upsilon_{\mathsf \eta}\,{\mathsf \eta}_i\,\Upsilon^{-1}_{\mathsf \eta}\,=\,{\mathsf \eta}_{i-1}^{-1}\,.
\label{UpsilonInter}
\end{equation}  

The relation \eqref{r++andR++} simply follows from \eqref{RppviaUniR}, \eqref{similarity},  \eqref{NUps} and 
\eqref{R++sol2}.
\qed

\subsubsection{$\tilde{\mathsf{r}}^{++}$ satisfies the Yang-Baxter equation}

We have seen that the intertwining relation \eqref{LLRpluplus} 
does not fix $\tilde{\mathsf{r}}^{++}$ uniquely. Here we adress the natural question of 
whether the Yang-Baxter equation for  $\tilde{\mathsf{r}}^{++}$ 
fixes this ambiguity. 
A solution of \eqref{LLRpluplus} is given by 
\begin{equation}
 \mathsf{r}^{++}_{\,a\,b}(\lambda_a,\lambda_b)\,=\,\mathbb{P}_{a\,b}\,\rho_z(\mathsf{w}_{ab})\,f_z(\eta_{ab})\,,
\qquad 
z=\lambda^{-1}_a\lambda_b^{}\,.
\label{frhoform}
\end{equation}
Here $\mathsf{w}_{ab}$ and  $\eta_{ab}$ are defined in \eqref{wdef}, \eqref{etadef},
 moreover  $\rho_z(\mathsf{w}_{ab})$ is defined by  \eqref{R++sol2} and $f_z(\eta_{ab})$ is any function of $\eta$.
The Yang-Baxter equation for  $\mathsf{r}^{++}$ can be brought to the braid-type form
\begin{equation}\label{braodFrho}
\begin{aligned}
& f_z(\eta_{1})\,\rho_z(\mathsf{w}_{1})\, f_{zw}(\eta_{2})\,\rho_{zw}(\mathsf{w}_{2})\,
 f_w(\eta_{1})\,\rho_w(\mathsf{w}_{1})=
 \\
&\hspace{4cm}
=f_w(\eta_{2})\,\rho_w(\mathsf{w}_{2})\, f_{zw}(\eta_{1})\,\rho_{zw}(\mathsf{w}_{1})\, f_z(\eta_{2})\,\rho_z(\mathsf{w}_{2})\,,
\end{aligned}\end{equation}
where $\eta_{1}=\eta_{ba}$, $\mathsf{w}_{1}=\mathsf{w}_{ba}$, $\eta_{2}=\eta_{cb}$, $\mathsf{w}_{2}=\mathsf{w}_{cb}$.
The important observation to be made is that
 $\eta_{\alpha,i}\,\mathsf{w}_{\beta,j}=\mathsf{w}_{\beta,j}\,\eta_{\alpha,i}$, where $\alpha,\beta=1,2$.
For this reason the braid relation above can be disentangled into two relations
\begin{align}
& f_z(\eta_{1})\, f_{zw}(\eta_{2})\, f_w(\eta_{1})\,=\,
 f_w(\eta_{2})\, f_{zw}(\eta_{1})\, f_z(\eta_{2})\,,
\label{braidf}\\
&\label{braidrho}
\rho_z(\mathsf{w}_{1})\,\rho_{zw}(\mathsf{w}_{2})\,\rho_w(\mathsf{w}_{1})\,=\,
\rho_w(\mathsf{w}_{2})\,\rho_{zw}(\mathsf{w}_{1})\,\rho_z(\mathsf{w}_{2})\,.
\end{align}
We conclude that a solution to \eqref{LLRpluplus} of the form
 \eqref{frhoform} satisfies the Yang-Baxter equation provided
 that $f_z$  and $\rho_z$  satisfy the braid relations \eqref{braidf}, \eqref{braidrho}.
In the discussion above we took  $f_w(\eta)$ to be either $1$ 
or proportional to $\Upsilon_{\eta}$, see \eqref{UpsilonInter}. 
One may observe that 
 $\Upsilon_{\eta_1}\Upsilon_{\eta_2}\Upsilon_{\eta_1}=\Upsilon_{\eta_2}\Upsilon_{\eta_1}\Upsilon_{\eta_2}$.
In Appendix \ref{app:TBforRpp} we verify 
 \eqref{braidrho}  directly.

The considerations above imply in particular that the normalised R-operator $\sr^{++}$ satisfies the 
same Yang-Baxter equation as the R-operator $\sr'{}^{++}$ coming from the universal R-matrix.

\subsection{Another useful expression for $\check{\mathsf{r}}^{++}$}


We are now going to derive another expression for the operator $\check{\mathsf{r}}^{++}_{\,a\,b}$ that will be
very useful for deriving representations as integral operators. The operator  $\check{\mathsf{r}}^{++}_{\,a\,b}$ 
can be represented as
\begin{equation}\label{R++toKERNELTEXT} 
 \check{\mathsf{r}}^{++}_{\,a\,b}\,=\,
\int d \mathbf{s}\;\tilde{\rho}_z^{++}(\mathbf{s})\,{\mathsf w}(\mathbf{s})\,,
\,\,\,\,\,\,\,\,\,
 d\mathbf{s}=\,\delta(s_{\text{tot}})\prod_{i=1}^Mds_i\,,
\qquad
{\mathsf w}(\mathbf{s})\,:=\,e^{\sum_{i=1}^M\,s_i\,\log  {\mathsf w}_i}\,,
\end{equation}
where $s_{\text{tot}}=\sum_{i=1}^M\,s_i$, $z=e^{2\pi b v}$, $q=e^{-i \pi b^2}$ and 
\begin{equation}
\tilde{\rho}_z^{++}(\mathbf{s})=N_z
\,\prod_{k=1}^{M}\mathbf{s}_b(ibs_{k,k+1}-v+c_b)\,,\qquad N_z:=\frac{e^{\pi i v^2 \frac{M(M-1)}{2}}}{\mathbf{s}_b(c_b-Mv)}\,,
\end{equation}
using the notations $s_{i,j}=s_i-s_j$, $v=(2\pi b)^{-1}{\log\,z}$. The special function 
$\mathbf{s}_b(x)$ is a close relative of 
$\mathbf{e}_b(x)$   defined as
\begin{equation}
\mathbf{s}_b(x)\,=\,\zeta_b^{-\frac{1}{2}}\,e^{-\frac{i \pi }{2}x^2}\,\mathbf{e}_b(x)\,.
\end{equation}


In order to  derive \eqref{R++toKERNELTEXT},
let us introduce the notation 
$\mathsf{x}_k=(2\pi b)^{-1}{\log \mathsf{w}_k}$ and 
$\mathsf{x}_{1\_\,k}:=({2\pi b})^{-1}{\log \mathsf{w}_{1\_\,k}}=\mathsf{x}_1+\dots+\mathsf{x}_k$.
From the exchange relations \eqref{wwexch} 
it follows that
\begin{equation}
  [\mathsf{x}_i,\mathsf{x}_j]\,=\,\frac{1}{2 \pi i}\,(\delta_{i+1,j}-\delta_{i,j+1})\,,
\end{equation}
where the indices are taken modulo $M$. 
 Consider \eqref{R++sol} and rewrite each term using \cite{FKV}
\begin{align}
\label{subright}
& \frac{h(z\,\mathsf{w}^{-1}_{1\_\,k})}{\theta(\mathsf{w}_{1\_\,k})}\,=\,
\frac{e^{i\pi v^2}e^{-2\pi i v\,\mathsf{x}_{1\_\,k}}}{\mathbf{e}_b(\mathsf{x}_{1\_\,k}-v)}
\,=\,
\zeta_0^{-1}\,
  \int_{\mathbb{R}}du_k\;e^{i \pi u_k^2}\, \left(\mathbf{e}_b(v-u_k-c_b)\right)^{-1}
 e^{-2\pi i u_k \mathsf{x}_{1\_\,k}}\,,\\
&h(z^k\,\mathsf{w}_{1\_\,k})\,=\,
\mathbf{e}_b(kv+\mathsf{x}_{1\_\,k})\,=\,
\zeta_0^{+1}\, \int_{\mathbb{R}}dv_k\,e^{-i \pi v^2_k}\;
\mathbf{e}_b(c_b+v_k)\,e^{-2\pi i v_k(kv+\mathsf{x}_{1\_\,k})}\,,
\end{align}
where $\zeta_0\,=\,e^{i \pi (1-4c_b^2)/12}$.
The next step is to group the non-commuting exponentials (using
the relation $e^A\,e^B=e^{A+B}\,e^{\frac{1}{2}[A,B]}$ when $[A,B]$ is central)
as follows
\begin{equation}
\begin{aligned}\left(e^{-2\pi i\,\mathsf{x}_1\, L_1}\!\!\dots
e^{-2\pi i \,\mathsf{x}_{1\_\,M-1}\,L_{M-1}} \right)& 
\left(e^{-2\pi i \,\mathsf{x}_{1\_\,M-1}\,R_{M-1}}\!\!\dots
e^{-2\pi i \,\mathsf{x}_{1}\,R_1} \right)=\\
&\qquad\qquad=e^{-2 \pi\,b\sum_{k=1}^{M-1}z_i\,s_{i+1,M} }
\,\mathsf{w}(\mathbf{s})\,,
\end{aligned}\end{equation}
where $z_i=(L_i-R_i)/2$ and  $L_i+R_i=ib(s_i-s_{i+1})$ with $\sum_{i=1}^Ms_i=0$ and 
$s_{i,j}=s_{i}-s_j$.
With this change of variables we rewrite
\begin{equation}\label{R++toKERNEL1}
\text{\eqref{R++sol}}\,=\,\int\prod_{i=1}^Mds_i\delta(s_{\text{tot}})\,\mathsf{w}(\mathbf{s})\,
\prod_{k=1}^{M-1}
\bigg{[}
e^{-2\pi i k\,v\,(a_k-c_b)}
\int_{\mathbb{R}}dz\,
\frac{\mathbf{e}_b(a_k-z)}{\mathbf{e}_b(b_k-z)}\,e^{-2 \pi i z\,\rho_k}
\bigg{]}\,,
\end{equation}
where
\begin{equation}
 a_k=c_b+i\frac{b}{2}\,s_{k,k+1}\,,
\qquad
 b_k=v-i\frac{b}{2}\,s_{k,k+1}-c_b\,,
\qquad
\rho_k\,=\,- i b\,s_{k,M}-k\,v\,.
\end{equation}
Notice that the integration has decomposed 
into the integration over $\mathbf{s}$ and 
the product of $M-1$ integrals over the variable $z$.
These integrals can be done using \cite{FKV,ByT}
\begin{equation}\label{forzintAPP}
\int_{\mathbb{R}}dy\;
\frac{\mathbf{e}_b(v-y)}{\mathbf{e}_b(u-y)}\,e^{- 2 \pi i w\,y}\,=\,
\frac{\mathbf{s}_b(v-u-c_b)\mathbf{s}_b(w+c_b)}{\mathbf{s}_b(v-u+w-c_b)}\,e^{ -\pi i w(u+v)}\,.
\end{equation}
 Note that the term in parenthesis in \eqref{R++toKERNEL1} can be simplified by 
using the identity
\begin{equation}\label{APPkernelslast}
\prod_{k=1}^{M-1}\frac{\mathbf{S}_b(b\,s_{k,k+1}+iv)\mathbf{S}_b(b\,s_{M,k}+ikv)}{\mathbf{S}_b(b\,s_{M,k+1}+i(k+1)v)}\,=\,
\frac{1}{\mathbf{S}_b(i M v)}\,
\prod_{k=1}^{M}\mathbf{S}_b(b\,s_{k,k+1}+iv)\,,
\end{equation}
which holds for any function $\mathbf{S}_b(x)$.  
The relation \eqref{R++toKERNELTEXT} immediately follows upon setting $\mathbf{S}_b(x):=\mathbf{s}_b(i x +c_b)$.



Let  us finally note that \eqref{R++toKERNELTEXT}  gives us a convenient way 
to re-prove that $\check{\mathsf{r}}^{++}_{\,a\,b}$ satisfies 
\eqref{eqforR++}. 
Inserting \eqref{R++toKERNELTEXT}  into \eqref{eqforR++} one finds that 
\eqref{eqforR++} will hold if 
$\mathcal{K}_z(\mathbf{s}):=
q^{-\frac{1}{2}\sum_{i,j}s_i\hat{A}_{ij}s_j}\tilde{\rho}_z^{++}(\mathbf{s})$ satisfies 
\begin{equation}
0=\mathcal{K}_z(\mathbf{s})(z\,\mathsf{t}_i^2-1)+
\mathcal{K}_z(\mathbf{s}-\delta_i+v_0)
\,(1-z\,q^{2}\,\mathsf{t}^2_{i-1})\,,\qquad \mathsf{t}_i:=q^{ s_i-s_{i+1}}\,,
\label{OS}
\end{equation}
where
$v_0=\frac{1}{M}(1,1,\dots,1)$
and $\delta_i$ is a $M$-vector with zero everywhere exept at position $i$.
In deriving \eqref{OS}  we made use of the following property: 
${\mathsf w}(\mathbf{s}+\alpha\,v_0)\,=\,{\mathsf w}(\mathbf{s})$ for an an arbitrary constant $\alpha$.
This will be the case if
\begin{equation}\label{SolutionFourier}
 \mathcal{K}_z(\mathbf{s})\,=\,
 \prod_{i=1}^M\,f_z(\mathsf{t}_{i}^2)\,,
\end{equation}
provided that $f_z(x)$ satisfies 
$
{f_z(q^{-2}\,x)}=(1-zx){f_z(x)}
$, as is clearly the case when $f_z(x)$ is chosen as
$f_z(\mathsf{t}_{i}^2)\,=\,
\left(\mathcal{E}_{b^2}(-qz\,\mathsf{t}_i^2)\right)^{-1}.
$



\subsection{The twisted story}\label{twiststory}

We had previously observed the possibility to modify the universal R-matrices by using Drinfeld twists.
It is natural to ask what is the  integrable lattice model constructed from the twisted R-matrices\footnote{
The same twist has been used in 
\cite{IS03}.
}.
 We will  consider 
the simple twist introduced in Section \ref{Twists}.
Some remarkable simplifications will later emerge in this
case. 

Let us first consider $(\pi^{\rm f}\ot\pi^{+})(\tilde{\CR})$,  and 
$(\pi^{\rm f}\ot\pi^{\dot{-}})(\tilde{\CR})$.
The resulting Lax matrices are 
\begin{align}
&\tilde{\mathsf{L}}^{+}_{2a-1}(\mu)\,=\,\Lambda(\tilde{\su})\bigg(1+
q^{\frac{M-1}{M}}\frac{1}{\mu}\sum_{i=1}^M\,\sv_{i,2a-1}\,\SE_{ii+1} \bigg)\,, \\
&\tilde{\mathsf{L}}^{\dot{-}}_{2a}(\bar\mu)\,=(1-q^{-1}\bar{\mu}^{M})
\bigg(1-q^{-\frac{1}{M}}\bar\mu
\sum_{i=1}^M\,\sv_{i,2a}\,\SE_{i+1i} \bigg)^{-1}\Lambda(\tilde{\su}')\,,
\end{align}
where $\tilde{\su}$ and $\tilde{\su}'$ are defined as
\begin{equation}
\tilde{\su}_{i,2a-1}\,=\,\prod_{j}(\su_{j,2a-1})^{-X_{ij}}\,,\qquad
\tilde{\su}_{i,2a}'\,=\,\prod_j(\su_{j,2a})^{-X_{ji}}\,.
\end{equation}
and $X_{ij}$ is given in \eqref{f-defn}.
The only non-trivial commutation relations involving the variables above are given by
\begin{equation}
\tilde{\su}_{i,2a-1}\,\sv_{j,2a-1}\,=\,q^{\frac{2}{M}-2\de_{ij}}\sv_{j,2a-1}\,\tilde{\su}_{i,2a-1}\,,\qquad
\tilde{\su}_{i,2a}'\,\sv_{j,2a}\,=\,q^{2\de_{i,j+1}-\frac{2}{M}}\sv_{j,2a}\,\tilde{\su}_{i,2a}'\,.
\end{equation}
The algebra generated by the matrix elements of $\tilde{\CL}(\bar\mu,\mu)=\tilde{\mathsf{L}}^{\dot{-}}(\bar\mu)
\tilde{\mathsf{L}}^{+}(\mu)$ has generators
\begin{equation}
\begin{aligned}
&\tilde\SV_{i,a}^2\,=\,\su_{i,2a}^{2}\sv_{i,2a}^{}\,\sv_{i,2a-1}^{}\su_{i+1,2a-1}^{-2}\,,\\
&\tilde\SC_{i,a}^2\,=\,\sv_{i,2a}^{}\,\sv_{i+1,2a-1}^{}\,,
\end{aligned}\qquad
\tilde\SU_{i,a}\,=\,\tilde{\su}_{i,2a}'\tilde{\su}_{i,2a-1}\,.
\end{equation}
The physical degrees of freedom are conveniently represented by
\begin{equation}
\tilde{\chi}_{i,2a}=\sv_{i,2a+1}\sv_{i,2a}\,,\qquad
\tilde{\chi}_{i,2a-1}\,=\,
\su_{i,2a}^{2}\sv_{i,2a}^{}\,\sv_{i,2a-1}^{}\su_{i+1,2a-1}^{-2}\,,
\end{equation}
they satisfy the same algebra as $\chi_{i,m}$.
The light-cone evolution operators are now found to be
\begin{equation}
\tilde{\SU}^{+}_{\ka}\,=\,\SC_{\rm\sst odd}\,\prod_{a=1}^N q^{2\sf_{2a,2a+1}}\,
\prod_{i=1}^M\CJ_{\kappa}(\tilde{\chi}_{2a})\,,\qquad
\tilde{\SU}^{-}_{\ka}\,=\,\tilde{\SU}^{+}_{\ka}\,\SC^{-1}\,,
\end{equation}
where 
\begin{equation}
\sf_{ab}:=-\frac{1}{2(\pi b^2)^2}\sum_{i,j}\log(\su_{i,a})\,X_{ij}\,\log(\su_{j,b})\,.
\end{equation}
The equations of motion \eqref{equationofmotiondiscrete} are unchanged.
This means that the integrable lattice model constructed from the 
twisted universal R-matrices is as good as a regularisation of the affine Toda theory as the 
original one. 

In order to clarify how much replacing the universal R-matrix $\mathscr{R}$ by 
$\tilde{\mathscr{R}}$ modifies the integrable lattice models constructed using these
universal R-matrices
let us temporarily consider more general twist elements of the form 
$\mathscr{F}=\si(q^{-f})$ with matrix $X_{ij}$ appearing in \rf{f-defn} left arbitrary. 
The Lax matrices obtained from  $\tilde{\mathscr{R}}$ always take the form
\begin{equation}
\begin{aligned}
\tilde{\mathsf{L}}^+(\mu)\,&=\,\Lambda(\mathsf{y}^L)\,\mathsf{\ell}\,\Lambda(\mathsf{y}^R)=
\sum_{i=1}^M\left(\mathsf{a}^+_i\,\SE_{i,i}\,+\,\mathsf{b}^+_i\,\SE_{i,i+1}\right)\,,\\
\tilde{\mathsf{L}}^{\dot{-}}(\bar{\mu})\,&=\Lambda(\bar{\mathsf{y}}^L)\,\bar{\mathsf{\ell}}\,\Lambda(\bar{\mathsf{y}}^R)=
(1-q^{-1}\bar{\mu}^{M})
\left[\sum_{i=1}^M\left((\mathsf{a}^-_i)^{-1}\,\SE_{i,i}\,-\,\mathsf{b}^-_i\,\SE_{i+1,i}\right)\right]^{-1}\,,
\end{aligned}
\end{equation}
where 
\begin{equation}
\mathsf{\ell}=1+q^{\frac{M-1}{M}}\frac{1}{\mu}\sum_{i=1}^M\,\SE_{i,i+1}\,,
\qquad
 \bar{\mathsf{\ell}}=(1-q^{-1}\bar{\mu}^{M})\left( 1-q^{-\frac{1}{M}}\bar{\mu}\sum_{i=1}^M\,\SE_{i+1,i}\right)^{-1}\,.
\end{equation}
The dependence on the twist is encoded in  the form of the variables  $\mathsf{y}_i^{L}$, $\mathsf{y}_i^{R}$, 
$\bar{\mathsf{y}}_i^{L}$, $\bar{\mathsf{y}}_i^{R}$ in terms of  $\su_i$, $\sv_i$. 
The explicit expressions will not be used in the following.
The gauge invariant combinations are
\begin{equation}
\begin{aligned}
\tilde{\chi}_{i,2a-1}\,&=\,
\left(\frac{1}{\bar{\mathsf{y}}^{R}_{i+1}}\,\bar{\mathsf{y}}^{R}_{i}\right)_{2a}
\left(\mathsf{y}^{L}_{i}\frac{1}{\mathsf{y}^{L}_{i+1}}\right)_{2a-1}\,
\sim
\Big{(}\mathsf{b}_i^-\,\mathsf{a}^-_i\Big{)}_{2a}
\Big{(}\mathsf{b}_i^+\,\frac{1}{\mathsf{a}^+_{i+1}}\Big{)}_{2a-1}\,,\\
\tilde{\chi}_{i,2a}\,&=\,
\left(\bar{\mathsf{y}}^{L}_{i+1}\,\frac{1}{\bar{\mathsf{y}}^{L}_{i}}\right)_{2a}
\left(\frac{1}{\mathsf{y}^{R}_{i}}\,\mathsf{y}^{R}_{i+1}\right)_{2a+1}\,\sim
\Big{(}\mathsf{a}_{i+1}^-\,\mathsf{b}^-_i\Big{)}_{2a}
\Big{(}\frac{1}{\mathsf{a}^+_{i}}\,\mathsf{b}_i^+\,\Big{)}_{2a+1}\,.
\end{aligned}
\end{equation}
It is not hard to see that the algebraic relations and the discrete equation of 
motion satisfied by the $\tilde{\chi}_{i,k}$ 
are independent of $X_{ij}$.
Let us furthermore note that for generic  $X_{ij}$  we have
\begin{equation}\label{TtildeT}
\widetilde{\mathsf{T}}(\la)\,=\,\mathsf{T}(\la)\big{|}_{\chi\,\mapsto\,\tilde{\chi}}\,
\end{equation}
where $\widetilde{\mathsf{T}}(\la)$ is the monodromy matrix defined from $\tilde{\mathsf{L}}^{\ep}(\mu)$
in the same way as ${\mathsf{T}}(\la)$ is built from ${\mathsf{L}}^{\epsilon}(\mu)$. 
In order to verify \rf{TtildeT} it suffices to note that 
\begin{equation}
\widetilde{\mathcal{M}}(\la)\,=\,
\Lambda(\mathsf{y}_1^R)^{-1}\,
\mathcal{M}'(\la)\,
\Lambda(\mathsf{y}_1^R)\,,
\qquad
\mathcal{M}'(\la)=\Lambda_{2N}\,\bar{\mathsf{\ell}}\,\Lambda_{2N-1}\,\ell\,\dots\,\Lambda_{1}\ell\,,
\end{equation}
where
\begin{equation}
\Lambda_{2a}:=\Lambda(\mathsf{y}_{2a+1}^R\,\bar{\mathsf{y}}_{2a}^L)\,,
\qquad
\Lambda_{2a-1}:=\Lambda(\bar{\mathsf{y}}_{2a}^R\,\mathsf{y}_{2a-1}^L)\,.
\end{equation}
Notice that the matrices $\Lambda_{k}$ contain only gauge invariant combinations.
Moreover, one can verify that the effect of the similarity transform $\Lambda(\mathsf{y}_1^R)$ on the transfer matrix 
is the same for any value of the twist.
We conclude that the twist only modifies the way 
the variables $\tilde{\chi}_{i,k}$ are constructed out of the basic 
variables $\su_{i,k}$ and $\sv_{i,k}$. It will turn out, however, that 
some choices of $X_{ij}$ are more convenient to work with than others.

\subsection{Assembling the fundamental R-operators}\label{assembly}

\subsubsection{Preparations}

We had previously observed that the Lax-matrices of our interest
can be represented in a factorised form, 
${\CL}_{A}(\la)=\bar\SL^{{-}}_{\bar{a}}(\bar\mu)
\SL^{+}_{{a}}(\mu)$.
We are using the notation $A=(\bar{a},a)$ and will denote
the Hilbert space the matrix elements of ${\CL}_{A}(\la)$
are realised on by $\CH_{A}=\CH_{a}\ot\CH_{\bar{a}}$.
It follows that the corresponding fundamental R-operators 
can be obtained from 
\begin{align}\label{R_AT_Sec6}
&\mathcal{R}_{AB}(\bar\mu,\mu;\bar\nu,\nu)\,=\,
\sr_{a\bar{b}}^{+\dot{-}}(\bar{\nu}/\mu)\,\sr_{ab}^{++}({\nu}/\mu)\,
\sr_{\bar{a}\bar{b}}^{\dot{-}\dot{-}}(\bar{\nu}/\bar\mu)\,\sr_{\bar{a}{b}}^{\dot{-}+}(\nu/\bar\mu)\,.
\end{align}
Our goal is to find more explicit representations for the operators $\mathcal{R}_{AB}$.
We begin by displaying the structure of the ingredients in 
a convenient form:
\begin{subequations}\label{ingredients}
\begin{equation}
\begin{aligned}
&\sr_{a\bar{b}}^{+\dot{-}}(\nu)\,=\, 
(\vartheta_{1/\nu}^{}(\check{\chi}_{\bar{b}a}))^{-1}\,q^{-\mst_{\bar{b}a}}\,,\\
&\sr_{\bar{a}b}^{\dot{-}+}(\nu)\,=\,q^{\mst_{\bar{a}b}}\,\vartheta_{\nu}^{}(\check{\chi}_{\bar{a}b})\,,
\end{aligned}\qquad
\begin{aligned}
&\sr_{ab}^{++}(\nu)\,=\,\BP_{ab}^{}\,\rho_{\nu}^{++}(\sw_{ab})\,,\\
&\sr_{\bar{a}\bar{b}}^{\dot{-}\dot{-}}(\nu)\,=\,\BP_{\bar{a}\bar{b}}^{}\,\rho_{\nu}^{\dot{-}\dot{-}}(\bar\sw_{\bar a\bar b})\,,
\end{aligned}
\end{equation}
where $\check{\chi}_{\bar{a}b}$, $\sw_{ab}$ and $\bar\sw_{\bar a\bar b}$ denote
the collection of operators
\begin{equation}\begin{aligned}
&\sw_{ab}^{i}\,=\,
\left(\sv^{}_{i}\,\su^{-1}_{i+1}\right)_{a}\,
\left(\sv_{i}^{-1}\,\su_{i}^{}\right)_{b}\,, \\ &
\bar\sw_{\bar a\bar b}^{i}\,=\,
\left(\sv_i^{}\,\su^{}_{i+1}\right)_{\bar{a}}\,
\left(\sv_{i}^{-1}\,\su_{i}^{-1}\right)_{\bar b}\,,
\end{aligned}
\qquad
\check{\chi}^{i}_{\bar{a}{b}}:=\,
\left(\su_{i+1}^{}\,\sv_{i}^{}\right)_{\bar a}\,
\left(\sv_{i}^{}\,\su_{i}^{-1}\right)_{{b}}\,.
\end{equation}
\end{subequations}
We may thereby represent $\mathcal{R}_{AB}$ in the following form:
\begin{align}\label{RATwith_xyz}
\mathcal{R}_{AB}(\bar\mu,\mu;\bar\nu,\nu)\,&=\,
\SP_{AB}\cdot\big[\vartheta^{}_{\mu/\bar\nu}(\sz_{AB})\big]^{-1}\!
\cdot
\rho^{++}_{\nu/\mu}(\sx_{AB})\rho^{--}_{\bar\nu/\bar\mu}({{\sy}}_{AB})
\cdot\vartheta^{}_{\bar\nu/\mu}(\sz_{AB})
\,,
\end{align}
where
\begin{equation}
\begin{aligned}
&\sx_{AB}^{i}:=\,q^{-\mst_{\bar{a}b}}\cdot \sw_{ab}^{i}\cdot q^{+\mst_{\bar{a}b}}\,=\,
\left(\sv_i^{}\,\su_{i+1}^{-1}\right)_a\,
\left(\su_{i+1}^{-1}\,\su^{}_i\right)_{\bar{a}}
\left(\sv_i^{-1}\,\su^{}_i\right)_b\,,
\\
&{\sy}_{AB}^{i}:=\,q^{-\mst_{\bar{a}b}}\cdot \bar\sw_{\bar{a}\bar{b}}^{i}\cdot q^{+\mst_{\bar{a}b}}\,=\,
\left(\sv^{}_i\,\su_{i+1}^{}\right)_{\bar a}
\left(\su_{i+1}\,\su_i^{-1}\right)_{b}
\left(\su_i^{}\,\sv_i^{-1}\right)_{\bar{b}} \,.
\end{aligned}
\end{equation}
and $\sz_{AB}=\check{\chi}^{i}_{\bar{a}{b}}$.
It will be useful to observe that the operators $\sx_{AB}^{i}$, ${\sy}_{AB}^{i}$ and $\sz_{AB}^i$ 
can be expressed in terms 
of operators $\SU^i_A$, $\hat\SV^i_A$, $\SU^i_B$ and $\check\SV^i_B$ defined as
\begin{equation}
\label{UVClater}
\begin{aligned}
&(\SU^i_A)^2=\,\left(\su_i^{}\,\sv^{}_i\right)_{\bar{a}}
\left(\sv_i^{}\,\su^{-1}_{i+1}\right)_a\,,\\
&(\SC^i_A)^2=\,
\left(
\sv_i^{-1}\,\su_i^{}\right)_{\bar{a}}
\left(\su_{i+1}^{}\,\sv_i^{}\right)_a\,,
\end{aligned}\qquad
\begin{aligned}
&\hat\SV^{i}_A=\,(\SC^{i}_A)^{-1}\,\SV_A^i\,,\\
&\check\SV^i_B=\,(\SC^{i+1}_B)^{-1}\,\SV_B^i\,,
\end{aligned}
\qquad
\SV_A^i=\,\su_{i+1,\bar{a}}\,\su_{i+1,a}\,.
\end{equation}
Notice that the operators $\SC^i_A$ are central in the algebra generated by the combinations \eqref{UVClater},
while $\SU^i_A$, $\SV^i_A$ satisfy the defining relations of the algebra $\mathcal{W}_M$.

The result is most conveniently expressed in the form 
\begin{equation}\label{xyz-XYZ}
\begin{aligned}
& 
\sx_{AB}^i=(\sz^i_{AB})^{-\frac{1}{2}}\,(\SU^i_{A})^{+2}\,(\sz^i_{AB})^{-\frac{1}{2}}\,,\\
&
\sy_{AB}^i=(\sz^i_{AB})^{+\frac{1}{2}}\,(\SU^{i}_{B})^{-2}\,(\sz^i_{AB})^{+\frac{1}{2}}\,,
\end{aligned}\qquad
\sz^i_{AB}=\SU_A^i\hat\SV_A^{i}\,(\check\SV_B^{i-1})^{-1}\SU_B^{i}\,.
\end{equation}
This representation makes clear that the operator $\SP_{AB}\mathcal{R}_{AB}$ commutes with $\SC_A^i$ and 
$\SC_B^i$. Noting that $\SC_A^i\equiv \SC_{2a-1}^i$ if $A=(\bar{a},a)\leadsto(2a,2a-1)$, it becomes easy to
see that the fundamental transfer matrices $\mathbf{T}(\bar\mu,\mu;\bar\nu,\nu)$ defined as  
$\mathbf{T}(\bar\mu,\mu;\bar\nu,\nu)=\mathsf{C}\cdot\mathcal{T}(\bar\mu,\mu;\bar\nu,\nu)$, where $\mathsf{C}$
is the shift operator,  commute with 
$\SC_{2a-1}^i$. 

In order to show that the fundamental transfer matrices $\mathbf{T}(\bar\mu,\mu;\bar\nu,\nu)$ also commute with 
$\SC_{2a}^i$ let us note that the cyclic symmetry of the trace allows us to rewrite the definition
of $\mathcal{T}(\bar\mu,\mu;\bar\nu,\nu)$ in terms of the fundamental R-operators $\mathcal{R}_{AB}'$
associated to the Lax-matrices ${\CL}_{A}'(\la)=
\SL^{+}_{{a}}(\mu)\bar\SL^{{-}}_{\bar{a}}(\bar\mu)$. The corresponding fundamental R-operator
$\mathcal{R}_{AB}'$
may be represented as 
\begin{align}
&\mathcal{R}_{AB}'(\mu,\bar\mu;\nu,\bar\nu)\,=\,
\sr_{\bar{a}{b}}^{\dot{-}+}(\bar{\nu}/\mu)\,\sr_{ab}^{++}({\nu}/\mu)\,
\sr_{\bar{a}\bar{b}}^{\dot{-}\dot{-}}(\bar{\nu}/\bar\mu)\,\sr_{{a}\bar{b}}^{+\dot{-}}(\nu/\bar\mu)\,.
\end{align}
A straightforward generalisation of the analysis above leads to the conclusion that 
$\SP_{AB}\mathcal{R}_{AB}'$ commutes with $\SC_{A}^{i'}$ and $\SC_{B}^{i'}$, defined
as 
\begin{equation}
\SC_{A}^{i'}:=\,\left(\sv^{}_i\,\su_i^{}\right)_a\,\left(\sv^{-1}_i\,\su^{}_{i+1}\right)_{\bar{a}}\,,\qquad
\SC_{B}^{i'}:=\,\left(\sv^{}_i\,\su_i^{}\right)_b\,\left(\sv_i^{-1}\,\su^{}_{i+1}\right)_{\bar{b}}\,.
\end{equation} 
Noting that $\SC_{A}^{i'}\equiv\SC_{2a}^i$ if $A=(a,\bar{a})\leadsto(2a+1,2a)$ 
leads to the conclusion  that $\mathbf{T}(\bar\mu,\mu;\bar\nu,\nu)$ commutes with 
$\SC_{2a}^i$. Taken together we have shown that the fundamental transfer matrix
is a physical observable.

\paragraph{XXZ-type chains.}
One may also consider 
$\mathcal{R}^{\rm\sst XXZ}_{AB}$  defined as 
\begin{equation}
\mathcal{R}^{\rm\sst XXZ}_{AB}(\bar\mu,\mu;\bar\nu,\nu)\,=\,
\sr_{a\bar{b}}^{-\dot{-}}(\bar{\nu}/\mu)\,\sr_{ab}^{--}(\nu/\mu)\,
\sr_{\bar{a}\bar{b}}^{\dot{-}\dot{-}}(\bar{\nu}/\bar\mu)\,\sr_{\bar{a}{b}}^{\dot{-}-}(\nu/\bar\mu)\,,
\end{equation}
with $\sr_{ab}^{--}(1)=\mathbb{P}_{ab}$, $\sr_{\bar{a}\bar{b}}^{\dot{-}\dot{-}}(1)=\mathbb{P}_{\bar{a}\bar{b}}$.
The operator
$\mathcal{R}^{\rm\sst XXZ}_{AB}$ is 
related to the lattice  affine Toda fundamental R-operator
$\mathcal{R}^{\rm\sst AT}_{AB}=\mathcal{R}_{AB}$ via\footnote{
This equation differs from \eqref{RfunfromRXXZ} by a similarity transform originating form the definition of $\mathcal{L}^{\rm XXZ}$,
see \eqref{Tdefsec3}.
}
\begin{equation}
\mathcal{R}^{\rm\sst XXZ}_{AB}(\bar\mu,\mu;\bar\nu,\nu)\,=\,
\Omega_A^{}\cdot(\CF_a\CF_b)^{-1}\!\cdot\mathcal{R}^{\rm\sst AT}_{AB}(\bar\mu,\mu;\bar\nu,\nu)
\cdot(\CF_a\CF_b)\cdot\Omega_B^{-1}\,.
\end{equation}
It follows that $\mathcal{R}^{\rm\sst XXZ}_{AB}$ takes the form \eqref{RATwith_xyz} with 
\begin{equation}
\left(\sx_{AB},\sy_{AB},\sz_{AB}\right)^{\rm\sst XXZ}\,=\,
\mathcal{S}\,\left(\sx_{AB},\sy_{AB},\sz_{AB}\right)^{\rm\sst AT}\,
\mathcal{S}^{-1}\,,
\qquad
\mathcal{S}:=\Omega_B^{}\cdot(\CF_a\CF_b)^{-1}\,.
\end{equation}


\subsubsection{Twisted lattice affine Toda}

One may easily carry out the same analysis for the R-operators coming from the twisted universal 
R-matrices $\tilde\CR^{\pm}$, see \eqref{twistedRpm}.
 The formulae for the ingredients are very similar
\begin{subequations}\label{tw-ingredients}
\begin{equation}
\begin{aligned}
&\tilde{\sr}_{a\bar{b}}^{+\dot{-}}(\nu)\,=\, 
(\vartheta_{1/\nu}^{}(\check{\chi}_{\bar{b}a}))^{-1}\,q^{-2\sf_{\bar{b}a}}\,,\\
&\tilde{\sr}_{\bar{a}b}^{\dot{-}+}(\nu)\,=\,q^{2\sf_{\bar{a}b}}\,\vartheta_{\nu}^{}(\check{\chi}_{\bar{a}b})\,,
\end{aligned}
\qquad
\begin{aligned}
&\tilde{\sr}_{ab}^{++}(\nu)\,=\,\BP_{ab}^{}\,\rho_{\nu}^{++}(\sw_{ab})\,,\\
&\tilde{\sr}_{\bar{a}\bar{b}}^{\dot{-}\dot{-}}(\nu)\,=\,\BP_{\bar{a}\bar{b}}^{}\,\rho_{\nu}^{--}(\bar\sw_{\bar a\bar b})\,,
\end{aligned}\end{equation}
where $\check{\chi}_{\bar{a}b}$, $\sw_{ab}$ and $\bar\sw_{\bar a\bar b}$ are now given by the expressions
\begin{equation}
\check{\chi}^{i}_{\bar{a}{b}}:=\,\sv^{}_{i,\bar a}\,
\sv^{}_{i,{b}}\,,
\qquad
\begin{aligned}
&\sw_{ab}^{i}\,=\,\left(\su_{i+1}^{-1}\,\sv^{}_{i}\,\su^{-1}_{i+1}\right)_a\,
\sv_{i,b}^{-1}\,, \\ &
\bar\sw_{\bar a\bar b}^{i}\,=\,\sv_{i,\bar{a}}^{}\,
\left(\su_{i}^{}\,\sv_{i}^{}\,\su^{}_{i}\right)_{\bar b}^{-1}\,.
\end{aligned}
\end{equation}
\end{subequations}
The rest of the analysis proceeds as before. The resulting formula 
for the operator $\tilde\SR_{AB}(\bar\mu,\mu;\bar\nu,\nu)$ is very similar to
formula \rf{xyz-XYZ}, the only changes being that 
one needs to replace the expression for 
$\sz^i_{AB}$ in \rf{xyz-XYZ} by 
\begin{equation}
\tilde\sz^i_{AB}
\,=\,\hat\SV_A^{i}\,(\check\SV_B^{i-1})^{-1}\,,
\end{equation}
and that one now needs to define
\begin{equation}\label{twistUV}
\begin{aligned}
&(\SU^i_A)^2=\,\left(\su_i^2\,\sv^{}_i\right)_{\bar{a}}\left(\sv^{}_i\,\su^{-2}_{i+1}\right)_a\,,\\
&(\SC^i_A)^2=\,\sv_{i,\bar{a}}^{-1}\,\sv^{}_{i+1,a}\,,
\end{aligned}\qquad
\begin{aligned}
&\hat\SV^{i}_A=\,\sv_{i,\bar{a}}^{}\,,\\
&\check\SV^i_B=\,\sv_{i+1,b}^{-1}\,.
\end{aligned}
\end{equation}
This innocent-looking modification has important consequences.
For application to integrable lattice models it is helpful to have a formula for the kernel
of $\mathcal{R}_{AB}(\bar\mu,\mu;\bar\nu,\nu)$ that is as simple as possible. Such a formula 
will be derived shortly for the operator $\tilde{\mathcal{R}}_{AB}(\bar\mu,\mu;\bar\nu,\nu)$
obtained from the twisted R-matrices $\tilde{\CR}_\pm$, taking advantage of the fact that
$\tilde\sz^i_{AB}$ is diagonal in representations where $\SV_A^i$ and $\SV_B^i$ are diagonal.


We may observe, on the other hand, that it is impossible to diagonalise the families of 
operators $\{\SU_A^i\SV_A^{i}; i=1,\dots,M\}$ 
and $\{(\SV_B^{i})^{-1}\SU_B^{i+1};i=1,\dots M\}$  simultaneously
as the operators in these families do not  mutually commute for different values of the index $i$. 
This means that it will be much more convenient to work with integrable lattice models build from the
twisted universal $R$-matrices $\tilde{\CR}_{\pm}$ rather than the original ones.

\subsubsection{Factorization from the universal R-matrix.}

In all the cases above were able to express the R-operators in terms of the 
operators $\SU_{i,R}$. $\SV_{i,R}$, $\SC_{i,R}$, $R=A,B$ generating a sub-algebra
of the algebra of all operators acting on $\CH_A\ot\CH_B$ which has a center generated by
the operators $\SC_{i,R}$. We will now see that this phenomenon has a 
natural representation-theoretic explanation.

We had observed in Section \ref{sec:decoupling} that the tensor product 
$\pi^-_{\la_1}\otimes {\pi}^{\dot{-}}_{\la_2}$ is isomorphic to the tensor product of 
a representation of evaluation type with a trivial representation. A similar statement
holds for the tensor product $\pi^+_{\la_1}\otimes {\pi}^{\dot{+}}_{\la_2}$. The precise
statement is 
\begin{equation}\label{factortriv2}
  \big(\pi^+_{\la_1}\otimes {\pi}^{\dot{+}}_{\la_2}\big)\Delta(a)\,=\,
\left(\pi^{\text{min}}_{\la,\ka}\ot\,\pi^{\text{triv}}_{}\right)\Delta(a)\,,
\qquad a\in\Borelm\,,
\end{equation}
where $\pi^{\text{min}}_{\la,\ka}$ and $\pi^{\text{triv}}_{}$ are defined via
\begin{align}
&\pi^{\text{min}}_{\la,\ka}(f_i)\,=\,\frac{\la}{q-q^{-1}}
\,{\mathsf{V}}_{i}^{\frac{1}{2}}\,
\big(\ka^{}\,{\mathsf{U}}^{-1}_{i+1}+\ka^{-1}{\mathsf{U}}_{i+1}^{}\big)\,
{\mathsf{V}}_{i}^{\frac{1}{2}}\,,
\qquad
\pi^{\text{min}}_{\la,\ka}(q^{\bar{\epsilon}_i})\,=\,
{\mathsf{U}}_i^{}\,,\\
& \pi^{\text{triv}}_{}(f_i)\,=\,0\,,\qquad
\pi^{\text{triv}}_{}(q^{\bar{\epsilon}_i})\,=\,{\mathsf{C}}_{i}^{}\,,
\end{align}
provided that we define
\begin{align}
& {\mathsf{U}}^2_i\,=\,\su_i^{}\sv^{-1}_{i-1}\ot\su_i^{}\,\sv_{i-1}^{}\,,
\qquad
{\mathsf{V}}_{i}^{}\,=\,\su_i^{-1}\ot\sv_{i}\cdot\SC_{i+1}^{}\,,
\qquad
{\mathsf{C}}^2_i\,=\,\su_i^{}\sv^{}_{i-1}\ot\su_i^{}\,\sv_{i-1}^{-1}\,,
\\
%
 \label{sim2BIS}
& \la:=\,q^{\frac{1}{2}}\left(\la_1\la_2\right)^{\frac{1}{2}}
\qquad
\ka:=\,(\la_1^{}\la_2^{-1})^{\frac{1}{2}}\,.
\end{align}

The isomorphisms \rf{factortriv} and \rf{factortriv2} implies, upon assuming the validity of \eqref{eqnUnivROther},
 that the R-operator
$\mathcal{R}'{}_{AB}^{\rm\sst XXZ}(\bar\mu,\mu;\bar\nu,\nu)$ has besides 
\rf{RXXZfactorrep} another factorisation of the schematic form
\begin{equation}\label{otherfact}
\mathcal{R}'{}_{AB}^{\rm\sst XXZ}\,=\,
\mathcal{R}{}_{AB}^{\rm(min)\ot(min)}\,\mathcal{R}{}_{AB}^{\rm (min)\ot(triv)}
\mathcal{R}{}_{AB}^{\rm (triv)\ot(min)}\, \mathcal{R}{}_{AB}^{\rm (triv)\ot(triv)}\,.
\end{equation}
Rewriting $\mathcal{R}'{}_{AB}^{\rm\sst XXZ}$ in the form \rf{otherfact} will allow us to 
extract $\mathcal{R}{}_{AB}^{\rm(min)\ot(min)}$ from $\mathcal{R}'{}_{AB}^{\rm\sst XXZ}$.

This is done as follows.
Let us  start from 
\eqref{RXXZfactorrep}, repeated here for convenience:
\begin{equation}\label{rrrrsec6}
\mathcal{R}'{}_{AB}^{\rm\sst XXZ}\,=\,\sr_{a,\bar{b}}^{+\dot{-}}(\mu,\bar\nu)\,
\sr_{a,b}^{+-}(\mu,\nu)\,\sr_{\bar{a},\bar{b}}^{\dot{+}\dot{-}}(\bar\mu,\bar\nu)\,
\sr_{\bar{a},b}^{\dot{+}-}(\bmu,\nu)\,.
\end{equation}
Introducing the notation 
\begin{equation}
\sr^{\epsilon_1 \epsilon_2}_{x,y}(\la,\mu)\,=\,
\bar{\mathsf{r}}^{\epsilon_1\epsilon_2}_{\la/\mu}(\{\mathsf{z}_{x,y}^{\epsilon_1\epsilon_2}\})\,q^{-\mst_{xy}}\,,
\qquad
\mathsf{z}_i^{\epsilon_1\epsilon_2}=-\tau_q^2\,\left(\pi^{\epsilon_1}(f_i)\ot \pi^{\epsilon_2}(e_i)\right)\,,
\end{equation}
and moving the factors $q^{-\mst_{xy}}$ to the right we see that 
$\mathcal{R}'{}_{AB}^{\rm\sst XXZ}$ can indeed be written in the form \rf{otherfact}
with
\begin{equation}
\mathcal{R}{}_{AB}^{\rm(min)\ot(min)}
=
\bar{\mathsf{r}}_{\mu/\bar{\nu}}^{+\dot{-}}\big(\,\mathsf{f}^+_A\,\mathsf{e}^{\dot{-}}_B\,\big)\;
\bar{\mathsf{r}}_{\mu/\nu}^{+-}\big(\,\mathsf{f}^+_A\,\mathsf{e}^{-}_B\,\big)\;
\bar{\mathsf{r}}_{\bar{\mu}/\bar{\nu}}^{\dot{+}\dot{-}}\big(\,\mathsf{f}^{\dot{+}}_A\,\mathsf{e}^{\dot{-}}_B\,\big)\;
\bar{\mathsf{r}}^{+-}_{\bar{\mu}/\nu}\big(\,\mathsf{f}^+_A\,\mathsf{e}^{-}_B\,\big)\,
q^{-\mst_{AB}}
\end{equation}
where we used the notation 
\begin{equation}
\mathsf{f}^+_{i,A}:=\SV_{i,A}^{}\SU_{i+1,A}^{-1}\,,
\quad
\mathsf{f}^{\dot{+}}_{i,A}:=\SU_{i+1,A}^{}\SV_{i,A}^{}\,,
\quad
\mathsf{e}^{\dot{-}}_{i,B}:=\SU^{}_{i,B}\SV^{}_{i,B}\,,
\quad
\mathsf{e}^{-}_{i,B}:=\SV^{}_{i,B}\SU^{-1}_{i,B}\,,
\end{equation}
and the relation
\begin{equation}
q^{-\left(\mst_{a\bar{b}}+\mst_{a b}+\mst_{\bar{a}\bar{b}}+\mst_{\bar{a}b}\right)}\,=\,
q^{-\mst_{AB}}\,
\mathcal{R}{}_{AB}^{\rm (min)\ot(triv)}
\mathcal{R}{}_{AB}^{\rm (triv)\ot(min)}\, \mathcal{R}{}_{AB}^{\rm (triv)\ot(triv)}\,.
\end{equation}

\subsection{Representation as integral operators}\label{kernels}

The generalized Baxter equation to be derived in the next section 
becomes an efficient tool for the calculation of the spectrum of the affine
Toda theories once it is supplemented by certain informations 
about the analytic properties of the Q-operators. 
In order to derive this information
it will be useful to represent the Q-operators
as integral operators, which will allow us to deduce 
the relevant information
from the analytic properties of the kernels representing $\SQ(\la)$, as
was done in \cite{ByT} for the Sinh-Gordon case.

Our first goal is therefore to present a representation of the 
fundamental R-operator $\SR_{AB}^{}(\bar\mu,\mu;\bar\nu,\nu)$ 
as an integral operator.

\subsubsection{Kernel of fundamental R-operator}

We shall now compute the kernel of $\tilde{\mathcal{R}}_{AB}^{\rm\sst XXZ}(\bar\mu,\mu;\bar\nu,\nu)$. 
This operator may be represented as in
\rf{RATwith_xyz}, where now
\begin{equation}
\hat\sz^i_{AB}
\,=\,\hat\SV_A^{i}\,(\check\SV_B^{i})^{-1}\,.
\end{equation}
Let $\check{\mathcal{R}}_{AB}^{\rm\sst XXZ}(\bar\mu,\mu;\bar\nu,\nu):=
\SP_{AB}\tilde{\mathcal{R}}_{AB}^{\rm\sst XXZ}(\bar\mu,\mu;\bar\nu,\nu)$.
As $\check{\mathcal{R}}_{AB}^{\rm\sst XXZ}$ commutes with $\SC_A^i$ and $\SC_B^i$ 
it suffices to consider the operator $\check{\mathsf{R}}_{AB}$ obtained 
from $\check{\mathcal{R}}_{AB}^{\rm\sst XXZ}$ by replacing the representation for the 
operators $\SU_A^i$, $\SU_B^i$,
$\hat\SV_A^{i}$ and $\check\SV_B^{i}$ following from \rf{twistUV} 
by a representation where these operators act on a Hilbert space 
spanned by 
states $\langle\, x,x'\,|$ such that
\begin{equation}
\langle\, x,y\,|\,\hat\SV_A^{i}\,=\,\langle \,x,y\,|\,e^{2\pi bx_{i,i+1}} \,,\qquad
\langle\, x,y\,|\,\check\SV_B^{i}\,=\,\langle \,x,y\,|\,e^{2\pi by_{i,i+1}}\,,
\end{equation}
using the notations $x_{i,j}:=x_i-x_j$. Our 
task is thereby reduced to the calculation of 
the matrix elements of the operator
$\rho^{++}_{\mu/\nu}(\tilde\sx_{AB}^{})\rho^{--}_{\bar\mu/\bar\nu}(\tilde{\sy}_{AB}^{})$,
where 
\begin{equation}
\begin{aligned}
& \tilde\sx_{AB}^i\,=\,(\tilde\sz_{AB}^i)^{-\frac{1}{2}}\,(\SU_A^i)^{+2}\,(\tilde\sz_{AB}^i)^{-\frac{1}{2}}\,,\\
& \tilde\sy_{AB}^{i}\,=\,(\tilde\sz_{AB}^{i})^{+\frac{1}{2}}\,(\SU^{i+1}_B)^{-2}\,(\tilde\sz_{AB}^{i})^{+\frac{1}{2}}\,.
\end{aligned}
\end{equation}

It is useful to represent the operators $\rho^{}_{\mu/\nu}(\tilde\sx_{AB}^{})
$ and $\rho^{}_{\bar\mu/\bar\nu}(\tilde{\sy}_{AB}^{})$ using a 
non-commutative generalisation of the Fourier transformation in the form
\begin{equation}
\begin{aligned}
&\rho^{++}_{\la}(\tilde{\sx}_{AB}^{})\,=\,
\int_{\BR^M} d\mu(s)\;\tilde{\rho}_\la^{++}(s)\;\SX(s)\,,
\qquad \SX(s):=
\exp\bigg(\frac{\mathrm i}{b}\sum_{i=1}^M {s_i}\log\tilde{\sx}_{AB}^{}\bigg)\,,\\
&\rho^{--}_{\la}(\tilde{\sy}_{AB}^{})\,=\,
\int_{\BR^M} d\mu(t)\;\tilde{\rho}_\la^{--}(t)\;\SY(t)\,,
\qquad \SY(t):=\exp
\bigg(\frac{\mathrm i}{b}\sum_{i=1}^M t_i\log\tilde{\sy}_{AB}^{}\bigg)\,,
\end{aligned}
\end{equation}
using the notation $d\mu(s)=\prod_{i=1}^Mds_i\,\de(\sum_{i=1}^M s_i)$.
Working in a representation where $\SU_A^i$ and $\SU_B^i$ are represented as operators
generating shifts of $x$ and $y$, respectively, 
leads to the following form for the matrix elements of  $\SX(s)\SY(t)$,
\begin{align}\label{SWmatel}
&\langle\,x,y\,|\,\SX(s)\SY(t)\,|\,x',y'\,\rangle\,=\,\\&\qquad\,=\,
\prod_i\de(x_i^{}-s_i^{}-x_i')\de(y_i^{}+t_i^{}-y_i')\notag\\[-1ex]
&\qquad\qquad\quad \times e^{-\pi{\mathrm i}s_i(x-y)_{i,i+1}}\, e^{-\pi{\mathrm i}s_i(x'-y')_{i,i+1}}
\,e^{+\pi{\mathrm i}t_i(x-y)_{i-1,i}}\,e^{+\pi{\mathrm i}t_i(x'-y')_{i-1,i}}\notag\\
&\qquad\,=\,\prod_i\de(x_i^{}+s_i^{}-x_i')\de(y_i^{}+t_i^{}-y_i')\notag\\[-1ex]
&\qquad\qquad\quad \times e^{-\pi{\mathrm i}(x-x')_i(x+x')_{i,i+1}}\,e^{\pi{\mathrm i}(y-y')_i(y+y')_{i-1,i}}\,
e^{2\pi{\mathrm i}x_i^{}y_{i,i+1}^{}}\,e^{-2\pi{\mathrm i}x_i'y'_{i,i+1}}\,.
\notag\end{align}
Thanks to the delta-functions in \rf{SWmatel}, 
the kernel of the operator $\check{\mathsf{R}}_{AB}(\bar\mu,\mu;\bar\nu,\nu)$, defined as
\begin{equation}
\check{R}_{\bar\mu,\mu;\bar\nu,\nu}(x,y|x',y'):=
\langle\, x,y\,|\,\check{\mathsf{R}}_{AB}(\bar\mu,\mu;\bar\nu,\nu)\,
|\,x',y'\,\rangle\,,
\end{equation}
becomes fully factorised,
\begin{align}\label{R-kernel1}
& \check{R}_{\bar\mu,\mu;\bar\nu,\nu}(x,y|x',y')=
\\
&\qquad\quad=\de(\bar{x}-\bar{x}')\de(\bar{y}-\bar{y}')\,
W^{^{\sst +-}}_{\bar\nu/\mu}(x,y)\,{W}^{^{\sst++}}_{\nu/\mu}(x,x')\,{W}^{^{\sst --}}_{\bar\nu/\bar\mu}(y,y')\,
W^{^{\sst -+}}_{\nu/\bar\mu}(x',y')\,,
\notag\end{align}
using the notation $\bar{x}=\sum_{i=1}^Mx_i$ for the sum of the components of a
vector $x\in \BR^M$, and 
\begin{subequations}
\begin{align}
&W^{^{\sst ++}}_\la(x,x')=W^{^{\sst--}}_\la(x',x)= e^{\pi{\mathrm i}P(x,x')}\bar{V}_{w}^{}(x-x')\,,\\
&W^{^{\sst -+}}_\la(x,y)=\big(W^{^{\sst +-}}_{1/\la}(x,y)\big)^{-1}=e^{\pi{\mathrm i}P(x,y)}{V}_{w}^{}(x-y)\,;
\end{align}
\end{subequations}
We are using the notation $P(x,y)=\sum_{i=1}^M(x_iy_{i+1}-y_ix_{i+1})$ and $w=\frac{1}{2\pi b}\log\la$.
The explicit formulae for the functions appearing in these expressions are
\begin{subequations}
\begin{align}
&{V}_w^{}(s)\,=\,e^{-\frac{\pi {\mathrm i}}{2}Mw^2}
\prod_{i=1}^M\frac{1}{\mathbf{s}_b(s_{i,i+1}+w)}\,,\\
&\bar{V}_w^{}(s)\,=\,N_w
\prod_{i=1}^M\mathbf{s}_b(w-s_{i,i+1}+c_b)\,.
\end{align}
\end{subequations}
The resulting expression resembles the one found for the generalised chiral Potts models found in \cite{BKMS,DJMM}.


Using \rf{RfunfromRXXZ} it is easy to get the kernel of $\mathcal{R}_{AB}(\bar\mu,\mu;\bar\nu,\nu)$
from the kernel of $\mathcal{R}_{AB}^{\rm\sst XXZ}(\bar\mu,\mu;\bar\nu,\nu)$.

\subsubsection{Fundamental transfer matrices}
\label{sec:physdofandFACT}

Having the kernel $\SR_{AB}(\bar\mu,\mu;\bar\nu,\nu)$
it is straightforward to compute the kernel representing the fundamental 
transfer matrices $\mathcal{T}(\bar\mu,\mu;\bar\nu,\nu)$ in an auxiliary representation 
for $\CH=\bigotimes_{a=1}^N \CH_{2a}\ot \CH_{{2a-1}}$
 that is defined as follows.
Let us introduce the operators ${\mathsf U}_{i,a}=\su_{i,2a}\su_{i,2a-1}$ commuting with 
$\mathsf{C}^{}_{i,a}\equiv\mathsf{c}^{}_{i,2a-1}$.
The operators  ${\mathsf U}_{i,a}$ and  ${\mathsf V}_{i,a}\equiv(\chi_{i,2a-1}^{})^{\frac{1}{2}}$ 
satisfy the defining relations
of the algebra $\mathcal{W}$. We may furthermore introduce the operators 
$\mathsf{D}_{i,a}:=(\su_{i,2a})^{-1}\su_{i,2a-1}$
commuting with ${\mathsf U}_{i,a}$ and  ${\mathsf V}_{i,a}$.
The representation of $\CW_M\ot\CW_M$ 
defined on a Hilbert-space $\hat{\CH}_a\simeq L^2(\BR^{2M})$ in terms of the operators $\su_{i,2a}$, $\sv_{i,2a}$,
$\su_{i,2a-1}$ and $\sv_{i,2a-1}$ is then unitarily equivalent 
to a representation on a Hilbert space represented by wave-functions 
$\psi(y_a,c_a)\in L^2(\BR^{2M})$ such that
${\mathsf U}_{i,a}$, ${\mathsf V}_{i,a}$   $\mathsf{C}_{i,a}$ and $\mathsf{D}_{i,a}$ are represented as
\begin{equation}\label{auxrep}
\begin{aligned}
&{\mathsf U}_{i,a}\,\psi(y_a,c_a)=\psi(y_a+\mathrm{ i}b\ep_{i},c_a)\,,\\
&{\mathsf V}_{i,a}\,\psi(y_a,c_a)=e^{\pi b(y_{i,a}-y_{i+1,a})}\,\psi(y_a,c_a)\,,
\end{aligned}
\qquad
\begin{aligned}
&{\mathsf C}_{i,a}\,\psi(y_a,c_a)=c_{i,a}\,\psi(y_a,c_a)\,,\\
&{\mathsf D}_{i,a}\,\psi(y_a,c_a)=\psi(y_a,c_a+\mathrm{i}b\ep_i)\,,
\end{aligned}\end{equation}
where $\ep_i$ is the vector in $\BR^M$ with $j$-th component being $\de_{i,j}-\frac{1}{M}$.
The vectors  in $\CH=\bigotimes_{a=1}^N \CH_{2a}\ot \CH_{{2a-1}}$ will accordingly be represented 
by wave-functions $\Psi(y,c)\in L^2(\BR^{2MN})$, where $y=(y_1,\dots,y_N)$, $c=(c_1,\dots,c_N)$.

If $\check{R}_{\bar\mu,\mu;\bar\nu,\nu}(x,y|x',y')$ is the kernel representing 
$\check{\mathcal{R}}(\bar\mu,\mu;\bar\nu,\nu)$ we may represent the fundamental transfer matrix
${\mathcal{T}}(\bar\mu,\mu;\bar\nu,\nu)$ as an
integral operator of the form
\begin{equation}
\big({\mathcal{T}}(\bar\mu,\mu;\bar\nu,\nu)\Psi\big)(y,c)=
\int d\mu_N^{}(y)\;T_{\bar\mu,\mu;\bar\nu,\nu}(y,y')\,\Psi(y,c)\,,
\end{equation}
with $d\mu_N^{}(y)=\prod_{a=1}^N d\mu(y_a)$, and the kernel $T_{\bar\mu,\mu;\bar\nu,\nu}(y,y')$ given as
\begin{equation}
T_{\bar\mu,\mu;\bar\nu,\nu}(y,y')=\int d\mu_N(x)\;
\prod_{a=1}^N \check{R}_{\bar\mu,\mu;\bar\nu,\nu}(x_{a+1}^{},y_a^{}|x'_a,y'_a)\,.
\end{equation}

It is finally not hard to see that the same kernel $T_{\bar\mu,\mu;\bar\nu,\nu}(y,y')$ can be used to represent 
the projection  $\mathsf{T}(\bar\mu,\mu;\bar\nu,\nu)$
of $\mathcal{T}(\bar\mu,\mu;\bar\nu,\nu)$ to the physical Hilbert space defined in Section
\ref{Phys-Obs}. Indeed, $\mathcal{T}(\bar\mu,\mu;\bar\nu,\nu)$ is a physical observable and there exists a
representation of the form \rf{intop}.  Such a representation is related to the representation defined
above in \rf{auxrep} by a gauge transformation $\Psi'(y,c)=e^{i\eta(y,c)}\Psi(y,c)$, in general. Such a 
gauge transformation modifies the kernel
$T_{\bar\mu,\mu;\bar\nu,\nu}(y,y')$ into 
$T_{\bar\mu,\mu;\bar\nu,\nu}(y,y')e^{i(\eta(y',c)-\eta(y,c))}$. The projection defined in Section
\ref{Phys-Obs} then has kernel 
$T_{\bar\mu,\mu;\bar\nu,\nu}(y,y')e^{i(\eta(y',\mathbf{1})-\eta(y,\mathbf{1}))}$.
The factor $e^{i(\eta(y',\mathbf{1})-\eta(y,\mathbf{1}))}$ can be removed by another gauge-transformation,
if necessary.

\section{Imaginary roots and functional relations I}
\label{Imrenorm-2}

Let us now consider the definition of the
imaginary root contributions to the  
R-matrices.
This turns out to be  more delicate than the case of the real
root contributions. The formula \rf{rpd} does not seem to have a natural renormalized 
counterpart at first sight. We are going to argue 
that the decisive requirement determining a canonical renormalisation of the 
imaginary root contributions will be the consistency with taking tensor products, 
or equivalently the validity of the conditions 
\begin{equation}\label{tensorR}
\begin{aligned}
&\SR_{V_1\ot V_2,V_3}\,=\,\SR_{V_1,V_3}\,\SR_{V_2,V_3}\,,\\
&\SR_{V_1, V_2\ot V_3}\,=\,\SR_{V_1,V_3}\,\SR_{V_1,V_2}\,.
\end{aligned}
\qquad
\SR_{V,W}:=(\pi_V\ot\pi_W)(\CR)\,,
\end{equation}
obtained by evaluating
the representation $\pi_{V_1}\ot\pi_{V_2}\ot\pi_{V_3}$ on 
$(\De\otimes\id)(\CR)=\CR_{13} \CR_{23}$ and
$(\id\otimes\De)(\CR)=\CR_{13} \CR_{12}$, respectively.

Our renormalisation prescription can be directly applied to both sides in \rf{tensorR} whenever
the infinite products representing the universal R-matrices truncate to finite ones in the given 
representations.
%
%
 This happens when one of the representations applied to the universal R-matrix
is of prefundamental type. 
A natural  strategy to construct families of operators $\SR_{V,W}$ satisfying \rf{tensorR} 
is of course to start by identifying a class of basic representations
from which more general ones may be constructed by taking tensor products and quotients.
Having defined $\SR_{V,W}$ for $V$, $W$ taken from the class of basic representations
one may simply use \rf{tensorR} recursively to extend the definition to more general representations. 
Whenever our renormalisation prescription can be applied to define all representations 
appearing in \rf{tensorR}
one needs to check explicitly that the relations following from \rf{tensorR}
 are satisfied.

We will apply this strategy using as basic representations the prefundamental representations of 
modular double type on the one hand, and the finite-dimensional representations on the other 
hand. It turns out, in particular, that the renormalisation prescriptions for the basic representations
are strongly constrained by the already chosen definitions for the real root contributions. 
The co-product mixes real and imaginary roots. This implies that
a part of the imaginary root contributions 
in $\SR_{V_1\ot V_2,V_3}$ is given by the real root contributions in 
$\SR_{V_1,V_3}$ and $\SR_{V_2,V_3}$, and similarly for $\SR_{V_1, V_2\ot V_3}$. 
The renormalisation prescriptions for real and imaginary roots
must therefore be related to each other. Consideration of tensor products of finite- and infinite-dimensional 
representations similarly implies relations between the prescriptions adopted in the two types of 
representations, respectively.

It may furthermore happen, for example, that the tensor product of representations
becomes reducible for certain values of the relevant parameters, containing basic representations
in sub-representations or quotients. Whenever this happens, it implies relations between
the imaginary root contributions to the respective R-matrices, as will be shown explicitly in 
some relevant examples. These relations take the form of certain functional relations 
restricting possible renormalisation prescriptions for the imaginary root contributions 
considerably. 

These considerations will lead us to a uniform and unambiguous prescription for the renormalisation 
of the imaginary root contributions for the whole family of representations of our interest. 
Most important for applications to integrable lattice models is the observation that 
the proper treatment of the imaginary root contributions provides the basis for 
the representation theoretic derivation of the Baxter equation, generalising 
the approach of \cite{BLZ3,AF} to the case of representations without extremal weight.

In order to make the overall logic transparent we will in this section restrict attention to the 
case of $\CU_q(\widehat{\fsl}_2)$. In the general case of $\CU_q(\widehat{\fsl}_M)$ one is 
facing a higher algebraic complexity which will be dealt with in the next section.


\subsection{Imaginary roots for basic representations}

To begin with, we shall compute the imaginary root contributions for the basic 
representations of finite-dimensional or prefundamental  type. 

\subsubsection{Prefundamental representations}\label{Imroot-pref}

As a warm-up, let us consider the case $M=2$, where the imaginary root contribution
to the universal R-matrix, see \eqref{rpd}, \eqref{umij}, simplifies to  the following form
\begin{equation}
\mathscr{R}^-_{\sim \delta} = \exp \bigg( -(q - q^{-1}) \sum_{k =1}^\infty
\frac{k}{[2k]_{q}}\, f^{(1)}_{k \delta} \otimes e^{(1)}_{k \delta} \bigg),
\label{rimM=2}
\end{equation}
An important feature of 
the representations $\pi_\la^\pm$  is the fact that
$e^{(1)}_{k\delta} $ and $f^{(1)}_{k \delta}$ 
get represented by central elements. The corresponding currents take the form
\begin{align}\label{E1prefund}
& \pi_\la^-(1+E_1'(z))\,=\,1+\la^{-2} z^{-1}\quad\Leftrightarrow \quad 
\pi_\la^-(e^{(1)}_{k\de})
\,=\,\frac{(-1)^{k+1}}{q^{-1}-q^{}}\frac{\la^{-2k}}{k}\,,\\
& \pi_\la^+(1+F_1'(z))\,=\,1+\la^{+2} z^{-1}\quad\Leftrightarrow \quad 
\pi_{\la}^+(f^{(1)}_{k\de})\,=\,\frac{(-1)^{k+1}}{q-q^{-1}}\frac{\la^{+2k}}{k}\,.
\label{F1prefund}
\end{align}
These equations follow straightforwardly from the definitions \eqref{piplus}, \eqref{piminus} and the 
iterative construction of imaginary root vectors given in  Section \ref{sec:rootgen}.
For $|q|\neq 1$  we therefore get
\begin{equation}
\begin{aligned}\label{rhopm}
\rho^{+-}(\la\mu^{-1}): &=(\pi^+_\la\ot\pi^-_\mu)(\mathscr{R}^-_{\sim \delta})
=\exp \Bigg( -\sum_{k =1}^\infty \frac{1}{k}\,
\frac{(-1)^{k+1}}{q^{2k}-q^{-2k}}\,
\bigg(-\frac{\la^2}{\mu^2}\bigg)^{k}\Bigg)\\
&=(\vep_{q^2}(-\la^2/\la^2))^{-1} \,,
\end{aligned}
\end{equation}
compare to \eqref{thetadef}.
Following the discussion in Section \ref{canren} we can immediately
suggest the following renormalized version of this special
function,
\begin{equation}\label{rhodef}
\rho^{+-}_{\rm ren}(\la\mu^{-1})\equiv
[(\pi^+_\la\ot\pi^-_\mu)(\mathscr{R}^-_{\sim \delta})]_{\rm ren}^{}:=
(\CE_{2b^2}(-\la^2/\mu^2))^{-1}\,,
\end{equation}
where
\begin{equation}\label{Eb-def}
\CE_{\hbar}(w):=\exp\bigg(\int_{\BR+i0}\frac{dt}{4t}\,
\frac{w^{-\frac{i}{\pi}t}}{\sinh(\hbar t)\sinh(t)}\bigg)\,.
\end{equation}
Note that $\CE_{\hbar}(w)$ is not single-valued in $w$, it is better understood
as a function of $\log(w)$. The definition \rf{rhodef} 
therefore needs to be supplemented by a choice of the logarithm of 
$-\la^2/\mu^2$. This 
is a subtle issue that will be resolved in Section \ref{branch} below.

\subsubsection{Evaluation representations}

By means of straightforward computations one may show that
the image of imaginary root currents under the evaluation map introduced in 
Section \ref{evHOMOsl2} takes the form 
\begin{subequations}
\begin{align}\label{evimaginaryrootsSL2}
\text{ev}_\la(1+E'_1(z))\,
&=\,
\frac{(1+q\,z^{-1}\,\la^{-2}\,q^{+2x})(1+q\,z^{-1}\,\la^{-2}\,q^{-2x})}
{(1+q^{-1}\,z^{-1}\,\la^{-2}\,q\,\SK^2)(1+q^{+1}\,z^{-1}\,\la^{-2}\,q\,\SK^2)}\,,\\
\text{ev}_\la(1+F'_1(z))\,
&=\,
\frac{(1+q^{-1}\,z^{-1}\,\la^{2}\,q^{+2x})(1+q^{-1}\,z^{-1}\,\la^{2}\,q^{-2x})}
{(1+q^{-1}\,z^{-1}\,\la^{2}\,q^{-1}\,\SK^{-2})(1+q^{+1}\,z^{-1}\,\la^{2}\,q^{-1}\,\SK^{-2})}\,.
\end{align}
\end{subequations}
We recall that $q^{2x}$,  where $x$ is defined up to a sign, parametrizes the 
$\CU_q(\fsl_2)$ Casimir as in    \eqref{sl2casimir}.


Considering finite dimensional representations of evaluation type one may note that
the imaginary root  currents for $\pi^{\rm f.d.}_{\la,j}$
 take the form \eqref{evimaginaryrootsSL2} 
 with $q^{2x}$ and  $\mathsf{K}$ replaced by  $q^{2j+1}$  and 
 $\mathsf{k}_j:=\text{diag}\left(q^{j},q^{j-1},\dots,q^{-j+1},q^{-j}\right)$ respectively.
Taking the second tensor factor to be $\pi^-_{\mu}$,  one could proceed along the lines
of Section \ref{Imroot-pref},  leading to 
\begin{equation}\label{imrootmixed}
[(\pi^{\rm f.d.}_{\la,j}\ot\pi^-_\mu)(\mathscr{R}^-_{\sim \delta})]_{\rm ren}^{}\,=\,
\frac{\CE_{2b^2}(-q^{-2}\,\mathsf{k}^{-2}_j\,\la^2/\mu^2)}{\CE_{2b^2}(-q^{-2j-2}\la^2/\mu^2)}
\frac{\CE_{2b^2}(-\mathsf{k}_j^{-2}\,\la^2/\mu^2)}{\CE_{2b^2}(-q^{2j}\la^2/\mu^2)}\,.
\end{equation}
If we further specialize \eqref{imrootmixed} to the case of spin $j=1/2$ we find
\begin{equation}\label{pifpi-IMAG}
[(\pi^{\rm f}_{\la}\ot\pi^-_\mu)(\mathscr{R}^-_{\sim \delta})]_{\rm ren}^{}=
\begin{pmatrix}
\frac{\CE_{2b^2}(-q^{-1}\,\la^2/\mu^2)}{\CE_{2b^2}(-q^{+1}\,\la^2/\mu^2)}\,
&
0
\\
0 &
\frac{\CE_{2b^2}(-q^{-1}\,\la^2/\mu^2)}{\CE_{2b^2}(-q^{-3}\la^2/\mu^2)}
\end{pmatrix}=
\theta(\la/\mu)
\begin{pmatrix}
1 & 0 \\
0 &  1-q^{-1}\la^2\mu^{-2}
\end{pmatrix}.
\end{equation}
Apart from defining the special function $\theta(z)$, the second 
equality in this equation follows from the relation 
$\CE_{2b^2}(q^2x)=(1+x)\,\CE_{2b^2}(q^{-2}x)$  applied in the case when 
$x=-q^{-1}\la^2\,\mu^{-2}$.
Let us observe that
\begin{equation}\label{thetaandrhopm}
\theta(\la)\,=\,
\frac{\rho^{+-}_{\text{ren}}(q^{+\frac{1}{2}}\la)}
{\rho^{+-}_{\text{ren}}(q^{-\frac{1}{2}}\la)}\,,
\end{equation}
where $\rho^{+-}_{\text{ren}}(\la)$ is given in 
\eqref{rhopm}.
%
%
Another example that will be useful in the following is
\begin{equation}\label{rdeltaevSL2}
[(\pi^{\rm f}_{\la}\ot\text{ev}_{\mu})(\mathscr{R}^-_{\sim \delta})]_{\rm ren}^{}=
\rho_{\text{ev}}(\la\mu^{-1})
\begin{pmatrix}
1-\la^2\mu^{-2}\,q\,\mathsf{K}^2 & 0 \\
0 & \frac{(1-\la^2\mu^{-2}\,q^{+2x} )(1-\la^2\mu^{-2}\,q^{-2x})}
{(1-\la^2\mu^{-2}\,q^{-1}\,\mathsf{K}^{2} )}
\end{pmatrix}
\end{equation}
where
\begin{equation}\label{rhoevaluationSL2}
\rho_{\text{ev}}(\la\mu^{-1})\,=\,
\frac{\mathcal{E}_{2b^2}(-\la^2\mu^{-2}\, q^{2x})\mathcal{E}_{2b^2}(-\la^2\mu^{-2}\, q^{-2x})}
{\mathcal{E}_{2b^2}(-\la^2\mu^{-2}\, q^{2+2x})\mathcal{E}_{2b^2}(-\la^2\mu^{-2}\, q^{2-2x})}\,=\,
\theta(q^{\frac{1}{2}+x}\la\mu^{-1})\,\theta(q^{\frac{1}{2}-x}\la\mu^{-1})\,\,.
\end{equation}
This result can be easily specialized to the modular double case as 
$\pi^{\text{m.d.}}_s(q^{\pm x})\,=\,-e^{\pm \pi b\,s}$.

\subsubsection{$\mathsf{L}^{\pm}(\la)$ from the renormalized universal R-matrix}
\label{sec:LpmfromuniSL2}

Let us now complete the derivation started in Section \ref{prefunddef} to obtain 
$\mathsf{L}^{\pm}(\la)$ from the renormalized product formula for the universal R-matrix.

As explained in the following section, the  real root contribution is not affected by renormalization 
in this case as the corresponding root vectors are realized by nilpotent operator in one tensor factor. 
The observations in Section \ref{prefunddef}  
 together with the calculation \eqref{pifpi-IMAG} then gives
\begin{align}\label{L-factor0}
\left[(\pi^{\rm f}_{\la}\ot\pi^{-}_1)(\CR^-)\right]_{\rm ren}^{}&=
\vep_{q}(-\tau_q^2\,
\la\,
\big(\begin{smallmatrix} 0 & 0 \\ 1 & 0 
\end{smallmatrix}\big)
\,\se_1)\cdot[(\pi^{\rm f}_{\la}\ot\pi^{-}_1)(\CR^-_{\sim\de})]^{}_{\rm ren}\cdot
\vep_q(-\tau_q^2\,\la\,
\big(\begin{smallmatrix} 0 & 1 \\ 0 & 0 
\end{smallmatrix}\big)
\,\se_0)\cdot q^{-\mst}\notag \\ 
& =\,\theta(\la)\,
\bigg(\begin{matrix}
1  & 0               \\ \la\,\sv\,\su^{-1} & 1
\end{matrix}\bigg)\bigg(\begin{matrix}
1   & 0                \\ 0                  &  1-q^{-1}\la^2
\end{matrix}\bigg)\bigg(\begin{matrix}
1  & \la\,\sv^{-1}\su \\ 0            & 1
\end{matrix}\bigg)\bigg(\,\begin{matrix} \su & 0 \\ 0  & \su^{-1}\end{matrix}\bigg)\\
& =\,\theta(\la)
\bigg(\begin{matrix}
\su  & \la\,\sv^{-1}               \\ \la\,\sv\, & \su^{-1}
\end{matrix}\bigg)\,.
\notag\end{align}
 For completeness let us recall the evaluation of the universal R-matrix for $\pi^{\rm f}_{\la}\ot\text{ev}_{\mu}$
 and how it is affected by the regularization.
 The infinite product of real root vectors gives
 \begin{equation}
(\pi^{\rm f}_{\la}\ot\text{ev}_{\mu})(\mathscr{R}^-_{\prec \delta})=
\begin{pmatrix}
1& 0 \\
\frac{\la\mu^{-1} (q^{-1}-q)}{1-\la^2\mu^{-2}\,q^{-1}\,\mathsf{K}^2 }\,q^{\frac{1}{2}}\mathsf{K}^{-1}\mathsf{E} & 1
\end{pmatrix}\,,
\end{equation}
\begin{equation}
(\pi^{\rm f}_{\la}\ot\text{ev}_{\mu})(\mathscr{R}^-_{\succ \delta})=
\begin{pmatrix}
1& 
\frac{\la\mu^{-1} (q^{-1}-q)}{1-\la^2\mu^{-2}\,q^{+1}\,\mathsf{K}^2 }\,
q^{\frac{1}{2}}\mathsf{K}^{+1}\mathsf{F} 
\\
0& 1
\end{pmatrix}\,.
\end{equation}
Together with \eqref{rdeltaevSL2} this implies that
\begin{equation}\label{LevSL2form}
[(\pi^{\rm f}_{\la}\ot\text{ev}_{\mu})(\mathscr{R}^-)]_{\rm ren}^{}=
\rho_{\text{ev}}(\la\mu^{-1})\,
\begin{pmatrix}
\mathsf{K}^{-1}-\la^2\mu^{-2}\,q\,\mathsf{K}^{+1} & \la\mu^{-1}(q^{-1}-q)\,q^{+\frac{1}{2}}\,\widetilde{\mathsf{F}} \\
\la\mu^{-1}(q^{-1}-q)\,q^{+\frac{1}{2}}\,\widetilde{\mathsf{E}} & \mathsf{K}^{+1}-\la^2\mu^{-2}\,q\,\mathsf{K}^{-1} \end{pmatrix}
\end{equation}
where $\widetilde{\mathsf{E}} =\mathsf{K}^{-1}\,\mathsf{E}\,\mathsf{K}^{-1}$,
$\widetilde{\mathsf{F}} =\mathsf{K}^{+1}\,\mathsf{F}\,\mathsf{K}^{+1}$
and  $\rho_{\text{ev}}(\la\mu^{-1})$ is given in 
\eqref{rhoevaluationSL2}.

\subsection{Rationality of currents}
\label{sec:ImrootRATIONAL_sl2}

The examples above lead us to a useful observation: 
An important role is played by the generating functions $1+E_1'(z)$ and $1+F_1'(z)$ 
that will be called currents. The currents generate a commutative algebra for 
the level zero representations we are considering. Whenever the 
currents get represented by rational functions
of $z$ there exists  a natural prescription for turning the formal series following from 
\rf{rimM=2} into well-defined operators. We are now going to show that the operators 
representing $1+E'_1(z)$ 
and $1+F'_1(z)$ will be
rational functions of $z$ for all representations of our interest. 
More precisely we shall show that for any tensor product $\boldsymbol{\pi}^-=\pi_1\otimes\dots\otimes\pi_N$
of basic representations of $\Borelp$ 
we have 
\begin{align}\label{E'rational}
\boldsymbol{\pi}^-(1+E'_1(z))\,&=\, 
\frac{
\prod_{\ell=1}^{n_-}
(1+z^{-1}\,\mathsf{N}_{\ell}^-)
}{\prod_{\ell=1}^{d_-}
(1+z^{-1}\,\mathsf{D}_{\ell}^-)}
\end{align}
 where 
$\mathsf{N}_{\ell}^-$, $\mathsf{D}_{\ell}^-$ are mutually commutative operators.
 A very similar statement holds
for tensor products of basic representations of $\Borelm$.

In order to derive \rf{E'rational}, let us consider 
the monodromy matrix
\begin{equation}
\SM(\la):=(\pi^{\rm f}_{\la}\ot\boldsymbol{\pi}^-)(\CR^-)\,.
\end{equation} 
It follows
from the product formula for $\CR^-$ that we may represent $\SM(\la)$ in the form
\begin{equation}
\SM(\la)\,=\,
\bigg(\,\begin{matrix} 1 & 0 \\ \SF(\la) & 1 \end{matrix}\,\bigg)
\bigg(\,\begin{matrix} \SK_+(\la) & 0 \\ 0 & \SK_{-}(\la) \end{matrix}\,\bigg)
\bigg(\,\begin{matrix} 1 & \SE(\la)\\ 0 & 1 \end{matrix}\,\bigg)
\bigg(\,\begin{matrix} \sk^{-1} & 0 \\ 0  & \sk \end{matrix}\,\bigg)\,,
\end{equation}
where $\SK_\pm(\la)$ are the eigenvalues of 
$(\pi^{\rm f}_{\la}\ot\boldsymbol{\pi}^-)(\CR^-_{\sim\delta})$ on $u_\pm$,
\begin{equation}
\SK_\pm(\la) = \exp \bigg( (q^{-1} - q^{+1}) 
\sum_{m \in \BZ_+} u_{m, 11} \, F^{(1)}_{m \delta, \pm} 
\,\pi_+(e^{(1)}_{m \delta}) \bigg),
\end{equation}
where 
the numbers $F^{(1)}_{m \delta, \pm}$ 
are the eigenvalues of $\pi^{\rm f}_{\la}(f^{(1)}_{m \delta})$ on the two basis 
vectors $u_\pm$ of $\BC^2$, 
$\pi^{\rm f}_{\la}(f^{(1)}_{m \delta})\,u_\pm = F^{(1)}_{m \delta,\pm}\,u_\pm$.
It follows straightforwardly from \rf{evimaginaryrootsSL2}
that  
\begin{equation}
F^{(1)}_{m \delta,-}-F^{(1)}_{m \delta,+}\,=\,(u_{m, 11})^{-1}
(-q^{-1}\la^2)^m\,.
\end{equation}
This implies that
\begin{equation}\label{currentKpm}
\boldsymbol{\pi}^-\big(1+E'_1(-q\la^{-2})\big)\,=\,
\frac{\SK_-(\la)}{\SK_{+}(\la)}\,.
\end{equation}
We may note, on the other hand, 
that for any basic representation 
$\pi_k$ the matrix 
$L_n(\la)=(\pi^{\rm f}_{\la}\ot\pi_n)(\CR^-)$ takes the form $\rho_n(\la){L}_n'(\la)$,
with $L'(\la)$ polynomial in $\la$. It follows that
$\SM(\la)={\SM}'(\la)\prod_{n=1}^N\rho_n(\la)$, where 
\begin{equation}
{\SM}'(\la):={L}_N'(\la){L}_{N-1}'(\la)\cdots {L}_1'(\la)\,=\,\bigg(\begin{matrix} \SA(\la) & \SB(\la) \\ \SC(\la) & \SD(\la) 
\end{matrix}\bigg)\,,
\end{equation}
is a matrix of polynomials in $\la$ such that
$\SA(\la)=\sk^{-1}+\CO(\la)$, $\SD(\la)=\sk+\CO(\la)$,  
$\SB(\la)=\CO(\la)$, $\SC(\la)=\CO(\la)$.

It remains to observe that both $\SK_+(\la)$ and 
$\SK_-(\la)$ can be expressed
as a rational function of the matrix elements of $\SM(\la)$, leading to the expression
\begin{equation}\label{KABCD}
\frac{\SK_-(\la)}{\SK_+(\la)}\,=\,\frac{\text{q-det}\big({\SM}'(q^{-\frac{1}{2}}\la)\big)}{\SA(\la)\,\SA(q^{-1}\la)}\,\sk^{-2}\,,
\end{equation}
where $\text{q-det}({\SM}'(\la))$ is defined as
\begin{equation}
\text{q-det}({\SM}'(\la))\,=\,
\SA(q^{-\frac{1}{2}}\la)\SD(q^{\frac{1}{2}}\la)-
\SC(q^{-\frac{1}{2}}\la)\SB(q^{\frac{1}{2}}\la)\,.
\end{equation}
In order to obtain formula \rf{KABCD} we used the commutation relations satisfied by  the matrix entries of ${\SM}'(\la)$.
Equations \rf{KABCD} and
\rf{currentKpm} imply that 
$\boldsymbol{\pi}^-(1+E'_1(z))$ is a rational function of $\la$ of the form
claimed in \rf{E'rational}. 
\hfill$\square$

 With these observations in mind, 
let us formulate the prescription:
for representation $\boldsymbol{\pi}^{\pm}$  of $\Borelmp$ such that
\begin{equation}\label{formpart1SL2}
\boldsymbol{\pi}^{+}(1+F_1'(z))=
\frac{
\prod_{\ell=1}^{n_+}
(1+z^{-1}\,\mathsf{N}_{\ell}^+)
}{\prod_{\ell=1}^{d_+}
(1+z^{-1}\,\mathsf{D}_{\ell}^+)}\,,
\quad
\boldsymbol{\pi}^{-}(1+E_1'(z))=
\frac{
\prod_{\ell=1}^{n_-}
(1+z^{-1}\,\mathsf{N}_{\ell}^-)
}{\prod_{\ell=1}^{d_-}
(1+z^{-1}\,\mathsf{D}_{\ell}^-)}
\end{equation}
let us set
\begin{equation}\label{formpart2SL2}
\big{[}
\left(\boldsymbol{\pi}^{+}\ot \boldsymbol{\pi}^{-}\right)
\CR^-_{\sim\de}
\big{]}_{\text{ren}}
=
\frac{
\prod_{\ell=1}^{d_+}
\prod_{\ell'=1}^{n_-}
\CE_{2b^2}(-\mathsf{D}_{\ell}^+\ot \mathsf{N}_{\ell'}^-)
}
{
\prod_{\ell=1}^{n_+}
\prod_{\ell'=1}^{n_-}
\CE_{2b^2}(-\mathsf{N}_{\ell}^+\ot \mathsf{N}_{\ell'}^-)
}
\frac
{
\prod_{\ell=1}^{n_+}
\prod_{\ell'=1}^{d_-}
\CE_{2b^2}(-\mathsf{N}_{\ell}^+\ot \mathsf{D}_{\ell'}^-)
}
{
\prod_{\ell=1}^{d_+}
\prod_{\ell'=1}^{d_-}
\CE_{2b^2}(-\mathsf{D}_{\ell}^+\ot \mathsf{D}_{\ell'}^-)
}\,,
\end{equation}
where  $\CE_{2b^2}(w)$ is defined in \eqref{Eb-def} below.
Notice that the unrenormalized version of \eqref{formpart2SL2}
is the same expression with $\CE_{2b^2}(w)$ replaced by $\varepsilon_{q^2}(w)$.
Above we  used the notation  $\boldsymbol{\pi}^{\pm}$ 
in order to avoid confusion with  the prefundamental representations
$\pi^{\pm}_\la$, which are a special case of $\boldsymbol{\pi}^{\pm}$.

As we will see,
after we fix a prescription  of the form \eqref{formpart1SL2}, \eqref{formpart2SL2}
 for the prefundamental representations $\pi^{\pm}_{\la}$,
the validity of the relations following from \rf{tensorR}
implies that the same prescription \eqref{formpart1SL2}, \eqref{formpart2SL2} 
needs to be used for representations obtained by taking tensor products.
The fact that this is a consistent prescription  is not obvious.  
We will show in all relevant cases that consistency follows
from  the  basic functional relations satisfied by $\CE_{b^2}(w)$.





\subsection{Co-product of imaginary roots}
\label{sec:Compatsub}

We have proposed a definition for the imaginary root contributions to the universal R-matrix
for the basic representations of our interest. We will now start analysing in some detail
if this definition is compatible with the relations \rf{tensorR}. To this aim 
we will now derive useful identities, formulae \rf{BEid} and \rf{BEidminus} below, 
satisfied by a generating function for the imaginary root 
generators from the basic relations
$(\id\ot\De)(\CR^-)=\CR^-_{13}\CR^-_{12}$
and $(\De\ot\id)(\CR^-)=\CR^-_{13}\CR^-_{23}$.

As a useful generating function for the imaginary root generators let us introduce
$\mathscr{M}^+_{\sim\de}(\la)$ via
\begin{equation}\label{Mdeltaplusdef}
1\otimes \mathscr{M}^+_{\sim\de}(\la)\,
:=\,(\pi^+_\la\ot \id)(\CR^-_{\sim\de})\,.
\end{equation}
This definition makes sense as $\pi^+_{\la}(f^{(1)}_{k\de})$ are complex numbers. For the time being
we shall continue to work with formal series in $\la$.
We are going to prove the identity
\begin{equation}\label{BEid}
\De(\mathscr{M}^+_{\sim\de}(\la))\,
=\,\left(1\ot\mathscr{M}^+_{\sim\de}(\la)\right)\, 
\varepsilon_q\big((\tau_q\la)^{2}e_1\ot e_0k_1\big)\,
\left(\mathscr{M}^+_{\sim\de}(\la)\ot 1\right)\,,
\end{equation}
giving a useful representation of the co-product of the imaginary root 
generators. The contribution containing real root generators is clearly visible in the 
argument of the function $\varepsilon_q(x)$.

As a preparation let us note that $\mathscr{M}^+_{\sim\de}(\la)$ appears in 
\begin{equation}\label{R-factor1}
(\pi^+_\la\ot \id)(\CR^-)\,=\,
\varepsilon_{q}(-\tau_q^2\,\sf_1\ot e_1)\,
\left(1\otimes \mathscr{M}^+_{\sim\de}(\la)\right)\,
\varepsilon_q(-\tau_q^2\,\sf_0\ot e_0)\,
\Lambda^{-1}(\su)\,,
\end{equation}
where $\Lambda^{-1}(\su)=e^{-\log\su\ot (\bar{\epsilon}_1-\bar{\epsilon}_2)}$,
$\sf_i:=\pi^+_{\la}(f_i)$,
$\tau_q=q-q^{-1}$. This may be rewritten  as
\begin{equation}\label{R-factor1BIS}
(\pi^+_\la\ot \id)(\CR^-)=
\Lambda(\mathsf{y})
\left(1\ot
\mathscr{M}^+(\la)\right)
\Lambda^{-1}(\mathsf{y})
\Lambda^{-1}(\mathsf{u})\,,
\end{equation}
\begin{equation}\label{DefMplusFULL}
\mathscr{M}^+(\la)\,:=\,
\varepsilon_{q}({\la}' e_1)\,
 \mathscr{M}^+_{\sim\de}(\la)
\varepsilon_q({\la}' e_0)
\end{equation}
where $\mathsf{y}:=\su^{-\frac{1}{2}}\sv\,\su^{-\frac{1}{2}}$, 
$\Lambda(\mathsf{y})=
e^{\frac{1}{2}\log\mathsf{y}\otimes (\bar{\epsilon}_1-\bar{\epsilon}_2)}$,
and ${\la}':=-\tau_q\,q^{\frac{1}{2}}\,\la$. 
It seems remarkable that there is a similarity transform $\Lambda(\mathsf{y})$
so that the first tensor factor in \eqref{R-factor1BIS} is the identity.
This follows from the identities
\begin{equation}
(q-q^{-1})\,\,\pi^{+}_\la(f_i)\ot e_i\,=\,
\la\,q^{\frac{1}{2}}\,
\Lambda(\mathsf{y})\left(1\ot e_i\right)\Lambda^{-1}(\mathsf{y})\,.
\end{equation}
 The rewriting \eqref{R-factor1BIS}
will be particularly useful in the higher rank case 
discussed in Section \ref{sec:ImaginaryrootMORE}.

\noindent \emph{Proof of \rf{BEid}.} 
The starting point of our derivation is the identity 
\begin{equation}\label{tensorRBIS}
(\pi^+_\la\otimes\De)(\CR^-)\,=\,
(\pi^+_\la\otimes1)(\CR^-_{13})\,
(\pi^+_\la\otimes1)(\CR^-_{12})\,.
\end{equation}
Inserting the form \eqref{R-factor1BIS}  into this equation 
and simplifying the $\Lambda$ factors we obtain
\begin{equation}\label{tensorRTRIS}
\Delta\left(\mathscr{M}^+(\la)\right)\,=\,
\left(1\ot \mathscr{M}^+(\la)\right)
q^{t}\,
\left(\mathscr{M}^+(\la)\ot 1\right)
q^{-t}\,,
\end{equation}
where we have cancelled the first tensor factor being proportional to the identity.
The contribution $q^t$ originates from reordering the factors of $\Lambda$. 
It acts as $q^t(e_i\ot 1)q^{-t}=e_i\ot k_i$.
The left hand side of \eqref{tensorRTRIS} 
contains terms 
$\varepsilon_{q}({\la}'\Delta(e_i))$,
 $i=0,1$, $\Delta(e_i))=e_i\ot k_i+1\ot e_i$
  which may be further factorized using 
\begin{equation}\label{qexp-prop1}
\varepsilon_{q}(U)\,\varepsilon_{q}(V)\,=\,\varepsilon_{q}(U+V)\,,
\end{equation}
if $UV=q^{-2}VU$.
Using \rf{qexp-prop1} we rewrite \eqref{tensorRTRIS} as 
\begin{equation}
\De(\mathscr{M}^+_{\sim\de}(\la))
=(1\ot\mathscr{M}^+_{\sim\de}(\la))\,\Theta\,
(\mathscr{M}^+_{\sim\de}(\la)\ot 1)\,,
\end{equation} where
\begin{equation}
\Theta:=
\frac{1}{\varepsilon_{q}({\la}'e_1\ot k_1)}\,
\varepsilon_{q}(1\ot{\la}'e_0)\,
\varepsilon_{q}({\la}'e_1\ot k_1)\,
\frac{1}{\varepsilon_{q}({\la}'e_1\ot k_1)}\,.
\end{equation}
This expression can be simplified using the pentagon relation
\begin{equation}\label{qexp-prop2}
\varepsilon_{q}(V)\,\varepsilon_{q}(U)
\,=\,\varepsilon_{q}(U)\,\varepsilon_{q}(qUV)\,\varepsilon_{q}(V)\,,
\end{equation}
and noting that $q^{-1}({\la}')^2=(\la \tau_q)^{2}$.
The resulting formula is \rf{BEid}, as claimed.\hfill $\square$

Applying $(\id \ot \pi^-_{\nu})$  to the second equation  in \eqref{tensorR}
a similar analysis shows that 
\begin{equation}\label{BEidminus}
\De(\mathscr{M}^-_{\sim\de}(\nu))\,
=\,\left(\mathscr{M}^-_{\sim\de}(\nu)\ot 1\right)\, 
\varepsilon_q\big((\tau_q\nu^{-1})^{2}\,k_0f_0\ot f_1\big)\,
\left(1\ot\mathscr{M}^-_{\sim\de}(\nu)\right)\,.
\end{equation}
where
\begin{equation}\label{Mdeltaminusdef}
 \mathscr{M}^-_{\sim\de}(\nu)\ot 1\,
:=\,(\id \ot \pi^-_\nu)(\CR^-_{\sim\de})\,.
\end{equation}
It is worth to observe that
\begin{equation}\label{DELTAf1deltaSL2}
\Delta(f^{(1)}_{\delta})=
f^{(1)}_{\delta}\ot 1+
1\ot f^{(1)}_{\delta}+
\tau_q\,[2]_q\,k_0f_0\ot f_1\,,
\end{equation}
where we have used the iterative definition 
$f^{(1)}_{\delta}=f_0f_1-q^{-2}f_1f_0$ given in 
\eqref{iterativeIMrootsandmore}.
Before regularization,   \eqref{BEidminus} can be considered
as an equality  of formal power series in $\nu^{-2}$.
In this interpretation \eqref{DELTAf1deltaSL2} corresponds to the 
term of order $\nu^{-2}$.
It is remarkable that the coproduct of all the  imaginary root vectors can be brought 
to the  simple form \eqref{BEidminus}.

Let us finally note that
our derivation of \rf{BEid}, \eqref{BEidminus} was based on the identities
\rf{qexp-prop1} and \rf{qexp-prop2}. As noted earlier, these identities are 
satisfied by the special function $\CE_{b^2}(w)$ whenever the arguments are 
replaced by positive self-adjoint operators \cite{F99,FKV,Vo}. This observation may be used to 
reduce the verification of \rf{tensorR} to the verification of the consequences of 
\rf{BEid} and \rf{BEidminus} in the representations of interest.

\subsection{Consistency}

The mixing between real and imaginary root generators
under the action of the co-product expressed in \rf{BEid}, \rf{BEidminus} implies that the 
renormalisation prescriptions adopted for the contributions of real and imaginary
root generators in the product formula must be related.
Let us first state the proposed renormalisation prescription 
of the real root contribution to the universal R-matrix.
We define 
\begin{equation}\label{realrootDEF}
\mathscr{E}_q(\sx)\,:=\,\left[\varepsilon_{q}(x)\right]_{\text{ren}}
\,=\,\left\{
\begin{aligned}
\varepsilon_{q}(\sx) & \;\;\text{if $\sx$ is a nilpotent operator}\\
\mathcal{E}_{b^2}(\sx) &\;\; \text{if $\sx$ is a positive self-adjoint operator}
\end{aligned}\right.
\end{equation}
where these special functions are defined in \eqref{thetadef} and \eqref{Eb-def}. 
We will now
verify that our proposed prescription for  the renormalisation of the imaginary root 
contributions is compatible with the definition \rf{realrootDEF} and the consequences of 
\rf{BEid}, \rf{BEidminus}.

\subsubsection{Check of compatibility for  $\pi^{-}_{\la_1\ot\la_2}:=(\pi^-_{\la_1}\ot\pi^-_{\la_2})\Delta$}
\label{sec:ExplicitSL2fundminus}

The image of the  left hand side  of \eqref{BEid}  under $\pi^{-}_{\la_1\ot\la_2}$
can be computed using the explicit form of the imaginary root currents
\begin{equation}\label{Ecurrentssl2}
(\pi^-_{\la_1}\ot\pi^-_{\la_2})\Delta
\left(1+E'_1(z)\right)\,=\,
\frac{\left(1+z^{-1}\,q^{2x}\,p\right)\left(1+z^{-1}\,q^{-2x}\,p\right)}
{\left(1-z^{-1}\,q^{+1}\,\mathsf{Z}\,p\right)\left(1-z^{-1}\,q^{-1}\,\mathsf{Z}\,p\right)}\,,
\end{equation}
where $q^{2x}=\la_1/\la_2$, $p=(\la_1\la_2)^{-1}$, $\mathsf{Z}=\sv\su^{-1}\ot \sv^{-1}\su^{-1}$.
Following the prescription outlined by equations \eqref{formpart1SL2} and \eqref{formpart2SL2}
this implies
\begin{equation}\label{tensorimroot2}
\left[\left(\pi^+_\la\ot\pi^{-}_{\la_1\ot\la_2}\right)\mathscr{R}^-_{\sim \delta}\right]_{\rm ren}
=(\pi^-_{\la_1}\ot\pi^-_{\la_2})\Delta
\left(\mathscr{M}^-_{\de}(\la)\right)=
\frac{\CE_{2b^2}\big(q^{+1}\frac{\la^2}{\la_1\la_2}\mathsf{Z}\big)
\CE_{2b^2}\big(q^{-1}\frac{\la^2}{\la_1\la_2}\mathsf{Z}\big)}
{\CE_{2b^2}\big(-\frac{\la^2}{\la_1^2}\big)\,\CE_{2b^2}\big(-\frac{\la^2}{\la_2^2}\big)}\,.
\end{equation}
On the other hand, applying $\pi^-_{\la_1}\ot\pi^-_{\la_2}$ to the right hand side of 
\eqref{BEid}   and using the definition \eqref{realrootDEF} we obtain 
\begin{equation}\label{tensorimroot2BIS}
\frac{\CE_{b^2}\big(\frac{\la^2}{\la_1\la_2}\mathsf{Z}\big)}
{\CE_{2b^2}\big(-\frac{\la^2}{\la_1^2}\big)\,\CE_{2b^2}\big(-\frac{\la^2}{\la_2^2}\big)}\,.
\end{equation}
The compatibility under tensor product,
encoded in  \eqref{BEid},  states that \eqref{tensorimroot2}
has to be equal to \eqref{tensorimroot2BIS}. This is so 
 provided that  the functional relation
\begin{equation}
\CE_{2b^2}(qw)\CE_{2b^2}(q^{-1}w)\,=\,\CE_{b^{2}}(w)\,,
\end{equation}
holds. This is indeed a simple consequence of the integral representation 
\rf{Eb-def}.


\subsubsection{Tensor products of finite- and 
infinite-dimensional representations, I}

For the derivation of the Baxter equation we will also need to consider tensor products of
finite- and infinite-dimensional representations such as $\pi^{\rm f}_{\zeta'}\ot \pi^+_\zeta$. 
Let us first generalise our renormalisation prescription in a way that will allow us to cover
cases involving such mixed tensor products. Let 
$\sx$ be an operator on a Hilbert-space of the form $\CH\ot V$ with $V$ being $n$-dimensional
that can be 
diagonalised by means of a similarity transform $\SS$ in the sense that
$\sx=\SS\cdot{\rm diag}(\la_1\sx_1,\dots,\la_n\sx_n)\cdot\SS^{-1}$, 
where $\la_k\in\BC^*$ and $\sx_k$,  $k=1,\dots,n$ are positive-selfadjoint operators. 
For such operators $\sx$ 
it is natural to define
\begin{equation}
\mathscr{E}_q(\sx)\,=\,\SS\cdot 
{\rm diag}\big(\mathcal{E}_{b^2}(\la_1\sx_1),\dots,\mathcal{E}_{b^2}(\la_n\sx_n)\big)\cdot\SS^{-1}
\,.
\end{equation}
This definition allows us to define
\begin{equation}
\sr^{(vw)-}_{(12)3}(\zeta',\zeta):=\big[(\pi^{v}_{\zeta'}\ot\pi^w_\zeta\ot \pi^-_1)
\big((\De\ot 1)(\CR^-)\big)\big]_{\rm\sst ren}^{}\,,
\end{equation}
for $v,w\in\{+,{\rm f}\}$, 
keeping in mind  that the infinite product over real root contributions 
truncates to a finite product whenever $\pi^-_\la$ is applied to the second tensor factor. 

Important for the derivation of the Baxter equation will be
the identities
\begin{equation}\label{DeRren}
\sr^{(vw)-}_{(12)3}(\zeta',\zeta)\,=\,
\sr^{v-}_{13}(\zeta')\sr^{w-}_{23}(\zeta)\,,\qquad
\begin{aligned}& \sr^{v-}_{13}(\zeta)\,=\,\big[(\pi^{v}_{\zeta}\ot \pi^-_1)
(\CR^-)\big]_{\rm\sst ren}^{}\,,\\
&\sr^{w-}_{23}(\zeta)\,=\,\big[(\pi^w_\zeta\ot \pi^-_1)
(\CR^-)\big]_{\rm\sst ren}^{}\,.
\end{aligned}
\end{equation}
The proof of these identities can follow almost literally the 
proof of \rf{BEidminus} provided that the identities 
\eqref{qexp-prop1}, \eqref{qexp-prop2} used in this calculation are preserved
by our renormalisation prescription. We need to verify that 
\begin{equation}\label{toshowSM}
\mathscr{E}_{q}(\SU)\,\mathscr{E}_{q}(\SV)\,=\,\mathscr{E}_{q}(\SU+\SV)\,,
\qquad
\mathscr{E}_{q}(\SV)\,\mathscr{E}_{q}(\SU)
\,=\,\mathscr{E}_{q}(\SU)\,\mathscr{E}_{q}(q\SU\SV)\,\mathscr{E}_{q}(\SV)\,,
\end{equation}
when
\begin{equation}
\SU=z(\pi^{\rm f}_\la\ot \pi^+_\mu)( f_i\ot 1)\,,
\qquad
\SV=z(\pi^{\rm f}_\la\ot \pi^+_\mu)(k_i^{-1}\ot f_i)\,.
\end{equation}
Let us start from the first equation in \eqref{toshowSM} for $i=1$. 
The case $i=0$ is similar.
First  notice that
\begin{equation}
\SU+\SV=
\begin{pmatrix}
q^{-1}\,x& 0 \\
z\,\la &  q^{+1}\,x 
\end{pmatrix} =
S\begin{pmatrix}
q^{-1}\,x & 0 \\
0 & q^{+1}\,x 
\end{pmatrix} S^{-1}=S\,\SV\,S^{-1}\,,
\end{equation}
where $x=z\mu\tau_q^{-1}\,\su^{-1}\sv$ and 
$S=\big(\begin{smallmatrix} 1 & 0 \\ t & 1 
\end{smallmatrix}\big)$ with $t=-\la\mu^{-1}\,\sv^{-1}\su$.
We thus have that $\SU+\SV$ is similar to $\SV$ which is self-adjoint and the prescription \eqref{realrootDEF} gives
\begin{equation}\label{first}
\mathscr{E}_{q}(\SU+\SV)=S \,\mathcal{E}_{b^2}(\SV) S^{-1}\,=\,
\begin{pmatrix}
\mathcal{E}_{b^2}(q^{-1}\,x) & 0 \\
t\left(\mathcal{E}_{b^2}(q^{-1}\,x)-\mathcal{E}_{b^2}(q^{+1}\,x)\right) & \mathcal{E}_{b^2}(q^{+1}\,x)
\end{pmatrix} \,,
\end{equation}
On the other hand $\SU$ is a nilpotent operator and the same prescription gives 
$\mathscr{E}_{q}(\SU)=\varepsilon_{q}(\SU)=1+\tau_q^{-1}\SU$ so that
\begin{equation}\label{second}
\mathscr{E}_{q}(\SU)\,
\mathscr{E}_{q}(\SV)=
\begin{pmatrix}
1& 0 \\
z\la \tau_q^{-1}& 1
\end{pmatrix}
\begin{pmatrix}
\mathcal{E}_{b^2}(q^{-1}\,x) & 0 \\
 0 & \mathcal{E}_{b^2}(q^{+1}\,x)
\end{pmatrix} \,.
\end{equation}
The equality between \eqref{first} and \eqref{second} 
follows from the functional relation 
$\mathcal{E}_{b^2}(q^{+1}x)=(1+x)\mathcal{E}_{b^2}(q^{-1}x)$
and the identity $z\la\tau_q^{-1}=-t\,x$.

Let us turn to the second relation in \eqref{toshowSM}.
 Using the nilpotency of $\SU$ and upon simplifying the the $\SU^0$ term it reduces to
\begin{equation}\label{third}
\mathcal{E}_{b^2}(\SV)\,\SU=\SU\,(1+q\SV)\,
\mathcal{E}_{b^2}(\SV)\,.
\end{equation}
Let us focus on the case $i=1$. The matrix $\SU$ is proportional to 
$\big(\begin{smallmatrix} 0 & 0 \\ 1 & 0 
\end{smallmatrix}\big)$ so that only the lower left
 entry of  \eqref{third} is non-trivial and reduses to the identity
$\mathcal{E}_{b^2}(q^{+1}x)=(1+x)\mathcal{E}_{b^2}(q^{-1}x)$.\hfill $\square$

\subsubsection{Tensor products of finite- and infinite-dimensional representations, II}
\label{sec:tensfinite-inf_SL2}

In order to verify that the consistency condition \eqref{BEidminus} 
holds after we apply the representations $\pi^{\rm f}_{\la}\ot \pi^+_\mu$ 
we first need to spell out the form of the imaginary root vectors.  
Concerning $\pi^{\rm f}$ and $\pi^-$, they are given
as a specialization of \eqref{evimaginaryrootsSL2}   
and \eqref{F1prefund} respectively.
The current of  imaginary roots for this tensor product
on the other hand takes the 
compact form
\begin{equation}\label{equalitycurrentsSL2}
\left[ 
\left(\pi^{\rm f}_{\la}\otimes \pi^+_\mu\right)\Delta\left(1+F_1'(z) \right)\right]
\,=\,
\tilde{\mathcal{S}}^{-1}\left[
\pi^{\rm f}_{\la}\left(1+F_1'(z) \right)
\otimes
 \pi^+_\mu\left(1+F_1'(z) \right)\right]
\tilde{\mathcal{S}}\,,
\end{equation}
where $\tilde{\mathcal{S}}=\mathcal{S}\Lambda^{-1}(\mathsf{y})$
 with  
$\mathcal{S}=1+q^{\frac{1}{2}}\mu\la^{-1}\big(\begin{smallmatrix} 0 & 1 \\0 & 0 
\end{smallmatrix}\big)$. Moreover, according to the definition below 
\eqref{R-factor1BIS} one has for the fundamental representation 
$\Lambda(\mathsf{y})=\mathsf{y}^{\frac{1}{2}} 
\big(\begin{smallmatrix} 1& 0 \\ 0 & \mathsf{y}^{-1}
\end{smallmatrix}\big)$.
%
The  equality \eqref{equalitycurrentsSL2} can be verified by lengthy calculations using the iterative
construction of root vectors given in Section \ref{sec:rootgen}.
The reader might be satisfied checking the first order in $z$ 
corresponding to the equality \eqref{DELTAf1deltaSL2}.
We will discuss \eqref{equalitycurrentsSL2} in Section \ref{sec:tensprodandDRcurr} in more details.

The relation \eqref{equalitycurrentsSL2} with 
 the  renormalization prescription \eqref{formpart1SL2}, \eqref{formpart2SL2}
 implies that the left hand side of   \eqref{BEidminus}
 reads 
\begin{equation}\label{LHSforfinINF}
\left[ 
\left(\pi_{\la}^{\rm f}
\otimes \pi_{\mu}^+\right)
\Delta\left(\mathscr{M}^-_{\sim\delta}(\nu)\right)\right]\,=\,
\tilde{\mathcal{S}}^{-1}
\left[ 
\pi_{\la}^{\rm f}\left(\mathscr{M}^-_{\sim\delta}(\nu)\right)
\otimes \pi_{\mu}^+\left(\mathscr{M}^-_{\sim\delta}(\nu)\right)\right]
\tilde{\mathcal{S}}\,.
\end{equation}
We are going to verify that this is equal to the right hand side of 
\eqref{BEidminus} given by 
\begin{equation}\label{RHSforfinINF}
\left(\pi_{\la}^{\rm f}\left(\mathscr{M}^-_{\sim\delta}(\nu)\right)
\otimes 1\right)
\left(1+\big(\begin{smallmatrix} 0 & \mathsf{t}\\ 0 & 0 
\end{smallmatrix}\big)\right)
\left(1\ot \pi_{\mu}^+\left(\mathscr{M}^-_{\sim\delta}(\nu)\right)\right)\,,
\end{equation}
where $\mathsf{t}=\,q^{-\frac{1}{2}}\,\frac{\la\,\mu}{\nu^2}\,\mathsf{y}$.
This formula is simply obtained recalling that
$\tau_q\pi^+_\mu(f_1)=q^{\frac{1}{2}}\mu\,\mathsf{y}$ and
$\pi_{\la}^{\rm f} (k_0\,f_0)=q^{-1}\la
\big(\begin{smallmatrix} 0 & 1\\ 0 & 0 
\end{smallmatrix}\big)$. 
Recall that the contribution $\pi_{\la}^{\rm f}(\mathscr{M}^-_{\sim\delta}(\nu))$
is given in \eqref{pifpi-IMAG}. 
As $\rho^-_{\rm f}(x)$ in  \eqref{pifpi-IMAG} and 
$\pi^{-}(\mathscr{M}^-_{\sim\delta}(\nu))$ 
are central, the equality between \eqref{LHSforfinINF} and  \eqref{RHSforfinINF}
reduces to 
\begin{equation}
\tilde{\mathcal{S}}^{-1}
\begin{pmatrix}
1 & 0 \\
0 &  1-q^{-1}\la^2\nu^{-2}
\end{pmatrix}
\tilde{\mathcal{S}}\,=\,
\begin{pmatrix}
1 & 0 \\
0 &  1-q^{-1}\la^2\nu^{-2}
\end{pmatrix}
\begin{pmatrix}
1 & \mathsf{t} \\
0 &  1
\end{pmatrix}\,.
\end{equation}
This is verified using the definition of $\mathsf{t}$ and
$\tilde{\mathcal{S}}$ given above.

\subsection{Reducibility of tensor products}

Other issues arise whenever tensor products of representations contain sub-representations
or quotients isomorphic to one of the basic representations. The renormalisation of the imaginary
root contributions must be compatible with the existence of such relations. This will be seen
to imply functional relations between the special functions appearing in the imaginary root
contributions.

\subsubsection{Highest weight representations}\label{sec:BAX2}

As a warm-up let us consider a 
representation of the  Weyl-algebra $\su\sv=q^{-1}\sv\su$ realised on vector spaces 
with basis $v_j$, $j\in\BZ$ by means of 
\begin{equation}
\sv\,v_j\,=\,v_{j+1}\,,\qquad\su\,v_j\,=\,q^{-j}v_j\,.
\end{equation}
It is possible to supplement the definition of $\pi^+_\zeta$ by a lowest- 
or highest weight condition, restricting the values of $j$ to a semi-infinite
subset of $\BZ$. 
It was first noted in \cite{AF} that the tensor product of 
representations $\pi_\zeta^+\ot\pi_{\zeta'}^{\rm f}$ contains for $\zeta'=q^{\frac{1}{2}}\zeta$ 
a subrepresentation
isomorphic to $\pi_{q\zeta}^+$, 
and that the quotient 
$\pi_\zeta^+\ot\pi_\zeta^f/\pi_{q\zeta}^+$ is isomorphic 
to $\pi_{q^{-1}\zeta}^+$. 

To see this, let us consider tensor products of the 
form $\pi_\zeta^+\ot\pi_{\zeta'}^{\rm f}$, 
and look for a sub-representation $\pi^+_{\zeta''}$
generated by vectors of the form
\begin{equation}
w_j:=\,a_jv_{j-1}\ot u_+
+ b_j
v_{j}\ot u_-\,,
\end{equation}
using the standard basis $u_+=\big(\begin{smallmatrix} 1\\ 0 
\end{smallmatrix}\big)$, $u_-=\big(\begin{smallmatrix} 0\\ 1 
\end{smallmatrix}\big)$ for 
$\BC^2$. A straightforward calculation shows that such a sub-representation
exists provided that $\zeta'$ and $\zeta$ are related as $\zeta'=q^{\frac{1}{2}}\zeta$.
The sub-representation  $\pi^+_{\zeta''}$ then has the parameter $\zeta''=q\zeta$.
It is furthermore straightforward to check that the quotient 
$\pi_\zeta^+\ot\pi_{\zeta'}^{\rm f}/\pi_{\zeta''}^+$ is isomorphic 
to $\pi_{q^{-1}\zeta}^+$ in this case.

Picking representatives $\bar{w}_j$ for the quotient
$\pi_\zeta^+\ot\pi_{\zeta'}^{\rm f}/\pi_{\zeta''}^+$ one gets a basis
for $\CH_-\ot\BC^2$ generated by 
vectors ${\mathbf w}_j=\big(\begin{smallmatrix} w_j\\ \bar{w}_j 
\end{smallmatrix}\big)$. The action of $\Borelm$, and therefore the 
representation of $ (\De\ot 1)(\CR^-)$ will be represented by
lower-triangular matrices 
with respect to this basis.

\subsubsection{Representations of modular double type}

We are now going to argue that this derivation can be generalised to
cases where the representation $\pi_\zeta^+$ is replaced by a representation 
of modular double type defined on the space $\CP$ of 
functions $f(p)$ which are entire, and
have a Fourier transformation that is entire as 
\begin{equation}
\su\, g(p)\,=\,e^{-\pi b \spp}g(p)\,,\qquad
\sv\, g(p)\,=\,g(p-ib)\,.
\end{equation}
The dual $\CP'$ of $\CP$ contains the complexified
delta-functionals $\de_p$ defined by $\langle \de_p,f\rangle=f(p)$ 
for all $f\in\CP$ and all $p\in\BC$. The dual representation $(\pi_\zeta^+)'$
will be realized on delta-functionals $\de_p$ in terms of the 
transpose operators 
\begin{equation}
\su'\, \de_p\,=\,e^{-\pi b \spp}\de_p\,,\qquad
\sv'\, \de_p\,=\,\de_{p-ib}\,.
\end{equation}

We claim that
the tensor product of representations $\pi_\zeta^+\ot\pi_\zeta^{\rm f}$
exhibits the same type of reducibility as observed in the previous 
subsection.
This is fairly easy to see:
We claim that the representation $\pi_\zeta^+\ot\pi_{\zeta'}^{\rm f}$
on $\CP\ot\BC^2$ becomes reducible for  $\zeta'=q^{\frac{1}{2}}\zeta$, 
containing the sub-representation $\pi_{q\zeta}^+$, 
and that the quotient $\pi_\zeta^+\ot\pi_{\zeta'}^{\rm f}/\pi_{q\zeta}^+$
is isomorphic to $\pi_{q^{-1}\zeta}^+$ in this case.

In order to verify this claim let us note that 
the tensor product $\pi_\zeta^+\ot\pi_\zeta^{\rm f}$
is realized on vector space $\CP\ot\BC^2$. 
Vectors in this space can be realised as 
vector-valued functions ${\mathbf v}(p)=f_+(p)u_++f_-(p) u_-$, 
where $f_\ep\in\CP$, $\ep=\pm$, and any basis $\{u_+,u_-\}$ for
$\BC^2$. The
dual $(\CP\ot\BC^2)'$ of $\CP\ot\BC^2$ is spanned by
elements of the form 
$d=d_+u_+'+d_-u_-'$, with $d_\pm\in\CP'$. 
$(\CP\ot\BC^2)'$ contains in particular elements 
of the form
\begin{equation}
w_+(p)\,=\,a(p)\de_{p+ib}\ot u_+
+ b(p)\de_{p}\ot u_-\,.
\end{equation}
One may check that there exist a choice for the coefficient
functions $a(p)$ and $b(p)$ such that the action of
$(\pi_\zeta^+\ot\pi_{\zeta'}^{\rm f})'$ 
on $w_+(p)$ becomes equivalent to $(\pi_{q\zeta}^+)'$. 
This boils down to the same calculation
as outlined in Section \ref{sec:BAX2}
using the identifications $q^j\equiv e^{\pi b p}$ and
$v_j\equiv \de_{-ibj}$. 
It follows that elements of $\CP\ot\BC^2$ of the form 
$\int dp \;g(p)w_+(p)$, $g\in\CP$, 
represented by the vector valued functions 
\begin{equation}\label{vgdef}
{\mathbf v}_g^+(p)\,=\,
g(p-ib)a(p-ib)u_++g(p)b(p) u_-\,,
\end{equation}
will generate a sub-representation $\pi_{q\zeta}^+$ 
in $\pi_\zeta^+\ot\pi_{\zeta'}^{\rm f}$ if $\zeta'=q^{\frac{1}{2}}\zeta$.
As before in Section \ref{sec:BAX2} one may check that
$\pi_\zeta^+\ot\pi_{\zeta'}^{\rm f}/\pi_{q\zeta}^+\simeq \pi_{q^{-1}\zeta}^+$.
As representatives for the quotient 
$\pi_\zeta^+\ot\pi_{\zeta'}^{\rm f}/\pi_{q\zeta}^+$ 
one may take vectors of the form 
${\mathbf v}_h^-(p)=e^{-\pi b p}h(p)u_-$, $h\in\CP$.  

Any vector ${\mathbf v}(p)$ in $\CP\ot\BC^2$ can be represented
in the form ${\mathbf v}_g^+(p)+{\mathbf v}_h^-(p)$
for suitable $g,h\in\CP$. This allows us to represent
any operator on $\CP\ot\BC^2$ in terms of a matrix of 
operators acting on the column vector 
$\big(\begin{smallmatrix} g\\ h \end{smallmatrix}\big)$. 
It follows that the matrix representing 
$[(\pi_\zeta^+\ot\pi_{\zeta'}^{\rm f}\ot\pi^-)(\CR^-)]_{\rm\sst ren}^{}$ 
will be lower triangular in such a representation, 
\begin{equation}\label{triangrep}
[(\pi_\zeta^+\ot\pi_{\zeta'}^{\rm f}\ot\pi^-)((\De\ot 1)(\CR^-))]_{\rm\sst ren}^{}\!=
\bigg(\begin{matrix} [(\pi_{q\zeta}^+\ot\pi^-)(\CR^-)]_{\rm\sst ren}^{} & 0 \!\!\!\!\! \\
*  &  \!\!\!\!\! [(\pi_{\zeta/q}^+\ot\pi^-)(\CR^-)]_{\rm\sst ren}^{}\end{matrix}\bigg).
\end{equation}
if $\zeta'=q^{\frac{1}{2}}\zeta$.
The existence of such a relation implies relations between the imaginary root contributions to the
R-matrices appearing in equation \rf{triangrep}. It is easy to see that
the relations following from \rf{triangrep} imply in particular equation
\rf{thetaandrhopm} that was previously observed to be satisfied by our renormalisation 
prescription.
\subsection{Relation to the Baxter equation}

Let us consider the Q-operator defined as
\begin{equation}\label{Qrepdef}
\hat{\SQ}(\zeta):={\rm tr}_{\CH^+}^{}
\Big{\{}\left(\Omega^{\ell}\ot 1\right)\big[(\pi_\zeta^+\ot\pi^q)\,\CR^-\big]_{\rm ren}^{}\Big{\}}\,, 
\end{equation}
together with the transfer matrix in the fundamental representation given by 
\begin{equation}\label{Ttransferdef}
\hat{\ST}(\zeta):={\rm tr}_{\BC^2}^{}
\Big{\{}\left(\Omega^{\ell}\ot 1\right)\big[(\pi_\zeta^{\rm f}\ot\pi^q)\,\CR^-\big]_{\rm ren}^{}\Big{\}}\,.
\end{equation}
The element $\Omega$, with $\ell\in \{0,1\}$ corresponds to the $\mathbb{Z}_2$ automorphism represented
 by the Pauil matrix $\sigma_1$
for $\hat{\ST}(\zeta)$ and by $\mathcal{F}^2$ for $\hat{\SQ}(\zeta)$, see \eqref{L+fromL-}.  
Introducing this factor is natural from the point of view of the quantum affine algebra $\mathcal{U}_q(\widehat{\mathfrak{sl}}_2)$
and it is necessary to discuss the  modular $XXZ$ magnet and lattice sinh-Gordon model on the same footing.

We are going to show that the validity of $\CR^-_{13}\CR^-_{23}=(\De\ot 1)(\CR^-)$
within representations of the form $\pi^+\ot\pi^{\rm f}\ot \pi^{q}$
implies the Baxter equation 
\begin{equation}\label{BAXn=2}
\hat{\ST}(q^{\frac{1}{2}}\zeta)\hat{\SQ}(\zeta)\,=\,\hat{\SQ}(q\zeta)+\hat{\SQ}(q^{-1}\zeta)\,.
\end{equation}
%
In order to derive \rf{BAXn=2}, let us note that
we may, on the one hand, represent $ \hat{\ST}(\zeta')$ and $\hat{\SQ}(\zeta)$ as
\begin{align*}
&\hat{\SQ}(\zeta)\,=\,{\rm tr}_{\CH_0}^{}\big[\,\sr_{0,2N}(\zeta)\cdots\sr_{0,1}(\zeta)\,\big]\,,\\
&\hat{\ST}(\zeta')\,=\,{\rm tr}_{\BC^2}^{}\big[\,\SL^-_{2N}(\zeta')\cdots\SL^-_1(\zeta')\,\big]\,.
\end{align*}
The right hand side of \rf{BAXn=2} may be represented as
\begin{align*}
\hat{\ST}(q^{\frac{1}{2}}\zeta)\hat{\SQ}(\zeta) ={\rm tr}_{\CH_0\ot\BC^2}^{}
\Big[\,\big[\sr_{0,2N}^{}(\zeta)\SL^-_{2N}(\zeta')\big]\;
\cdots\;\big[\sr_{0,1}(\zeta)\SL^-_1(\zeta')\big]\Big]_{\zeta'=q^{\frac{1}{2}}\zeta}^{}\,.
\end{align*}
Using identity 
\rf{DeRren} we may  represent each factor $\sr_{0,k}(\zeta)\SL^-_k(\zeta')$ in the trace
representing $\hat{\ST}(\zeta')\hat{\SQ}(\zeta)$
in terms of  $\big[(\pi^{+}_{\zeta'}\ot\pi^{\rm f}_\zeta\ot \pi^-_1)
\big((\De\ot 1)(\CR^-)\big)\big]_{\rm\sst ren}^{}$, which was found to have a lower triangular matrix
representation in \rf{triangrep}. It follows that the matrix representation 
of $\big[\sr_{0,2N}^{}(\zeta)\SL^-_{2N}(\zeta')\big]
\cdots\big[\sr_{0,1}(\zeta)\SL^-_1(\zeta')\big]$ will also be lower triangular. The Baxter equation
follows immediately from this observation.


\subsubsection{Baxter equation for XXZ-type spin chains}

It remains to show that the universal form of the Baxter equation \rf{BAXn=2}
reproduces previous forms of the Baxter equation appearing in the literature.

Let us look at the explicit form of \eqref{Qrepdef} and \eqref{Ttransferdef}.
To do so, recall that for each site of the spin-chain we have
\begin{equation}
\big[(\pi_\zeta^+\ot\pi^-_{\la})\,\CR^-\big]_{\rm ren}^{}\,=\,
\rho^{+-}_{\text{ren}}(\zeta/\la)\,\mathsf{r}^{+-}(\zeta/\la)\,,
\!\!\quad
\big[(\pi_\zeta^{\rm f }\ot\pi^-_{\la})\,\CR^-\big]_{\rm ren}^{}\,=\,
\theta(\zeta/\la)\,\mathsf{L}^-(\zeta/\la)\,.
\end{equation}
The normalization  
$\rho^{+-}_{\text{ren}}(x)$ 
and $\theta(x)$ are defined in \eqref{rhopm} and \eqref{pifpi-IMAG} 
respectively and the remaining operators $\mathsf{r}^{+-}(\zeta)$ and $\mathsf{L}^-(\zeta)$ 
are given in \rf{R-factor-ren} 
and \eqref{LpmfromR}, respectively.
The definitions \eqref{Qrepdef} and \eqref{Ttransferdef} will then reduce to 
\begin{equation}
\hat{\ST}(\zeta)=\Theta_\ka(\zeta)\,\ST^{\rm\sst XXZ}(\zeta)\,,\qquad 
\hat{\SQ}(\zeta)=\Xi_\ka(\zeta)\,\SQ^{\rm\sst XXZ}(\zeta)\,,
\end{equation}
where $\Theta_\ka(\zeta)=\prod_{n=1}^N \theta(\zeta/\ka_n)
\theta(\zeta/\bar{\ka}_n)$,
 $\Xi_\ka(\zeta)=\prod_{n=1}^N \rho^{+-}_{\text{ren}}(\zeta/\ka_n)
 \rho^{+-}_{\text{ren}}(\zeta/\bar{\ka}_n)$
and
\begin{align}
&\ST^{\rm\sst XXZ}(\zeta):=
{\rm tr}_{\BC^2}^{}\big[\,
\SL_{\bar{N}}^{-}(\zeta/\bar{\ka}_N)\SL_{N}^{-}(\zeta/{\ka}_N)\cdot\dots\cdot
\SL_{\bar{1}}^{-}(\zeta/\bar{\ka}_1)\SL_{1}^{-}(\zeta/\ka_1)\big]\,, \notag\\
&\SQ^{\rm\sst XXZ}(\zeta):={\rm tr}_{\CH_0}^{}\big[\,
\mathsf{r}_{0\bar{N}}^{+-}(\zeta/\bar{\ka}_N)\mathsf{r}_{0N}^{+-}(\zeta/{\ka}_N)\cdot\dots\cdot
\mathsf{r}_{0\bar{1}}^{+-}(\zeta/\bar{\ka}_1)\mathsf{r}_{01}^{+-}(\zeta/{\ka}_1)\big]\,.
\end{align}
We have set $\ell=0$ in  \eqref{Qrepdef} and \eqref{Ttransferdef}.
Using \eqref{thetaandrhopm} 
the Baxter equation  \rf{BAXn=2} is equivalent to 
\begin{equation}\label{BAX-QISM}
{\ST}^{\rm\sst XXZ}(q^{\frac{1}{2}}\zeta){\SQ}^{\rm\sst XXZ}(\zeta)\,=\,{\SQ}^{\rm\sst XXZ}(q\zeta)+\De(\zeta)\,
{\SQ}^{\rm\sst XXZ}(q^{-1}\zeta)\,,
\end{equation}
where 
\begin{equation}
\De(\zeta)\,=\,
\frac{\Xi_\ka(q^{-1}\zeta)}{\Theta_\ka(q^{\frac{1}{2}}\zeta)\Xi_\ka(\zeta)}
\,=\,
\frac{\Xi_\ka(q^{+1}\zeta)}{\Xi_\ka(q^{-1}\zeta)}\,=\,\prod_{n=1}^{N}
\bigg(1-\frac{\zeta^2}{\bar{\ka}_n^2}\bigg)\bigg(1-\frac{\zeta^2}{{\ka}_n^2}\bigg)\,.
\end{equation}
This is essentially the form of the Baxter equation for integrable spin chains of 
XXZ type, with $\De(\zeta)$ being the quantum determinant of the monodromy matrix.
Notice that in order to simplify $\De(\zeta)$ we used again \eqref{thetaandrhopm} 
and  the functional relation
 $\rho^{+-}_{\text{ren}}(q^{+1}\zeta/\ka)=(1-\zeta^2/\ka^2)\rho^{+-}_{\text{ren}}(q^{-1}\zeta/\ka)$.
%
%

\subsubsection{Baxter equation for the lattice Sinh-Gordon model}

Let us finally note that the Baxter equation for the lattice Sinh-Gordon model studied in 
\cite{ByT} is an easy consequence of \rf{BAX-QISM}. 
Using the relations 
\rf{Lpmrel} and \rf{r++fromr+-} it is straightforward to
deduce from \rf{BAX-QISM} that the operators
\begin{align}
&\ST^{\rm\sst SG}(\zeta):=
{\rm tr}_{\BC^2}^{}\big[\,
\SL_{\bar{N}}^{-}(1/\zeta\ka)\SL_{N}^{+}({\ka}/\zeta)\cdot\dots\cdot
\SL_{\bar{1}}^{-}(1/\zeta\ka)\SL_{1}^{+}({\ka}/\zeta)\big]\,, \notag\\
&\SQ^{\rm\sst SG}(\zeta):={\rm tr}_{\CH_0}^{}\big[\,
\sr_{0\bar{N}}^{+-}(\zeta{\ka})\sr_{0N}^{++}(\zeta/{\ka})\cdot\dots\cdot
\sr_{0\bar{1}}^{+-}(\zeta{\ka})\sr_{01}^{++}(\zeta/{\ka})\big]\,.
\end{align}
satisfy a Baxter equation of the form
\begin{equation}\label{BAX-SG}
{\ST}^{\rm\sst SG}(q^{\frac{1}{2}}\zeta){\SQ}^{\rm\sst SG}(\zeta)\,=\,
a^{\rm\sst SG}(\zeta)\,{\SQ}^{\rm\sst SG}(q^{-1}\zeta)
+d^{\rm\sst SG}(\zeta){\SQ}^{\rm\sst SG}(q\zeta)\,,
\end{equation}
where 
\begin{equation}
a^{\rm\sst SG}(\zeta)\,=\,(q^{\frac{1}{2}}\zeta/\ka)^{-N}
(1-{\zeta^2}/{\ka}^2)^N
(1-{\zeta^2}{\ka}^2)^N\,,\qquad
d^{\rm\sst SG}(\zeta)\,=\,(q^{\frac{1}{2}}\zeta/\ka)^{-N}\,.
\end{equation}
The equation \rf{BAX-SG} is equivalent to the Baxter equation derived  
previously in \cite{ByT}, as discussed in some detail in Appendix \ref{ShGapp}.

\subsubsection{Relation with previous representation-theoretic constructions of Q-operators}

Our definition \rf{Qrepdef} of Q-operators is in some respects similar, but not quite
identical to the definitions of Q-operators based on representations of the q-oscillator
algebra introduced in \cite{BLZ3}. The most important difference is that the representations
considered in \cite{BLZ3} have extremal weight vectors, which is not the case for the 
representations used in this paper.
In the rest of this subsection we will compare the 
two constructions in more detail.

Both type of representations are constructed starting from the following algebra homomorphism
\begin{equation}
\pi_{\la}(e_0)\,=\,\la^{-1}\,\mathbf{a}\,,
\qquad
\pi_{\la}(e_1)\,=\,\la^{-1}\,\bar{\mathbf{a}}\,,
\qquad
\pi_{\la}(k^{}_1)\,=\,\pi_{\la}(k_0^{-1})\,=q^{ 2\mathbf{h}}\,
\end{equation}
where  $\mathbf{a}$, $\bar{\mathbf{a}}$, $q^{\pm 2\mathbf{h}}$
 satisfy the defining  relations of the  q-oscillator algebra 
\begin{equation}\label{qOSCalg}
q\,\mathbf{a}\,\bar{\mathbf{a}}- q^{-1}\,\bar{\mathbf{a}}\,\mathbf{a}\,=\,\frac{1}{q-q^{-1}}\,,
\qquad
q^{+ 2\mathbf{h}}\,\bar{\mathbf{a}}\,q^{ -2\mathbf{h}}\,=\,q^{+2}\,\bar{\mathbf{a}}\,,
\qquad
q^{+2\mathbf{h}}\,\mathbf{a}\,q^{- 2\mathbf{h}}\,=\,q^{-2}\,\mathbf{a}\,.
\end{equation}
If $q$ is not a root of unity this algebra admits only infinite dimensional representations.
As observed in Section \ref{prefunddef} the relations \eqref{qOSCalg} imply that $\pi_{\la}(e_{\delta})$ is central, from 
which it quickly follows that 
\begin{equation}\label{piimrootnoOMEGA}
\pi_{\la}(1+E'(z))\,=\,1+\omega\,z^{-1}\,,
\qquad
\omega:=q\,\la^{-2}\,.
\end{equation}
Given any representation $\pi$ we can obtain a new one as $\pi\circ \Omega$ using the automorphims
$\Omega(e_{1})=e_0$,  $\Omega(e_{0})=e_1$. 
Applying this procedure to the  case above we find
\begin{equation}\label{piOMEGAimroot}
\pi_{\la}\circ \Omega(1+E'(z))\,=\
\frac{\left(1+\omega\,z^{-1}\right)}{
\left(1+q^{-1}\,\mathbf{C}\,q^{-2\mathbf{h}}\,\omega\,z^{-1}\right)
\left(1+q^{+1}\,\mathbf{C}\,q^{-2\mathbf{h}}\,\omega\,z^{-1}\right)}\,,
\end{equation}
where $\mathbf{C}$ generates the center of the q-oscillator algebra and is defined as
\begin{equation}
\mathbf{C}\,:=\,(q-q^{-1}\,)q^{2\,\mathbf{h}}\,(\mathbf{a}\,\bar{\mathbf{a}}-\bar{\mathbf{a}}\,\mathbf{a})\,.
\end{equation}
Notice that  if $\mathbf{C}\neq 0$ the imaginary root currents are not represented by central elements.

The representation  $\pi^-$ used in this paper, see \eqref{piminus}, corresponds to $\mathbf{C}=0$.
In this case $\mathbf{a}$ and $\bar{\mathbf{a}}$ are inverse of each other up to a constant 
and we conclude  that for  $\mathbf{C}=0$ the q-oscillator algebra is isomorphic to the Weyl algebra
generated by invertible elements $\su,\sv$ satisfying $\su\sv=q^{-1}\sv\su$.
In this case the representations $\pi_{\la}$ and  $\pi_{\la}\circ \Omega$ are equivalent.

The  representations considered  in \cite{BLZ3} 
are highest weight representations of  the q-oscillator algebra generated from the Fock vacuum 
$|0\rangle$ satisfying
$\mathbf{a}|0\rangle=0$.
Upon introducing the notation 
\begin{equation}
\pi_{\text{BLZ}}^+\,:=\,\pi_{\la}\,,
\qquad
\pi_{\text{BLZ}}^-\,:=\,\pi_{\la}\circ\Omega\,,
\end{equation}
we find that the eigenvalues of the currents  \eqref{piimrootnoOMEGA} \eqref{piOMEGAimroot}
on the highest weight state gives
\begin{equation}
\pi_{\text{BLZ}}^{\pm}(1+E'(z))\, |0\rangle\,=\
\left(1+q^{\pm1}\,\la^{-2}\,z^{-1}\right)^{\pm1}|0\rangle\,.
\end{equation}
In the equation above, as  in  \cite{BLZ3} and \cite{HJ}, the label $\pm$
 refers to simple pole or simple zero 
for the eigenvalue of the current $1+E'(z)$ on the highest weight state.
Such eigenvalues are rational expression in $z^{-1}$ for the category of representations 
introduced in  \cite{HJ}.
In our paper $\pm$ labels representations of the  two Borel halves. 
 
The representations considered in our paper do not have extremal weight vectors. It is unknown to us if
useful Q-operators can be constructed using highest weight type representations in auxiliary space if
the representations used in quantum space are of modular double type.  
 





\subsection{Choice of branch}\label{branch}

Let us finally return to the issue to fix a choice of branch for logarithm of the
argument of the  special functions $\CE_{\hbar}(w)$ used above to represent the 
imaginary root contributions. It will be fixed by the following reasoning:
It was shown in Section \ref{sec:decoupling} that the tensor product of two pre-fundamental representations
contains an evaluation representation of  modular double type. 
It will be observed below that the {\it dual} of such a representation contains 
representations of highest weight type. The rational function representing the eigenvalues
of the current on the highest weight vector simplifies
somewhat compared to the eigenvalues of a generic vector.
We demand that the eigenvalues 
of $[(\pi^+_\la\ot\pi^{\rm m.d.}_{\mu,s})(\mathscr{R}^-_{\sim \delta})]_{\rm ren}^{}$ on the highest weight
vectors of these sub-representations coincide with what is  obtained by applying our
renormalisation prescription to the eigenvalues of the current on the highest weight vector. 
This gives a natural way to fix the choice of branch of $\log(w)$ in the definition 
$\CE_{\hbar}(w)$, as will now be described in more detail.

\subsubsection{Highest weight representations in the dual of $\CP_s$}\label{dualhw}

The key observation is that the highest weight representations of $\CU_q(\fsl_2)$
are contained in the {\it dual} to the representations $\CP_s$. 
In order to see this, let us note that by simple changes of notation 
one may rewrite the representation 
defined in \rf{uqsl-rep} as
\begin{equation}\label{uqsl-rep2}
\begin{aligned}
&\SE_j\,\equiv\,\se_s\,=\,[j-m]_q\,\ST_-\,,\\
&\SF_j\,\equiv\,\sf_s\,=\,[j+m]_q\,\ST_+\,,
\end{aligned}
\qquad \SK_j\,=\,q^m\,,
\end{equation}
where $p=-ibm$, $\ST_\pm f(m)=f(m\pm 1)$, and the parameter $j$ is related to $s$ via
\begin{equation}\label{j-s}
j=\frac{i}{b}(s-c_b)\,,\quad\text{where}\quad 
c_b:=\frac{i}{2}(b+b^{-1})\,.
\end{equation}
The dual space $\CP'_s$ contains complexified 
delta-distrbutions $e^{j}_m:=\de_p$. By duality one gets
\begin{equation}\label{uqsl-rep3}
\begin{aligned}
&\SE_j'\,e^j_m=\,[j-m]_q\,e^j_{m-1}\,,\\
&\SF_j'\,e^j_m\,=\,[j+m]_q\,e^j_{m+1}\,,
\end{aligned}
\qquad \SK_j'\,e^j_m\,=\,q^m\,e^j_m\,.
\end{equation}
It follows that the distributions $\{e^j_m;m=j,j+1,\dots\}$ 
generate a Verma submodule 
$\CR_j$ within the dual $\CP_s'$ of $\CP_s$. 

  \subsubsection{Eigenvalues of currents on the highest weight vector}
  
  The form   of the imaginary root currents for representations of modular double type follows
 form the  first equation in \eqref{evimaginaryrootsSL2}  and  the 
 expression of the  \eqref{mdCASIMIR} to be 
\begin{align}\label{currentmdSL2}
 \pi^{\rm m.d.}_{\mu,s}(1+E'_1(z))\,&=\,
\frac{(1-q\,z^{-1}\,\mu^{-2}\,e^{+2\pi b s})(1+q\,z^{-1}\,\mu^{-2}\,e^{-2\pi b s})}
{(1+q^{-1}\,z^{-1}\,\mu^{-2}\,q\,\mathsf{k}_s^2)(1+q^{+1}\,z^{-1}\,\mu^{-2}\,q\,\mathsf{k}_s^2)}\,.
\end{align}
The prescription \eqref{formpart1SL2}, \eqref{formpart2SL2} with the currents
as in \eqref{E1prefund}, \eqref{currentmdSL2} then gives
\begin{equation}\label{imrooteval}
[(\pi^+_\la\ot\pi^{\rm m.d.}_{\mu,s})(\mathscr{R}^-_{\sim \delta})]_{\rm ren}^{}
=
\frac{\CE_{2b^2}(-q^2\mathsf{k}_s^2\la^2/\mu^2)\CE_{2b^2}(-\mathsf{k}_s^2\la^2/\mu^2)}
{\CE_{2b^2}(qe^{2\pi b s}\la^2/\mu^2)\CE_{2b^2}(qe^{-2\pi b s}\la^2/\mu^2)}
\,.
\end{equation}  
Let us now consider the dual action of $ \pi^{\rm m.d.}_{\mu,s}(1+E'_1(z))$ on 
$e^j_j$. Note that \rf{currentmdSL2} simplifies in this case, as a factor in the numerator
can be canceled against a factor in the denominator.

Requiring that our renormalisation prescription leading to \rf{imrooteval} is consistent with 
this fact finally fixes the choice of the branch of the logarithm in the definition 
of factors like $\CE_{2b^2}(-w)$: It should be such that 
the same cancellation takes place when  \rf{imrooteval} is evaluated on $e^j_j$. 
This will be the case when $\log(-w)=-\pi i +\log(w)$.

\subsection{Towards a "more universal" R-matrix}
\label{sec:moreUNI}

Our findings suggest that there should exist a generalisation of the 
universal R-matrix that not only makes sense for $|q|=1$, but which also extends the
class of representations in which it can be evaluated by an interesting class of
infinite-dimensional representations.
The representations of interest for us can all be found in the tensor products of two types
of representations, the prefundamental representations of modular double type on the 
one hand, and the finite-dimensional representations on the other hand.
We have defined renormalised versions of the image of the universal R-matrix
for the basic examples of such representations from which more general representations
can be constructed by taking tensor products. 

Note that we have not defined the renormalisation 
of the product formula for general tensor products yet. However, if we have a tensor product
$\pi_{ij}:=(\pi_i\ot\pi_j)\circ\De$ 
of two representation for which we have already
defined the image of the R-matrix, we may define the corresponding R-operators via
\rf{tensorR}.
One may thereby extend the definition of the renormalised universal R-matrix to the 
whole category of representations generated by taking tensor products of 
representations of prefundamental and finite-dimensional type.
This allows us, in particular, to construct 
\begin{equation}
\SR_{s_1s_2}(\la/\mu)\,=\,
\big[(\pi^{\rm ev}_{\la,s_1}\ot\pi^{\rm ev}_{\mu,s_2})(\CR^-)\big]_{\rm ren}^{}
\end{equation}
from the product of four operators 
$\sr^{+-}(\la/\mu)=[(\pi^+_{\la}\ot\pi^-_{\mu})(\CR^-)]_{\rm ren}^{}$,
as noted previously.
 
 We'd finally like to propose that the prescription for the renormalisation in the case
of finite-dimensional representations
is related to the one for the case of infinite-dimensional representations even more deeply.
We are going to argue that the latter implies the former. 

In Section \ref{dualhw} we discussed the dual of the representations $\CP_s$. It is clear
that the action of $\SR_{s_1s_2}(\la/\mu)$ on 
$\CP_{s_1}\ot\CP_{s_2}$ defines the dual action on
$(\CP_{s_1}\ot\CP_{s_2})'$. As the latter contains highest weight representations
$\CR_{j_1}\ot\CR_{j_2}$ with $j_i$ related to $s_i$ via \rf{j-s} for $i=1,2$,
we get an action of $(\SR_{s_1s_2}(\la/\mu))^t$ on $\CR_{j_1}\ot\CR_{j_2}$.
We are using the notation $\SO^t$ for the transpose (dual) of an operator $\SO$.
We conjecture that this action coincides with the action of the R-matrix
obtained from the 
universal R-matrix 
using the renormalisation prescription introduced above,
\begin{equation}\label{hw-conj}
(\SR_{s_1s_2}(\la/\mu))^t\cdot e_1\ot e_2\,=\,
[(\pi^{\rm ev}_{\la,j_1}\ot\pi^{\rm ev}_{\mu,j_2}(\CR^-)]_{\rm ren}^{}\cdot e_1\ot e_2\,,
\end{equation}
where 
$e_1\in\CR_{j_1}$, $e_2\in\CR_{j_2}$. A result in this direction
was obtained in \cite{ByT2}: A formula like
\rf{hw-conj} holds if $\SR_{s_1s_2}(\la/\mu)$ is replaced by the 
spectral parameter independent R-matrix $\SR_{s_1s_2}$
acting on the tensor product of two representations of the modular 
double. We believe that a proof should be possible
for example using the alternative representation of the 
operator $\SR_{s_1s_2}(\la/\mu)$ derived in \cite[Appendix D]{ByT}.

The validity of the conjecture \rf{hw-conj} would underline 
in which sense the renormalised version of the universal R-matrix is 
a "more universal" R-matrix: It can not only be used for 
infinite-dimensional representations of modular double type, it also 
defines the action of the R-matrices on finite-dimensional representations
in a way that automatically ensures compatibility with the structure 
of the enlarged
category of representation generated from both finite-dimensional 
representations and the infinite-dimensional representations of modular double type.


\section{Imaginary roots and functional relations II}
\label{sec:ImrootsSLM}


In this section we shall begin by deriving a universal form of the Baxter equation
for models with $\CU_q(\widehat{\fgl}_M)$ quantum group symmetry. 
A new feature in our derivation is the use of a fermionic representation $\pi^{\CF}$ containing 
all fundamental representations of $\CU_q(\widehat{\fgl}_M)$ as sub-representations.
Being reducible, it admits a collection of spectral parameters $\mu=(\mu_0,\dots,\mu_M)$, 
one for each fundamental representation $\CV_k$ contained in $\CF$.
The Baxter equation will follow from the reducibility of the tensor products $\pi^{\CF}\ot\pi^+$ 
at certain values of the spectral parameters.

The proof of the universal Baxter equation will be valid for the infinite-dimensional representations
of our interest if the renormalised R-matrices
satisfy the relations $\SR_{V_1\ot V_2,V_3}=\SR_{V_1,V_3}\SR_{V_2,V_3}$ and $
\SR_{V_1, V_2\ot V_3}=\SR_{V_1,V_3}\SR_{V_1,V_2}$.
We verify that this  is the case for the representations of our interest.
This will again follow from a delicate interplay between the contributions 
associated to real and imaginary roots in the product formula.


\subsection{Universal Baxter equation}
\label{sec:universalBaxter}

We are now going to prove the following  universal form of the Baxter equation:
\begin{equation}\label{TQstart}
 \sum_{k=0}^M\,(-1)^k\,\mathbb{T}^{(k)}(q^{\frac{k}{M}}\zeta)\,
\mathsf{Q}^+(-\omega \,q^{\frac{2k-M}{M}}\zeta)\,=\,0\,,
\end{equation}
where $\omega$ is an $M$-th root of unity $\omega^M=1$.
This equation reduces to \eqref{BAXn=2} for $M=2$.
The  "universal" 
Baxter operator  $\mathsf{Q}^+(\lambda)$ 
is  defined as 
\begin{equation}
\mathsf{Q}^{+}(\lambda)\,:=\,
\tr_{\mathcal{H}}^{}
\big\{ \!\left(\Omega^{\ell}\otimes 1\right)\,
\left[\left(\pi^{+}_{\lambda}\otimes \pi_{q}\right)\CR^-\right]_{\text{ren}}
\big\}\,.
\end{equation}
The representation $\pi^{+}_{\lambda}$ corresponding to  the auxiliary space 
$\mathcal{H}$ is given in \eqref{repplusTEXT}.  
The trace is  twisted by the $\ell$-th power of  the $\mathbb{Z}_M$ automorphism  $\Omega$ given in 
\eqref{OmegaonUq}.
The choice of the representation in the quantum space, denoted by $\pi_{q}$,
 will  only be restricted by the condition that the trace should exist.
The (higher)
transfer matrices $\mathbb{T}^{(k)}(\la)$ are similarly defined
as traces
\begin{equation}\label{TkLambda}
\mathbb{T}^{(k)}(\la)\,:=\,
\tr_{\mathcal{V}_k}
\big\{ \!\left(\Omega^{\ell}\otimes 1\right)\,
\big[\big(\pi^{(k)}_{\lambda}\otimes \pi_{q}\big)\CR^-\big]_{\text{ren}}
\big\}\,.
\end{equation}
over certain finite-dimensional irreducible representations $\mathcal{V}_k$
that we descibe in the following.
It will be very useful for us to observe that the representations
$\pi_{\lambda}^{(k)}$ relevant for the  formulation of the Baxter equation
\rf{TQstart} appear as irreducible components in a reducible 
representation 
constructed from fermionic creation- and annihiliation 
operators $\bar{\mathbf{c}}_i$, $\mathbf{c}_{i}$, $i=1,\dots,M$ 
which satisfy
\begin{equation}\label{fermoscalgebra}
 \{\mathbf{c}_{i},\bar{\mathbf{c}}_{j}\}\,=\,\delta_{ij}\,,\qquad 
\{\mathbf{c}_{i},\mathbf{c}_{j}\}\,=\,0\,,\qquad 
\{\bar{\mathbf{c}}_{i},\bar{\mathbf{c}}_{j}\}\,=\,0\,.
\end{equation}
Let $\mathcal{F}$ denote the fermionic Fock space.
 The representation   $\pi_{\mathcal{F}}$ is defined via
\begin{equation}\label{PiFdefTEXT}
 \pi_{\lambda}^{\mathcal{F}}(e_i)\,=\,\lambda^{-1}\,\bar{\mathbf{c}}_i\,\mathbf{c}_{i+1}\,,
\qquad
 \pi_{\lambda}^{\mathcal{F}}(f_i)\,=\,\lambda\,\bar{\mathbf{c}}_{i+1}\,\mathbf{c}_{i}\,,
\qquad
 \pi_{\lambda}^{\mathcal{F}}(k_i)\,=\,q^{\mathfrak{n}_i-\mathfrak{n}_{i+1}}\,,
\end{equation}
where $\mathfrak{n}_i:=\bar{\mathbf{c}}_{i}\mathbf{c}_{i}$.
Notice that this is a representation of the full $\mathcal{U}_q(\widehat{\mathfrak{gl}}_M)$.
 It is easy  to see that the total fermion  number operator $\mathfrak{n}:=\sum_{i=1}^M\mathfrak{n}_i$
is in the center of the representation $\pi^{\mathcal{F}}$.
The  eigenspaces $\mathcal{V}_k \simeq \mathbb{C}^{{M}\choose{k}}$
 of $\mathfrak{n}$ associated to the eigenvalue $k$  are irreducible. 
Each $\mathcal{V}_k$ corresponds to the  $k$-th fundamental representation.

\begin{rem}
The $M$-th root of unity $\omega$ appearing explicitly in \eqref{TQstart}
 will turn out to play an important role 
for the  integrable model studied in this paper.
It is not hard to see from the definition above that 
$ \mathbb{T}^{(k)}(\omega\zeta)= \mathbb{T}^{(k)}(\zeta)$,
so that the Baxter equation posses a $\mathbb{Z}_M$ symmetry.
 We will see in Section 
\ref{sec:moddualandWrons} 
 that this symmetry acts non-trivially on
 the solution $\mathsf{Q}(\zeta)$
 for  the choice of quantum space relevant for this paper.
 \end{rem}

\begin{rem}
In Section \ref{FunR/Q} we introduced two Q-operators $\mathcal{Q}^{\pm}(\la)$, they
correspond to  the two Q-operators $\mathsf{Q}^+(\la)$, $\bar{\mathsf{Q}}^+(\la)$.
These are constructed using the representations  $\pi^+_{\la}$ and $\bar{\pi}^+_{\la}$
given in \eqref{repplusSectionBAXTER} and \eqref{TQstartBIS}
and the renormalized universal R-matrix.
The operator $\bar{\mathsf{Q}}^+(\la)$ satisfies the Baxter equation \eqref{TQstartBAR}
\end{rem}

\subsubsection{Preliminaries}

In order to show \eqref{TQstart}, 
let us start with a simple observation:
Operators as the one appearing in \eqref{TQstart} 
can be represented as traces over the tensor product 
$\CF\ot\mathcal{H}$ in the following way
\begin{equation}\label{TQastrace}
 \sum_{k=0}^M\,(-1)^k\,\mathbb{T}^{(k)}(\mu_k)\,
\mathsf{Q}^+(\lambda_k)\,=\,
\tr_{\CF\ot\mathcal{H}}
\big\{ \!\left(\Omega^{\ell}\otimes 1\right)\,
\big[
(-1)^{\mathfrak{n}}\,
\big(\pi_{\mathcal{F}\otimes \mathcal{H}}^{\mu_{\mathfrak{n}},\la_{\mathfrak{n}}}
\ot
 \pi_{q}\big)\CR^-\big]_{\text{ren}}
\big\}\,.
\end{equation}
where the operators $\la_{\mathfrak{n}}$ on $\CF\ot\CH$ are for given 
 $(\la_0,\dots,\la_M)\in\BC^{M+1}$ defined 
as the operators multiplying each vector in $\CV_k$ by $\la_k$, respectively. 
The tensor product of representations is defined using the coproduct as
\begin{equation}\label{tensorproddef}
\pi_{\mathcal{F}\otimes \mathcal{H}}^{\mu_{\mathfrak{n}},\lambda_{\mathfrak{n}}}\,
:=\,
\left(\pi_{\mathcal{F}}^{\mu_{\mathfrak{n}}}\otimes \pi_+^{\lambda_{\mathfrak{n}}}\right)\Delta\,.
\end{equation}
The action of $\Omega\in \text{End}\left(\CF\ot\mathcal{H}\right)$ in auxiliary space is understood.
The identity \eqref{TQastrace} follows from the decomposition of the fermionic representation
$\pi_\CF$ into irreducible finite-dimensional representations and from the following property of the universal R-matrix:
\begin{equation}\label{RenormRRR}
\left[
\left(\pi^{\mu}_{\CF}\otimes 1\otimes \pi_{q}\right)\CR^-_{13}
\right]_{\text{ren}}
\left[\left(1\otimes \pi_+^{\lambda}\otimes \pi_{q}\right)\CR^-_{23}\right]_{\text{ren}}=
\left[\left(\pi^{\mu}_{\CF}\otimes \pi_+^{\lambda}\otimes \pi_{q}\right)\left(\Delta\otimes 1\right)\CR^-\right]_{\text{ren}}
\end{equation}
This relation is crucial for the derivation of the Baxter equation. 
We will show in Section \ref{sec:ImaginaryrootMORE_baxt}
 that the renormalization of the universal R-matrix proposed in this paper preserves this property.

\subsubsection{Block triangular structure of $\pi^{\mu_{\mathfrak{n}}}_{\CF} \otimes  \pi^{\lambda_{\mathfrak{n}}}_+$}
\label{sec:blocktriang}

The following observation will be the key to the derivation of the Baxter equation \eqref{TQstart}.
There exist special values of the spectral parameter
\begin{equation}\label{specialValueslambdaN}
\lambda_{k}\,=\,
\hat{\lambda}_{k,\ell}\,:=\,
-\omega_\ell\,q^{\frac{2k-M}{M}}\,\zeta\,,
\qquad
\mu_{k}\,=\,
\hat{\mu}_{k}\,:=\,
q^{\frac{k}{M}}\,\zeta\,,
\qquad
\omega_\ell=e^{-\frac{2\pi i}{M}\ell }\,,
\end{equation}
such that the tensor product representation \eqref{tensorproddef} 
has the following triangular structure:
For any $\chi\in \mathcal{U}_q(\mathfrak{b}_-)$ there exist 
orthogonal projectors $\Pi_1^{(\ell)}$, $\Pi_2^{(\ell)}$ and an operator $\widetilde{\pi}_{\text{new}}\left(\chi\right)$
such that
\begin{subequations}\label{pipi}
\begin{align}\label{pipi0}
& \Pi_1^{(\ell)}\,\big(\pi^{\hat{\mu}_{\mathfrak{n}}}_{\mathcal{F}}\otimes 
\pi^{\hat{\lambda}_{\mathfrak{n},\ell}}_{+} \big)\Delta(\chi)\,\Pi_2^{(\ell)}\,=\,0\,,\\
\label{pi1pi1}
&\Pi_1^{(\ell)}\,\big(\pi^{\hat{\mu}_{\mathfrak{n}}}_{\mathcal{F}}\otimes
 \pi^{\hat{\lambda}_{\mathfrak{n},\ell}}_{+} \big)\Delta(\chi)\,\Pi_1^{(\ell)}
\,=\,\widetilde{\pi}_{\text{new}}\left(\chi\right)\,\Pi_1^{(\ell)}\,,\\
\label{pi2pi2}
& \Pi_2^{(\ell)}\,\big(\pi^{\hat{\mu}_{\mathfrak{n}}}_{\mathcal{F}}\otimes 
\pi^{\hat{\lambda}_{\mathfrak{n},\ell}}_{+} \big)\Delta(\chi)\,\Pi_2^{(\ell)}
\,=\,\widetilde{\pi}_{\text{new}}\left(\chi\right)\,\Pi_2^{(\ell)}\,.
\end{align}
\end{subequations}
The  projectors $\Pi_1^{(\ell)}$ and $\Pi_2^{(\ell)}$ determine a ($\ell$ dependent) decomposition
\begin{equation}\label{FHdecomposition}
\CF\ot\mathcal{H}\,
\simeq\,
\mathbf{V}_1\oplus \mathbf{V}_2\,=\,\mathbb{C}^2\otimes \mathbf{V}\,,
\end{equation}
where in order to write the second equality we used 
$\mathbf{V}_1\simeq \mathbf{V}_2\simeq \mathbf{V}$.
We will show that $\widetilde{\pi}_{\text{new}}(\chi)\simeq
\big(\begin{smallmatrix} 1 & 0\\ 0 & 1 
\end{smallmatrix}\big)\otimes\pi_{\text{new}}(\chi)$
and 
$(-1)^{\mathfrak{n}}\simeq
\big(\begin{smallmatrix} 1 & 0\\ 0 & -1 
\end{smallmatrix}\big)\otimes (-1)^{\mathfrak{n}'}$
with respect to the decomposition $\mathbb{C}^2\otimes \mathbf{V}$.
The relation  \eqref{pipi} can thus be rewritten  in block matrix form as
\begin{equation}\label{triang}
(-1)^{\mathfrak{n}}
\big(\pi^{\hat{\mu}_{\mathfrak{n}}}_{\mathcal{F}}\otimes \pi^{\hat{\lambda}_{\mathfrak{n},\ell}}_{+} \big)\Delta(\chi)
\,\simeq\,
\left(\begin{matrix} 
(-1)^{\mathfrak{n}'}\,\pi_{\text{new}}(\chi) & \ast \\
0 & -(-1)^{\mathfrak{n}'}\,\pi_{\text{new}}(\chi)
\end{matrix}\right)\,.
\end{equation}
This is an operator acting on  
$\mathbb{C}^2\otimes \mathbf{V}$
where each block acts on $\mathbf{V}$.

\emph{Proof of \eqref{pipi}.}
 To prove this fact it is enough to show that it holds for the generators $f_i$, $k_i$.
To do so, it is convenient to rewrite the representation $\pi^\la_+$ in terms of new variables $\mathsf{y}_i$ 
that are defined such that
\begin{equation}\label{repplusSectionBAXTER}
 \pi^{\lambda}_+(f_i)\,=\,\frac{\lambda}{q-q^{-1}}\,\su_i^{-1}\,\sv_i\,=\,
\frac{q^{\frac{M-1}{M}}\,\lambda}{q-q^{-1}}\,\mathsf{y}^{-1}_{i}\,\mathsf{y}^{}_{i+1}\,,\qquad
 \pi^{\lambda}_+(k_i)\,=\,\su^{}_i\,\su^{-1}_{i+1}\,.
\end{equation}
It is not hard to see that \rf{repplusSectionBAXTER} will hold provided that 
\begin{equation}
\log\mathsf{y}_i\,=\,\frac{1}{4}\,\sum_{i}\,X_{ij}\,\log\big(\su_i^{-1}\sv_i^2\su_i^{-1}\big)\,,
\end{equation}
where $X_{ij}$ was defined in \rf{f-defn}.
The variables $\mathsf{y}_i$ satisfy the following exchange relations
\begin{equation}\label{yalgebra}
 \mathsf{y}_i\,\mathsf{y}_j\,=\,q^{Y_{ij}}\,\mathsf{y}_j\,\mathsf{y}_i\,,\qquad
 \su_i\,\mathsf{y}_j=q^{\delta_{ij}-\frac{1}{M}}\,\mathsf{y}_j\,\su_i\,,
\end{equation}
where $Y_{ij}\,=\, \delta_{ij}-1+\frac{2}{M}(i-j)_{\text{mod-}M}$.
One of the advantages of introducing  $\mathsf{y}_i$'s is that they will allow us to 
simplify the study of tensor products involving $\pi_+$ by use of the following formulas
\begin{equation}\label{onetimespiplus}
 \left(1\otimes \pi^{\lambda}_+\right)\Delta(f_i)\,=
\,\Lambda^{-1}(\mathsf{y})\cdot
q^{\frac{1-\bar{\epsilon}}{M}}
\bigg(\hat{f}_i\,+\,\frac{\la\,q^{\frac{M-\bar{\epsilon}}{M}}}{q^2-1}\,q^{2\bar{\epsilon}_{i+1}}\bigg)
\otimes \mathsf{y}^{}_{i+1}\,\mathsf{y}_i^{-1}
\cdot \Lambda(\mathsf{y})\,,
\end{equation}
\begin{equation}
 \left(1\otimes \pi^{\lambda}_+\right)\Delta(q^{\bar{\epsilon}_i})\,=
\,\Lambda^{-1}(\mathsf{y})
\cdot
q^{\frac{\bar{\epsilon}}{M}}\otimes \su_i
\cdot \Lambda(\mathsf{y})\,,
\end{equation}
where $\bar{\epsilon}\equiv\bar{\epsilon}_{\text{tot}}$, 
 $\hat{f}_i=q^{\frac{1}{2}(\bar{\epsilon}_i+\bar{\epsilon}_{i+1}-1)}f_i$ 
 and 
 $\Lambda(\mathsf{y}):=e^{\sum_{i=1}^M\,\bar{\epsilon}_i\otimes\log\,\mathsf{y}_i}.$

{\it Proof of \eqref{onetimespiplus}:} It is straightforward to check that
\begin{equation}\label{id-1}
 \Lambda^{-1}(\mathsf{y})\cdot
(f_i \otimes 1) \cdot
\,\Lambda(\mathsf{y})\,=\,
f_i\,q^{+\frac{1}{2}E_i}\otimes \mathsf{y}^{-1}_{i+1}\,\mathsf{y}_{i}^{}\,,
\end{equation}
where 
\begin{equation}
E_i\,=\,
\left(\bar{\epsilon}_i+\bar{\epsilon}_{i+1}-1\right)+
\frac{2}{M}\left(1-\bar{\epsilon}_{}\right)\,.
\end{equation}
It is furthermore easy to verify that  $\Lambda(\mathsf{y})^{-1}\,(1\otimes \mathsf{y}_i)\,\Lambda(\mathsf{y})\,=\,q^{-\sum_{k=1}^M\bar{\epsilon}_kY_{ki}}\otimes \mathsf{\sy}_i$.
Noting that
\begin{equation}
 Y_{i,j}-Y_{i+1,j}\,=\,-\frac{2}{M}\,,
\qquad
\text{if $\,\,\,j\,\neq\,i,i+1$}\,,
\label{YmY}
\end{equation}
one finds that
\begin{equation}\label{id-2}
\Lambda^{-1}(\mathsf{y})\,
\left(1\otimes \sy_{i+1}^{}\sy_i^{-1}\right)\,\Lambda(\mathsf{y})\,=\,
q^{-\frac{2}{M}(1-\bar{\epsilon}_{})}\,
q^{-(\bar{\epsilon}_i+\bar{\epsilon}_{i+1}-1)}\otimes \sy_i^{-1}\sy_{i+1}^{}\,.
\end{equation}
The identity \rf{onetimespiplus} follows easily by combining \rf{id-1} and \rf{id-2}.
\qed

For the fermionic Fock space representation \eqref{PiFdefTEXT}, using 
$q^{2\mathfrak{n}_{i}-1}=(q-q^{-1})\mathfrak{n}_{i}+q^{-1}$,
the identity \eqref{onetimespiplus}  can be rewritten in the following way
\begin{equation}\label{pipidelta}
\begin{aligned}
&\big(\pi^{\mu_\mathfrak{n}}_{\mathcal{F}}\otimes \pi^{\lambda_{\mathfrak{n}}}_{+} \big)\Delta(f_i)\,=\,\\
&\qquad=\Lambda^{-1}(\mathsf{y})\cdot
\mu_\mathfrak{n}\,q^{\frac{1-\mathfrak{n}}{M}}
\left(\bar{\mathbf{c}}_{i+1}(\mathbf{c}_{i}-g_{\mathfrak{n}}\,\mathbf{c}_{i+1})-\frac{g_{\mathfrak{n}}}{q^2-1}\right)
\otimes 
\,\mathsf{y}_{i+1}\,\mathsf{y}_i^{-1}
\cdot
\Lambda(\mathsf{y})\,,
\end{aligned}
\end{equation}
where $g_{\mathfrak{n}}:=-q^{\frac{M-\mathfrak{n}}{M}}\,\mu_\mathfrak{n}^{-1}\lambda^{}_{\mathfrak{n}}$.
The triangular structure \eqref{pipi} will follow easily from \rf{pipidelta}.
This is best seen by   performing a discrete Fourier transform along the affine Dynkin diagram as follows
\begin{equation}\label{MOMosc}
\bar{\mathbf{c}}(p)\,:=\,\frac{1}{\sqrt{M}}\,\sum_{\ell=1}^M\,e^{\frac{2\pi i p}{M}\ell}\,\bar{\mathbf{c}}_{\ell}\,
\qquad
\mathbf{c}(p)\,:=\,\frac{1}{\sqrt{M}}\,\sum_{\ell=1}^M\,e^{-\frac{2\pi i p}{M}\ell}\,\mathbf{c}_{\ell}\,.
\end{equation}
This transformation preserves the anti-commutation relations \eqref{fermoscalgebra}.
We are going to show that \eqref{pipi} holds with projectors 
\begin{equation}\label{proj12}
\Pi^{(\ell)}_1:=\Lambda^{-1}(\mathsf{y})\,
\overline{\mathbf{N}}(\ell)
\,\Lambda(\mathsf{y})\,,
\qquad
\Pi^{(\ell)}_2:=\Lambda^{-1}(\mathsf{y})\,
\mathbf{N}(\ell)
\,\Lambda(\mathsf{y})\,,
\end{equation}
where  $\mathbf{N}(p)=\bar{\mathbf{c}}(p)\mathbf{c}(p)$
and $\overline{\mathbf{N}}(p)=\mathbf{c}(p)\bar{\mathbf{c}}(p)$.
Indeed, using  \eqref{pipidelta}  and   \eqref{proj12}, 
the relation  \eqref{pipi} for $\chi=f_i$ is rewritten as
\begin{subequations}\label{pipiBIS}
\begin{align}
& \overline{\mathbf{N}}(\ell)\,
\left[\bar{\mathbf{c}}_{j+1}(\mathbf{c}_{j}-\omega_\ell\,\mathbf{c}_{j+1})\right]\,
\mathbf{N}(\ell)\,=\,0\,,\\
\label{pi1pi1BIS}
&\overline{\mathbf{N}}(\ell)\,
\left[\bar{\mathbf{c}}_{j+1}(\mathbf{c}_{j}-\omega_\ell\,\mathbf{c}_{j+1})\right]\,
\overline{\mathbf{N}}(\ell)
\,=\,\mathbf{M}_{\ell,j}\,\,\overline{\mathbf{N}}(\ell)\,,\\
\label{pi2pi2BIS}
& \mathbf{N}(\ell)\,
\left[\bar{\mathbf{c}}_{j+1}(\mathbf{c}_{j}-\omega_\ell\,\mathbf{c}_{j+1})\right]\,
\mathbf{N}(\ell)
\,=\,\mathbf{M}_{\ell,j}\,\,\mathbf{N}(\ell)\,,
\end{align}
\end{subequations}
where $\omega_\ell=e^{-\frac{2\pi i}{M}\ell }$ and $\mathbf{M}_{\ell,j}$ is the same in the last two lines.
Notice that the term proportional to the identity 
in the first tensor factor of \eqref{pipidelta} 
has already been simplified.
The interested reader can find the specialization of the formulae above to the case $M=2$
in Appendix \ref{App:M=2}.

In order to prove \eqref{pipiBIS}  let us
rewrite the relevant combination entering \eqref{pipidelta} in terms of 
momentum space oscillators as
\begin{equation}\label{FouriertransfSUM}
\bar{\mathbf{c}}_{j+1}(\mathbf{c}_{j}-g_{\mathfrak{n}}\,\mathbf{c}_{j+1})\,=\,
\frac{1}{M}\,\sum_{p,k=0}^{M-1}\,e^{\frac{2 \pi i}{M}(p-k)(j+1)}\,
\big(e^{-\frac{2 \pi i}{M}\,p}-g_{\mathfrak{n}} \big)\bar{\mathbf{c}}(k)\mathbf{c}(p)\,.
\end{equation} 
The projectors $\overline{\mathbf{N}}(\ell)$, $\mathbf{N}(\ell)$ act in a simple way on Fourier
transformed fermionic oscillators
\begin{subequations}\label{pipiTRIS}
\begin{align}
& \overline{\mathbf{N}}(\ell)\,
\bar{\mathbf{c}}(k)\mathbf{c}(p)\,
\mathbf{N}(\ell)\,=\,
\delta_{\ell,p}\left(1-\delta_{\ell,k}\right)\,\bar{\mathbf{c}}(k)\mathbf{c}(p)\,,\\
\label{pi1pibis1}
&\overline{\mathbf{N}}(\ell)\,
\bar{\mathbf{c}}(k)\mathbf{c}(p)\,
\overline{\mathbf{N}}(\ell)
\,=\,
\left(1-\delta_{\ell,p}\right)\,\left(1-\delta_{\ell,k}\right)\,\bar{\mathbf{c}}(k)\mathbf{c}(p)
\,\,\overline{\mathbf{N}}(\ell)\,,\\
\label{pi2pibis2}
& \mathbf{N}(\ell)\,
\bar{\mathbf{c}}(k)\mathbf{c}(p)\,
\mathbf{N}(\ell)
\,=\,
\big{[}\left(1-\delta_{\ell,p}\right)\,
\left(1-\delta_{\ell,k}\right)
+\delta_{\ell,k}\delta_{\ell,p}\big{]}\,
\bar{\mathbf{c}}(k)\mathbf{c}(p)\,\,\mathbf{N}(\ell)\,.
\end{align}
\end{subequations}
Applying these  relations to  \eqref{FouriertransfSUM} with $g_{\mathfrak{n}}=\omega_{\ell}$,
relation \eqref{pipiBIS} follows with $\mathbf{M}_{\ell,j}$ given as
\begin{equation}
\mathbf{M}_{\ell,j}\,=\,
 \frac{1}{M}\,
\sum_{p,k\neq \ell}
\,e^{\frac{2 \pi i}{M}(p-k)(j+1)}\,
\big(e^{-\frac{2 \pi i}{M}\,p}-\omega_{\ell} \big)
\bar{\mathbf{c}}(k)\mathbf{c}(p)\,.
\end{equation}

Notice that the oscillator of "momentum" $\ell$ does not appear in this expression.
We have thereby completed the proof of the triangular structure \eqref{pipi}.
\qed

%
It is worth to emphasize that while for \eqref{pipi0}
to hold it is enough to have $g_{\mathfrak{n}}=\omega_{\ell}$, the relations 
\eqref{pi1pi1}, \eqref{pi2pi2} further require that $\mu_{k}q^{-\frac{k}{M}}$
is independent of $k$, see \eqref{pipidelta}.
The values \eqref{specialValueslambdaN} follows from these requirements.
From the explicit form of the projectors the decomposition \eqref{FHdecomposition} is easy to interpret:
up to the similarity transform $\Lambda(\mathsf{y})$  one has
\begin{equation}
\mathbf{V}_1\,\simeq\,\mathcal{F}_1\otimes \mathcal{H}\,,
\qquad
\mathbf{V}_2\,\simeq\,\mathcal{F}_2\otimes \mathcal{H}\,,
\end{equation}
where $\mathcal{F}_1$ and $\mathcal{F}_2$ corresponds to the subspaces of the Fock
space $\mathcal{F}$ where the $\ell$-th mode oscillators is respectively absent or present.
They are clearly isomorphic and their total number operator $\mathfrak{n}$ differs by one unit.

It is clear that the Baxter equation \rf{TQstart} will immediately follow
from our preliminary observation \eqref{TQastrace} combined with the triangular structure \rf{triang}.
This is so as the operators
appearing in the diagonal elements of the matrix in \rf{triang} coincide
up to a sign, from which the vanishing of traces over
$\CF\ot\mathcal{H}$ follows.

\begin{rem}
The form of projectors \eqref{proj12}, the similarity transform $\Lambda(\mathsf{y})$
and the introduction of the fermionic oscillators in \eqref{MOMosc}  is motivated by the study of 
$(\pi^{+}\ot \pi^{\mathcal{F}})\mathscr{R}^-$. Indeed,
the triangular structure of $(\pi^{+}\ot \pi^{\mathcal{F}})\Delta$ for special values of the spectral
parameter is related to values of the spectral parameter for which the operator 
$(\pi^{+}\ot \pi^{\mathcal{F}})\mathscr{R}^-$ has a non-trivial kernel.
\end{rem}
\begin{rem}
 A form of the Baxter equation similar to \eqref{TQstart}
 was derived in \cite{Hi01} for $M=3$ using different techniques.
In the language of this paper the model considered in
\cite{Hi01} corresponds to the quantum space to be 
$\left(\pi^-_{\ka}\ot\dots\ot\pi^-_{\ka}\right)\De^{(N)}$. 
\end{rem}
\begin{rem}
One may notice that for any $a\,\in\,\Borelm$ there exist $\Psi(a)$ such that 
\begin{equation}\label{Qexactness}
\big(\pi^{\hat{\mu}_\mathfrak{n}}_{\mathcal{F}}\otimes \pi^{\hat{\lambda}_{\mathfrak{n},\ell}}_{+} \big)\Delta(a)\,=\,
\Lambda^{-1}(\mathsf{y})\,
\{\bar{\mathbf{c}}(\ell)\otimes 1,\Psi(a)\}\,\Lambda^{+1}(\mathsf{y})\,,
\end{equation}
where $\{a,b\}:=a\,b+b\,a$. The explicit form of $\Psi(q^{\bar{\epsilon}_i})$  and $\Psi(f_i)$ is easily obtained from the discussion above,
the existence of  $\Psi(a)$ follows. 
\end{rem}


\subsubsection{Tensor products and Drinfeld's currents}
\label{sec:tensprodandDRcurr}

It is instructive to spell out explicitly what happens to the imaginary root vectors when 
taking the tensor product 
$(\pi_{\mathcal{F}}^{\mu_{\mathfrak{n}}}\otimes \pi_+^{\lambda_{\mathfrak{n}}})\Delta$ as in 
\eqref{tensorproddef}.
We will use these observations in Section \ref{sec:ImaginaryrootMORE_baxt}
to show that \eqref{RenormRRR}  holds for the choice of quantum space  studied in this paper.

The imaginary root vectors are encoded 
in the generating currents $1+F_i'(z)$, $i=1,\dots, M-1$ 
defined in \eqref{Imrootscurrents}.
Their image under  $\pi_{\mathcal{F}}$ and $\pi_+$ is given by
\begin{subequations}\label{summaryDricurrents1}
\begin{align}\label{currentFERM_form1}
& \pi^{\mu}_{\mathcal{F}}\left(1+F_i'(z) \right)\,=\,
\frac{1-\bar{\kappa}_i\,z^{-1}\,q^{2(\mathfrak{n}_{i} -\mathfrak{n}_{i+1})}}{1-\bar{\kappa}_i\,z^{-1}}\,,\\
& \pi^{\lambda}_{+}\left(1+F_i'(z) \right)\,=\,
1+\delta_{i,M-1}\,\lambda^M\,z^{-1}\,.
\end{align}
\end{subequations}
The first expression is derived in Appendix \ref{app:CWbasisFERMfs}, the second is equivalent to \eqref{piplusimaginary}.
These are rational expressions in $z$.
An important feature of the imaginary root currents is that
 in many cases their (generalized) eigenvalues behave multiplicatively
under tensor product. We will return to this observation in Section 
\ref{sec:checkcoproducMdelta_prefund}
 where we will also present  some new interesting counter examples.
For now, let us see explicitly how this works in the case relevant for the Baxter equation. 

The form of the imaginary root currents for the tensor
product of these representation is encoded in the following relation 
\begin{equation}\label{equalitycurrents}
\Lambda(\mathsf{y})
\left[ 
\left(\pi^{\mu_{\mathfrak{n}}}_{\mathcal{F}}
\!\otimes\!
 \pi^{\lambda_{\mathfrak{n}}}_+\right)\Delta\left(1+F_i'(z) \right)\right]
\Lambda(\mathsf{y})^{-1}\!\!=
\mathcal{S}^{-1}
\left[
\pi^{\mu_{\mathfrak{n}}}_{\mathcal{F}}\!\left(1+F_i'(z) \right)
\otimes
\pi^{\lambda_{\mathfrak{n}}}_+\!\left(1+F_i'(z) \right)
\right]
\mathcal{S}
\end{equation}
where
\begin{equation}\label{SdefcurrentFermpiplus}
\mathcal{S}=\left(1-g_{\mathfrak{n}}\,\bar{\mathbf{c}}_{M-1}\mathbf{c}_M\right)
\cdots
\left(1-g_{\mathfrak{n}}\,\bar{\mathbf{c}}_{2}\mathbf{c}_3\right)
\left(1-g_{\mathfrak{n}}\,\bar{\mathbf{c}}_{1}\mathbf{c}_2\right)
\qquad
g_{\mathfrak{n}}=-\lambda^{}_{\mathfrak{n}}\mu_{\mathfrak{n}}^{-1}q^{\frac{M-\mathfrak{n}}{M}}\,.
\end{equation}
Notice that $\mathcal{S}$ is invertible for any value of $g_{\mathfrak{n}}$.
The  equality \eqref{equalitycurrents} can be verified by lengthy calculations using the iterative
construction of root vectors given in Section \ref{sec:rootgen}.
It also follows from  Theorem  8.1 of \cite{KT94}.

It is manifest from \eqref{equalitycurrents}  and \eqref{summaryDricurrents1} 
that the  tensor product  
$(\pi_{\mathcal{F}}\otimes \pi_+)\Delta(1+F_i'(z))$
is a rational expression in $z$.
If we rewrite \eqref{currentFERM_form1} for $i=M-1$ as follows
\begin{equation}\label{currFERMMminus1}
 \pi^{\mu_{\mathfrak{n}}}_{\mathcal{F}}\left(1+F_{M-1}'(z) \right)\,=\,
\frac{1+(-q^{-1}\mu_{\mathfrak{n}})^{M}\,q^{\mathfrak{n}}\,q^{2(\mathfrak{n}_{M-1} -\mathfrak{n}_{M})}\,z^{-1}}
{1+(-q^{-1}\mu_{\mathfrak{n}})^{M}\,q^{\mathfrak{n}}\,z^{-1}}\,,
\end{equation}
it is then clear that for 
\begin{equation}\label{polezerocancellation}
\left(\lambda_{\mathfrak{n}}\right)^M\,=\,\big(-q^{\frac{\mathfrak{n}-M}{M}}\,\mu_{\mathfrak{n}}\big)^M\,,
\end{equation}
the zero 
of $\pi_+(1+F_{M-1}'(z))$ cancels with the pole of $\pi_{\mathcal{F}}(1+F_{M-1}'(z))$.  
This mechanism  signals the reducibility of the tensor product. Indeed, the condition 
\eqref{polezerocancellation} follows from \eqref{specialValueslambdaN}.

\subsubsection{The representation  $\bar{\pi}_+$ and the Baxter equation}

There is a second representation that can be used in auxiliary space to construct Baxter Q-operators:
\begin{equation}\label{TQstartBIS}
\bar{\pi}_+^{\lambda}(f_i)\,=\,\frac{\lambda}{q-q^{-1}}\,\su_{i+1}\,\sv_i\,=\,
\frac{\lambda}{q-q^{-1}}\,
q^{\frac{1}{M}}\,
\bar{\mathsf{y}}_i^{-1}\,\bar{\mathsf{y}}_{i+1}^{}
\,.
\end{equation}
Following similar steps as the one given above for $\pi_+$,  we can show that
\begin{equation}\label{TQstartBAR}
 \sum_{k=0}^M\,(-1)^k\,\mathbb{T}^{(k)}(q^{\frac{M-k}{M}}\zeta)\,
\bar{\mathsf{Q}}^+(\omega\, q^{\frac{M-2k}{M}}\zeta)\,=\,0\,,
\end{equation}
where $\mathbb{T}^{(k)}$ are the same as in \eqref{TQstart}.
For $M=2$, $\pi^+=\bar{\pi}^+$ and one can show that the two Baxter equations 
\eqref{TQstart} and \eqref{TQstartBIS} are indeed equivalent by noticing that 
$\mathbb{T}^{(0)}(\la)=\mathbb{T}^{(M)}(\la)=1$ and $-\omega$
squares to one when $\omega$ does.

 We collect some of the relevant formulae used in the derivation
\begin{equation}\label{onetimespibarplus}
 \left(1\otimes \bar{\pi}^{\lambda}_+\right)\Delta(f_i)\,=
\,\Lambda^{-1}(\bar{\mathsf{y}})\cdot
q^{\frac{\bar{\epsilon}-1}{M}}
\bigg(\check{f}_i\,+\,\frac{\la\,q^{\frac{\bar{\epsilon}}{M}}}{q-q^{-1}}\,q^{1-2\bar{\epsilon}_{i}}\bigg)
\otimes \bar{\mathsf{y}}_{i+1}\,\bar{\mathsf{y}}_i^{-1}
\cdot\Lambda(\bar{\mathsf{y}})\,,
\end{equation}
 where $\check{f}_i=q^{-\frac{1}{2}(\bar{\epsilon}_i+\bar{\epsilon}_{i+1}-1)}f_i$.
From the equality above the analog of \eqref{pipidelta}  follows
\begin{equation}
\big(\pi^{\mu_{\mathfrak{n}}}_{\mathcal{F}}\otimes \bar{\pi}^{\lambda_{\mathfrak{n}}}_{+} \big)\Delta(f_i)\,=\,
\Lambda^{-1}(\bar{\mathsf{y}})\cdot
\mu_{\mathfrak{n}}\,q^{\frac{\mathfrak{n}-1}{M}}
\left((\bar{\mathbf{c}}_{i+1}-\bar{g}_{\mathfrak{n}}\,\bar{\mathbf{c}}_{i})\mathbf{c}_{i}
+\frac{\bar{g}_{\mathfrak{n}}\, q}{q-q^{-1}}\right)
\otimes 
\,\bar{\mathsf{y}}_{i+1}\,\bar{\mathsf{y}}_i^{-1}
\cdot
\Lambda(\bar{\mathsf{y}})\,,
\end{equation}
where $\bar{g}_{\mathfrak{n}}:=q^{\frac{\mathfrak{n}}{M}}\,\mu_{\mathfrak{n}}^{-1}\lambda_{\mathfrak{n}}$.
The tensor product representation exhibit triangular structure for $\bar{g}_{\mathfrak{n}}=\omega_{\ell}$.
Together with the condition that $\mu_{\mathfrak{n}}\,q^{\frac{\mathfrak{n}-1}{M}}$ is independent of $\mathfrak{n}$
this implies that 
$\la_{\mathfrak{n}}=\omega_\ell\,q^{\frac{M-2{\mathfrak{n}}}{M}}\zeta$ and 
$\mu_{\mathfrak{n}}=q^{\frac{M-{\mathfrak{n}}}{M}}\zeta$.

Let us finally quote the formulae for the 
Drinfeld currents relevant for this case.  We have
\begin{subequations}\label{summaryDricurrentsforpibarplus}
\begin{align}\label{currentFERM1}
&  \bar{\pi}^{\lambda}_{+}\left(1+F_i'(z) \right)\,=\,
1+\delta_{i,1}\,\lambda^M\,z^{-1}\,.\\
\label{currFERM1}
&  \pi^{\mu}_{\mathcal{F}}\left(1+F_{1}'(z) \right)\,=\,
\frac{1+\mu^{M}\,q^{-\mathfrak{n}}\,q^{2(\mathfrak{n}_{1} -\mathfrak{n}_{2})}\,z^{-1}}
{1+\mu^{M}\,q^{-\mathfrak{n}}\,z^{-1}}\,,.
\end{align}
\end{subequations}
The poles in the tensor product 
$(\pi_{\mathcal{F}}^{\mu_{\mathfrak{n}}}\otimes  \bar{\pi}_+^{\lambda_{\mathfrak{n}}})\Delta$
 cancels under the condition that
$(\lambda_{\mathfrak{n}})^M\,=\,q^{-\mathfrak{n}}\,(\mu_{\mathfrak{n}})^M$.
In the special case $M=2$ the representations  $\pi_+$ and $\bar{\pi}_+$ are manifestly the same
and the current \eqref{currFERMMminus1} coincides with  \eqref{currFERM1}.


\subsection{Renormalization of the imaginary root contribution 
to the universal R-matrix} 
\label{sec:ImaginaryrootMORE}

We had previously observed that the imaginary root contributions play
a key role for the validity of the identity \eqref{RenormRRR} underlying 
the derivation of the  Baxter equation \eqref{TQstart} presented in Section 
\ref{sec:universalBaxter}. As a preparation for the verification of \eqref{RenormRRR} we shall now
introduce a prescription for renormalising the imaginary root contribution 
to the universal R-matrix


\subsubsection{Renormalization prescription for the imaginary root contributions
}
In order to formulate our prescription it is necessary to spell out the structure of the imaginary root currents first.
As in the case of  $\mathcal{U}_q(\widehat{\mathfrak{gl}}_2)$ imaginary root currents form a commutative algebra.
We will restrict our attention to 
 representations in which the currents are represented by rational functions of the form 
\begin{equation}\label{formpart1SLM}
\boldsymbol{\pi}^{+}(1+F_i'(z))=
\frac{
\prod_{\ell=1}^{n_{i,+}}
(1+z^{-1}\,\mathsf{N}_{\ell,i}^{+})
}{\prod_{\ell=1}^{d_{i,+}}
(1+z^{-1}\,\mathsf{D}_{\ell,i}^+)}\,,
\quad
\boldsymbol{\pi}^{-}(1+E_i'(z))=
\frac{
\prod_{\ell=1}^{n_{i,-}}
(1+z^{-1}\,\mathsf{N}_{\ell,i}^-)
}{\prod_{\ell=1}^{d_{i,-}}
(1+z^{-1}\,\mathsf{D}_{\ell,i}^-)}\,.
\end{equation}
It will be shown  in 
Section \ref{sec:UqglMandiso} below that this condition holds for a large class of representation including
the ones we are interested in. 
Moreover this property is preserved by taking 
tensor products.

Next notice that the coefficients $u_{m,ij}$ given  in   \eqref{umij} that enter the imaginary root contributions to the universal R-matrix
\eqref{rpd}, 
can be rewritten using
\begin{equation}
(-1)^{m(i-j)} \,
[M-\text{max}(i,j)]_{q^m}\,[\text{min}(i,j)]_{q^m}\,=\,
(-1)^{mM}
\sum_{s=1}^{k_{ij}}
\,
(-q)^{m(k_{ij}-2s+1)}\,
\gamma_{i,j}^{(s)}\,,
\end{equation}
where  $k_{ij}:=M-|i-j|-1$, 
and $\gamma_{i,j}^{(s)}=\sum_{a=1}^{M-\text{max}(i,j)}\sum_{b=1}^{\text{min}(i,j)}\delta_{s,a+b-1}$.
In order to derive this relation one rewrites $[n]_q=\sum_{s=1}^{n}\,q^{n-2s+1}$.

With this observations in mind 
 it is clear that, before renormalization, the contribution 
of imaginary roots for given representations takes the form of a finite product
$\prod_{\alpha}\varepsilon_{q^M}(w_{\alpha})$, where $\varepsilon_{q}(w)$
is defined in \eqref{thetadef}.
Our renormalization prescriptions 
consists 
in  replacing $\varepsilon_{q^M}(w)$
with   $\CE_{Mb^2}(w)$  defined in \eqref{Eb-def}. For convenience we report the definition here
\begin{equation}\label{Eb-def_sect8}
\CE_{Mb^2}(w):=\exp\bigg(\int_{\BR+i0}\frac{dt}{4t}\,
\frac{w^{-\frac{i}{\pi}t}}{\sinh({Mb^2t)\sinh(t)}}\bigg)\,.
\end{equation}
The prescription above can be formulated more explicitly as follows
\begin{equation}\label{renIM_SLM}
\begin{aligned}
\big{[}
\left(\boldsymbol{\pi}^{+}\ot
\boldsymbol{\pi}^{-}\right)& 
\CR^-_{\sim\de}
\big{]}_{\text{ren}}
=\\
&
\prod_{i,j=1}^{M-1}\,\,
\frac{
\prod_{\ell=1}^{d_{i,+}}
\prod_{\ell'=1}^{n_{j,-}}\,
\mathcal{G}_{ij}(\mathsf{D}_{\ell,i}^+\ot \mathsf{N}_{\ell',j}^-)
}
{
\prod_{\ell=1}^{n_{i,+}}
\prod_{\ell'=1}^{n_{j,-}}\,
\mathcal{G}_{ij}(\mathsf{N}_{\ell,i}^+\ot \mathsf{N}_{\ell',j}^-)
}\,
\frac
{
\prod_{\ell=1}^{n_{i,+}}
\prod_{\ell'=1}^{d_{j,-}}\,
\mathcal{G}_{ij}(\mathsf{N}_{\ell,i}^+\ot \mathsf{D}_{\ell',j}^-)
}
{
\prod_{\ell=1}^{d_{i,+}}
\prod_{\ell'=1}^{d_{j,-}}\,
\mathcal{G}_{ij}(\mathsf{D}_{\ell,i}^+\ot \mathsf{D}_{\ell',j}^-)
}\,,
\end{aligned}
\end{equation}
where the image of the imaginary root currents under $\boldsymbol{\pi}^{\pm}$ 
is given in  \eqref{formpart1SLM} and 
\begin{equation}\label{Gij_def}
\mathcal{G}_{ij}(x):=
\prod_{s=1}^{k_{ij}}\Big(\mathcal{E}_{Mb^2}\left((-1)^{M-1}(-q)^{2\rho_s(k_{ij})}\,x\right)\Big)^{\gamma^{(s)}_{ij}}\,,
\end{equation}
using the notation $\rho_s(k):=\frac{k-2s+1}{2}$.

\subsubsection{Examples of renormalized imaginary root contributions}
\label{sec:examplerenIMroot}

In this section we calculate the currents and 
formulate the resulting prescription \eqref{renIM_SLM}
for the renormalization of imaginary root contributions 
for the basic representations of our interest. 
Let us start recalling the form of imaginary root currents for prefundamental representations
\begin{subequations}
\begin{align}
& \pi_{\lambda}^{+}\left(1+F_i'(z) \right)\,=\,
1+\delta_{i,M-1}\,\lambda^{+M}\,z^{-1}\,,\\
& \bar{\pi}_{\lambda}^{ + }\left(1+F_i'(z) \right)\,=\,
1+\,\,\,\,\,\delta_{i,1}\,\,\,\,\lambda^{+M}\,z^{-1}\,,\\
& 
\label{piminlater}\pi_{\lambda}^{-}\left(1+E_i'(z) \right)\,=\,
1+\delta_{i,M-1}\,\lambda^{-M}\,z^{-1}\,,\\
&\bar{\pi}_{\lambda}^{-}\left(1+E_i'(z) \right)\,=\,
1+\,\,\,\,\,\delta_{i,1}\,\,\,\,\lambda^{-M}\,z^{-1}\,.
\end{align}
\end{subequations}
These equations are collected from \eqref{summaryDricurrents1},  \eqref{summaryDricurrentsforpibarplus},
\eqref{eiIMpiminus} and \eqref{eiimbarminus}.
Let us define 
\begin{equation}
\rho^{\sigma^+\sigma^-}(\la\mu^{-1})\,:=\,
 \big[\big({\pi^{\sigma^+}_{\lambda}}\!\otimes {\pi^{\sigma^-}_{\mu}} \big)\,\mathscr{R}^-_{\sim\delta}\big]_{\text{ren}}\,,
\qquad\,\,
\sigma^{\pm}\,\in\,\{\pm,\dot{\pm}\}\,,
\end{equation}
compare to \eqref{srSEC6_all}.
Following the prescription given in \eqref{renIM_SLM} one obtains
\begin{subequations}
\label{rhopmsec8}
\begin{align}
& \rho^{+-}(\zeta)=\rho^{\dot{+}\dot{-}}(\zeta)=\prod_{s=1}^{M-1}\frac{1}{\mathcal{E}_{Mb^2}\left(-q^{M-2s}\,\zeta^M\right)}\,,\\
& \rho^{+\dot{-}}(\zeta)=\rho^{\dot{+}-}(\zeta)=\,\,\,\,\,\,\,\,\frac{1}{\mathcal{E}_{Mb^2}\left((-1)^{M-1}\,\zeta^M\right)}\,.
\end{align}
\end{subequations}
Notice that for $M=2$, these two expressions coincide and are equal to \eqref{rhodef}.

The next example is the renormalization of $(\pi^{\mathcal{F}}\otimes \pi^-)\mathscr{R}^-_{\sim\delta}$.
In this case, the prescription \eqref{renIM_SLM} for the currents  \eqref{summaryDricurrents1}, and \eqref{piminlater} gives
\begin{equation}
\left[\left(\pi^{\mathcal{F}}_{\mu_{\mathfrak{n}}}\otimes \pi^-_{\nu}\right)\,\mathscr{R}^-_{\sim\delta}\right]_{\text{ren}}\,=\,
\frac{\mathcal{E}_{Mb^2}
\left(
-q^{M-2\mathfrak{n}}\,g_-^M
\right)}{
\mathcal{E}_{Mb^2}
\left(
-q^{M}\,g_-^M\,q^{-2M\,\mathfrak{n}_M}
\right)
}\,,
\qquad
g_-:=-q^{\frac{\mathfrak{n}-M}{M}}\,\mu_{\mathfrak{n}}\,\nu^{-1}\,.
\end{equation}
This equality results after a cancellation of terms in  \eqref{renIM_SLM}.
The simplification does not rely on any special property of the function $\mathcal{E}_{Mb^2}(\omega)$
and uses the fact that each $\mathfrak{n}_i$ takes the values $\{0,1\}$.
At this point one can use the property 
$\mathcal{E}_{Mb^2}(q^{+M}x)=(1+x)\mathcal{E}_{Mb^2}(q^{-M}x)$  to rewrite 
\begin{equation}\label{Thetadef}
\left[\left(\pi^{\mathcal{F}}_{\mu_{\mathfrak{n}}}\otimes \pi^-_{\nu}\right)\,\mathscr{R}^-_{\sim\delta}\right]_{\text{ren}}=
\theta^-_{\mathcal{F}}(g_-)\,
\left(1-g_-^M\,\mathfrak{n}_M\right)\,,
\quad\,
\theta^-_{\mathcal{F}}(g_-):=\frac{\mathcal{E}_{Mb^2}
\left(
-q^{M-2\mathfrak{n}}\,g_-^M
\right)}{
\mathcal{E}_{Mb^2}
\left(
-q^{M}\,g_-^M
\right)
}\,.
\end{equation}
A similar analysis gives
\begin{equation}\label{Thetabardef}
\left[\left(\pi^{\mathcal{F}}_{\mu_{\mathfrak{n}}}\otimes \bar{\pi}^-_{\nu}\right)\,\mathscr{R}^-_{\sim\delta}\right]_{\text{ren}}\,=\,
\bar{\theta}^-_{\mathcal{F}}(\bar{g}_-)\,
\left(1-\bar{g}_-^M\,\bar{\mathfrak{n}}_1\right)\,,
\qquad\,
\bar{\theta}^-_{\mathcal{F}}(\bar{g}_-):=\frac{\mathcal{E}_{Mb^2}
\left(
-q^{M-2\bar{\mathfrak{n}}}\,\bar{g}_-^M
\right)}{
\mathcal{E}_{Mb^2}
\left(
-q^{M}\,\bar{g}_-^M
\right)
}\,,
\end{equation}
where
\begin{equation}
\bar{g}_-:=q^{\frac{\bar{\mathfrak{n}}-M}{M}}\,\mu_{\mathfrak{n}}\,\nu^{-1}\,,
\qquad
\bar{\mathfrak{n}}:=M-\mathfrak{n}\,,
\qquad
\bar{\mathfrak{n}}_i:=1-\mathfrak{n}_i\,.
\end{equation}
More examples of renormalization of imaginary root contributions are presented in the following section and Appendix \ref{sec:imageimGTalg}.

\paragraph{Lax operators for $\mathbb{T}(\la)$.}Using  the results \eqref{Thetadef}
\eqref{Thetabardef} we can write down the explicit expression obtained from the renormalized universal R-matrix
 for the Lax operators entering the tranfer matrices \eqref{TkLambda} with quantum space 
\eqref{quantumspacefinal}.
\begin{equation}
\left[\left(\pi_{\mu_{\mathfrak{n}}}^{\mathcal{F}} \otimes \pi_{\nu}^{-}\right)\,\mathscr{R}^-\right]_{\text{ren}} \,=\,
\theta^-_{\mathcal{F}}(g_-)\,\,\Lambda(\mathsf{y})
\left[\prod_{p=1}^M\,\left(1-g_-\,e^{-\frac{2\pi i p}{M}}\,\mathbf{N}(p)\right)\right]\,
\Lambda^{-1}(\mathsf{y})\,\Lambda(\su)\,,
\label{LminusF}
\end{equation}
\begin{equation}
\left[\left(\pi_{\mu_{\mathfrak{n}}}^{\mathcal{F}} \otimes \bar{\pi}_{\nu}^{-}\right)\,\mathscr{R}^-\right]_{\text{ren}} \,=\,
\bar{\theta}^-_{\mathcal{F}}(\bar{g}_-)\,\,\Lambda(\bar{\mathsf{y}})
\left[\prod_{p=1}^M\,\left(1-\bar{g}_-\,e^{-\frac{2\pi i p}{M}}\,\overline{\mathbf{N}}(p)\right)\right]\,
\Lambda^{-1}(\bar{\mathsf{y}})\,\Lambda(\su)\,.
\label{LminusFbar}
\end{equation}
The variables $\mathsf{y}_i$ and $\bar{\mathsf{y}}_i$ entering the expressions above are introduced in 
\eqref{eipiminus}, \eqref{introduceybar}  (equivalently in \eqref{repplusSectionBAXTER}, \eqref{TQstartBIS}) and 
the fermionic number operators
 $\mathbf{N}(p)=\bar{\mathbf{c}}(p)\mathbf{c}(p)$ and $\overline{\mathbf{N}}(p)=\mathbf{c}(p)\bar{\mathbf{c}}(p)$ are defined in terms of the 
fermionic oscillators \eqref{MOMosc} in "momentum space" conjugated to the Dynkin diagram circle.
The main steps of the dervation are left to Appendix \ref{app:stepsfotLminusF}.
The Lax operators \eqref{LminusUNIR}, \eqref{LpmanduniR}
 can be recovered from these expressions upon acting on the subspace of the fermionic Fock space where the total number operator 
 $\mathfrak{n}$ has eigenvalue $1$.




\subsubsection{ Rationality of the imaginary root currents}
\label{sec:UqglMandiso}

It remains to show that the currents are indeed represented by rational functions 
of the form \rf{formpart1SLM} in the representations of our interest. To this aim
we need to generalise the proof of the rationality of the currents described in Section 
\ref{sec:ImrootRATIONAL_sl2} for the case of $\mathcal{U}_q(\widehat{\mathfrak{gl}}_2)$
to $\mathcal{U}_q(\widehat{\mathfrak{gl}}_M)$. This turns out to be somewhat more involved.
We will outline the proof below, leaving some technical details to appendices.


It will be useful to consider the so-called universal Lax matrix
\begin{equation}
 \mathscr{L}(\la)\,:=\,\left(\pi^{\text{f}}_{\la}\,\otimes\,1\right)\mathscr{R}^-\,,
\label{uniL}
\end{equation}
where $\pi^{\text{f}}_{\la}$ is the fundamental representation of
 $\mathcal{U}_q(\widehat{\mathfrak{sl}}_M)$ defined in \eqref{fundrepTEXT}.
It follows from the universal Yang-Baxter equation 
\eqref{YBE} that $\mathscr{L}(\la)$ satisfies the quadratic relations \eqref{RLLLLR}. 
The product formula for the universal R-matrices yields a
triangular decomposition of the form
\begin{equation}\label{triangFORevaluation}
\mathscr{L}(\la)\,=\,
\left( 1\,+\,\sum_{i>j}\,\ell_{ji}(\la)\,\SE_{ij}\right)\,
\left(\sum_{i=1}^M\,a_i(\la)\,\SE_{ii}\right)\,
\left(1\,+\,\sum_{i<j}\,\ell_{ji}(\la)\,\SE_{ij}\right)\,,
\end{equation}
where $\SE_{ij}$ are the matrix units, as before.
It can be shown, see Appendix \ref{app:minors} for details, that for any matrix 
$\mathscr{L}(\la)$ that satisfies the relations \eqref{RLLLLR} the following relations hold
\begin{equation}\label{apfromqdet}
a_p(\la)\,=\,\frac{A_{p}(q^{-\frac{p-1}{M}}\,\la)}
{A_{p-1}(q^{-\frac{p}{M}}\,\la)}\,,\qquad 
A_{p}(\la)\,:=\,
\text{q-det}\big(\mathscr{L}^{[p]}(\la^{\frac{M}{p}})\big)\,,
\qquad A_0(\la)\,:=\,1\,,
\end{equation}
where $\rho_{k}=\frac{m-2k+1}{2}$ and the $p\times p$ matrices $\mathscr{L}^{[p]}(\la)$
 are defined as 
 \begin{equation}
 \big(\mathscr{L}^{[p]}(\lambda^{\frac{M}{p}})\big)_{ij}\,:=\,
\lambda^{\frac{M-p}{p}(i-j)}\,\left(\mathscr{L}(\lambda)\right)_{ij}\,,
\qquad
i,j=1,2,\dots,p\,.
\label{Lp}
\end{equation}
The quantum determinant $\text{q-det}\left(\mathscr{L}(\la)\right)$ in \rf{apfromqdet}
is defined by an expression of the form
\begin{align}
 \label{qdetdef-text}
 \text{q-det}& \left(\mathscr{L}(\la)\right)\,=\\
&=\sum_{\sigma\,\in\,\mathfrak{S}_M}
\,c_{\sigma}(q)\,\mathscr{L}_{\sigma(1),1}(q^{-\frac{2}{M}\rho_1}\la)\,
\mathscr{L}_{\sigma(2),2}(q^{-\frac{2}{M}\rho_2}\la)\,\dots\,\mathscr{L}_{\sigma(M),M}(q^{-\frac{2}{M}\rho_M}\la)\,,
\notag \end{align}
The summation in \rf{qdetdef-text} is extended over all permutations $\sigma$ of $M$ elements. An explicit  
formula for the coefficients $c_{\sigma}(q)$ in \rf{qdetdef-text} can be found in \rf{csigmaexpl}. 
Note that $[A_p(\la),A_q(\mu)]=0$.

We are interested in 
the contributions of the imaginary root generators to the universal Lax matrix contained in 
generating functions $k_i(\la)$ defined 
via
\begin{equation}
 \left(\pi^{\text{f}}_\la\,\otimes\,1\right)\mathscr{R}^-_{\sim\delta}\,=\,
\sum_{i=1}^M\,k_i(\la)\,\mathsf{E}_{ii}\,.
\label{Imrootfundpart}
\end{equation}
The explicit form of  $k_i(\la)$ can be obtained using  the definition
$\mathscr{R}^-_{\sim\delta}$, see \eqref{rpd} with \eqref{umij}, and the explicit formula 
for $\pi^{\rm f}_{\la}(f^{(i)}_{m\delta})$
given in Appendix \ref{app:CWbasisFERMfs} .
One can verify by direct comparison that  $k_i(\la)$
satisfy the following  relations
\begin{equation}
  \frac{k_{i+1}(\la)}{ k_i(\la)}\,=\,1+E_i'((-q)^{i} \la^{-M})\,,
\qquad
\prod_{i=1}^M  k_i(q^{-\frac{2}{M}\rho_i}\la)\,=\,1\,,
\label{kkgivesE}
\end{equation}
where $1+E_i'(z)$ is defined in \eqref{Imrootscurrents} and $\rho_i=\frac{M-2i+1}{2}$ are the components of 
the Weyl vector.
Combining this observation with \eqref{apfromqdet} and
\begin{equation}
 \left(\pi^{\text{f}}_\lambda\,\otimes\,1\right)q^{-t}\,=\,
\sum_{i=1}^M\,\,\SE_{ii}\ot
 q^{\frac{\bar{\epsilon}}{M}-\bar{\epsilon}_i}\,.
\label{fundqt}
\end{equation}
 we obtain
\begin{equation}\label{Epfromqdet}
1+E_i'((-1)^{i} \la^{-M})\,=\,
\frac{A_{i+1}(\la)A_{i-1}(\la)}{A_{i}(q^{-\frac{1}{M}}\la)A_{i}(q^{+\frac{1}{M}}\la)}
\,q^{\bar{\epsilon}_{i+1}-\bar{\epsilon}_{i}}\,,
\end{equation}
where $A_i(\la)$ are defined in \eqref{apfromqdet}.
Notice that this combination remains unchanged if we rescale the matrix $\mathscr{L}(\la)$ 
by an overall function of $\la$.
Formula \rf{Epfromqdet} allows us to complete  the proof of rationality of the currents 
for the representations of interest along the lines of Section \ref{sec:ImrootRATIONAL_sl2}.
It suffices to note that
the generating functions  $A_{i}(\la)$ get represented, up to an $i$-independent factor proportional 
to the identity, by polynomials in $\la$. We have checked this fact explicitly 
for the basic representations of our interest, and it will continue to hold for any tensor product
of these representations.


\begin{rem}
In Section \ref{QuaAffAlg} we presented a realization of the quantum affine algebra
 $\mathcal{U}_q(\widehat{\mathfrak{g}})$ in terms of 
$3\,r$ generators. This presentation is due to Drinfel'd and Jimbo \cite{Dr1,J}.
 There is an other realization known as Drinfel'd
second realization \cite{Dr87}. This realization involves certain currents which, as explained in \cite{KT2},  are 
directly connected to the root vectors defined in Section \ref{sec:rootgen}. 
The isomorphism between the realization of 
  Drinfel'd and Jimbo and the  Drinfel'd second realization has been proven 
  in \cite{Beck1}. 

In the case $\widehat{\mathfrak{g}}=\widehat{\mathfrak{sl}}_M$
 there is yet an other presentation of the quantum affine algebra following
the Leningrad school, see \cite{FaReTa,ReSe}. 
The isomorphism between this realization and the Drinfel'd second realization 
was establshed in \cite{DF}. We may note that
the universal Lax matrix introduced above 
contains (half) of the generators of $\mathcal{U}_q(\widehat{\mathfrak{sl}}_M)$ 
in the presentation of \cite{FaReTa,ReSe}. 
The proof above therefore combines elements of all three realisations.
\end{rem}



\subsection{Co-product of imaginary root generators}
\label{sec:coprIM_root_slM}

In Section \ref{sec:Compatsub} we had found the  useful identity \rf{BEid} expressing the mixing between
real and imaginary roots under co-product in the case of $\CU_q(\widehat{\fsl}_2)$.  
It allowed us to analyse possible consistency conditions on the renormalisation of the imaginary root
contributions that might arise from this mixing. 
We shall now describe the generalisation of the identity \rf{BEid} to the case of
$\CU_q(\widehat{\fsl}_M)$. 
As a useful generating function we shall again consider 
\begin{equation}
 \mathscr{M}^-_{\sim\delta}(\nu)\,\otimes 1 :=
\left(1\otimes\pi_{\nu}^-\right)(\mathscr{R}^-_{\sim \delta})\,,
\end{equation}
The explicit expression of  $\mathscr{M}^-_{\sim\delta}(\nu)$ follows from the definitions
\eqref{Mdeltaminus},  \eqref{rpd}
 and 
the form of the 
 imaginary root vectors given in \eqref{eiIMpiminus}:
\begin{equation}\label{miminusdeltaexpl1}
 \mathscr{M}^-_{\sim\delta}(\nu)\,=\,
\,\exp\left(\sum_{m=1}^{\infty}\frac{(-1)^{m+1}}{m}\,\frac{\nu^{-mM}}{q^{mM}-q^{-mM}}\,
\tilde{f}^{(M-1)}_{m\delta}\right)
,
\end{equation}
with 
\begin{equation}\label{VmandXidef2}
\tilde{f}^{(j)}_{m\delta}:=
 (q^{mM}-q^{-mM})
\sum_{i=1}^{M-1}
\, u_{m,j\,i}\,f^{(i)}_{m\delta}\,,
\end{equation}
 and $u_{m,ij}$  given in \eqref{umij}. We are going to show that
the co-product of $ \mathscr{M}^-_{\sim\delta}(\nu)$  takes the form
\begin{equation}\label{coproductimSLM}
 \Delta\left(\mathscr{M}^-_{\sim \delta }(\nu)\right)\,=\,
\left(\mathscr{M}^-_{\sim \delta}(\nu)\otimes 1\right) \,\varepsilon_q\left(\nu^{-M}\,\Xi\right)\,
\left(1\otimes \mathscr{M}^-_{\sim \delta}(\nu)\right)\,,
\end{equation}
generalising \rf{BEid} to the cases with $M>2$.
We are using the notation $\tau_q=q-q^{-1}$ and 
\begin{equation}\label{Mdeltaminus}
\Xi:=
\tau_q^2\,\sum_{j=1}^{M-1}\,q^{\bar{\epsilon}_M-\bar{\epsilon}_j}\,
f^{}_{\delta-(\epsilon_j-\epsilon_M)}\otimes f^{\text{op}}_{\epsilon_j-\epsilon_M}\,,
\end{equation}
is the combination of real root generators appearing in the co-product of $ \mathscr{M}^-_{\sim\delta}(\nu)$.
In the definition of $\Xi$ the terms in the second tensor factor $f^{\text{op}}_{\gamma}$ are
 constructed using the opposite root ordering
 compared to the one defined in Appendix \ref{sec:rootSLM} which is used for the construction of 
 $f_{\gamma}$. 
Their explicit expression can be found in  \eqref{fOPPexplicit}.


In the following we will report the main ideas that enter the derivation of \eqref{coproductimSLM}
leaving most of the technical details to Appendix \ref{app:derivationIMROOTPART}.
The first observation is the following
\begin{equation}
\left(1\otimes\pi_{\nu}^-\right)(\bar{\mathscr{R}}^-)\,=
\,\Lambda(\mathsf{y})\, \left(\mathscr{M}^-(\nu)\otimes 1\right)\,\Lambda^{-1}(\mathsf{y})\,,
\qquad
\Lambda(\mathsf{y})\,:=\,e^{\sum_{i=1}^M\,\bar{\epsilon}_i \otimes \log\mathsf{y}_i}\,,
\end{equation}
where $\bar{\epsilon}_i$ are the Cartan generators \eqref{introduceebar} 
and  the variables $\mathsf{y}_i$ are introduced in \eqref{repplusSectionBAXTER}.
Notice that we have already used the operators $\mathsf{y}_i$ and the similarity transform $\Lambda(\mathsf{y})$
to simplify the study of tensor products involving $\pi_+$ in Section \eqref{sec:blocktriang}.
The explicit expression of  $\mathscr{M}^-(\nu)$ follows from the product formula of the universal R-matrix \eqref{KT}  
and the  form of $(1\otimes \pi^-)(f_{\gamma}\otimes e_{\gamma})$ for $\gamma$ a real root given in 
\eqref{Posrootanypibarminus}. It takes the form
\begin{equation}\label{Mminusexplicit}
\mathscr{M}^-(\nu)=
\mathscr{M}^-_{\prec}(\nu)
 \,\mathscr{M}^-_{\sim\delta}(\nu)
\,\mathscr{M}^-_{\succ}(\nu)\,,
\qquad
\begin{aligned}
\mathscr{M}^-_{\prec}(\nu)& =\,
\varepsilon_q(\mathsf{X}_1^{\prec})\dots \varepsilon_q(\mathsf{X}_{M-1}^{\prec})\,,
\\
\mathscr{M}^-_{\succ}(\nu)& =\,
\varepsilon_q(\mathsf{X}_{M-1}^{\succ})\dots \varepsilon_q(\mathsf{X}_{1}^{\succ})\,,
\end{aligned}
\end{equation}
where
\begin{equation}\label{Xpreccsuccdef}
\mathsf{X}_i^{\prec}:=\tau_q\,\check{\nu}^{-1}\,\check{f}_i\,,\quad
\mathsf{X}_i^{\succ}:=\tau_q\,\check{\nu}^{-i}\,[\text{\footnotesize{$\cdots$}}[\check{f}_0,\check{f}_1],\dots,\check{f}_{i-1}]\,,
\quad
\check{f}_i\,:=\,q^{-\frac{1}{2}\,\left(\bar{\epsilon}_i+\bar{\epsilon}_{i+1}-1\right)}\, f_{i}\,,
\end{equation}
with $\check{\nu}= \nu\,q^{\frac{M-\bar{\epsilon}}{M}}$ and 
$\tau_q=q-q^{-1}$.
The fact that only finitely many real roots contribute to the product formula \eqref{KT}  
is due to the special property of $\pi^-$ spelled out  in Section \ref{sec:Imagerootvectors2}.
Notice that the  nested commutator in the definition of 
$\mathsf{X}_i^{\succ}$ is
\begin{equation}\label{fromftoXsucc}
\,[\text{\footnotesize{$\cdots$}}[\check{f}_0,\check{f}_1],\dots,\check{f}_{i-1}]\,=\,
q^{-\sum_{k=1}^{i-1}(\bar{\epsilon}_k-1)} q^{-\frac{1}{2}\,\left(\bar{\epsilon}_i+\bar{\epsilon}_{M}-1\right)}\,
f_{\delta-(\epsilon_i-\epsilon_M)} \,.
\end{equation}
The commutation relations and coproduct formulae for the elements \eqref{Xpreccsuccdef} are collected 
in Appendix \ref{app:derivationIMROOTPART}.

The second ingredient used in derivation of \eqref{coproductimSLM}
 are certain  identities  satisfied by
$\varepsilon_q(\mathsf{X})$.
 In addition to the relations \eqref{qexp-prop1}, \eqref{qexp-prop2}
 used in Section \ref{sec:Compatsub} in the case $M=2$, 
the following generalized pentagon equation holds
\begin{equation}\label{pentagonwithSerre}
 \varepsilon_q(V)\varepsilon_q(U)=
\varepsilon_q(U)\,\varepsilon_q\!\left(\frac{VU-UV}{q-q^{-1}}\right)\,\varepsilon_q(V)\,,
\end{equation}
if
\begin{align}
 q^{-1}\,V^2\,U+(q+q^{-1})\,V\,U\,V+q^{+1}\,U\,V^2\,=\,0\,,
\label{serreUV1}\\
 q^{+1}\,U^2\,V+(q+q^{-1})\,U\,V\,U+q^{-1}\,V\,U^2\,=\,0\,.
\label{serreUV2}
\end{align}
Notice that the identity  \eqref{qexp-prop2} is a special
 case of \eqref{pentagonwithSerre} for 
$U\,V\,=\,q^{-2}\,V\,U$.
%
The two basic identities \eqref{qexp-prop1} and \eqref{pentagonwithSerre} are known to be satisfied by
$\varepsilon_q(x)$. 

The last important observation used in the derivation  is that 
\begin{equation}\label{Vmcommuteswithfj}
[f_{j}, \tilde{f}^{(M-1)}_{m\delta}]=0\,,\qquad \text{for}\,\,\,\,\,j\,=\,1,\dots,\,M-2\,,
\end{equation}
where $\tilde{f}^{(M-1)}_{m\delta}$ are defined in \eqref{VmandXidef2}.
This follow  from the definition \eqref{VmandXidef2} and the
commutation relations\footnote{It is actually obtained by applying the
 Cartan anti-involution \eqref{DefZeta} to \eqref{eecomm_gives_t}} \eqref{eecomm_gives_t}.

\begin{rem}
For future use let us note that 
the relations obtained from \rf{pentagonwithSerre} by replacing $\varepsilon_q(x)$ by
$\mathscr{E}_q(x)$ and $U$, $V$ by positive self-adjoint operators 
are also satisfied, see e.g.~\cite{Ip12} for a derivation. 
The identities obtained by using our renormalisation prescription 
to define the evaluation of $ \Delta\left(\mathscr{M}^-_{\sim \delta }(\nu)\right)$ in representations of modular
double type will therefore also be valid.
\end{rem}

\begin{rem}
The identity \eqref{coproductimSLM} is understood
as an equality of formal power series in the spectral parameter.
One may notice that the the first non-trivial term in this expansion reads
\begin{equation}
\Delta(\tilde{f}^{(M-1)}_{\delta})-\tilde{f}^{(M-1)}_{\delta}\ot 1- 1\ot \tilde{f}^{(M-1)}_{\delta}\,=\,
[M]_{q}\,\Xi\,.
\end{equation}
Within this interpretation, the relation \eqref{coproductimSLM}
 provides  a compact expression for the coproduct of imaginary root vectors.  
This should be compared with known expressions in the literature from \cite{Dam2} and \cite{KT94}.
In \cite{KT94} an explicit twist that maps the coproduct defined in this paper, to the so-called Drinfeld coproduct,
with respect to which imaginary roots are primitive elements,
is constructed. This form is not of direct use when both tensor factors correspond to representations 
of $\mathcal{U}_q(\mathfrak{b}_+)$ that cannot be extended to representations of the full
$\mathcal{U}_q(\widehat{\mathfrak{sl}}_M)$.
\end{rem}

\begin{rem}
The  quantity
 $\mathscr{M}^-_{\sim\delta}$ defined in \eqref{Mdeltaminus} appeared also 
in  \cite{FH} (Section 7.2), where it is called $T_{i=M-1}(z)$.
\end{rem}

\subsection{Checks of compatibility}

In the previous section we had verified in the case of $\mathcal{U}_q(\widehat{\mathfrak{sl}}_2)$
that the proposed renormalisation prescription  preserves all the basic properties of the 
universal R-matrices. This was found to be a consequence of the fact that  the 
function $\mathscr{E}_q(x)$ used to define
the renormalisation of the real root contributions
satisfies the same functional relations \eqref{qexp-prop1}, \eqref{qexp-prop2}, and  
\eqref{pentagonwithSerre} as are satisfied by the function $\varepsilon_q(x)$ appearing in the
product formula. In the following we will outline how to 
generalise this discussion to the case of $\mathcal{U}_q(\widehat{\mathfrak{sl}}_M)$. 

It will furthermore be explained how the consequences of the identity 
\rf{coproductimSLM} are consistent with the renormalisation prescription
\begin{equation}\label{piplusMdelta}
\boldsymbol{\pi}^{+} \left(\mathscr{M}^-_{\sim\delta}(\nu)\right)\,=\, \prod_{i=1}^{M-1}\,
\frac{\prod_{\ell=1}^{d_{i,+}}\,\mathcal{G}_{i,M-1}(\nu^{-M}\,\mathsf{D}_{\ell,i}^+)}%
{\prod_{\ell=1}^{n_{i,+}}\,\mathcal{G}_{i,M-1}(\nu^{-M}\,\mathsf{N}_{\ell,i}^+)}\,.
\end{equation}
This will again be a consequence of the functional equations satisfied by the special function
$\mathscr{E}_q(x)$.


\subsubsection{The case of  $\pi_{\mathcal{F}}\otimes \pi_+$ and the Baxter equation }
\label{sec:ImaginaryrootMORE_baxt}

In the following we verify \eqref{coproductimSLM}  when the first two tensor factors are chosen as 
$\pi_{\mathcal{F}}\otimes \pi_+$.
We leave the proof of the identity involving real root contributions, generalizing the one presented in Section \ref{sec:tensfinite-inf_SL2},
to Appendix \ref{APP:formixedpentagon}. 
This is a prototypical example of tensor products involving finite dimensional representations and modular double type representations.
This verification, supplemented with a similar analysis where $\pi^-$ is replaced by $\bar{\pi}^-$
that goes along the same lines, allows to  complete  the proof of the Baxter equation.





\paragraph{Explicit verification of  $\pi^{\mathcal{F}}\otimes \pi^+$ 
applied to  \eqref{coproductimSLM}}
The verification of  \eqref{coproductimSLM} in this case is greatly simplified by the
analysis of imaginary root currents given in Section
\ref{sec:tensprodandDRcurr}. More specifically, the relation 
 \eqref{equalitycurrents} implies that the 
left hand side of \eqref{coproductimSLM}  can be rewritten using 
\begin{equation}\label{forpiFpipuslIMA}
\Lambda(\mathsf{y})
\left[ 
\left(\pi_{\mu_{\mathfrak{n}}}^{\mathcal{F}}
\otimes \pi_{\lambda_{\mathfrak{n}}}^+\right)
\Delta\left(\mathscr{M}^-_{\sim\delta}(\nu)\right)\right]
\Lambda(\mathsf{y})^{-1}=
\mathcal{S}^{-1}
\left[ 
\pi_{\mu_{\mathfrak{n}}}^{\mathcal{F}}\left(\mathscr{M}^-_{\sim\delta}(\nu)\right)
\otimes \pi_{\lambda_{\mathfrak{n}}}^+\left(\mathscr{M}^-_{\sim\delta}(\nu)\right)\right]
\mathcal{S}\,.
\end{equation}
where $\mathcal{S}$  is given in \eqref{SdefcurrentFermpiplus}.
Concerning the right hand side,  the following holds:
\begin{equation}\label{item1}
\pi_{\mu_{\mathfrak{n}}}^{\mathcal{F}}\left(\mathscr{M}^-_{\sim\delta}(\nu)\right)\,=\,
\theta^-_{\mathcal{F}}(g_-)\,\left(1-g_-^M\,\mathfrak{n}_M\right)\,,
\end{equation}
\begin{equation}\label{item2}
\Lambda(\mathsf{y})
\left[ 
\left(\pi^{\mu_{\mathfrak{n}}}_{\mathcal{F}}
\otimes \pi_{\lambda_{\mathfrak{n}}}^+\right)
\Xi\right]
\Lambda(\mathsf{y})^{-1}=\Xi'\otimes 1
\,,\quad
\Xi'=
-\tau_q\,\la_{\mathfrak{n}}^M
\left(\sum_{k=1}^{M-1}\,g_{\mathfrak{n}}^{-k}\, \bar{\mathbf{c}}_k\right)\mathbf{c}_M\,,
\end{equation}
where  $\theta^-_{\mathcal{F}}(x)$ is defined in \eqref{Thetadef}
 and the operator $\pi_+(\mathscr{M}^-_{\sim\delta})$ is central.
The equality  \eqref{item2} follows from  \eqref{MdeltaEQ1FERM} and the definition
 \eqref{Mdeltaminus}.

It follows from these observations and the prescription \eqref{realrootDEF}
 for $\mathscr{E}_q(\nu^{-M}\Xi'\otimes 1)$
that  \eqref{coproductimSLM} reduces to 
\begin{equation}\label{stepinpiFpiplus}
\mathcal{S}^{-1}\left(1-g_-^M\,\mathfrak{n}_M\right)\mathcal{S}\,=\,
\left(1-g_-^M\,\mathfrak{n}_M\right)\left(1+\tau_q^{-1}\,\nu^{-M}\,\,\Xi'\right)\,.
\end{equation}
This simple equality of operators acting on the fermionic Fock space holds
as a consequence of  
\begin{equation}\label{stepinpiFpiplus2}
\mathcal{S}^{-1}\bar{\mathbf{c}}_M\,\mathcal{S}\,=\,
\bar{\mathbf{c}}_M+\sum_{k=1}^{M-1}\,g_{\mathfrak{n}}^{M-k}\,\,\bar{\mathbf{c}}_k\,.
\end{equation}
In order to reduce \eqref{stepinpiFpiplus} to \eqref{stepinpiFpiplus2},
one can use the explicit form of $\Xi'$ and the following relations:
 $\mathfrak{n}_M\,\mathbf{c}_M=0$,  $\mathcal{S}$ commutes with $\mathbf{c}_M$
and $\nu^{-M}\,(\la_{\mathfrak{n}})^{M}(g_-)^{-M}=(g_{\mathfrak{n}})^{M}$.
The identity  \eqref{stepinpiFpiplus2} is easy to show.  \qed

\subsubsection{Checks of \eqref{coproductimSLM} evaluated on prefundamental representations}
\label{sec:checkcoproducMdelta_prefund}

This section contains an explicit verification of  the identity that follows from  \eqref{coproductimSLM} 
after applying  $\pi^{+}\otimes \bar{\pi}^+$ or $\pi^{+}\otimes \pi^+$ to it.
The verification requires a careful study of the the image of imaginary root currents 
under the tensor product representations
 $(\pi^{+}\otimes \bar{\pi}^+)\Delta$ or $(\pi^{+}\otimes \pi^+)\Delta$.
 These are representations of $\Borelm$.
We will see that in this case
the (generalized) eigenvalues of imaginary root currents do not behave multiplicatively under 
tensor product, see \eqref{Imrootpiplusplus} and \eqref{pibarpluspiplusimaginary} below.
This should be compared to a rather general result, which is a corollary of Theorem  8.1 of \cite{KT94},
which states the following:

\noindent
Let $\boldsymbol{\pi}_{\text{full}}$ be a representation of $\mathcal{U}_q(\widehat{\mathfrak{g}})$
and $\boldsymbol{\pi}^{+}$ a representation of $\Borelm$, then 
the generalized eigenvalues of $(\boldsymbol{\pi}_{\text{full}}\otimes \boldsymbol{\pi}^{+})\Delta\left(1+F'_i(z)\right)$
and  $(\boldsymbol{\pi}^{+}\ot \boldsymbol{\pi}_{\text{full}})\Delta\left(1+F'_i(z)\right)$
are equal to the eigenvalues of $\boldsymbol{\pi}^{+}\left(1+F'_i(z)\right)\times \boldsymbol{\pi}_{\text{full}}\left(1+F'_i(z)\right)$. Notice that this result, once supplemented by the information that any finite dimensional representation 
of  $\Borelm$ can be extended to a finite dimensional representation of 
$\mathcal{U}_q(\widehat{\mathfrak{g}})$,
 implies the result of Proposition 1 in \cite{FrR}.

\paragraph{Explicit verification of  $\pi^{+}\otimes \pi^+$ 
applied to  \eqref{coproductimSLM}.}
As in the example in Section \ref{sec:ImaginaryrootMORE_baxt},
 in order to verify \eqref{coproductimSLM}, we need to evaluate two basic quantities:
(1) the coproduct of imaginary root currents, (2)  the element $\Xi$ defined in \eqref{Mdeltaminus}.
Let us proceed in order.
On the one hand 
the currents of imaginary root vectors for  the tensor product of two 
prefundamental representations $\pi^+$ take a particularly simple form
\begin{equation}\label{Imrootpiplusplus}
\left(\pi^+_{\lambda_1}\ot\pi^+_{\lambda_2}\right)\Delta
\left(1+F'_i(z)\right)\,=\,
\begin{cases}
1 & i\neq M-2,M-1 \\
1+\mathbf{r}\,z^{-1} & i=M-2 \\
\frac{\left(1+\la^M_{1}\,z^{-1}\right)
\left(1+\la_{2}^M\,z^{-1}\right)}
{\left(1-q^{+1}\,\mathbf{r}\,z^{-1}\right)\left(1-q^{-1}\,\mathbf{r}\,z^{-1}\right)}  & i=M-1
\end{cases}
\end{equation}
 where $\tau_q=q-q^{-1}$ and the operator $\mathbf{r}$ is given below.
The result \eqref{Imrootpiplusplus} follows from a  straightforward but lengthy calculation.
The form \eqref{Imrootpiplusplus} is not too surprising if we recall that, 
in the special case of $\mathcal{U}_q(\widehat{\mathfrak{sl}}_M)$,
 the imaginary root currents can be  computed using the formula  \eqref{Epfromqdet}
with $\mathscr{L}$ replaced by $\mathsf{L}^+\mathsf{L}^+$. 
It follows  from  \eqref{Imrootpiplusplus}
 that the linear combination of imaginary roots defined in  \eqref{VmandXidef2} 
 satisfies the relation
\begin{equation}\label{tildefpiplus2}
\left(\pi^+_{\lambda_1}\ot\pi^+_{\lambda_2}\right)\Big{[}
\Delta(\tilde{f}^{(M-1)}_{m\delta})-\tilde{f}^{(M-1)}_{m\delta}\ot 1- 1\ot \tilde{f}^{(M-1)}_{m\delta}
\Big{]}\,=\,
[M]_{q^m}\,\mathsf{r}^m\,.
\end{equation}
The result  \eqref{Imrootpiplusplus} with the definition \eqref{Mdeltaminus} implies that
\begin{equation}\label{coproductimSLMplusplus}
\frac{
\left(\pi^+_{\lambda_1}\ot\pi^+_{\lambda_2}\right)
\!\Delta\left(\mathscr{M}^-_{\sim \delta }(\nu)\right)
}{
\left(\pi^+_{\lambda_1}\ot\pi^+_{\lambda_2}\right)
\left(\mathscr{M}^-_{\sim \delta}(\nu)\otimes \mathscr{M}^-_{\sim \delta}(\nu)\right)}
\,=\,
\mathcal{E}_{b^2}\left(\nu^{-M}\,\mathbf{r}\right)\,
\,
\end{equation}
In writing the left hand side of this expression we have used the fact that the denominator is represented by central elements. The identity \eqref{coproductimSLMplusplus} is then obtained by first computing \eqref{tildefpiplus2} and then  applying  the renormalization  prescription to the expression  \eqref{miminusdeltaexpl1}.
It is instructive to rederive  \eqref{coproductimSLMplusplus} from the general formula \eqref{piplusMdelta} with $\boldsymbol{\pi}^{+}\mapsto (\pi^+_{\lambda_1}\ot\pi^+_{\lambda_2})\Delta$.
From this point of view \eqref{coproductimSLMplusplus} 
holds as a consequence of the following identity
\begin{equation}\label{Gijsimplifies}
\frac{\mathcal{G}_{M-1,M-1}(-q^{+1}\omega)\,\mathcal{G}_{M-1,M-1}(-q^{-1}\omega)}{\mathcal{G}_{M-2,M-1}(\omega)}\,=\,
\prod_{s=1}^M\,\mathcal{E}_{Mb^2}(q^{2\rho_s(M)}\omega)\,=\,
\mathcal{E}_{b^2}(\omega)\,,
\end{equation}
with $\omega=\mathsf{r}$. The first equality in \eqref{Gijsimplifies}  follows from the definition of $\mathcal{G}_{ij}(x)$ given in \eqref{Gij_def}
and does not use any property of $\mathcal{E}_{Mb^2}(x)$. 
The second equality in  \eqref{Gijsimplifies}  is a simple consequence of the definition \eqref{Eb-def_sect8}.
 
 In order to complete the verification 
that  \eqref{coproductimSLM} 
holds when we  apply the representation 
 $\pi^{+}\otimes \pi^+$, 
 we need to evaluate the image of 
 $\Xi$  defined in \eqref{Mdeltaminus}.
 A simple calculation shows that 
\begin{equation}
\left(\pi^+_{\lambda_1}\ot\pi^+_{\lambda_2}\right)\Xi\,=\,
(q-q^{-1})^M\,
\sum_{j=1}^{M-1}\,\su_M^{}\su_j^{-1}
\left(\mathsf{f}_{j-1}\dots \mathsf{f}_1\,\mathsf{f}_0\right)
\ot
\left(\mathsf{f}_{M-1}\dots \mathsf{f}_{j+1}\,\mathsf{f}_j\right)\,=\,
\mathbf{r}\,,
\end{equation}
with $\pi^+(f_{\delta-(\epsilon_j-\epsilon_M)})$  
and $\pi^+(f^{\text{op}}_{\epsilon_i-\epsilon_M})$  given in
\eqref{eisuccpiplus} and below \eqref{MdeltaEQ1} respecively.
Above we used the by now standard notation  
$(\pi^+_{\lambda_1}\ot\pi^+_{\lambda_2})(f_i\ot f_j)=\mathsf{f}_i\ot \mathsf{f}_j$.
The operator $\mathsf{r}$ is the same as the one appearing in the currents \eqref{Imrootpiplusplus}.
This conclude the check in this case.

\paragraph{Explicit verification of  $\bar{\pi}_{+}\otimes \pi_+$ 
applied to  \eqref{coproductimSLM}.}
The steps are the same as in the previous paragraph with important
structural differences.
The imaginary root currents take the form 
\begin{equation}\label{pibarpluspiplusimaginary}
\left(\bar{\pi}^+_{\lambda_1}\ot\pi^+_{\lambda_2}\right)\Delta(1+F'_i(z))\,=\,
\frac{\left(1+z^{-1}\,X_{i-1}\right)\left(1+z^{-1}\,X_{i+1}\right)}
{\left(1-q^{-1}\,z^{-1}\,X_{i}\right)\left(1-q^{+1}\,z^{-1}\,X_{i}\right)}\,,
\end{equation}
where 
\begin{equation}
X_i\,=\,\la_1^{M-i}\,\la_2^i\,
\left(t_1\cdots t_i \right)^{-1}
\left(t_{i+1} \cdots t_M \right)^{}
\qquad
t_i\,=\,
\left(q\,\su_i^{}\sv_i^{}\ot \sv_i^{-1}\su_i^{}\right)^{\frac{1}{2}}\,.
\end{equation}
Notice that 
$t_i$ are commuting operators and satisfy $t_1\dots t_M=1$.
%
%
The linear combination of imaginary roots defined in \eqref{VmandXidef2} 
 satisfies the relation
\begin{equation}\label{tildefpiplus2next}
\left(\bar{\pi}^+_{\lambda_1}\ot\pi^+_{\lambda_2}\right)\Big{[}
\Delta(\tilde{f}^{(M-1)}_{m\delta})-\tilde{f}^{(M-1)}_{m\delta}\ot 1- 1\ot \tilde{f}^{(M-1)}_{m\delta}
\Big{]}\,=\,
[M]_{q^m}\,\left(\la_1\la_2^{M-1}\,t_M^2\right)^m\,.
\end{equation}
To obtain this expression it is useful to observe that most of the terms in the sum 
\eqref{VmandXidef2}  cancel with each other due to the form \eqref{pibarpluspiplusimaginary}
and the identity 
$[i+1]_{q^k}+
[i-1]_{q^k}-
[i]_{q^k}(q^{k}+q^{-k})=0$.
 By a similar mechanism as in \eqref{coproductimSLMplusplus}, this implies that
\begin{equation}\label{coproductimSLMBIS}
\frac{
\left(\bar{\pi}^+_{\lambda_1}\ot\pi^+_{\lambda_2}\right)
\!\Delta\left(\mathscr{M}^-_{\sim \delta }(\nu)\right)
}{
\left(\bar{\pi}^+_{\lambda_1}\ot\pi^+_{\lambda_2}\right)
\left(\mathscr{M}^-_{\sim \delta}(\nu)\otimes \mathscr{M}^-_{\sim \delta}(\nu)\right)}
\,=\,
\mathcal{E}_{b^2}\left(\nu^{-M}\,\la_1\la_2^{M-1}\,t_M^2\right)\,
\,.
\end{equation}
It is  instructive to rederive  \eqref{coproductimSLMBIS}
 from the general formula \eqref{piplusMdelta} with
  $\boldsymbol{\pi}^{+}\mapsto (\bar{\pi}^+_{\lambda_1}\ot\pi^+_{\lambda_2})\Delta$.
From this point of view  \eqref{coproductimSLMBIS} holds as a consequence of  
\begin{equation}\label{GsimpleSECOND}
\prod_{i=1}^{M-1}\,
\frac{\mathcal{G}_{iM-1}(-q^{-1}\omega_i)\,\mathcal{G}_{iM-1}(-q^{+1}\omega_i)}
{\mathcal{G}_{iM-1}(\omega_{i+1})\,\mathcal{G}_{iM-1}(\omega_{i+1})}=
\frac{\prod_{i=1}^M\,\mathcal{E}_{Mb^2}(q^{2\rho_s(M)}\omega_{M-1})}{\rho^{\dot{+}-}(\omega_0)\rho^{+-}(\omega_M)}=
\frac{\mathcal{E}_{b^2}(\omega_{M-1})}{\rho^{\dot{+}-}(\omega_0)\rho^{+-}(\omega_M)}\,,
\end{equation}
with $\omega_i=\nu^{-M}\,\mathsf{X}_i$ and $\rho^{\epsilon_1\epsilon_2}(\omega)$ defined in 
\eqref{rhopmsec8}. The first equality in \eqref{GsimpleSECOND} does not uses any property of the special function $\mathcal{E}_{Mb^2}(\omega)$.
The second equality is the same as in \eqref{Gijsimplifies}.
For the right hand side of \eqref{coproductimSLM} one finds that 
\begin{equation}
\left(\bar{\pi}^+_{\lambda_1}\ot\pi^+_{\lambda_2}\right)\Xi\,=\,
(q-q^{-1})^M\,
\,\su_M^{}\su_1^{-1}\,
\bar{\mathsf{f}}_{0}
\ot
\left(\mathsf{f}_{M-1}\dots \mathsf{f}_{2}\,\mathsf{f}_1\right)=
\la_1\la_2^{M-1}\,t_M^2\,,
\end{equation}
where $\Xi$ is defined in \eqref{Mdeltaminus}
The form of
 $\bar{\pi}^+(f_{\delta-(\epsilon_j-\epsilon_M)})=\delta_{j,1}\,\bar{\mathsf{f}}_0$  
follows from the definition \eqref{TQstartBIS} and the iterative construction of root vectors,
the second tensor factor
$\pi^+(f^{\text{op}}_{\epsilon_i-\epsilon_M})$  is given 
below \eqref{MdeltaEQ1}.
 This concludes the verification of \eqref{coproductimSLM} in this case.




\subsection{Modular duality and quantum Wronskian relations}
\label{sec:moddualandWrons}

By dividing the Q-operators by the scalar factors coming from the imaginary roots one
obtains Q-operators that are manifestly self-dual under $b\ra b^{-1}$. We are now going 
to show that this has important consequences, leading to functional relations 
among the Q-operators of quantum Wronskian type. In the case $M=2$ it has been observed
in \cite{Z00} that such functional relations can be solved to express the eigenvalues of Q-operators
in terms of solutions to certain nonlinear difference equations of thermodynamic Bethe ansatz (TBA) type.

\subsubsection{Rewriting the Baxter equations}
When the quantum space is taken  as  
\begin{equation}\label{quantumspacefinal}
 \pi_{q}(a)\,=\,\left(\pi^-_{\kappa_N}\otimes\bar{\pi}_{\bar{\kappa}_N}^-\otimes \dots \otimes 
 \pi^-_{\kappa_1}\otimes\bar{\pi}^-_{\bar{\kappa}_1} \right)\Delta^{(2N)}(a)
\qquad
a\,\in\,\mathcal{U}_q(\mathfrak{b}^+)\,.
\end{equation}
 the transfer matrices entering the Baxter equation
  \eqref{TQstart} can be rewritten as follows
\begin{equation}
\label{TandQreduced}
\mathsf{Q}^+(\zeta)\,=\,
\Xi(\zeta)\,\mathsf{q}^+(\zeta)\,,
\qquad
\mathbb{T}^{(k)}(\zeta)\,=\,
\Theta^{(k)}(\zeta)\,\mathsf{t}_k(\zeta)\,,
\qquad
k=1,\dots,M-1\,,
\end{equation}
where
\begin{equation}
\Xi(\zeta)\,:=\,
\prod_{a=1}^N\,
\rho^{+-}(\zeta\,\ka_a^{-1})\,
\rho^{+\dot{-}}(\zeta\,\bar{\ka}_a^{-1})\,,
\quad
\Theta^{(k)}(\zeta)
\,:=\,
\prod_{a=1}^N\,
\theta^-_k(\zeta\,\ka_a^{-1})\,
\theta_k^{\dot{-}}(\zeta\,\bar{\ka}_a^{-1})\,.
\end{equation}
The function
$\rho^{\epsilon_1\epsilon_2}(\zeta)$ are given in \eqref{rhopmsec8} and the form of 
$\theta^{\epsilon}_k(\zeta)$ follows from 
  \eqref{Thetadef}, \eqref{Thetabardef} to be
  \begin{equation}
\theta^-_k(\zeta)=
\frac{\mathcal{E}_{Mb^2}((-1)^{M-1}\,q^{-k}\,\zeta^M)}
{\mathcal{E}_{Mb^2}((-1)^{M-1}\,q^{+k}\,\zeta^M)}\,,
\qquad\,\,
\theta^{\dot{-}}_k(\zeta)=\theta^{-}_{M-k}(-\zeta)\,.
  \end{equation}
  The remaining transfer matrices involved in  \eqref{TQstart}  are simply given by
      $\mathbb{T}^{(1)}(\zeta)=\mathbb{T}^{(M)}(\zeta)=\mathbf{1}$.
The rewriting above is convenient because the transfer matrices 
 $\mathsf{t}_k(\zeta)$ and $\mathsf{q}^+(\zeta)$ have simpler analytic properties as functions of the 
 spectral parameter compared to their ancestors.
  
Inserting \eqref{TandQreduced} in the Baxter equation \eqref{TQstart} and dividing 
 by $\Xi(-q\,\zeta)$ we obtain
\begin{equation}\label{BaxterAbitMore_expl}
\begin{aligned}
 \sum_{k=1}^{M-1}\,(-1)^{k-1}&\,c_k(\zeta)\,\mathsf{t}_k(q^{\frac{k}{M}}\zeta)\,
\mathsf{q}^+(-\omega \,q^{\frac{2k-M}{M}}\zeta)
\,=\,\\
&\,=\,\Delta(\zeta)\,\mathsf{q}^+(-\omega \,q^{-1}\zeta)+
(-1)^{M}\,\mathsf{q}^+(-\omega \,q^{+1}\zeta)\,,
\end{aligned}\end{equation}
where
\begin{equation}
\label{DeltaSLM}
\Delta(\zeta):=\frac{\Xi(-q^{-1}\,\zeta)}{\Xi(-q^{+1}\,\zeta)}\,=\,
\prod_{a=1}^N\,
\left[
\left(1-\frac{\zeta^M}{\bar{\ka}_a^M}\right)
\prod_{s=1}^{M-1}\,\left(1-q^{-2s}\,\left(-q\frac{\zeta}{\ka_a}\right)^M\right)
\right]\,,
\end{equation}
\begin{equation}
c_k(\zeta):=\frac{\Xi(-q^{\frac{2k-M}{M}}\,\zeta)}{\Xi(-q^{+1}\,\zeta)}\,\Theta^{(k)}(q^{\frac{k}{M}}\zeta)\,=\,
\prod_{a=1}^N\,\prod_{s=1}^{M-k-1}\,\left(1+(-1)^{M-1}\,q^{M-2s}\,\frac{\zeta^M}{\ka_a^M}\right)\,.
\end{equation}
Notice that compared to \eqref{TQstart}  we reabsorbed the $M$-th root of unity  $\omega$
in the definition of $\zeta$.
In order to derive \eqref{DeltaSLM} it is useful to notice that
\begin{subequations}
\begin{align}
& \frac{\rho^{+-}(q^{-1}\,\la)}{\rho^{+-}(q^{+1}\,\la)}\,=\,
\prod_{s=1}^{M-1}\,\left(1+q^{M-2s}\,\la^M\right)\,=\,\text{q-det}\big(\mathsf{L}^{\dot{-}}(\la)\big)\,,\\
&  \frac{\rho^{+\dot{-}}(q^{-1}\,\la)}{\rho^{+\dot{-}}(q^{+1}\,\la)}\,=\,\,
\,\,\left(1+(-1)^{M-1}\,\la^M\right)\,\,=\,\text{q-det}\big(\mathsf{L}^{-}(\la)\big)\,.
\end{align}
\end{subequations}

\subsubsection{Elementary properties of functional difference equations}

Consider the M-th order  functional difference equation for $q(\la)$
\begin{equation}\label{MainDifferenceEquation}
\sum_{k=0}^M\,(-1)^k\,t_k(\la)\,q^{[k]}(\la)\,=\,0\,,
\end{equation}
where $f^{[k]}(\la)$ means to shift the argument of $f(\la)$ in certain units, e.g.~$f^{[k]}(\la):=f^{}(p^k\la)$.
We set $t_0(\la)=t_M(\la)=1$. This is the generic situation as they can be reintroduced by rescaling the equation
\eqref{MainDifferenceEquation}  with $t_0(\la)$ and by redefining $q(\la)$.
Let us recall two elementary facts about functional difference relations:

\noindent{\bf 1.}
Let $q_1(\la),\dots,q_M(\la)$ be $M$ solutions of  \eqref{MainDifferenceEquation} then 
the quantum Wronskian 
\begin{equation}\label{wronskian}
W(\la)\,:=\,\det_{1\leq\, a,b\, \leq M}\left(q^{[a-1]}_{b}(\la)\right)\,,
\end{equation}
is a quasiconstant, i.e.~$W(\la)=W^{[1]}(\la)$.

\noindent{\bf 2.}
 Let $q_1(\la),\dots,q_{M-1}(\la)$ be $M-1$ solutions of  \eqref{MainDifferenceEquation} then 
\begin{equation}\label{qbarq}
\bar{q}^{}(\la)\,:=\,
\det_{1\leq\, a,b\, \leq M-1}\left(q^{[a-1]}_{b}(\la)\right)\,,
\end{equation}
satisfies the conjugate Baxter equation
\begin{equation}\label{MainDifferenceEquation2}
\sum_{k=0}^M\,(-1)^k\,\bar{t}_k(\la)\,\bar{q}^{[k]}(\la)\,=\,0\,,
\qquad
\bar{t}_k(\la)\,:=\, t^{[k-1]}_{M-k}(\la)\,.
\end{equation}

The statements can strengthened considerable provided one is dealing with Q-operators
that are self-dual under $b\ra b^{-1}$.

\subsubsection{Modular duality}
It is manifest from its explicit expression that $\mathsf{q}^+(\zeta)$ is invariant upon replacing 
$b$ with $b^{-1}$. This means that $\mathsf{q}^+(\zeta)$ satisfies a \emph{dual} Baxter equation obtained  by replacing $b$ with $b^{-1}$. In order to make the behaviour under $b\ra b^{-1}$ more
visible let us introduce
$u:=\frac{M}{2\pi b}\,\log \zeta$ along with $s_a:=\frac{M}{2\pi b}\,\log \kappa_a$
and
$\bar{s}_a:=\frac{M}{2\pi b}\,\log \bar\kappa_a$.
Multiplication by  $q^{\frac{2}{M}}$ and $e^{-\pi i \frac{2}{M}}$ in the
 $\zeta$-plane
translates into shifts by $-{i}b^{+1}$ and $-{i}b^{-1}$ in the $u$-plane.

We have already observed in the remark below \eqref{PiFdefTEXT} that one can obtain $M$ solutions to the Baxter equation  \eqref{TQstart} by shifting the argument of the Q-operator  as follows $\mathsf{Q}^+(\omega_{\ell}\,\zeta)$ with $\omega_{\ell}=e^{2\pi i \ell/M}$. 
The dual Baxter equation guarantees that these are linear independent.
It will be argued that the following relations hold
\begin{equation}\label{wronskianproposal2}
\det_{1\leq\, k, \ell \,\leq M}\mathsf{q}^+\big(u-{i}(k\,b^{+1}+\ell\,b^{-1})\big)\,=\,\SF(u-Mc_b)\,,
\end{equation}
where the operator $\SF(u)$ is determined up to a  $u$-independent operator 
as
\begin{equation}\label{SFsolution}
\SF(u)\,=\,\SF_0
\prod_{a=1}^N\,
\bigg[
\mathbf{e}_b^{}\big(u-\bar{s}_a-c_b\big)
\prod_{s=1}^{M-1}\,\mathbf{e}_b\big(u-s_a+(2s-M-1)c_b\big)
\bigg]\,.
\end{equation}
We had noted above that the Baxter equation implies quasi-constancy of 
$\SF(u)$, more precisely we find in our case the functional equation
\begin{equation}
\SF(u-Mc_b)\,=\,\De(-qe^{\frac{2\pi b}{M}u})\SF(u+ib-Mc_b)\,.
\end{equation}
The dual Baxter equation obtained by replacing $b\ra b^{-1}$ in the coefficients 
implies that $F(u)$ must satisfy a very similar difference equation with 
$b$ replaced by $b^{-1}$. 
These equations posses the manifestly self-dual solution \rf{SFsolution}. 
Taken together these two difference equations determine $\SF(u)$ up to 
a constant operator $\SF_0$. This operator can be determined by studying 
the asymptotics of $\mathsf{q}^+(\zeta)$ for $\zeta\ra \infty$, as
was done for $M=2$ in \cite{ByT}.
We intend to return to this question elsewhere. 


\begin{rem}
It was observed in Remark \ref{highertensor} above that the tensor product
$\pi^+_{\la_1}\,\otimes \dots\otimes \pi^+_{\la_M}$ contains for generic values of  $\{\la_s\}$
an irreducible representations of evaluation type, as expressed more precisely in 
equation \rf{evaluationgeneric}.
Formal reasoning indicates that for certain values of  $\{\la_s\}$ there may exist
invariant subspaces in the dual of $\pi^+_{\la_1}\,\otimes \dots\otimes \pi^+_{\la_M}$.
In particular for $\la_s=q^{\frac{2s}{M}}\la$ there seems to exist a sub-representation 
isomorphic to the trivial representation. Similar observations have been used in the 
case of highest weight representations to derive functional relations similar to 
\rf{wronskianproposal2} using resolutions of the identity representation
of Bernstein-Gel'fand-Gel'fand (BGG)-type
 \cite{BLZ3,BHK,BFLMS,DM}. It would be interesting to know if a similar approach can 
 be used to derive functional equations in the case of representations that do not have
 extremal weight vectors as considered in our paper. A more systematic analysis of  the  tensor products 
 $(\pi^+_{\la_1}\,\otimes \dots\otimes \pi^+_{\la_{\ell}})\Delta^{(\ell)}$ 
 and their connections with the functional relations involving Q-operators may be an interesting
 project for the future.

%
\end{rem}

\appendix

\section*{Appendices}






\section{Quantum minors and triangular decomposition of $\mathscr{L}(x)$}
\label{app:minors}

\paragraph{The quantum determininant.}
In this appendix we introduce the quantum determinant, see 
\cite{KuSk81}, \cite{Mol}, \cite{Tar92}.
It follows from the relation \eqref{RLLLLR}  that 
\begin{equation}\label{PiL}
 \Pi^{-}_{12\dots m}\,
\mathscr{L}_1(q^{-\frac{2}{M}\rho_1}\lambda)\, 
 \dots  \, \mathscr{L}_m(q^{-\frac{2}{M}\rho_m}\lambda)\,=\,
\mathscr{L}_m(q^{-\frac{2}{M}\rho_m}\lambda)\, 
 \dots  \, \mathscr{L}_1(q^{-\frac{2}{M}\rho_1}\lambda)\,\Pi^{-}_{12\dots m}\,,
\end{equation}
where
\begin{equation}
\Pi^-_{12\dots m}\,:=\,
\left(R_{m-1,m}\right)
\left(R_{m-2,m}R^-_{m-2,m-1}\right)
\dots
\left(R_{1,m}\cdots R^-_{1,2}\right)
\,\in\,\text{End}\left((\mathbb{C}^M)^{\ot m}\right)\,,
\end{equation}
and
\begin{equation}
R_{a,b}=\frac{q^{-1}\la_b^M-q\la_a^M}{q-q^{-1}}\mathsf{R}_{a,b}(\la_a,\la_b)\,,
\qquad
\la_a=q^{-\frac{2}{M}\rho_a}\lambda\,,
\end{equation}
with $\mathsf{R}(\la,\mu)$  given in \eqref{RfundfundTEXT}
and  $\rho_{a}=\frac{m-2a+1}{2}$. The 
 indices $a,b$ in $R_{a,b}$ and  $\mathscr{L}_a(\la)$ entering \eqref{PiL} denotes the $a$-th ($b$-th) copy of 
$\mathbb{C}^M$ in $(\mathbb{C}^M)^{\ot m}$.
One can show that $\Pi^{-}_{12\dots m}$ projects into the totally antisymmetric part of 
$\mathbb{C}^M$ in $(\mathbb{C}^M)^{\ot m}$.
%
%
%
The case $m=M$ plays a distinguished role. On the one hand
 \begin{equation}
 \Pi^{-}_{12\dots M}\,
\mathscr{L}_1(q^{-\frac{2}{M}\rho_1}\lambda)\, 
 \dots  \, \mathscr{L}_M(q^{-\frac{2}{M}\rho_M}\lambda) \Pi^{-}_{12\dots M}\,=\,
\text{q-det}\left(\mathscr{L}(\la)\right)\,
 \Pi^{-}_{12\dots M}\,,
\end{equation}
where $\text{q-det}\left(\mathscr{L}(\la)\right)$ acts as a scalar in $(\mathbb{C}^M)^{\ot M}$ and takes the form 
\begin{equation}
  \text{q-det}\left(\mathscr{L}(\la)\right)=
\sum_{\sigma\,\in\,\mathfrak{S}_M}
\,c_{\sigma}(q)\,\mathscr{L}_{\sigma(1),1}(q^{-\frac{2}{M}\rho_1}\la)\,
\mathscr{L}_{\sigma(2),2}(q^{-\frac{2}{M}\rho_2}\la)\,\dots\,\mathscr{L}_{\sigma(M),M}(q^{-\frac{2}{M}\rho_M}\la)\,,
\label{qdetdef}
 \end{equation}
where $\rho_{k}=\frac{M-2k+1}{2}$. The coefficients $c_{\sigma}(q)$ are determined by the relation
\begin{equation}
 \Pi^{-}_{12\dots M}\,\left(e_{\sigma(1)}\ot e_{\sigma(2)}\dots\ot e_{\sigma(M)}\right)=
c_{\sigma}(q)\,
 \Pi^{-}_{12\dots M}\,\left(e_1\ot e_2\dots\ot e_M\right)\,,
\end{equation}
where $e_i$ denote the canonical basis of $\mathbb{C}^M$, see \cite{Mol}. 
With a little inspection one finds that 
where
\begin{equation}\label{csigmaexpl}
c_{\sigma}(q)=(-q)^{\ell(\sigma)}q^{f(\sigma)}\,,\qquad
f(\sigma)=-\frac{2}{M}\sum_{k=1}^M(k-1)(k-\sigma(k))\,.
\end{equation}
One the other hand one can show via the fusion procedure that $[\text{q-det}\left(\mathscr{L}(\la)\right),\mathscr{L}(\mu)]=0$.

\noindent \emph{Examples}:
The definition above produce
\begin{equation}
\text{q-det}(\mathsf{L}^{-}(\la))\,=\,1+(-1)^{M-1}\,\la^M\,,
\end{equation}
where $\mathsf{L}^{-}(\la)$
is defined in \eqref{Lminsaux}.
Notice that only two permutation contributes to the expression for the quantum determinant given above: 
$\sigma=\text{id}$
and  $\sigma=\omega:=(2,3,\dots,M,1)$.
The coefficient in the quantum determinant are computed 
recalling that $\ell(\omega)=M-1$ and $f(\omega)=1-M$.

An other relevant example is 
\begin{align}
\pi^-_{\mu=1}\left(A_i(\la)\right)&=
\su_1\dots\su_{i}\,(1-\delta_{i,M-1}\,\la^M(-1)^M)\,,
\\
\bar{\pi}^-_{\mu=1}\left(A_i(\la)\right)&=
\su_1\dots\su_{i}\,\prod_{s=1}^{i-1}(1-\la^Mq^{i-2s})
\end{align}
where $A_i(\la)$ are  defined in \eqref{apfromqdet}.
Notice that for $M=2$ the two expressions above coincide.

\paragraph{Definition.}
It is convenient to define
\begin{equation}
 \left(\mathscr{L}^{[p]}(\lambda^{\frac{M}{p}})\right)_{ij}\,:=\,
\lambda^{\frac{M-p}{p}(i-j)}\,\left(\mathscr{L}(\lambda)\right)_{ij}\,,
\qquad
i,j=1,2,\dots,p\,.
\end{equation}
This definition is motivated by the fact that $\mathscr{L}^{[p]}(\lambda)$ satisfies the same relations as 
$\mathscr{L}(\lambda)$ with $M$ replaced by $p$. The expression for the quantum determinant of $\mathscr{L}^{[p]}(\lambda)$
is understood as \eqref{qdetdef} with $M$ replaced by $k$.

\paragraph{The quantum comatrix.}
Let us define the quantum comatrix $\overline{\mathscr{L}}(\lambda)$ of $\mathscr{L}(\lambda)$ by
\begin{equation}\label{comatrprop}
 \mathscr{L}(q^{\frac{M-1}{M}}\lambda)\,\overline{\mathscr{L}}(q^{-\frac{1}{M}}\lambda)\,=\, 
 \text{q-det}\left(\mathscr{L}(\la)\right)\,.
\end{equation}
The matrix entries of $\overline{\mathscr{L}}(\lambda)$ can be expressed in terms of quantum minors of $\mathscr{L}(\lambda)$.
In the following we will need only the last diagonal elements given by 
\begin{equation}\label{comaLMM}
\left(\overline{\mathscr{L}}(\lambda)\right)_{M,M}\,=\,
\text{q-det}\left(\mathscr{L}^{[M-1]}(\lambda^{\frac{M}{M-1}}) \right)
\end{equation}
where $\mathscr{L}^{[p]}(\la)$ is defined in \eqref{Lp}
\paragraph{Triangular decomposition of $\mathscr{L}(x)$.}
Consider the  triangular decomposition of the type \eqref{triangFORevaluation}  
of a  matrix $X$ with non-commutative entries $X_{ij}$.
One has
\begin{equation}\label{ap}
a_p\,=\,\left(\left(\left(X^{[p]}\right)^{-1}\right)_{pp}\right)^{-1}\,,
\qquad 
\left(X^{[p]}\right)_{ij}\,:=\,X_{ij}\,,
\qquad 
i,j=1,\dots, p\,.
\end{equation}
The derivation of this fact is elementary, see e.g.~\cite{Ioh} for its application in a similar context.
If $X$ is replaced by $\mathscr{L}(\la)$,  one finds a simple expression for \eqref{ap} as follows from 
\eqref{comatrprop}  combined with \eqref{comaLMM}.
The relation   \eqref{apfromqdet} follows.






\section{On the evaluation representation}

In this appendix we review the definition of evaluation representation.
Along the way we will obtain explicit formulae for the image of imaginary root currents under the evaluation homomorphism.
We could not find such expressions in the literature.
These formulae allow  to compute the image of  the universal R-matrix   under $\pi^{\rm f}\ot\text{ev}$,
filling  an apparent gap in the literature. 




\subsection{Jimbo evaluation homomorphism }
\label{sec:IntroduceLev}

In \cite{J85} Jimbo introduced an homomorphism,
 usually called \emph{evaluation homomorphism} and denoted by $\text{ev}$,
from  $\mathcal{U}_q(\widehat{\mathfrak{gl}}_M)$ to $\mathcal{U}_q(\mathfrak{gl}_M)$. 
This homomorphism can be given in terms of the generators 
$\{e_i,f_i,q^{\bar{\epsilon}}_i\}$ of  $\mathcal{U}_q(\widehat{\mathfrak{gl}}_M)$
 and $\mathcal{U}_q(\mathfrak{gl}_M)$ respectively, see e.g.~\cite{CP}.
For the purposes of this section it is more convenient
to exploit this homomorphism using 
\begin{equation}
\mathbb{L}_{\text{ev}}(\la)\,:=\,
\frac{1}{\rho_{\text{ev}}(\la)}\,
 \left(\pi^{\text{f}}_\la\,\otimes\,\text{ev}\right)\mathscr{R}^-\,.
\label{evfromUNI}
\end{equation}
It can be shown that, upon choosing the scalar factor
 $\rho_{\text{ev}}(\la)$ appropriately (see below), one has
\begin{equation}
\label{Lev1}
 \mathbb{L}_{\text{ev}}(\lambda)\,=\,
\sum_{i=1}^M\,\SE_{ii}\otimes \left(q^{+\mathcal{H}_i}+\lambda^{M}\,q^{\gamma-\mathcal{H}_i}\right)
+\sum_{i\neq j}\,\lambda^{(i-j)_M}\,\SE_{ij}\otimes\mathcal{E}_{ji} \,.
\end{equation}
It follows from the universal Yang-Baxter equation \eqref{YBE} 
that this Lax operator satisfies the quadratic relations \eqref{RLLLLR}.
These relations, together with the specific dependence of 
$\mathbb{L}_{\text{ev}}(\lambda)$ on the spectral parameter $\lambda$,
provides a definition of $\mathcal{U}_q(\mathfrak{gl}_M)$ 
in terms of the generators $\{q^{\mathcal{H}_{i}}\}_{i=1,\dots, M}$, $\{\mathcal{E}_{ij}\}_{i\neq j}$.
The fact that the definition \eqref{evfromUNI} gives rise to
 a Lax operator of the form \eqref{Lev1} follows from the interwining property
 \eqref{eqnUnivRIntertwiner} of the universal R-matrix.
It is shown in Appendix \ref{app:interforev} that this is the case upon defining
\begin{equation}
\text{ev}(e_i)\,=\,\frac{1}{q^{-1}-q}\, \mathcal{E}_{i,i+1}\,q^{-\mathcal{H}_i}\,,
\qquad
\text{ev}(f_i)\,=\,\frac{1}{q-q^{-1}}\,q^{\,\mathcal{H}_i-\gamma}\, \mathcal{E}_{i+1,i}\,,
\label{evdef1APP}
\end{equation}
\begin{equation}
 \text{ev}\left(q^{\bar{\epsilon}_i-\frac{1}{M}\bar{\epsilon}}\right)\,=\,
q^{-\mathcal{H}_i}\,.
\label{evdef2APP}
\end{equation}

A direct calculation of \eqref{evfromUNI} using the infinite
 product formula for the universal R-matrix has been done for
 $\mathcal{U}_q(\widehat{\mathfrak{gl}}_2)$ in \cite{KST94}, see also Section \ref{sec:LpmfromuniSL2},
 and $\mathcal{U}_q(\widehat{\mathfrak{gl}}_3)$ in \cite{Raz13}.
As opposed to the derivation based on \eqref{eqnUnivRIntertwiner},
 the direct calculation of the product formula determines 
the scalar factor $\rho_{\text{ev}}(\la)$ as well.
In the following section we determine the image of the imaginary root vectors
 under the evaluation homomorphism and, as a byproduct, the factor $\rho_{\text{ev}}(\la)$.

\begin{rem} One may consider fixing the spectral parameter 
dependence of some Lax operator to be that  of a degree
 $k$ polynomial in $\lambda^{-1}$ for $k\leq M$.
 The case $k=M$ corresponds to  \eqref{Lev1}.
An identification  of the type \eqref{evfromUNI} 
would then provide an homomorphisms form 
$\mathcal{U}_q(\mathfrak{b}_+)$ to some 
algebra whose commutation  relations are dictated by the \eqref{RLLLLR}.
The case $k=1$ will produce $\mathsf{L}^-(\lambda)$ defined in \eqref{Lminsaux}.
\end{rem}

\begin{rem}  The R-matrix in \eqref{RfundfundTEXT} is related to \eqref{Lev1} as follows 
\begin{equation}
\pi^{\rm f}_{\mu=1}\left( \mathbb{L}_{\text{ev}}(\lambda)\right)=
q^{\frac{1}{M}}(q^{-1}-q^{+1}\la^M)\,\mathsf{R}(\la,1)\,,
\end{equation}
upon setting $q^{\gamma}=-q^{\frac{2}{M}}$ in the left hand side.
Moreover,
the expression \eqref{LevSL2form} coincide with  \eqref{Lev1} 
in the special case $M=2$,  upon identifying  $q^{\gamma}=-q$.
\end{rem} 


\subsection{Intertwining properties for  $\mathbb{L}_{\text{ev}}(\la)$}
\label{app:interforev}

It follows from the definition  \eqref{evfromUNI}
that $\mathbb{L}_{\text{ev}}(\lambda)$ satisfies the intertwining property
\begin{equation}\label{intertwiningforEV}
\mathbb{L}_{\text{ev}}(\lambda)\,\left(\pi_{\lambda}^{\text{f}}\otimes \text{ev}\right)\Delta(a)\,
=\,\left(\pi_{\lambda}^{\text{f}}\otimes \text{ev}\right)\Delta^{\text{op}}(a)\,
\mathbb{L}_{\text{ev}}(\lambda)\,,
\qquad
\forall \,a\,\in\,\mathcal{U}_q(\widehat{\mathfrak{sl}}_M)\,.
\end{equation}
In the following we will study the implications of  \eqref{intertwiningforEV} 
where  $\mathbb{L}_{\text{ev}}(\lambda)$ is taken to be  of the form
\begin{equation}\label{Lev1APP}
 \mathbb{L}_{\text{ev}}(\lambda)\,=\,
\sum_{i=1}^M\,\SE_{ii}\otimes \left(q^{\,\mathcal{N}_i}+
\lambda^{M}\,q^{\,\overline{\mathcal{N}}_i}\right)
+\sum_{i\neq j}\,\lambda^{(i-j)_M}\,\SE_{ij} \otimes \mathcal{E}_{ji}\,.
\end{equation}
One can argue that  
the solution of \eqref{intertwiningforEV} is unique up to  multiplication by an element of the form 
$1\otimes \rho(\lambda)$ where $\rho(\lambda)$ belongs to the center of
  $\mathcal{U}_q(\mathfrak{gl}_M)$.
In order for this to be the case it is important that \eqref{intertwiningforEV} holds for the full 
$\mathcal{U}_q(\widehat{\mathfrak{sl}}_M)$ and not just a Borel half. 
The fact that we can find a solution of the interwining property of the form  \eqref{Lev1APP} 
 thus  provides a  proof of \eqref{evfromUNI}.

Let us proceed with the analysis.
Using the form \eqref{Lev1APP} and taking $a=q^{\bar{\epsilon}_i}$, 
 the intertwining property implies
\begin{equation}
 q^{\,\mathcal{N}_i}\,\text{ev}(q^{\bar{\epsilon}_j})\,=\,\text{ev}(q^{\bar{\epsilon}_j})\, 
 q^{\,\mathcal{N}_i}\,,
 \qquad
q^{\,\overline{\mathcal{N}}_i}\,\text{ev}(q^{\bar{\epsilon}_j})\,=\,
\text{ev}(q^{\bar{\epsilon}_j})\,q^{\,\overline{\mathcal{N}}_i}\,,
\end{equation}
\begin{equation}\label{EEqe}
 \mathcal{E}_{ij}\,\text{ev}(q^{\bar{\epsilon}_k})\,=\,
q^{\delta_{jk}-\delta_{ik}}\,\text{ev}(q^{\bar{\epsilon}_k})\,\mathcal{E}_{ij}\,.
\end{equation}
Next, consider the intertwining property for $a=f_i$.
The $\lambda^{M+1}$ term of these equations immediately implies that 
\begin{equation}\label{ident1}
\text{ev}\left(q^{\bar{\epsilon}_i}\right)\,=\,q^{\bar{x}}\,q^{\,\overline{\mathcal{N}}_i}\,,
\end{equation}
for some constant $\bar{x}$. Using this identification and \eqref{EEqe}, the 
$\lambda^{M}$ terms of the same equations give
\begin{equation}\label{ident2}
 \mathcal{E}_{i+1,i}\,=\,(q-q^{-1})\,
 q^{\,\overline{\mathcal{N}}_i}\,
\text{ev}(f_i)\,.
\end{equation}
Let us turn to the case $a=e_i$ in \eqref{intertwiningforEV}.
A similar analysis applied to the terms of order $\lambda^{-1}$ and $\lambda^0$ 
shows that 
\begin{equation}\label{ident3}
\text{ev}\left(q^{\bar{\epsilon}_i}\right)\,=\,q^{x}\,q^{-\mathcal{N}_i}\,,
\qquad
 \mathcal{E}_{i,i+1}\,=\,
 (q^{-1}-q)\,
\text{ev}(e_i)\,
 q^{\mathcal{N}_i}\,,
\end{equation}
for some constant $x$. 
The equations 
 \eqref{ident1}, \eqref{ident2}, \eqref{ident3} 
give the identification between the generators of 
$\mathcal{U}_q(\widehat{\mathfrak{gl}}_M)$ and 
$\mathcal{U}_q(\mathfrak{gl}_M)$.
The constants $x$ and $\bar{x}$ correspond to the  freedom 
of overall  rescaling of $\mathbb{L}_{\text{ev}}$ and
introducing the  spectral parameter for $\text{ev}$.
To obtain \eqref{Lev1} we demand that the leading term in the $\lambda$ expansion is 
\begin{equation}
(\pi^{\rm f}\ot \text{ev})q^{-t}=
\sum_{i=1}^M\,\mathsf{E}_{i,i}\,\ot\,q^{\mathcal{H}_i}\,,
\end{equation}
where $q^t$ is given in \eqref{qtdefSLM} , $\pi^{\rm f}(q^{\bar{\epsilon}_i})=q^{\mathsf{E}_{ii}}$
and $q^{-\mathcal{H}_i}=\text{ev}(q^{\bar{\epsilon}_i-\frac{1}{M}\bar{\epsilon}})$. 
Notice that $\prod_i\,q^{-\mathcal{H}_i}=1$.
This requirement implies that $q^{\mathcal{N}_i}=q^{\mathcal{H}_i}$
and $q^{\overline{\mathcal{N}}_i}=q^{\gamma-\mathcal{H}_i}$.
The remaining equation contained in \eqref{intertwiningforEV} prescribe how 
to express $\mathcal{E}_{ij}$ in terms of these generators.
The equivalence  between
different looking expressions 
for $\mathcal{E}_{ij}$ is equivalent to the Serre relations.

\subsubsection{Image of imaginary root vectors  and Gelfand-Tsetlin algebra}
\label{sec:imageimGTalg}

The image of the imaginary root vectors under the 
evaluation homomorphism can be obtained by applying 
the procedure explained in 
Section \ref{sec:rootgen}. 
As this  procedure is quite involved we will use a shortcut based
 on the observations presented in Section \ref{sec:UqglMandiso}. 
The expression \eqref{Epfromqdet} for the imaginary root currents 
$1+E_i'(z)$ in terms of quantum minors is independent of a rescaling of  
$\mathscr{L}(\la)$ by an arbitrary function of $\la$.
For this reason the quantum minors of  $\mathbb{L}_{\text{ev}}(\la)$ 
given in \eqref{Lev1} can be directly used to obtain 
 $\text{ev}\left(1+E_i'(z)\right)$.   
It is not hard to see
 that the relevant quantum minors take the form
\begin{equation}
 \mathbb{G}_p(\la^M)\,:=\,\text{q-det}\,\mathbb{L}^{[p]}_{\text{ev}}(\la^{\frac{M}{p}})\,=\,
q^{\sum_{s=1}^p\,\mathcal{H}_s}\,\prod_{s=1}^p\,\left(1+\la^{M}\,q^{2\,\nu_{p,s}+\gamma}\right)\,,
\label{qdetev}
\end{equation}
with $\sum_{s=1}^p\,\nu_{p,s}=-\sum_{s=1}^p\,\mathcal{H}_s$.
These quantum minors commute  $[\mathbb{G}_p(\la),\, \mathbb{G}_q(\mu)]=0$ 
and generate a  maximally commutative subalgebra
 of $\mathcal{U}_q(\mathfrak{gl}_M)$
known as Gelfand-Tsetlin algebra, see e.g.~\cite{NaTa}.
This algebra can be descibed as follows.
Let 
$\mathcal{Z}\left(\mathcal{U}_q(\mathfrak{gl}_M)\right)$ be the center of $\mathcal{U}_q(\mathfrak{gl}_M)$ and
$\mathcal{U}_q(\mathfrak{gl}_p)$ be the subalgebra generated by
 $\{q^{\mathcal{H}_{i}}\}_{i=1,\dots, p}$, $\{\mathcal{E}_{ij}\}_{1 \leq i\neq j\leq p}$.
 The subalgebra of $\mathcal{U}_q(\mathfrak{gl}_M)$ generated by 
$\mathcal{Z}\left(\mathcal{U}_q(\mathfrak{gl}_1)\right)$, $\mathcal{Z}\left(\mathcal{U}_q(\mathfrak{gl}_2)\right)$,  \dots,  $\mathcal{Z}\left(\mathcal{U}_q(\mathfrak{gl}_M)\right)$
is evidently commutative. This is what is called Gelfand-Tsetlin algebra. 
From \eqref{qdetev} and \eqref{Epfromqdet}
we conclude that
\begin{equation}
 \text{ev}\left(1+E'_p((-1)^p z)\right)\,=\
\frac{\prod_{s=1}^{p+1}\,\left(1+z^{-1}\,q^{2\,\nu_{p+1,s}+\gamma}\right)\,
\prod_{s=1}^{p-1}\,\left(1+\,z^{-1}\,q^{2\,\nu_{p-1,s}+\gamma}\right)}
{\prod_{s=1}^p\,\left(1+q^{-1}\,z^{-1}\,q^{2\,\nu_{p,s}+\gamma}\right)\,
\prod_{s=1}^p\,\left(1+q^{+1}\,z^{-1}\,q^{2\,\nu_{p,s}+\gamma}\right)}
\label{imfromGT}
\end{equation}
or equivalently 
\begin{equation}
 \text{ev}\left(e_{k\delta}^{(p)}\right)\,=\,
\frac{1}{k}\,
\frac{\left((-1)^{p+1}q^{\gamma}\right)^k}{q-q^{-1}}\,
\left(t^{(k)}_{p+1}+
t^{(k)}_{p-1}-
[2]_{q^k}\,t^{(k)}_{p}\right)\,,
\,\,\,\,\,\,
t^{(k)}_p\,:=\,
\sum_{s=1}^p\,q^{2k\,\nu_{p,s}}\,.
\end{equation}
Using this formula for the imaginary root vectors we can obtain the scalar factor in \eqref{evfromUNI} to be
\begin{equation}
 \rho_{\text{ev}}(\la)\,=\,
\exp\left(\sum_{m=1}^{\infty}\,\frac{(-x)^m}{m}\,\frac{t_M^{(m)}}{\,\,\,[M]_{q^{m}}}\right)
\,=\,
\prod_{s=1}^M\,
\frac{\varepsilon_{q^{M}}(q^{2\nu_{M,s}-1}\,x)}{\varepsilon_{q^{M}}(q^{2\nu_{M,s}+1}\,x)}\,,
\label{rhoev}
\end{equation}
where $x:=q^{\gamma}\,\la^M$ and $\varepsilon_{q}(x)$ is defined in \eqref{thetadef}.

\section{Evaluation of the Universal R-matrix}

\subsection{Cartan-Weyl basis for $\mathcal{U}_q(\widehat{\mathfrak{sl}}_M)$}

\subsubsection{Choice of convex order for  $\mathcal{U}_q(\widehat{\mathfrak{sl}}_M)$}
\label{sec:convexorderSLM}

Recall that the simple roots of $\mathfrak{sl}_M$ are
 $\alpha_i\,=\,\epsilon_i-\epsilon_{i+1}$ with $i=1,\dots,M-1$ and 
\begin{equation}
 \Delta_+(\mathfrak{sl}_M)\,=\,\{\epsilon_i-\epsilon_j\,,\,1\leq i<j\leq M\}\,.
\end{equation}
The highest root $\theta=\alpha_1+\dots+\alpha_{M-1}=\epsilon_1-\epsilon_M$
 and the remaining simple root of $\widehat{\mathfrak{sl}}_M$ is $\alpha_0=\delta-\theta$.
The set $\Delta_+(\widehat{\mathfrak{sl}}_M)$ is given in \eqref{affinerootRE}, \eqref{affinerootIM}.
We endow this set with a convex (normal) order, see \eqref{convexorddef} for the definition, as follows
\begin{equation}
\widehat{\mathbb{A}}_1\prec\, 
\widehat{\mathbb{A}}_2\prec \dots\prec\, \widehat{\mathbb{A}}_{M-1}\,\prec\,\,
\mathbb{Z}_{>0}\,\delta\,\prec \widehat{\mathbb{B}}_{M-1}\prec\,
 \widehat{\mathbb{B}}_{M-2}\prec \dots\prec\, \widehat{\mathbb{B}}_{1}\,,
\label{rootorder}
\end{equation}
compare to \eqref{precsucc}.
The ordered sets of real positive roots $\widehat{\mathbb{A}}_i$ and $\widehat{\mathbb{B}}_i$ are defined as
\begin{equation}
 \widehat{\mathbb{A}}_i\,:=\,\mathbb{A}_i\prec\mathbb{A}_i+\delta\prec \mathbb{A}_i+2\,\delta\prec\dots\,,
\qquad 
\mathbb{A}_i\,:=\,\epsilon_i-\epsilon_{i+1}\prec \epsilon_i-\epsilon_{i+2}\prec\dots\prec\epsilon_i-\epsilon_{M}\,,
\label{setA}
\end{equation}
\begin{equation}
 \widehat{\mathbb{B}}_i\,:=\,\dots\prec \mathbb{B}_i+2\,\delta \prec\mathbb{B}_i+\delta\prec \mathbb{B}_i\,,
\qquad 
\mathbb{B}_i\,:=\,\delta-\left(\epsilon_i-\epsilon_{i+1}\right)\prec\dots\prec\delta-\left(\epsilon_i-\epsilon_{M}\right)\,,
\label{setB}
\end{equation}
A similar root ordering appears in relation to the universal R-matrix for the Yangian in \cite{Stu}.
We remark that the ordering above can be obtained in the framework of \cite{Ito}, as an ordering of ''M-raw type``,  
using the action of the extended affine Weyl group. According to theorem 2.3 in \cite{Tol2} any convex order
 can be obtained form any other by composition of so called elementary inversions.

\subsubsection{Explicit construction of root vectors for $\mathcal{U}_q(\widehat{\mathfrak{sl}}_M)$}
\label{sec:rootSLM}

\paragraph{Root vectors $e_{\gamma}$ where  $\gamma \in \Delta_+(\mathfrak{sl}_M)$.}

\begin{equation}
 e_{\alpha_i+\alpha_{i+1}}\,:=\,\left[e_{\alpha_i},e_{\alpha_{i+1}}\right]_{q^{-1}}\,,
\end{equation}
\begin{equation}
 e_{\alpha_i+\alpha_{i+1}+\alpha_{i+2}}\,:=\,\left[e_{\alpha_i},e_{\alpha_{i+1}+\alpha_{i+2}}\right]_{q^{-1}}\,,
\end{equation}
and so on.
\paragraph{Root vectors $e_{\delta-\gamma}$ where  $\gamma \in \Delta_+(\mathfrak{sl}_M)$.}

There are $M-1$ steps in the consrtuction.
One has the following $M-1$ definitions (first step)
\begin{equation}
 e_{\delta-\theta}\,:=\,e_{\alpha_0}\,,
\end{equation}
\begin{equation}
 e_{\delta-\theta+\alpha_{M-1}}\,:=\,\left[e_{\alpha_{M-1}},e_{\delta-\theta}\right]_{q^{-1}}\,,
\end{equation}
\begin{equation}
 e_{\delta-\theta+\alpha_{M-2}+\alpha_{M-1}}\,:=\,\left[e_{\alpha_{M-2}},e_{\delta-\theta+\alpha_{M-1}}\right]_{q^{-1}}\,,
\end{equation}
\begin{equation}
 e_{\delta-\theta+\alpha_{M-3}+\alpha_{M-2}+\alpha_{M-1}}\,:=\,\left[e_{\alpha_{M-3}},e_{\delta-\theta+\alpha_{M-2}+\alpha_{M-1}}\right]_{q^{-1}}\,,
\end{equation}
\begin{equation}
\vdots
\end{equation}
\begin{equation}
 e_{\delta-\alpha_{1}}\,:=\,\left[e_{\alpha_{2}},e_{\delta-\alpha_1-\alpha_{2}}\right]_{q^{-1}}\,,
\end{equation}
One has the following $M-2$ definitions
\begin{equation}
 e_{\delta+\alpha_1-\theta}\,:=\,\left[e_{\alpha_1},e_{\delta-\theta}\right]_{q^{-1}}\,,
\end{equation}
\begin{equation}
 e_{\delta+\alpha_1-\theta+\alpha_{M-1}}\,:=\,\left[e_{\alpha_{M-1}},e_{\delta+\alpha_1-\theta}\right]_{q^{-1}}\,,
\end{equation}
\begin{equation}
\vdots
\end{equation}
\begin{equation}
 e_{\delta-\alpha_{2}}\,:=\,\left[e_{\alpha_{3}},e_{\delta-\alpha_2-\alpha_{3}}\right]_{q^{-1}}\,,
\end{equation}
 One has the following $M-3$ definitions
\begin{equation}
 e_{\delta+\alpha_1+\alpha_2-\theta}\,:=\,\left[e_{\alpha_2},e_{\delta+\alpha_1-\theta}\right]_{q^{-1}}\,,
\end{equation}
\begin{equation}
 e_{\delta+\alpha_1+\alpha_2-\theta+\alpha_{M-1}}\,:=\,\left[e_{\alpha_{M-1}},e_{\delta+\alpha_1+\alpha_2-\theta}\right]_{q^{-1}}\,,
\end{equation}
\begin{equation}
\vdots
\end{equation}
\begin{equation}
 e_{\delta-\alpha_{3}}\,:=\,\left[e_{\alpha_{4}},e_{\delta-\alpha_3-\alpha_{4}}\right]_{q^{-1}}\,,
\end{equation}
One has the following final definition ( step $M-1$)
\begin{equation}
 e_{\delta-\alpha_{M-1}}\,:=\,\left[e_{\alpha_{M-2}},e_{\delta-\alpha_{M-1}-\alpha_{M-2}}\right]_{q^{-1}}\,.
\end{equation}

\subsection{Fermionic Fock space representation}
\subsubsection{Fermionic Fock space representation: definition}
\label{App:piferm}

\begin{equation}
\pi^{\mu}_{\mathcal{F}}\,:\,\,\mathcal{U}_q(\widehat{\mathfrak{sl}}_{M})\,\,\,\rightarrow \,\,\mathcal{F}_{M}\, 
\end{equation} 
 \begin{equation}
 \pi^{\mu}_{\mathcal{F}}(e_i)\,=\,\mu^{-1}\,\bar{\mathbf{c}}_i\,\mathbf{c}_{i+1}\,
\qquad
 \pi^{\mu}_{\mathcal{F}}(f_i)\,=\,\mu\,\bar{\mathbf{c}}_{i+1}\,\mathbf{c}_{i}\,
\qquad
 \pi^{\mu}_{\mathcal{F}}(k_i)\,=\,q^{\mathfrak{n}_i-\mathfrak{n}_{i+1}}\,
\end{equation}
\begin{equation}
\mathcal{F}_{ M}\,:\,\,\,\
 \{\mathbf{c}_{i},\bar{\mathbf{c}}_{j}\}\,=\,\delta_{ij}\,\qquad 
\{\mathbf{c}_{i},\mathbf{c}_{j}\}\,=\,0\,\qquad 
\{\bar{\mathbf{c}}_{i},\bar{\mathbf{c}}_{j}\}\,=\,0\,
\qquad
\mathfrak{n}_i:=\bar{\mathbf{c}}_{i}\mathbf{c}_{i}\,
\end{equation}
where the indices $i,j,k,\dots$ are subject to cyclic identification: $i+M\sim i$.
This representation is not irreducible as $\mathfrak{n}_{\text{tot}}$ is central.
The fundamental representation corresponds to $\mathfrak{n}_{\text{tot}}=1$. In this case
\begin{equation}\label{fundrep}
 \pi_{\mu}(e_i)\,=\,\mu^{-1}\,\SE_{i,i+1}\,,\qquad
 \pi_{\mu}(f_i)\,=\,\mu\,\SE_{i+1,i}\,,\qquad
 \pi_{\mu}(h_i)\,=\,\SE_{i,i}-\SE_{i+1,i+1}\,,
\end{equation}
 and 
\begin{equation}
\SE_{ij}\,\SE_{kl}\,=\,\delta_{jk}\,\SE_{il}\,.
\end{equation}
\subsubsection{Fermionic Fock space representation: evaluation of root vectors}
\label{App:rootFerm}
Using the explicit definitions in Section \ref{sec:rootgen}  and Appendix \ref{sec:rootSLM} one obtains
\begin{itemize}
\item[1.] 
\begin{equation}
 \pi_{\mathcal{F}}(e_{\epsilon_i-\epsilon_j})\,=\,
\mu^{i-j}\,\bar{\mathbf{c}}_i\,\mathbf{c}_{j}\,q^{\left(\sum_{k=i+1}^{j-1}\mathfrak{n}_{k} \right)}\,,
\end{equation}
\begin{equation}
 \pi_{\mathcal{F}}(e_{\delta-\left( \epsilon_i-\epsilon_j\right)})\,=\,
(-q)^{i-1}\,\mu^{j-i-M}\,\bar{\mathbf{c}}_{j}\,\mathbf{c}_{i}
\,q^{\left(\sum_{k=j+1}^{M}\mathfrak{n}_{k} \right)-\left(\sum_{k=1}^{i-1}\mathfrak{n}_{k} \right)}\,.
\end{equation}
\item[2.]
\begin{equation}
 \pi_{\mathcal{F}}(e^{(i)}_{\delta})\,=\,
\kappa_i\,q^{\mathfrak{n}_{i+1} -\mathfrak{n}_{i}}
\,[\mathfrak{n}_{i+1}-\mathfrak{n}_{i}]_q
\,,
\end{equation}
\begin{equation}
\pi_{\mathcal{F}}(e_{\alpha_i+k\delta})\,=\,
\left(\kappa_i\right)^k\,\pi_{\mathcal{F}}(e_{\alpha_i})\,,
\end{equation}
\begin{equation}
\pi_{\mathcal{F}}(e_{(\delta-\alpha_i)+k\delta})\,=\,
\left(\kappa_i\right)^k\,\pi_{\mathcal{F}}(e_{(\delta-\alpha_i)})\,,
\end{equation}
\begin{equation}
 \pi_{\mathcal{F}}(e'^{(i)}_{k\delta})\,=\,
 \left(\kappa_i\right)^{k-1}\,\pi_{\mathcal{F}}(e^{(i)}_{\delta})\,.
\end{equation}
\begin{equation}
 \kappa_i\,:=\,\mu^{-M}\,(-q)^i\,q^{\left(\sum_{k=i+2}^{M}\mathfrak{n}_{k} \right)-\left(\sum_{k=1}^{i-1}\mathfrak{n}_{k} \right)}
\end{equation}
\item[3.] 
In the case of interest we do not need these generators.
\item[4.] It follows that
\begin{equation}
 \pi_{\mathcal{F}}\left(1+E_i'(z) \right)\,=\,
\frac{1-\kappa_i\,z^{-1}\,q^{2(\mathfrak{n}_{i+1} -\mathfrak{n}_{i})}}{1-\kappa_i\,z^{-1}}
\,,
\end{equation}
which upon Taylor expansion gives
\begin{equation}\label{piFekdelta}
 \pi_{\mathcal{F}}(e^{(i)}_{k\delta})\,=\,
\frac{1}{k}\,\left(\kappa_i \right)^k\,\,q^{k(\mathfrak{n}_{i+1} -\mathfrak{n}_{i})}
\,[k(\mathfrak{n}_{i+1}-\mathfrak{n}_{i})]_q
\,.
\end{equation}
\end{itemize}
{\bf Remark}. From the  formulas above one can easly obtain root vectors for the fundamental representation (and further include  step $3$):
\begin{equation}\label{relationfund1}
\pi^{\mu}_{\rm f}(e_{(\epsilon_i-\epsilon_j)+k\delta})\,=\,\mu^{i-j}\left(-q\,t_i\right)^k\,\SE_{i,j}\,,
\end{equation}
\begin{equation}\label{relationfund2}
\pi^{\mu}_{\rm f}(e_{(\delta-(\epsilon_i-\epsilon_j))+k\delta})\,=\,\,t_i\,\mu^{j-i}\left(-q\,t_i\right)^k\,\SE_{j,i}\,,
\end{equation}
\begin{equation}
 \pi^{\mu}_{\rm f}(e^{(i)}_{k\delta})\,=\,t_i^{k}\,
\frac{(-1)^{k+1}}{k}\,[k]_q\,\,
\Big(\SE_{i,i}\,-\,q^{2k}\,\SE_{i+1,i+1} \Big)\,,
\label{imrootfund}
\end{equation}
where $i<j$ and $t_i=\mu^{-M}\,(-q)^{i-1}$. 
\subsubsection{Fermionic Cartan-Weyl basis: second Borel half}
\label{app:CWbasisFERMfs} 

\begin{itemize}
\item[1.] 
\begin{equation}
 \pi_{\mathcal{F}}(f_{\epsilon_i-\epsilon_j})\,=\,
\mu^{j-i}\,\bar{\mathbf{c}}_j\,\mathbf{c}_{i}\,q^{-\left(\sum_{k=i+1}^{j-1}\mathfrak{n}_{k} \right)}\,,
\label{PiFfre+}
\end{equation}
\begin{equation}
 \pi_{\mathcal{F}}(f_{\delta-\left( \epsilon_i-\epsilon_j\right)})\,=\,
\left(-q^{-1}\right)^{i-1}\,\mu^{M-(j-i)}\,
\bar{\mathbf{c}}_{i}\,\mathbf{c}_{j}
\,q^{-\left(\sum_{k=j+1}^{M}\mathfrak{n}_{k} \right)+\left(\sum_{k=1}^{i-1}\mathfrak{n}_{k} \right)}\,.
\label{PiFfre+bis}
\end{equation}
\item[2.]
\begin{equation}
 \pi_{\mathcal{F}}(f^{(i)}_{\delta})\,=\,
\bar{\kappa}_i\,q^{\mathfrak{n}_{i} -\mathfrak{n}_{i+1}}
\,[\mathfrak{n}_{i+1}-\mathfrak{n}_{i}]_q
\,,
\end{equation}
\begin{equation}
\pi_{\mathcal{F}}(f_{\alpha_i+k\delta})\,=\,
\left(\bar{\kappa}_i\right)^k\,\pi_{\mathcal{F}}(f_{\alpha_i})\,,
\end{equation}
\begin{equation}
\pi_{\mathcal{F}}(f_{(\delta-\alpha_i)+k\delta})\,=\,
\left(\bar{\kappa}_i\right)^k\,\pi_{\mathcal{F}}(f_{(\delta-\alpha_i)})\,,
\end{equation}
\begin{equation}
 \pi_{\mathcal{F}}(f'^{(i)}_{k\delta})\,=\,
 \left(\bar{\kappa}_i\right)^{k-1}\,\pi_{\mathcal{F}}(f^{(i)}_{\delta})\,.
\end{equation}
\begin{equation}
 \bar{\kappa}_i\,:=\,\mu^{M}\,(-q^{-1})^i\,q^{-\left(\sum_{k=i+2}^{M}\mathfrak{n}_{k} \right)+\left(\sum_{k=1}^{i-1}\mathfrak{n}_{k} \right)}
\end{equation}
\item[3.] 
In the case of interest we do not need these generators.
\item[4.] 
Finally, notice that we just need to replace $q$ with $q^{-1}$ and $x$ with $x^{-1}$ so that
\begin{equation}
 \pi_{\mathcal{F}}\left(1+F_i'(z) \right)\,=\,
\frac{1-\bar{\kappa}_i\,z^{-1}\,q^{2(\mathfrak{n}_{i} -\mathfrak{n}_{i+1})}}{1-\bar{\kappa}_i\,z^{-1}}
\,,
\end{equation}
which upon Taylor expansion gives
\begin{equation}\label{piFfkdelta}
 \pi_{\mathcal{F}}(f^{(i)}_{k\delta})\,=\,
\frac{1}{k}\,\left(\bar{\kappa}_i \right)^k\,\,q^{k(\mathfrak{n}_{i} -\mathfrak{n}_{i+1})}
\,[k(\mathfrak{n}_{i+1}-\mathfrak{n}_{i})]_q
\,.
\end{equation}
\end{itemize}


\subsection{Minimal representations of $\mathcal{U}(\mathfrak{gl}_M)$}

Let us define the following representation of $\mathcal{U}_q(\mathfrak{n}^+)$
\begin{equation}\label{newrepAPP}
 \pi(e_i)\,=\,\frac{s_i}{q^{-1}-q}\,\,\mathsf{W}^{}_i\,\mathsf{W}^{-1}_{i+1}\,
\left(t\,\mathsf{Z}_{i}^{}+t^{-1}\mathsf{Z}_{i}^{-1}\right)\,,
\end{equation}
where
\begin{equation}
\mathsf{W}_{i}\,\mathsf{Z}_{j}\,=\,q^{\delta_{ij}}\,\mathsf{Z}_{j}\,\mathsf{W}_{i}\,,
\qquad
\mathsf{W}_{i}\,\mathsf{W}_{j}\,=\,\mathsf{W}_{j}\,\mathsf{W}_{i}\,,
\qquad
\mathsf{Z}_{i}\,\mathsf{Z}_{j}\,=\,\mathsf{Z}_{j}\,\mathsf{Z}_{i}\,,
\end{equation}
with $\prod_i\,\mathsf{Z}_{i}=\prod_i\,\mathsf{W}_{i}=1$
and $s_i$ and $t$ are complex numbers.
The goal of this appendix is to compute the image of the Cartan-Weyl generators under $\pi$.
We will see that image of infinitely many real roots is non zero.
Using the explicit iterative contruction presented in Section \ref{sec:rootgen} and Appendix \ref{sec:rootSLM}, 
one obtains
\begin{itemize}
 \item[1.]
\begin{equation}
\pi\left(e_{\epsilon_i-\epsilon_j} \right)\,=\,
 \frac{s_i}{q^{-1}-q}\,\left[\prod_{k=i+1}^{j-1}qts_k\mathsf{Z}^{}_k\right]\,\mathsf{W}^{}_i\,\mathsf{W}_{j}^{-1}\,
\left(t\,\mathsf{Z}_{i}^{}+t^{-1}\mathsf{Z}_{i}^{-1}\right)\,,
\end{equation}
\begin{equation}
\pi\left(e_{\delta-(\epsilon_i-\epsilon_j)} \right)\,=\,
 \frac{s_j}{q^{-1}-q}\,
\left[\prod_{k=1}^{i-1}t^{-1}s_k\mathsf{Z}^{-1}_k\right]
\left[\prod_{k=j+1}^{M}qts_k\mathsf{Z}^{}_k\right]\,
\mathsf{W}^{}_j\,\mathsf{W}_{i}^{-1}\,
\left(t\,\mathsf{Z}_{j}^{}+t^{-1}\mathsf{Z}_{j}^{-1}\right)\,,
\end{equation}
where
$ 1\leq i< j\leq M$.
\item[2.] Once we have constructed $\pi(e_{\delta-\alpha_i})$, we may notice that for each node $i$
 we have an evaluation type representation of $\mathcal{U}_q(\widehat{\mathfrak{sl}}_2)$.
To make this observation explicit we write
\begin{equation}
 \pi(e^{(i)}_{\delta})\,=\,
\frac{p_i}{q^{-1}-q}\,\left(
[2]_q\,\mathsf{k}_{i}\,+\,\left(q^{2\,x_i}+q^{-2\,x_i}\right)
\right)\,,
\end{equation}
where
\begin{equation}
\mathsf{k}_i:=\mathsf{Z}_{i+1}^{}\mathsf{Z}_{i}^{-1}\,,
\,\,\,\,
q^{2x_i}:=q \,t^2\mathsf{Z}_{i+1}^{}\mathsf{Z}_{i}^{}\,,
\quad
p_i:=q \,s_{\text{tot}}
\left[\prod_{k=1}^{i-1}t^{-1}\mathsf{Z}^{-1}_k\right]
\left[\prod_{k=i+2}^{M}qt\mathsf{Z}^{}_k\right]\,.
\end{equation}
It is easy to verify that $p_i$ and $q^{2x_i}$ commute with 
$\pi(e_{\alpha_i})$, $\pi(e_{\delta-\alpha_i})$ for fixed $i$.
With this observation in mind we evaluate the remaining root vectors associated to the node $i$ to be
\begin{equation}
 \pi(e_{\alpha_i+k\,\delta})\,=\,\left(q^{-1}p_i\,\mathsf{k}_{i}\right)^k\,\pi(e_{\alpha_i})\,,
\quad
 \pi(e_{(\delta-\alpha_i)+k\,\delta})\,=\,\left(q^{+1}p_i\,\mathsf{k}_{i}\right)^{k}
\,\pi(e_{\delta-\alpha_i})\,,
\end{equation}
\begin{equation}
 \pi(e'^{(i)}_{k\delta})\,=\,
\frac{(p_i)^k}{q^{-1}-q}\,
(\mathsf{k}_i)^{k-2}
\left(
[k+1]_q\,\mathsf{k}_{i}^{2}+[k]_q\left(q^{2\,x_i}+q^{-2\,x_i}\right)\mathsf{k}_i+[k-1]_q
\right)\,,
\end{equation}

\begin{equation}
\pi\left(1+E'_i(z)\right)\,=\,
\frac{\left(1+z^{-1}q^{+2x_i}\,p_i\right)\left(1+z^{-1}q^{-2x_i}\,p_i\right)}
{\left(1-z^{-1}\,q^{+1}\,\mathsf{k}_i\,p_i\right)\left(1-z^{+1}\,q^{+1}\,\mathsf{k}_i\,p_i\right)}\,.
\label{1plusE}
\end{equation}
\end{itemize}

 \paragraph{Comparison with the general form of the currents \eqref{imfromGT}.}
The imaginary root currents \eqref{1plusE} can be rewritten as
\begin{equation}\label{Ecurrents}
\pi\left(1+E'_i(z)\right)=
\frac{\left(1+z^{-1}\,X_{i-1}\right)\left(1+z^{-1}\,X_{i+1}\right)}
{\left(1-z^{-1}q^{+1}X_i\right)\left(1-z^{-1}q^{-1}X_i\right)}\,,
\end{equation}
 %
 The comparison with  \eqref{imfromGT} follows from the formula
\begin{equation}
\pi'\left(\prod_{s=1}^p\,\left(1+t\,q^{2\,\nu_{p,s}+\gamma}\right)\right)\,=\,
g_p(t)\,\left(1-(-1)^p\,t\,X_p\right)\,,,
\qquad
\pi=\pi'\circ \text{ev}\,,
\label{imageGTmin}
\end{equation}
where
\begin{equation}
g_p(t)\,=\,\prod_{s=1}^{p-1}\left(1-t\,\la_2^{-M}\,q^{p-2s}\right)\,.
\end{equation}
Notice that the contribution from $g_p(\lambda)$ cancel out (for $p>1$) in the combination \eqref{imfromGT} leaving 
a rational function with two zeroes and two poles in  $\lambda^{M}$.
We conclude that for these representations of $\mathcal{U}_q(\mathfrak{gl}_M)$,
 the image of the Gelfand-Tsetlin algebra coincides with the image of the Cartan 
subalgebra.

\section{Triangular decomposition of $\left(\pi^{\mathcal{F}}_{\mu}\ot \pi^{+}_{\la_{\mathfrak{n}}}\right)\Delta$
 for $M=2$}
\label{App:M=2}

It can be useful to present the main formulae of  Section \eqref{sec:blocktriang} in a more 
explicit form for the case of $M=2$. The relation \eqref{pipidelta} in this case reads
\begin{equation}
\left(\pi^{\mathcal{F}}_{\mu}\ot \pi^{+}_{\la_{\mathfrak{n}}}\right)\Delta(f_1)\,=\,
\Lambda(\mathsf{y})^{-1}\,\,
\left(
\frac{q^{\frac{1}{2}}\,\mathsf{y}^{-2}}{q-q^{-1}}
\right)\,
\begin{pmatrix}
\la_0 & 0  & 0 & 0 \\
0 & q^{-1}\,\la_1& 0 & 0 \\
0 & q^{-\frac{1}{2}}\, \tau_q\,\mu& q^{+1}\,\la_1 & 0 \\
0 & 0& 0 & \la_2
\end{pmatrix}
\Lambda(\mathsf{y})
\end{equation}
\begin{equation}
\left(\pi^{\mathcal{F}}_{\mu}\ot \pi^{+}_{\la_{\mathfrak{n}}}\right)\Delta(f_0)\,=\,
\Lambda(\mathsf{y})^{-1}\,\,
\left(
\frac{q^{\frac{1}{2}}\,\mathsf{y}^{+2}}{q-q^{-1}}
\right)\,
\begin{pmatrix}
\la_0 & 0  & 0 & 0 \\
0 & q^{+1}\,\la_1& q^{-\frac{1}{2}}\, \tau_q\,\mu & 0 \\
0 & 0& q^{-1}\,\la_1 & 0 \\
0 & 0& 0 & \la_2
\end{pmatrix}
\Lambda(\mathsf{y})
\end{equation}
where $\mathsf{y}^{}_1=\mathsf{y}_2^{-1}=\mathsf{y}$ and $\tau_q=q-q^{-1}$.
If $\mu=q^{\frac{1}{2}}\,\la_1$
one finds a block triangular structure\footnote{
The terminology refers to the following fact:
For an operator $\mathcal{O}$, we say that it has  a block triangular  structure 
if $\mathbf{P}_+\mathcal{O}\,\mathbf{P}_-=0$ and  $\mathbf{P}_-\mathcal{O}\,\mathbf{P}_+\neq0$ 
for orthogonal projectors $\mathbf{P}_\pm$.
}
 given by
\begin{equation}\label{blocktrianM=2part1}
\mathbf{P}_+\,\,
\Lambda(\mathsf{y})
\bigg{[}
\left(\pi^{\mathcal{F}}_{q^{\frac{1}{2}}\la_1}\!\!\!\ot \pi^{+}_{\la_{\mathfrak{n}}}\right)\Delta(f_i)
\bigg{]}
\Lambda(\mathsf{y})^{-1}\,\mathbf{P}_-
\,=\,
0
\end{equation}
\begin{equation}\label{blocktrianM=2part2}
\mathbf{P}_{\pm}\,
\Lambda(\mathsf{y})
\bigg{[}
\left(\pi^{\mathcal{F}}_{q^{\frac{1}{2}}\la_1}\!\!\!\ot \pi^{+}_{\la_{\mathfrak{n}}}\right)\Delta(f_i)
\bigg{]}
\Lambda(\mathsf{y})^{-1}\,\mathbf{P}_{\pm}
=
\left(
\frac{q^{\frac{1}{2}}\,\mathsf{y}^{+2\sigma_i}}{q-q^{-1}}
\right)
\begin{pmatrix}
\la_0 & 0  & 0 & 0 \\
0 & q^{\pm 1}\,\la_1&0  & 0 \\
0 & 0& q^{\pm 1}\,\la_1 & 0 \\
0 & 0& 0 & \la_2
\end{pmatrix}
\mathbf{P}_{\pm}
\end{equation}
where $\sigma_1=-1,\sigma_0=+1$ and
\begin{equation}
\mathbf{P}_+:=
\begin{pmatrix}
1 & 0  & 0 & 0 \\
0 & 1/2& 1/2& 0 \\
0 & 1/2& 1/2& 0 \\
0 & 0& 0 & 0
\end{pmatrix}\,=\,
\mathcal{S}\,
\left[
\begin{pmatrix}
1 & 0 \\
0 & 0  
\end{pmatrix}
\ot
\begin{pmatrix}
1 & 0 \\
0 & 1  
\end{pmatrix}
\right]
\mathcal{S}^{-1}\,,
\quad \qquad
\mathbf{P}_-:=1-\mathbf{P}_+\,.
\end{equation}
Where $\ot$ refers to the Kronecker product 
and the matrix $\mathcal{S}$ is easily worked out.
These relations reduce to \eqref{pipi} with 
$\Pi_1=\Lambda(\mathsf{y})^{-1}\mathbf{P}_+\Lambda(\mathsf{y})$
and
$\Pi_2=\Lambda(\mathsf{y})^{-1}\mathbf{P}_-\Lambda(\mathsf{y})$.
Using the similarity transform $\mathcal{S}$ we  rewrite \eqref{blocktrianM=2part2} as
\begin{equation}\label{tiangPplus}
\mathcal{S}^{-1}\mathbf{P}_{+}\,
\Lambda(\mathsf{y})
\bigg{[}
\left(\pi^{\mathcal{F}}_{q^{\frac{1}{2}}\la_1}\ot \pi^{+}_{\la_{\mathfrak{n}}}\right)\Delta(f_i)
\bigg{]}
\Lambda(\mathsf{y})^{-1}\,\mathbf{P}_{+}\,\mathcal{S}
=
\left(
\frac{q^{\frac{1}{2}}\,\mathsf{y}^{+2\sigma_i}}{q-q^{-1}}
\right)
\begin{pmatrix}
1 & 0 \\
0 & 0  
\end{pmatrix}
\ot
\begin{pmatrix}
\la_0 & 0 \\
0 & q^{+1}\,\la_1 
\end{pmatrix}
\end{equation}
\begin{equation}\label{tiangPminus}
\mathcal{S}^{-1}\mathbf{P}_{-}\,
\Lambda(\mathsf{y})
\bigg{[}
\left(\pi^{\mathcal{F}}_{q^{\frac{1}{2}}\la_1}\ot \pi^{+}_{\la_{\mathfrak{n}}}\right)\Delta(f_i)
\bigg{]}
\Lambda(\mathsf{y})^{-1}\,\mathbf{P}_{-}\,\mathcal{S}
=
\left(
\frac{q^{\frac{1}{2}}\,\mathsf{y}^{+2\sigma_i}}{q-q^{-1}}
\right)
\begin{pmatrix}
0 & 0 \\
0 & 1  
\end{pmatrix}
\ot
\begin{pmatrix}
q^{-1}\,\la_1 & 0 \\
0 & \la_2
\end{pmatrix}
\end{equation}
The statement expressed by \eqref{triang} is actually stronger then 
\eqref{blocktrianM=2part1} and \eqref{tiangPplus},  \eqref{tiangPminus}
as it states that the $2\times 2$ matrix in the right hand side of 
 \eqref{tiangPplus} and  \eqref{tiangPminus}
as to be the same, up to a similarity transform.
This  implies, up to exchange of $\la_0$ with $\la_2$, that   $\la_0=q^{-1}\la_1$ and $\la_2=q^{+1}\la_1$.


A similar analysis can be done  in the case of
 $\mu=-q^{\frac{1}{2}}\,\la_1$.

\section{Form of $\left(1\otimes \pi^-\right)\mathscr{R}^-$ 
and $\left(1\otimes \bar{\pi}^-\right)\mathscr{R}^-$ and action of the coproduct on the first tensor factor }

\subsection{Image of the universal R-matrix under $1\otimes \pi^-$ and $1\otimes \bar{\pi}^-$}


For the following analysis it is convenient  to rewrite 
\begin{equation}\label{repminusAPP}
 \pi_{\lambda}^-(e_i)\,=\,\frac{\lambda^{-1}}{q^{-1}-q^{}}\,\sv_i\,\su_i^{-1}\,=\,
\frac{\left(q^{\frac{1-M}{M}}\lambda\right)^{-1}}{q^{-1}-q^{}}\,\mathsf{y}_{i+1}\,\mathsf{y}^{-1}_{i}\,,\qquad
 \pi_{\lambda}^-(k_i)\,=\,\su^{-1}_i\,\su_{i+1}\,.
\end{equation}
The exchange relations of these variables are given in \eqref{yalgebra}.

%
%

\paragraph{$1\ot \pi^-$ on combinations of root vectors entering the universal R-matrix.}
Let $\gamma\in\,\Delta_+^{\text{re}}(\widehat{\mathfrak{sl}}_M)$, the relations \eqref{id-1}, \eqref{id-2},
 together with \eqref{eipiminus}, imply
\begin{equation}
\text{\small{$\Lambda^{-1}(\mathsf{y})\left[(1\otimes\pi_{\nu}^-)
(f_{\gamma}\otimes e_{\gamma})\right]\Lambda(\mathsf{y})
=\frac{1}{\bar{\tau}_q}\!$}}
\begin{cases}
\text{\small{$\check{\nu}^{-1}\!
\left(q^{-\frac{1}{2}\,\left(\bar{\epsilon}_i+\bar{\epsilon}_{i+1}-1\right)}\,
f_{\gamma}\otimes 1\right) \!$}}& 
\text{\footnotesize{$\gamma=\epsilon_i\!-\!\epsilon_{i+1}$}}\\
\text{\small{$\check{\nu}^{-i}\!
\left(q^{-\sum_{k=1}^{i-1}(\bar{\epsilon}_k-1)} q^{-\frac{1}{2}\,
\left(\bar{\epsilon}_i+\bar{\epsilon}_{M}-1\right)}\,f_\gamma\otimes 1\right) \!$}}&
\text{\footnotesize{$ \gamma=\delta\!-\!(\epsilon_i\!-\!\epsilon_M)$}}\\
\text{\small{$0$}} & \text{otherwise}
\end{cases}
\label{Posrootanypiminus}
\end{equation}
where $\bar{\tau}_q:=q^{-1}-q$ and 
$\check{\nu}^{-1}=q^{\frac{\bar{\epsilon}-M}{M}}\,\nu^{-1}$.
The contribution of the imaginary root  to the universal R-matrix is left unchanged by the action of $\Lambda(\mathsf{y})$. 
We conclude that the image of the reduced universal R-matrix can be written as
\begin{equation}\label{Mminusdef}
\left[\left(1\otimes\pi_{\nu}^-\right)\bar{\mathscr{R}}^-\right]_{\text{ren}}\,=\,
\Lambda(\mathsf{y})\left( \mathscr{M}^-(\nu)\otimes 1\right)\Lambda^{-1}(\mathsf{y})\,.
\end{equation}
The explicit expression for \eqref{Mminusexplicit} follows from the from \eqref{Posrootanypiminus}
and the product formula  \eqref{KT}, upon recalling that $\mathscr{E}_q(\tau_q\,x)=\left[\exp_{q^2}(x)\right]_{\text{ren}}$.

\paragraph{Intertwining relation for $\mathscr{M}^-$.}
The property \eqref{Rm1} of the universal R-matrix implies 
\begin{equation}
 [\mathscr{M}^-(\nu),\,\check{e}_i]\,=\,
\nu^{-1}\,q^{\frac{\bar{\epsilon}}{M}}\,
\frac{q^{-2 \bar{\epsilon}_{i+1}}\,\mathscr{M}^-(\nu)-\mathscr{M}^-(\nu)\,q^{-2 \bar{\epsilon}_{i}}}{q^{-1}-q}\,,
\qquad
\check{e}_i\,=\,q^{-\frac{1}{2}(\bar{\epsilon}_i+\bar{\epsilon}_{i+1}-1)}\,e_i\,.
\label{onetensorJimbo}
\end{equation}
\paragraph{The form of $\left(1\otimes \bar{\pi}^-\right)\mathscr{R}^-$}

Introduce $\bar{\mathsf{y}}_i$ via 
\begin{equation}
\bar{\pi}^-(e_i)=\bar{\mathsf{e}}_i\,=\,
\frac{
q^{-\frac{1}{M}}
\,\lambda^{-1}}{q^{-1}-q}\,\bar{\mathsf{y}}_{i+1}\,\bar{\mathsf{y}}^{-1}_i\,.
\label{introduceybar}
\end{equation}
The variables $\bar{\mathsf{y}}_i$ satisfy the same exchange relations as $\mathsf{y}_i$ with $q$ replaced by $q^{-1}$.
We can rewrite  \eqref{eipibarminus}  and \eqref{eisuccbarpiminus} as
\begin{equation}
\bar{\pi}^-(e_{\epsilon_i-\epsilon_j})\,=\,\frac{\left(q^{\frac{1}{M}}\,\lambda\right)^{i-j}}{q^{-1}-q}\,
\bar{\mathsf{y}}_{j}\,\bar{\mathsf{y}}^{-1}_i\,,
\qquad
\bar{\pi}^-(e_{\delta-(\epsilon_1-\epsilon_j)})\,=\,
\frac{\left(q^{\frac{1}{M}}\,\lambda\right)^{j-M-1}}{q^{-1}-q}\,
\bar{\mathsf{y}}_{1}\,\bar{\mathsf{y}}^{-1}_j\,,
\end{equation}
In analogy with \eqref{Posrootanypiminus}  we obtain 
\begin{equation}
\text{\small{$\Lambda^{-1}(\bar{\mathsf{y}})\left[(1\otimes\pi^-)(f_{\gamma}\otimes e_{\gamma})\right]\Lambda(\bar{\mathsf{y}})
=\frac{1}{\bar{\tau}_q}\!$}}
\begin{cases}
\text{\small{$\hat{\lambda}^{i-j}\!
\left(
q^{\sum_{k=i+1}^{j-1}\bar{\epsilon}_k}\,
q^{\frac{1}{2}\,\left(\bar{\epsilon}_i+\bar{\epsilon}_{j}-1\right)}\,
f_{\gamma} 
\otimes 1\right) \!$}}& 
\text{\footnotesize{$\gamma=\epsilon_i\!-\!\epsilon_{j}$}}\\
\text{\small{$
\hat{\lambda}^{j-M-1}\!
\left(
q^{\sum_{k=j+1}^{M}\bar{\epsilon}_k}
q^{\frac{1}{2}\,\left(\bar{\epsilon}_1+\bar{\epsilon}_{j}-1\right)}\,f_\gamma \otimes 1\right) \!$}}&
\text{\footnotesize{$ \gamma=\delta\!-\!(\epsilon_1\!-\!\epsilon_j)$}}\\
\text{\small{$0$}} & \text{otherwise}
\end{cases}
\label{Posrootanypibarminus}
\end{equation}
where $\bar{\tau}_q=q^{-1}-q$ and $\hat{\lambda}=\lambda\,q^{\frac{\bar{\epsilon}}{M}}$.
The asymmetry between $\pi^-$ and $\bar{\pi}^-$ is a consequence of the fact that we choose the same root ordering. 

In analogy with \eqref{onetensorJimbo}  
\begin{equation}
\left[\left(1\otimes\bar{\pi}^-\right)\bar{\mathscr{R}}^-\right]
\,=\,
\Lambda(\bar{\mathsf{y}})\left(\bar{\mathscr{M}}^-\otimes 1\right)\Lambda^{-1}(\bar{\mathsf{y}})\,,
\label{Mminusbardef}
\end{equation}
satisfy the interwining relation
\begin{equation}
 [\bar{\mathscr{M}}^-,\,\hat{e}_i]\,=\,\bar{\alpha}\,\frac{q^{2 \bar{\epsilon}_{i}}\,\bar{\mathscr{M}}^--\bar{\mathscr{M}}^-\,q^{2 \bar{\epsilon}_{i+1}}}{q^{-1}-q}\,,
\qquad
\hat{e}_i\,=\,q^{+\frac{1}{2}(\bar{\epsilon}_i+\bar{\epsilon}_{i+1}-1)}\,e_i\,,
\label{onetensorJimbobar}
\end{equation}
where $\bar{\alpha}=-q^{-1}\bar{\gamma}_-q^{\frac{1}{M}(1-\bar{\epsilon})}$.


\subsection{Some steps for the evaluation of $ \left( \pi^{\mathcal{F}}\otimes \pi^-\right)\,\mathscr{R}^{-}$ 
and $ \left( \pi^{\mathcal{F}}\otimes \bar{\pi}^-\right)\,\mathscr{R}^{-}$}
\label{app:stepsfotLminusF}

\paragraph{Computation of $ \left( \pi^{\mathcal{F}}\otimes \pi^-\right)\,\mathscr{R}^{-}$.}
Applying $\pi_\mathcal{F}$ to \eqref{Posrootanypiminus} and using \eqref{PiFfre+}, \eqref{PiFfre+bis} one obtains
\begin{equation}
\left(\pi^{\mathcal{F}}_{\mu_{\mathfrak{n}}}\otimes  \pi^-_{\nu}\right)\left(f_{\gamma}\otimes e_{\gamma}\right)\,=\,
\begin{cases}
\Lambda(\mathsf{y})\,\left(\frac{g_-}{q-q^{-1}}\,\bar{\mathbf{c}}_{i+1}\,\mathbf{c}_{i}\right)\,\Lambda^{-1}(\mathsf{y}) & \gamma=\epsilon_i-\epsilon_{i+1}\\
\Lambda(\mathsf{y})\,\left(\frac{g^i_-}{q-q^{-1}}\,\,\bar{\mathbf{c}}_{i}\,\,\,\mathbf{c}_{M}\right)\,\Lambda^{-1}(\mathsf{y})\,, & \gamma=\delta-(\epsilon_i-\epsilon_M) \\
0 & \text{otherwise}
\end{cases}
\end{equation}
 where $\gamma\in \Delta_+^{\text{re}}(\widehat{\mathfrak{sl}}_M)$
and 
and  $g_-$ is defined in \eqref{Thetadef}.
Next, one obtains
\begin{equation}\label{imaginaryFermminus}
 \left( \pi^{\mathcal{F}}_{\mu_{\mathfrak{n}}}\otimes \pi^-_{\nu}\right)\,\mathscr{R}^{-}_{\sim \delta}\,
=\,\tilde{\theta}_{\mathcal{F}}^{-}\,\left(1-g_-^M\,\mathfrak{n}_{M}\right)\,,
\qquad \qquad
\tilde{\theta}_{\mathcal{F}}^{-}\,=\,
\frac{(g_-^M\,q^{2M};q^{2M})_{\infty}}{(g_-^M\,q^{2(M-\mathfrak{n})};q^{2M})_{\infty}}\,.
\end{equation}
This calculation is the same as in \eqref{Thetadef}  before regularization.
The last non-trivial identity used in the derivation of \eqref{LminusF} is
 \begin{equation}
\underbrace{(1\!-\!g_-\bar{\mathbf{c}}_{2}\,\mathbf{c}_{1})\dots (1\!-\!g_-\bar{\mathbf{c}}_{M}\,\mathbf{c}_{M-1})}_{\text{from $\mathscr{R}_{\prec \delta}$}}\!
\left(1\!-\!g_-^M\,\mathfrak{n}_{M}\right)\!\!
\underbrace{\left(1\!-\!\sum_{i=1}^{M-1}\,g_-^i\bar{\mathbf{c}}_{i}\,\mathbf{c}_{M} \right)}_{\text{from $\mathscr{R}_{\succ \delta}$}}\!\!=\!\!
\prod_{p=1}^M\left(1\!-\!g_-e^{-\frac{2\pi i p}{M}}\mathbf{N}(p)\right)\,.
\label{UnifermM}
 \end{equation}
In particular notice that the cyclicity property, i.e.~the fact that it commutes with the internal shift operator, of this object is obscure
in the left hand side and totally manifest in the right hand side.


\paragraph{Computation of $ \left( \pi^{\mathcal{F}}\otimes \bar{\pi}^-\right)\,\mathscr{R}^{-}$}.
Applying $\pi_\mathcal{F}$ to \eqref{Posrootanypibarminus} and using \eqref{PiFfre+}, \eqref{PiFfre+bis} one obtains
\begin{equation}
\left(\pi^{\mathcal{F}}_{\mu_{\mathfrak{n}}}\otimes  \bar{\pi}^-_{\nu}\right)\left(f_{\gamma}\otimes e_{\gamma}\right)\,=\,
\begin{cases}
\Lambda(\bar{\mathsf{y}})\,\left(\frac{\bar{g}^{j-i}_-}{q^{-1}-q}\,\bar{\mathbf{c}}_{j}\,\mathbf{c}_{i}\right)\,\Lambda^{-1}(\bar{\mathsf{y}}) & \gamma=\epsilon_i-\epsilon_{j}\\
\Lambda(\bar{\mathsf{y}})\,\left(\frac{\bar{g}^{M-j+1}_-}{q^{-1}-q}\,\bar{\mathbf{c}}_{1}\,\mathbf{c}_{j}\right)\,\Lambda^{-1}(\bar{\mathsf{y}})\,, & \gamma=\delta-(\epsilon_1-\epsilon_j) \\
0 & \text{otherwise}
\end{cases}
\end{equation}
where 
$\bar{g}_-$ is given in \eqref{Thetabardef}.. The contribution of the imaginary roots is
\begin{equation}
 \left( \pi^{\mathcal{F}}_{\mu_{\mathfrak{n}}}\otimes \bar{\pi}^-_{\nu}\right)\,\mathscr{R}^{-}_{\sim \delta}\,=\,\tilde{\theta}_{\mathcal{F}}^{\dot{-}}\,\left(1-\bar{g}_-^M\,\overline{\mathfrak{n}}_{1}\right)\,,
\qquad \qquad
\tilde{\theta}_{\mathcal{F}}^{\dot{-}}\,=\,\frac{(\bar{g}_-^M\,q^{2M};q^{2M})_{\infty}}{(\bar{g}_-^M\,q^{2\mathfrak{n}};q^{2M})_{\infty}}\,,
\end{equation}
compare to \eqref{Thetabardef} 
The last identity we use to prove \eqref{LminusFbar} is
 \begin{equation}
\underbrace{(1\!+\!\mathsf{B}_1\mathbf{c}_{1})\dots (1\!+\!\mathsf{B}_{M-1}\mathbf{c}_{M-1})}_{\text{from $\mathscr{R}_{\prec \delta}$}}\!
\left(1\!-\!\bar{g}_-^M\,\overline{\mathfrak{n}}_{1}\right)\!\!
\underbrace{\left(1\!+\!\sum_{j=2}^{M}\,\bar{g}_-^{M-j+1}\bar{\mathbf{c}}_{1}\,\mathbf{c}_{j} \right)}_{\text{from $\mathscr{R}_{\succ \delta}$}}\!\!=\!\!
\prod_{p=1}^M\left(1\!-\!\bar{g}_-e^{-\frac{2\pi i p}{M}}\overline{\mathbf{N}}(p)\right)\,.
\label{UnifermMbar}
 \end{equation}
where
$\mathsf{B}_i=\sum_{j=i+1}^{M}\,\bar{g}_-^{j-i}\bar{\mathbf{c}}_{j}$ and $\overline{\mathfrak{n}}_i=1-\mathfrak{n}_i=\mathbf{c}\,\bar{\mathbf{c}}$.
As $(1-\alpha\,\overline{\mathfrak{n}}_i)(1-\alpha\,\mathfrak{n}_i)=1-\alpha$ this is the inverse matrix of \eqref{UnifermM}.

%

\subsubsection{Check of the Jimbo equation}

Let us  verify that \eqref{UnifermM} satisfies the relations 
\eqref{onetensorJimbo} via an explicit calculation.
We can rewrite \eqref{onetensorJimbo} as
\begin{equation}
\left[\pi_{\mathcal{F}}\left(\mathscr{M}^-\right),\bar{\mathbf{c}}_i\,\mathbf{c}_{i+1}\right]\,=\,
-g_-\,\left(\mathfrak{n}_{i+1}\,\pi_{\mathcal{F}}\left(\mathscr{M}^-\right)-\pi_{\mathcal{F}}\left(\mathscr{M}^-\right)\,\mathfrak{n}_i\right)\,,
\label{InterFerm}
\end{equation}
where we have used $q^{-2 \mathfrak{n}_i}=1-\mathfrak{n}_i\,q^{-1}(q-q^{-1})$ to simplify the right hand side 
and  introduced $g_-$ as in \eqref{Thetadef}.
It is easy to check that the relation \eqref{InterFerm} is satisfied if
\begin{align}
 \pi_{\mathcal{F}}\left(\mathscr{M}^-\right)\,\bar{\mathbf{c}}_i\,&=\,
 \left(\bar{\mathbf{c}}_i-g_-\bar{\mathbf{c}}_{i+1}\right)\,\pi_{\mathcal{F}}\left(\mathscr{M}^-\right)\,,
\\
\mathbf{c}_{i+1}\,\pi_{\mathcal{F}}\left(\mathscr{M}^-\right)\,&=\,
\pi_{\mathcal{F}}\left(\mathscr{M}^-\right)\, \left(\mathbf{c}_{i+1}-g_-\mathbf{c}_{i}\right)\,.
\end{align}
These equations are easy to solve upon Fourier transformation in the index $i$ and give the solution \eqref{UnifermM}.
$\pi_{\mathcal{F}}\left(\bar{\mathscr{M}}^-\right)$ satisfies the same equations as $\pi_{\mathcal{F}}\left(\mathscr{M}^-\right)^{-1}$
with $g_-$ replaced by $\bar{g}_-$.

\subsection{Derivation of \eqref{coproductimSLM}}
\label{app:derivationIMROOTPART}

\subsubsection{From $\func{\brac{\Delta \otimes \id}}{\CR} = \CR_{13} \CR_{23}$ to $\Delta(\mathscr{M}^-)$}

Applying $(1\otimes 1 \otimes \pi^-)$ to \eqref{eqnUnivROther} and using  \eqref{Mminusdef}, one obtains
\begin{equation}\label{DeltaMapp1}
 \Delta\left(\mathscr{M}^-\right)\,=\,
\left(\mathscr{F}_{12}^{-1}\,\mathscr{M}^-_1\,\mathscr{F}_{12}\right)\,
\left(q_{\infty}\,\mathscr{F}_{12}\,\mathscr{M}^-_2\,\mathscr{F}_{12}^{-1}\,q_{\infty}^{-1}\right)\,,
\end{equation}
 where
\begin{equation}
\mathscr{F}_{12}\,:=\,q^{-\frac{1}{2}\,\sum_{i,j=1}^M\,\left(\bar{\epsilon}_i\otimes \bar{\epsilon}_j\right)\,Y_{ij}}\,,
\qquad
q_{\infty}\,=\,q^{\sum_{i=1}^M\,\left(\bar{\epsilon}_i\otimes \bar{\epsilon}_i\right)}\,.
\end{equation}
This claim can be easily derived using $(1\otimes \pi^-)q^{-t}\,=\,\Lambda(\su)$ and
\begin{equation}
 \Delta\left(\Lambda(\mathsf{y})\right)^{-1}\,\Lambda_1(\mathsf{y})\,=\,\mathscr{F}_{12}^{-1}\,\Lambda_2(\mathsf{y})\,,
\end{equation}
\begin{equation}
 \Lambda_2(\mathsf{y})^{-1}\Lambda_2(\su)\,\Delta\left( \Lambda(\mathsf{y})^{-1}\Lambda(\su)\right)^{-1}\,=\,
q^{-\frac{c}{M}(\bar{\epsilon}\otimes\bar{\epsilon} )}\Lambda_1(\mathsf{u})^{-1}\Lambda_1(\mathsf{y})\mathscr{F}_{12}^{-1}\,q_{\infty}^{-1}\,,
\end{equation}
\begin{equation}
 \Lambda_2(\mathsf{y})\Lambda_1(\mathsf{y})^{-1}\Lambda_1(\su)\Lambda_2(\mathsf{y})\Lambda_1(\su)^{-1}\,\Lambda_1(\mathsf{y})\,=\,
q^{+\frac{c}{M}(\bar{\epsilon}\otimes\bar{\epsilon} )}q_{\infty}\,\mathscr{F}_{12}^{2}\,.
\end{equation}
These relations are derived using \eqref{yalgebra}.

\subsubsection{Preliminaries}
\paragraph{Commutation relations involving $\mathsf{X}^{\prec}_i$, $\mathsf{X}^{\succ}_i$  defined in \eqref{Xpreccsuccdef}.}
The following relations hold
\begin{equation}\label{Xsuccsucc}
 \mathsf{X}^{\succ}_i\,\mathsf{X}^{\succ}_j\,=\,q^{2\left(\delta_{i<j}-\delta_{i>j}\right)}\,
\mathsf{X}^{\succ}_j\,\mathsf{X}^{\succ}_i\,,
\qquad
i,j=1,\dots, M-1\,.
\end{equation}
\begin{equation}
\mathsf{X}^{\succ}_i\,\mathsf{X}^{\prec}_j\,=\,
\begin{cases}
 q^{2\delta_{i,j+1}}\,\mathsf{X}^{\prec}_j\,\mathsf{X}^{\succ}_i & i>j\\
\mathsf{X}^{\prec}_j\,\mathsf{X}^{\succ}_i & i<j\neq M-1
\end{cases}\,,
\,\,\,\,\,\qquad
\text{for}\,\,\, 1\leq i \neq j\leq M-1\,. 
\label{summaryeq}
\end{equation}
The case $i=j$ corresponds to the iterative definition
$\mathsf{X}^{\succ}_{i+1}\,=\,\tau_q^{-1}\,[\mathsf{X}^{\succ}_{i},\mathsf{X}^{\prec}_{i}]$, 
where $\tau_q=q-q^{-1}$ (compare to \eqref{Xpreccsuccdef}).

\noindent\emph{Proof:}
One may verify the relations above by direct calculations and inductive arguments.
In the following we will show how these relations arise as a consequence of \eqref{eeCONVEX} 
and the definitions \eqref{Xpreccsuccdef}, \eqref{fromftoXsucc}.
This  is a simple corollary of \eqref{eeCONVEX}:

\noindent
Let $\alpha,\,\beta\in \Delta_+(\widehat{\mathfrak{g}})$ with  $\alpha \prec \beta$ 
be such that the decomposition 
$\alpha+\beta\,=\,\sum_k\,n_k\,\gamma_k$
 with $n_k\in \mathbb{Z}_{>0}$ and  $\gamma_k\in \Delta_+(\widehat{\mathfrak{g}})$
is unique. Then
\begin{equation}
f_{\alpha}\,f_{\beta}\,=\,q^{-(\alpha,\beta)}\,f_{\beta}\,f_{\alpha}\,.
\label{falphabetaexchange}
\end{equation}

As an illustrative example let us show  how this corollary implies \eqref{Xsuccsucc} .
 The identity \eqref{summaryeq} is shown similarly.
It is easy to see that $\alpha=\delta-(\epsilon_i-\epsilon_M)$ and 
$\beta=\delta-(\epsilon_j-\epsilon_M)$ for 
 $i>j$ satisfy the conditions for \eqref{falphabetaexchange} to hold. 
We conclude that
\begin{equation}
f_{\delta-(\epsilon_i-\epsilon_M)}\,f_{\delta-(\epsilon_j-\epsilon_M)}
\,=\,q^{-1}
\,f_{\delta-(\epsilon_j-\epsilon_M)}\,f_{\delta-(\epsilon_i-\epsilon_M)}\,,
\qquad \qquad i>j\,.
\end{equation}
The relation \eqref{Xsuccsucc} easily follows from this identity together 
with the definitions \eqref{Xpreccsuccdef}, \eqref{fromftoXsucc} and 
the relation $q^{h}f_{\gamma}=q^{-\langle h ,\gamma \rangle}f_{\gamma} q^h$.
\qed
\paragraph{Coproducts of $\mathsf{X}^{\prec}_i$, $\mathsf{X}^{\succ}_i$  defined in \eqref{Xpreccsuccdef}.}
A simple calculation using the definition of the coproduct shows that
\begin{equation}
 \Delta\left(\mathsf{X}_i^{\prec}\right)\,=\, \mathsf{X}^{\prec}_i(1)\,+\, \mathsf{X}^{\prec}_i(2)\,,
\end{equation}
where
\begin{equation}\label{Xprec12def}
 \mathsf{X}^{\prec}_i(1):=\mathsf{X}_i^{\prec}\otimes \mathsf{a}_i\,,
\qquad
 \mathsf{X}^{\prec}_i(2):=\mathsf{a}_i\,k_i^{-1}\otimes \mathsf{X}_i^{\prec}\,,
\qquad
 \mathsf{a}_i
\,=\, q^{\frac{1}{M}\bar{\epsilon}}q^{-\frac{1}{2}\left(\bar{\epsilon}_i+\bar{\epsilon}_{i+1}\right)}\,.
\end{equation}
The coproduct of $\mathsf{X}^{\succ}_i$ defined in \eqref{Xpreccsuccdef} is more complicated.
Set 
\begin{equation}\label{deltaicopr}
 \delta_i\,:=\,\Delta(\mathsf{X}^{\succ}_{i})-\mathsf{X}^{\succ}_i(1)-\mathsf{X}^{\succ}_i(2)\,.
\end{equation}
where
\begin{equation}\label{Xsucc12def}
 \mathsf{X}^{\succ}_i(1):=\mathsf{X}_i^{\succ}\otimes \mathsf{b}_i\,,
\,\,\,\,
 \mathsf{X}^{\succ}_i(2):=\mathsf{b}_i\,q^{-\left(\bar{\epsilon}_M-\bar{\epsilon}_i\right)}\otimes \mathsf{X}_i^{\succ}\,,
\,\,\,\,
\mathsf{b}_i\,=\,
q^{\frac{i}{M}\bar{\epsilon}_{\text{tot}}}\,
q^{-\sum_{k=1}^{i-1}\,\bar{\epsilon}_k}\,q^{-\frac{1}{2}\left(\bar{\epsilon}_M+\bar{\epsilon}_{i}\right)}\,.
\end{equation}
Notice that $\mathsf{b}_i$ commutes with $f_{\delta-(\epsilon_i-\epsilon_M)}$.
The explicit expression of $\delta_i$ is given below. 

\paragraph{Remark}
\begin{equation}
 \mathscr{F}_{12}^{-1}\,\left(\mathsf{X}^{\prec}_i\otimes 1\right)\,\mathscr{F}_{12}
\,=\,\mathsf{X}^{\prec}_i(1)
\qquad
 q_{\infty}\mathscr{F}_{12}\,\left(1\otimes \mathsf{X}^{\prec}_i\right)\,\mathscr{F}^{-1}_{12}\,q^{-1}_{\infty}
\,=\,\mathsf{X}^{\prec}_i(2)\,,
\label{Fsim1}
\end{equation}
\begin{equation} \mathscr{F}_{12}^{-1}\,\left(\mathsf{X}^{\succ}_i\otimes 1\right)\,\mathscr{F}_{12}
\,=\,\mathsf{X}^{\succ}_i(1)
\qquad
 q_{\infty}\mathscr{F}_{12}\,\left(1\otimes \mathsf{X}^{\succ}_i\right)\,\mathscr{F}^{-1}_{12}\,q^{-1}_{\infty}
\,=\,\mathsf{X}^{{\succ}}_i(2)\,.
\label{Fsim2}
\end{equation}
\paragraph{More commutation relations.} It is a simple exercise to show that
the combinations defined in \eqref{Xprec12def}, \eqref{Xsucc12def} satisfy the following relations
\begin{equation}\label{exchXprecprec}
  \mathsf{X}^{\prec}_i(1)\, \mathsf{X}^{\prec}_j(2)\,=\,
 q^{-2\left(\delta_{i,j}-\delta_{i+1,j}\right)}\, \mathsf{X}^{\prec}_j(2)\,\mathsf{X}^{\prec}_i(1)
\end{equation}
\begin{equation}\label{exchXprecsucc}
  \mathsf{X}^{\prec}_i(1)\, \mathsf{X}^{\succ}_j(2)\,=\,
 q^{2\left(\delta_{i,M-1}-\delta_{i,j-1}\right)}\, \mathsf{X}^{\succ}_j(2)\,\mathsf{X}^{\prec}_i(1)
\end{equation}
\begin{equation}\label{exchXsuccprec}
  \mathsf{X}^{\succ}_i(1)\, \mathsf{X}^{\prec}_j(2)\,=\,
 q^{2\,\delta_{i,j}}\, \mathsf{X}^{\prec}_j(2)\,\mathsf{X}^{\succ}_i(1)
\end{equation}
\begin{equation}\label{exchXsuccsucc}
  \mathsf{X}^{\succ}_i(1)\, \mathsf{X}^{\succ}_j(2)\,=\,
 q^{-2\,\delta_{i\geq j}}\, \mathsf{X}^{\succ}_j(2)\,\mathsf{X}^{\succ}_i(1)
\end{equation}
The exhange relations  involving  $\mathsf{X}^{\star}_i(a)$,  $\mathsf{X}^{\star'}_j(a)$ 
with $a$ fixed are the same as  \eqref{Xsuccsucc} and \eqref{summaryeq}.

\paragraph{Explicit form of $\delta_i$.}
It follows from the definition  \eqref{deltaicopr} that
\begin{equation}\label{deltaellsum}
\delta_{\ell}\,=\,q\,\sum_{k=1}^{\ell-1}\,\tau_q^{(k-\ell+1)}\,
[\cdots[\mathsf{X}^{\prec}_k(2),\mathsf{X}^{\prec}_{k+1}(2)]\dots,\mathsf{X}^{\prec}_{\ell-1}(2)]\,
\mathsf{X}^{\succ}_k(1)\,,
\end{equation}
where $\tau_q=q-q^{-1}$.

\noindent \emph{Proof.} 
Upon applying the coproduct to the inductive definition 
$\mathsf{X}^{\succ}_{i+1}=\tau_q\,[\mathsf{X}^{\succ}_{i},\mathsf{X}^{\prec}_{i}]$ 
and using \eqref{exchXsuccprec}, \eqref{exchXprecsucc} one easily obtains
\begin{equation}
 \delta_{i+1}\,=\,q\,\mathsf{X}^{\prec}_i(2)\,\mathsf{X}^{\succ}_i(1)
+\tau_q^{-1}\,[\delta_i,\mathsf{X}^{\prec}_i(1)+\mathsf{X}^{\prec}_i(2)]\,.
\label{deltaip1}
\end{equation}
Further observe that 
\begin{equation}
[\delta_i,\mathsf{X}^{\prec}_k(1)]=0\,,
\qquad \,\,\,k=i,i+1,\dots, M-2\,.
\end{equation}
This can be easily shown by induction using \eqref{deltaip1} and the exchange properties given in the previous paragraph.
Equation  \eqref{deltaip1}  thus reduces to 
\begin{equation}
 \delta_{i+1}\,=\,q\,\mathsf{X}^{\prec}_i(2)\,\mathsf{X}^{\succ}_i(1)
+\tau_q^{-1}\,[\delta_i,\mathsf{X}^{\prec}_i(2)]\,,
\label{deltaip1BIS}
\end{equation}
from which the explicit form of $\delta_i$ given above follows.

 \qed

We notice that while $\delta_i$ was originally defined for 
$i=1,\dots, M-1$, we extend the definition to $i=M$
using the explicit formula  \eqref{deltaellsum}.

\paragraph{Some commutation relations involving $\delta_i$.}
We collect the following relations
\begin{equation}\label{deltaexch}
\delta_i\,\mathsf{X}_i^{\succ}(1)\,=\,q^2\,\mathsf{X}_i^{\succ}(1)\,\delta_i\,,
\qquad \qquad
\delta_i\,\mathsf{X}_k^{\succ}(2)\,=\,q^{-2\delta_{k,i}}\,\mathsf{X}_k^{\succ}(2)\,\delta_i\,,\qquad
k\geq i\,.
\end{equation}
\begin{equation}\label{deltaXprec2}
\delta_i\,\mathsf{X}^{\prec}_k(2)\,=\,
\mathsf{X}^{\prec}_k(2)\,\delta_i\,,
\qquad k=i+1,\dots,M-1\,.
\end{equation}
\begin{equation}\label{exchangeforlaststep0}
[\delta_i,\mathsf{X}^{\prec}_i(2)]\,\mathsf{X}^{\succ}_i(1)\,=\,
\mathsf{X}^{\succ}_i(1)\,[\delta_i,\mathsf{X}^{\prec}_i(2)]\,
\end{equation}
\begin{equation}\label{exchangeforlaststep}
[\delta_i,\mathsf{X}^{\prec}_i(2)]\,\left(\mathsf{X}^{\prec}_i(2)\mathsf{X}^{\succ}_i(1)\right)\,=\,
q^2\,\left(\mathsf{X}^{\prec}_i(2)\mathsf{X}^{\succ}_i(1)\right)\,[\delta_i,\mathsf{X}^{\prec}_i(2)]\,
\end{equation}
The last identity
follows from the Serre relations \eqref{Serreeasy1} (). 
Finally
\begin{equation}\label{Serrefordeltai}
\SV\,:=\,\delta_{\ell}\,,
\qquad
\SU\,:=\,\mathsf{X}^{\prec}_{\ell}(2)\,,
\qquad
\text{satisfy the (twisted) Serre relations \eqref{serreUV1}, \eqref{serreUV2}.}
\end{equation}
The relation \eqref{serreUV2}, which is linear in $\delta_{\ell}$,  
can be shown easily using the exchange relations collected above and 
the fact that  $\mathsf{X}^{\prec}_{\ell-1}(2)$, $\mathsf{X}^{\prec}_{\ell}(2)$ satisfy the 
(twisted) Serre relations \eqref{serreUV2}.
Showing \eqref{serreUV1} requires a bit of work. 
 It is not hard to see, using the explicit expression for $\delta_{\ell}$ given in \eqref{deltaellsum},
 that the equality
\begin{equation}\label{wwequations}
q^{-1}(w_m\,w_n+w_n\,w_m)\,\check{f}_{\ell}+q^{+1}\check{f}_{\ell}\,(w_m\,w_n+w_n\,w_m)\,=\,
(q+q^{-1})\,
\left(w_m\,\check{f}_{\ell}\,w_n+w_n\,\check{f}_{\ell}\,w_m\right)\,,
\end{equation}
where $n\neq m<\ell$ and $w_n=[\cdots[\check{f}_n,\check{f}_{n+1}],\dots, \check{f}_{\ell-1}]$
implies \eqref{serreUV2}. 
The relations \eqref{wwequations} can be shown as follows. 
Let $m>n$ and notice that $w_n=[x,w_m]$ where $[x,\check{f}_{\ell}]=0$,  
\eqref{wwequations}  is satisfied if the same equation holds for $w_n\mapsto w_m$.
The relation  \eqref{wwequations}  for $m=n$ is a consequence of this elementary fact:
If $(f_1,\dots,f_i,f_{i+1}\dots,f_M)$ satisfy the Serre relations of $\mathcal{U}_q(\widehat{\mathfrak{gl}}_M)$,
 then for any $i\,\in\mathbb{Z}/M\mathbb{Z}$ and choice of sign $\sigma$, the elements
$(f_1,\dots,f_i\,f_{i+1}-q^{\sigma}\,f_{i+1}\,f_{i}\dots,f_M)$ 
satisfy the Serre relations of $\mathcal{U}_q(\widehat{\mathfrak{gl}}_{M-1})$.

\paragraph{$\delta_i$ and the opposite root ordering.}
Let $f^{\text{op}}_{\epsilon_i-\epsilon_j}$ be root vectors constructed using the opposite root ordering,
 explicitly
\begin{equation}
f^{\text{op}}_{\epsilon_i-\epsilon_j}\,=\,
f_i\,f^{\text{op}}_{\epsilon_{i+1}-\epsilon_j}-q^{-1}\,
f^{\text{op}}_{\epsilon_{i+1}-\epsilon_j}\,f_i\,,
\qquad
j=i+2,\dots,M\,,
\end{equation}
with $f^{\text{op}}_{\epsilon_i-\epsilon_{i+1}}=f_{i}$. 
It is easy to inductively show that
\begin{equation}
f^{\text{op}}_{\epsilon_i-\epsilon_j}\,=\,
q^{\frac{i-j+1}{2}}\,
q^{\frac{1}{2}\sum_{k=i}^{j-1}(\bar{\epsilon}_k+\bar{\epsilon}_{k+1}-1)}\,
[\cdots[\check{f}_i,\check{f}_{i+1}],\dots,\check{f}_{j-1}]\,,
\qquad
1\leq i<j\ \leq M\,.
\end{equation}
In the special case of $j=M$ it may be rewritten as
\begin{equation}
\label{fOPPexplicit}
f^{\text{op}}_{\epsilon_i-\epsilon_M}\,=\, q^{\sum_{k=i+1}^{M-1}\,\left(\bar{\epsilon}_k-1\right)}\,q^{\frac{1}{2}\,\left(\bar{\epsilon}_i+\bar{\epsilon}_{M}-1\right)}\,
[\dots[[\check{f}_{i},\check{f}_{i+1}],\check{f}_{i+2}],\dots], \check{f}_{M-1}]\,.
\end{equation}

\subsubsection{From $\Delta(\mathscr{M}^-)$ to $\Delta(\mathscr{M}^-_{\sim \delta})$}

\paragraph{On coproduct of $\mathscr{M}^-_{\prec}$.}
The following identity holds
\begin{equation}\label{DeltaMprec}
 \Delta\left(\mathscr{M}^-_{\prec}\right)\,=\,
\left(\mathscr{F}_{12}^{-1}\,\mathscr{M}^-_{\prec,1}\,\mathscr{F}_{12}\right)\,
\left(q_{\infty}\,\mathscr{F}_{12}\,\mathscr{M}^-_{\prec,2}\,\mathscr{F}_{12}^{-1}\,q_{\infty}^{-1}\right)\,,
\end{equation}

\noindent 
\emph{Proof.}
 Recall the form of $\mathscr{M}^-$ from \eqref{Mminusexplicit}.
It follows from \eqref{qexp-prop1} and the exchange relation \eqref{exchXprecprec} that
\begin{equation}\label{idenaux1}
 \Delta\left(\mathscr{E}_q(\mathsf{X}_i^{\prec})\right)\,=\,
\mathscr{E}_q\left(\Delta(\mathsf{X}_i^{\prec})\right)\,=\,
\mathscr{E}_q(\mathsf{X}_i^{\prec}(1))\,\mathscr{E}_q(\mathsf{X}_i^{\prec}(2)).
\end{equation}
The identity \eqref{DeltaMprec}  follows from this relation together with 
\eqref{Fsim1}  and  the exchange relations \eqref{exchXprecprec}.
\paragraph{On coproduct of $\mathscr{M}^-_{\succ}$.}
Let us define $\mathscr{B}$ as follows
\begin{equation}\label{DeltaMsucc}
\Delta\left(\mathscr{M}^-_{\succ}\right)\,=\,
\mathscr{B}^{-1}\,\left(q_{\infty}\,\mathscr{F}_{12}\,\mathscr{M}^-_{\succ,2}\,\mathscr{F}_{12}^{-1}\,q_{\infty}^{-1}\right)\,.
\end{equation}
More explicitely, using the form of $\mathscr{M}^{\succ}$ given in \eqref{Mminusexplicit} and  \eqref{Fsim2} 
\begin{equation}\label{Bsecondform}
 \mathscr{B}\,=\,\left[\mathscr{E}_q(\mathsf{X}^{\succ}_{M-1}(2))\dots \mathscr{E}_q(\mathsf{X}^{\succ}_{1}(2))\right]\,
\left[
\frac{1}{\mathscr{E}_q(\Delta(\mathsf{X}^{\succ}_{1}))}
\dots \frac{1}{\mathscr{E}_q(\Delta(\mathsf{X}^{\succ}_{M-1}))}\right]\,.
\end{equation}
In order to simplify this expression we will use the following lemma.
\paragraph{Lemma.} 
This identity holds
\begin{equation}\label{idenaux2}
 \Delta\left(\mathscr{E}_q(\mathsf{X}_i^{\succ})\right)\,=\,
\mathscr{E}_q\left(\Delta(\mathsf{X}_i^{\succ})\right)\,=\,
\mathscr{E}_q(\mathsf{X}_i^{\succ}(1))\,\mathscr{E}_q(\delta_i)\,\mathscr{E}_q(\mathsf{X}_i^{\succ}(2)),
\end{equation}
where $\Delta(\mathsf{X}_i^{\succ})=\mathsf{X}_i^{\succ}(1)+\mathsf{X}_i^{\succ}(2)+\delta_i$, compare to \eqref{deltaicopr}.

\noindent
\emph{Proof}: \eqref{idenaux2} is derived using two simple observations
\begin{itemize}
\item[\emph{(i)}]
\begin{equation}
 \mathscr{E}_q( \mathsf{U}+ \mathsf{V}+ \mathsf{W})\,=\,
\mathscr{E}_q( \mathsf{U})\,\mathscr{E}_q( \mathsf{W})\,\mathscr{E}_q( \mathsf{V})\,,
\end{equation}
if 
\begin{equation}\label{UVW}
 \mathsf{U}\,\mathsf{V}\,=\,q^{-2}\,\mathsf{V}\,\mathsf{U}\,,
\qquad
\mathsf{U}\,\mathsf{W}\,=\,q^{-2}\,\mathsf{W}\,\mathsf{U}\,,
\qquad
\mathsf{V}\,\mathsf{W}\,=\,q^{+2}\,\mathsf{W}\,\mathsf{V}\,.
\end{equation}
\item[\emph{(ii)}]
The exchange relations \eqref{UVW} are satisfied by
\begin{equation}
 \SU\,=\,\mathsf{X}^{\succ}_i(1)\,,
\qquad
\SV\,=\,\mathsf{X}^{\succ}_i(2)\,,
\qquad
\mathsf{W}\,=\,\delta_i\,.
\end{equation}
\end{itemize}
Point \emph{(i)} is derived using \eqref{qexp-prop1} twice.
Point \emph{(ii)} uses the exchange relations \eqref{exchXsuccprec} and \eqref{deltaexch}.
 \qed

By applying this lemma to \eqref{Bsecondform}  and rearranging terms using the exchange relations 
\eqref{exchXsuccsucc}  and \eqref{deltaexch} we obtain
\begin{equation}\label{DeltaMsuccEuqality}
 \mathscr{B}\,=\,
\left(\frac{1}{\mathscr{E}_q(\mathsf{X}^{\succ}_1(1))}\right)\,
\left(\frac{1}{\mathscr{E}_q(\delta_2)}\frac{1}{\mathscr{E}_q(\mathsf{X}^{\succ}_2(1))}\right)
\dots 
\left(\frac{1}{\mathscr{E}_q(\delta_{M-1})}\frac{1}{\mathscr{E}_q(\mathsf{X}^{\succ}_{M-1}(1))}\right)\,.
\end{equation}
The second tensor factors of $\delta_i$ and  $\mathsf{X}^{\succ}_i(1)$ are written in terms of 
$\{f_k\}_{k\in\{1,\dots,i-1\}}$ and Cartan generators only.
This fact, combined with   the observation \eqref{Vmcommuteswithfj} 
and the explicit form  \eqref{DeltaMsuccEuqality}, makes it manifest that
\begin{equation}\label{MBcomm}
 \left(1\otimes \mathscr{M}_{\sim \delta}\right)\mathscr{B}\,=\,
\mathscr{B}\left(1\otimes \mathscr{M}_{\sim \delta}\right)\,.
\end{equation}
Using this relation and the explicit form of $\mathscr{B}$, 
we rewrite \eqref{DeltaMapp1}  as
\begin{equation}
 \Delta\left(\mathscr{M}^-_{\sim \delta}\right)\,=\,
\left(\mathscr{M}^-_{\sim \delta}\otimes 1\right) \,\mathscr{A}\mathscr{B}\,
\left(1\otimes \mathscr{M}^-_{\sim \delta}\right)\,,
\end{equation}
where
\begin{equation}\label{Afinal}
\text{\small{$
\mathscr{A}:=
\left(
\frac{1}{\mathscr{E}_q(\mathsf{X}_{M\!-\!1}^{\prec}(2))}
\dots 
\frac{1}{\mathscr{E}_q(\mathsf{X}_{1}^{\prec}(2))}\right)
\left(\mathscr{E}_q(\mathsf{X}_{M\!-\!1}^{\succ}(1))\dots\mathscr{E}_q(\mathsf{X}_{1}^{\succ}(1))\right)
\left(\mathscr{E}_q(\mathsf{X}_{1}^{\prec}(2))\dots\mathscr{E}_q(\mathsf{X}_{M\!-\!1}^{\prec}(2))\right)\,,$}}
\end{equation}
To derive this expression we also used  the fact that  
$\mathscr{A}$ and $( \mathscr{M}^-_{\sim \delta}\otimes 1)$  commute.

\paragraph{Completing the derivation.}
In the following we will show that
\begin{equation}
\mathscr{A}\mathscr{B}\,=\,
\mathscr{X}_{\ell}\,\mathscr{E}_q(\delta_{\ell})\,\mathscr{Y}_{\ell}\,,
\qquad
\ell=1,\dots,M\,.
\label{ABtoXYidentity}
\end{equation}
where
\begin{equation}
\mathscr{X}_{\ell}\,=\,
\left(
\frac{1}{\mathscr{E}_q(\mathsf{X}_{M\!-\!1}^{\prec}(2))}
\dots 
\frac{1}{\mathscr{E}_q(\mathsf{X}_{\ell}^{\prec}(2))}
\right)
\left(
\mathscr{E}_q(\mathsf{X}_{M\!-\!1}^{\succ}(1))\dots
\mathscr{E}_q(\mathsf{X}_{\ell}^{\succ}(1))\right)
\end{equation}
\begin{equation}
\mathscr{Y}_{\ell}\,=\,
\left(\mathscr{E}_q(\mathsf{X}_{\ell}^{\prec}(2))\dots\mathscr{E}_q(\mathsf{X}_{M\!-\!1}^{\prec}(2))\right)
\left[ 
\left(\frac{1}{\mathscr{E}_q(\delta_{\ell})}\,\frac{1}{\mathscr{E}_q(\mathsf{X}_{\ell}^{\succ}(1))}\right)
\dots
\left(\frac{1}{\mathscr{E}_q(\delta_{M\!-\!1})}\,\frac{1}{\mathscr{E}_q(\mathsf{X}_{M\!-\!1}^{\succ}(1))}\right)
\right]
\end{equation}
and $\delta_i$ are given in \eqref{deltaellsum}.

\noindent
\emph{Proof.}
For $\ell=1$ the identity trivially follows from the explicit form of $\mathscr{A}$, $\mathscr{B}$
 given in \eqref{Afinal}, \eqref{DeltaMsuccEuqality}
 and the fact that $\mathscr{E}_q(\delta_{1}=0)=1$. 
For $\ell=M$  one has $\mathscr{X}_{M}=\mathscr{Y}_{M}=1$ and
the identity \eqref{ABtoXYidentity} implies \eqref{coproductimSLM}. 
We will prove \eqref{ABtoXYidentity} by induction on $\ell$. First notice that
\begin{equation}
\mathscr{X}_{\ell}\,=\,
\mathscr{X}_{\ell+1}\,\,
\mathscr{E}_q(q\mathsf{X}_{\ell}^{\prec}(2)\mathsf{X}_{\ell}^{\succ}(1))\,\,
\mathscr{E}_q(\mathsf{X}_{\ell}^{\succ}(1))\,\,
\frac{1}{\mathscr{E}_q(\mathsf{X}_{\ell}^{\prec}(2))}
\end{equation}
\begin{equation}
\mathscr{Y}_{\ell}\,=\,
\mathscr{E}_q(\mathsf{X}_{\ell}^{\prec}(2))\,\,
\frac{1}{\mathscr{E}_q(\delta_{\ell})}\,\,
\frac{1}{\mathscr{E}_q(\mathsf{X}_{\ell}^{\succ}(1))}\,\,
\mathscr{Y}_{\ell+1}\,.
\end{equation}
The first identity easily follows from the exchange relations \eqref{exchXsuccprec} 
and the pentagon relation \eqref{qexp-prop2}.
The second identity follows from the exchange relations \eqref{deltaXprec2} 
and \eqref{exchXsuccprec}.  The crucial observation  is that as a consequence 
of \eqref{Serrefordeltai} one can use \eqref{pentagonwithSerre} to rewrite
\begin{equation}\label{idenauxLAST}
\frac{1}{\mathscr{E}_q(\mathsf{X}_{\ell}^{\prec}(2))}\,\,
\mathscr{E}_q(\delta_{\ell})\,\,
\mathscr{E}_q(\mathsf{X}_{\ell}^{\prec}(2))\,\,
\frac{1}{\mathscr{E}_q(\delta_{\ell})}\,\,=\,
\mathscr{E}_q(\tau_q^{-1}[\delta_{\ell},\mathsf{X}_{\ell}^{\prec}(2)])\,.
\end{equation}
Finally \eqref{exchangeforlaststep0} and 
\eqref{exchangeforlaststep} with \eqref{qexp-prop1}
 imply the result.

\qed

\subsection{For mixed pentagon}
\label{APP:formixedpentagon}

The goal of this appendix it to show that
\eqref{idenaux1},
\eqref{idenaux2} and 
\eqref{idenauxLAST}
are satisfied when we apply
$\pi^{\mathcal{F}}\otimes \pi^+$. 

\subsubsection{Preliminaries}
The first step is to provide explicit expressions for the arguments of the special functions 
entering, \eqref{idenaux1},
\eqref{idenaux2} and 
\eqref{idenauxLAST} when we apply $\pi^{\mathcal{F}}\otimes \pi^+$.
\paragraph{Action of $1\ot \pi^+$ on $\mathsf{X}^{\prec}_{\ell}(1)$ and  $\mathsf{X}^{\prec}_{\ell}(2)$.}
Consider $\mathsf{X}^{\prec}_{\ell}(1)$ and $\mathsf{X}^{\prec}_{\ell}(2)$
 defined in \eqref{Xprec12def} and  \eqref{Xpreccsuccdef}.
They satisfy the following relations
\begin{align}
\Lambda(\mathsf{y})\,\left[
\left(1\ot \pi^+_{\la}\right)\mathsf{X}^{\prec}_{\ell}(1)
\right]\Lambda^{-1}(\mathsf{y})\,&=\,\tau_{q}\,
\la^{-1}\,
q^{-\frac{\bar{\epsilon}}{M}}\,
\hat{f}_\ell
\ot \mathsf{m}^{\prec}_{\ell}\,,
\label{1timespiplusXprec1}\\
\Lambda(\mathsf{y})\,\left[
\left(1\ot \pi^+_{\la}\right)\mathsf{X}^{\prec}_{\ell}(2)
\right]\Lambda^{-1}(\mathsf{y})\,&=\,
q^{2\left(\bar{\epsilon}_{\ell+1}-\frac{\bar{\epsilon}}{M}\right)}
\ot \mathsf{m}^{\prec}_{\ell}\,,
\label{1timespiplusXprec2}
\end{align}
where
\begin{equation}\label{mprecelldef}
\mathsf{m}^{\prec}_{\ell}:=q^{-\frac{1}{2}}\,
q^{\frac{1}{M}}\,\la^{}\,\nu^{-1}\,
\pi^+_{\la}(\mathsf{a}_{\ell})\,\mathsf{y}_{\ell+1}^{}\,\mathsf{y}_{\ell}^{-1}\,,
\end{equation}
and $\hat{f}_{\ell}=q^{\frac{1}{2}\left(\bar{\epsilon}_{\ell}+\bar{\epsilon}_{\ell}-1\right)}$,
compare to the definition below \eqref{onetimespiplus}.
These relation  follows from \eqref{id-1}, \eqref{id-2} and
$\Lambda(\mathsf{y})\left[
\left(1\ot \pi^+_{\la}(q^{\bar{\epsilon}_i})\right)
\right]\Lambda^{-1}(\mathsf{y})=q^{-\bar{\epsilon}_i+\frac{\bar{\epsilon}}{M}}\ot\pi^+_{\la}(q^{\bar{\epsilon}_i})$.
\paragraph{Applying $\pi^{\mathcal{F}}_\mu$ to the first tensor factor.}
From the identities above it follows that 
\begin{align}
\Lambda(\mathsf{y})\,\left[
\left(\pi^{\mathcal{F}}_\mu\ot \pi^+_{\la}\right)\mathsf{X}^{\prec}_{\ell}(1)
\right]\Lambda^{-1}(\mathsf{y})\,&=\,\tau_{q}\,
\mu\,\la^{-1}\,
q^{-\frac{\mathfrak{n}}{M}}\,
\bar{\mathbf{c}}_{\ell+1}
\mathbf{c}_{\ell}
\ot \mathsf{m}^{\prec}_{\ell}\,,
\label{1timespiplusXprec1FERM}\\
\Lambda(\mathsf{y})\,\left[
\left(\pi^{\mathcal{F}}_\mu\ot \pi^+_{\la}\right)\mathsf{X}^{\prec}_{\ell}(2)
\right]\Lambda^{-1}(\mathsf{y})\,&=\,
q^{2\left(\mathfrak{n}_{\ell+1}-\frac{\mathfrak{n}}{M}\right)}
\ot \mathsf{m}^{\prec}_{\ell}\,.
\label{1timespiplusXprec2FERM}
\end{align}

\paragraph{Rewriting of $\delta_{\ell}$ defined in \eqref{deltaellsum}.}
One can rewrite $\delta_{\ell}$ defined in \eqref{deltaellsum}.as
\begin{equation}\label{deltaellrewriting}
\delta_{\ell}\,=\,q^{\frac{1}{2}}\tau_q^2\,\nu^{-\ell}\,
\,\sum_{k=1}^{\ell-1}\,q^{k-\ell}\,
\mathsf{b}_\ell\,
q^{\bar{\epsilon}_{\ell}-\bar{\epsilon}_{k}}
f_{\delta-(\epsilon_k-\epsilon_M)}\ot
q^{\left(\frac{\ell-k}{M}\right)\bar{\epsilon}}\,
[\cdots[\check{f}_k,\check{f}_{k+1}]\dots,\check{f}_{\ell-1}]\,
\mathsf{b}_k\,,
\end{equation}
where $\mathsf{b}_{\ell}$ are defined in \eqref{Xsucc12def}.

\noindent
\emph{Derivation:}
It follows form the definitions
\eqref{Xprec12def} and  \eqref{Xpreccsuccdef}   that\footnote{
To derive this identity one may notice that
\begin{equation}
\mathsf{b}_{\ell}^{}\,\mathsf{b}_k^{-1}\,=\,
q^{\frac{\ell-k}{M}\bar{\epsilon}}\,
q^{-\sum_{s=k+1}^{\ell-1}\,\bar{\epsilon}_s}
\,q^{-\frac{1}{2}\,\left(\bar{\epsilon}_{\ell}+\bar{\epsilon}_{k}\right)}\,=\,
\prod_{s=k}^{\ell-1}\mathsf{a}_s\,.
\end{equation}
} 
\begin{equation}
[\cdots[\mathsf{X}^{\prec}_k(2),\mathsf{X}^{\prec}_{k+1}(2)]\dots,\mathsf{X}^{\prec}_{\ell-1}(2)]=
x^{\ell-k}\,
\mathsf{b}_{\ell}^{}\,\mathsf{b}_k^{-1}\,
q^{\bar{\epsilon}_{\ell}-\bar{\epsilon}_{k}}\ot
q^{\left(\frac{\ell-k}{M}\right)\bar{\epsilon}}\,
[\cdots[\check{f}_k,\check{f}_{k+1}]\dots,\check{f}_{\ell-1}]\,,
\end{equation}
where $x=q^{-1}\tau_q\,\nu^{-1}$.
If follows from the definitions \eqref{Xsucc12def}, 
 \eqref{Xpreccsuccdef} and the observation \eqref{fromftoXsucc} that
\begin{equation}\label{Xsucck1APP}
\mathsf{X}^{\succ}_{k}(1)
=q^{-\frac{1}{2}}\,\tau_q\,\nu^{-k}\,\mathsf{b}_{k}\,f_{\delta-(\epsilon_k-\epsilon_M)}\ot
\mathsf{b}_{k}
\end{equation}

\paragraph{Action of $1\ot \pi^+$ on $\delta_{\ell}$,  $\mathsf{X}^{\succ}_{\ell}(1)$ and  $\mathsf{X}^{\succ}_{\ell}(2)$.}
\begin{align}
\Lambda(\mathsf{y})\,\left[
\left(1\ot \pi^+_{\la}\right)\delta_{\ell}
\right]\Lambda^{-1}(\mathsf{y})\,&=\,\tau_q\,
\left(\sum_{k=1}^{\ell-1}\,\la^{-k}\,
q^{\bar{\epsilon}_\ell-\bar{\epsilon}_k}\,q^{-\frac{1}{2}}\,
\mathsf{b}_{\ell}^{-2}\,
\mathsf{b}_k^{}\,f_{\delta-(\epsilon_k-\epsilon_M)}\right)
\ot \mathsf{m}^{\succ}_{\ell}
\,,
\label{deltaonepiplusNEW}\\
\Lambda(\mathsf{y})\,\left[
\left(1\ot \pi^+_{\la}\right)\mathsf{X}^{\succ}_{\ell}(1)
\right]\Lambda^{-1}(\mathsf{y})\,&=\,\tau_{q}\,
\la^{-\ell}\,
q^{-\frac{1}{2}}\,
\mathsf{b}_{\ell}^{-1}\,
f_{\delta-(\epsilon_\ell-\epsilon_M)}
\ot \mathsf{m}^{\succ}_{\ell}\,,
\label{1timespiplusXsucc1}\\
\Lambda(\mathsf{y})\,\left[
\left(1\ot \pi^+_{\la}\right)\mathsf{X}^{\succ}_{\ell}(2)
\right]\Lambda^{-1}(\mathsf{y})\,&=\,
\mathsf{b}_{\ell}^{-2}\,
q^{\bar{\epsilon}_\ell-\bar{\epsilon}_M}
\ot \mathsf{m}^{\succ}_{\ell}\,,
\label{1timespiplusXsucc2}
\end{align}
where
\begin{equation}\label{msuccelldef}
\mathsf{m}^{\succ}_{\ell}:=q^{\frac{1}{2}}\,
q^{\frac{\ell-M}{M}}\,\la^{\ell}\,\nu^{-\ell}\,
\pi^+_{\la}(\mathsf{b}_{\ell})\,\mathsf{y}_{\ell}^{}\,\mathsf{y}_M^{-1}
\end{equation}

\noindent
\emph{Derivation:}
The relation \eqref{deltaonepiplusNEW} is obtained from \eqref{deltaellrewriting}
by applying the following
\begin{equation}
\pi^+_{\la}\left(
[\cdots[\check{f}_k,\check{f}_{k+1}]\dots,\check{f}_{\ell-1}]\right)=
q^{-\frac{1}{2}}\tau_q^{-1}\,
\left(q^{\frac{M+1}{M}}\la\right)^{\ell-k}\,
\pi^+_\la(\mathsf{b}_{\ell}^{}\mathsf{b}_{k}^{-1})\,
\mathsf{y}_{\ell}^{}\,\mathsf{y}_k^{-1}\,,
\label{piplusfchecks}
\end{equation}
\begin{equation}
\pi^+_{\la}(\mathsf{b}_k^{-1})\,
\mathsf{y}_{\ell}^{}\,\mathsf{y}_k^{-1}\,
\pi^+_{\la}(\mathsf{b}_k^{})\,=\,
q^{-\frac{1}{2}}\,
\mathsf{y}_{\ell}^{}\,\mathsf{y}_k^{-1}\,,
\end{equation}
\begin{equation}\label{LambdaonfdeltaminuskM}
\Lambda(\mathsf{y})\,\left(
f_{\delta-(\epsilon_k-\epsilon_M)}
\ot\, 1\right)
\Lambda^{-1}(\mathsf{y})\,=\,
q^{-\frac{1}{2}}q^{\frac{k}{M}}\,
f_{\delta-(\epsilon_k-\epsilon_M)}\,
\mathsf{b}_{k}^{-1}
\ot\, \mathsf{y}_{k}^{}\,\mathsf{y}_M^{-1}\,.
\end{equation}
\begin{equation}\label{Lambdaonyy}
\Lambda(\mathsf{y})\,\left(1
\ot\, \mathsf{y}_{\ell}^{}\,\mathsf{y}_k^{-1}\right)
\Lambda^{-1}(\mathsf{y})\,=\,
\mathsf{b}_{\ell}^{-2}\,\mathsf{b}_{k}^{+2}
\ot\, \mathsf{y}_{\ell}^{}\,\mathsf{y}_k^{-1}\,.
\end{equation}
\begin{equation}
\Lambda(\mathsf{y})\label{Lambdaonb}
\left(1\ot \pi^+_\la(\mathsf{b}^{}_{\ell})\right)
\Lambda^{-1}(\mathsf{y})\,=\,
\mathsf{b}^{-1}_{\ell}\ot \pi^+_\la(\mathsf{b}^{}_{\ell})\,.
\end{equation} 
The relation \eqref{1timespiplusXsucc1} follows from \eqref{Xsucck1APP}
with \eqref{LambdaonfdeltaminuskM} and \eqref{Lambdaonb}.
The relation \eqref{1timespiplusXsucc2} follows from 
\begin{equation}\label{piplusfsuccwithy}
\pi^+_{\la}(f_{\delta-(\epsilon_\ell-\epsilon_M)})\,=\,\frac{q^{\frac{\ell}{M}}\la^{\ell}}{q-q^{-1}}\,
\mathsf{y}_{\ell}^{}\,\mathsf{y}_M^{-1}\,.
\end{equation}
with  \eqref{Lambdaonyy} and \eqref{Lambdaonb}.

\paragraph{Applying $\pi^{\mathcal{F}}$ to the first tensor factor.}
We can apply $\pi^{\mathcal{F}}_{\mu}$ to the first tensor factor of
 \eqref{deltaonepiplusNEW},
\eqref{1timespiplusXsucc1}, \eqref{1timespiplusXsucc2}
 and use the 
expressions collected in Appendix \ref{app:CWbasisFERMfs} to obtain
\begin{align}
\Lambda(\mathsf{y})\,\left[
\left(\pi^{\mathcal{F}}_{\mu}\ot \pi^+_{\la}\right)\delta_{\ell}
\right]\Lambda^{-1}(\mathsf{y})\,&=\,
-\tau_q\,q^{-1}\,
t_{\ell}\,
\left(\sum_{k=1}^{\ell-1}\,g_{\mathfrak{n}}^{-k}\,
\bar{\mathbf{c}}_k\right)
\mathbf{c}_M
\ot \mathsf{m}^{\succ}_{\ell}
\,,
\label{deltaonepiplusNEWferm}\\
\Lambda(\mathsf{y})\,\left[
\left(\pi^{\mathcal{F}}_{\mu}\ot \pi^+_{\la}\right)\mathsf{X}^{\succ}_{\ell}(1)
\right]\Lambda^{-1}(\mathsf{y})\,&=\,
-\tau_{q}\,\,q^{-1}\,
t_{\ell}\,
g_{\mathfrak{n}}^{-\ell}\,\,
\bar{\mathbf{c}}_{\ell}\,
\mathbf{c}_M
\ot \mathsf{m}^{\succ}_{\ell}\,
\label{1timespiplusXsucc1ferm}\\
\Lambda(\mathsf{y})\,\left[
\left(\pi^{\mathcal{F}}_{\mu}\ot \pi^+_{\la}\right)\mathsf{X}^{\succ}_{\ell}(2)
\right]\Lambda^{-1}(\mathsf{y})\,&=\,
t_{\ell}
\ot \mathsf{m}^{\succ}_{\ell}\,,
\label{1timespiplusXsucc2ferm}
\end{align}
where
$t_{\ell}=q^{-\frac{2\ell}{M}\mathfrak{n}}q^{2\sum_{s=1}^{\ell}\mathfrak{n}_s}$ 
and
$g_{\mathfrak{n}}:=-q^{\frac{M-\mathfrak{n}}{M}}\mu^{-1}\la$.
To derive these relations recall that 
$\pi^{\mathcal{F}}_{\mu}(q^{-\frac{1}{2}}\,\mathsf{b}_k^{}\,f_{\delta-(\epsilon_k-\epsilon_M)})=
-(-q^{\frac{\mathfrak{n}-M}{M}}\mu)^k\,\bar{\mathbf{c}}_k\mathbf{c}_M$.

\paragraph{Action of $1\ot \pi^+$ on $\tau_{q}^{-1}[\delta_{\ell},\mathsf{X}^{\prec}_{\ell}(2)]$.}
The following holds
\begin{equation}
\Lambda(\mathsf{y})\,\left[
\left(1\ot \pi^+_{\la}\right)\tau_{q}^{-1}[\delta_{\ell},\mathsf{X}^{\prec}_{\ell}(2)]
\right]\Lambda^{-1}(\mathsf{y})
=\tau_q\,
\left(\sum_{k=1}^{\ell-1}\,\la^{-k}\,
q^{\bar{\epsilon}_{\ell+1}-\bar{\epsilon}_k-\frac{1}{2}}\,
\mathsf{b}_{\ell+1}^{-2}\,
\mathsf{b}_k^{}\,f_{\delta-(\epsilon_k-\epsilon_M)}\right)
\ot \mathsf{m}^{\succ}_{\ell+1}
\,.
\end{equation}

\noindent
\emph{Derivation:}
The starting point is \eqref{deltaip1BIS} with \eqref{1timespiplusXprec2}, \eqref{1timespiplusXsucc1}
 and \eqref{deltaonepiplusNEW}.
It follows from the definitions \eqref{mprecelldef} and \eqref{msuccelldef} that
$\mathsf{m}^{\succ}_{\ell+1}=q\,\mathsf{m}^{\prec}_{\ell}\,\mathsf{m}^{\succ}_{\ell}$.

\paragraph{Applying $\pi^{\mathcal{F}}_{\mu}$ to the first tensor factor.}
\begin{equation}\label{piFpipluscommutat}
\Lambda(\mathsf{y})\,\left[
\left(\pi^{\mathcal{F}}_{\mu}\ot \pi^+_{\la}\right)\tau_{q}^{-1}[\delta_{\ell},\mathsf{X}^{\prec}_{\ell}(2)]
\right]\Lambda^{-1}(\mathsf{y})\,=\,-q^{-1}\tau_q\,
t_{\ell+1}
\left(\sum_{k=1}^{\ell-1}\,\,
g_{\mathfrak{n}}^{-k}\,\bar{\mathbf{c}}_k\right)\mathbf{c}_M
\ot \mathsf{m}^{\succ}_{\ell+1}
\,.
\end{equation}

\subsubsection{Verifications of $\pi^{\mathcal{F}}\ot \pi^+$ on \eqref{idenaux1},
\eqref{idenaux2} and 
\eqref{idenauxLAST}}

\paragraph{Verification of $\pi^{\mathcal{F}}\ot \pi^+$ on \eqref{idenaux1}.}
In order to verify \eqref{idenaux1}  using the prescription \eqref{realrootDEF}, let us first observe that the image of the sum 
$\Delta(\mathsf{X}^{\prec}_{\ell})=\mathsf{X}^{\prec}_{\ell}(1)+\mathsf{X}^{\prec}_{\ell}(2)$ can be rewritten as
\begin{equation}
\Lambda(\mathsf{y})\,\left[
\left(\pi^{\mathcal{F}}_\mu\ot \pi^+_{\la}\right)\Delta(\mathsf{X}^{\prec}_{\ell})
\right]\Lambda^{-1}(\mathsf{y})\,=\,
\left(\mathcal{S}\ot 1\right)
\left(q^{2\mathfrak{n}_{\ell+1}}\ot 1\right)\omega_{\ell}
\left(\mathcal{S}\ot 1\right)^{-1}\,,
\end{equation}
where
$\omega_{\ell}:=q^{-2\frac{\mathfrak{n}}{M}}
\ot \mathsf{m}^{\prec}_{\ell}$.
To obtain this expression we used the relations \eqref{1timespiplusXprec1FERM} and 
\eqref{1timespiplusXprec2FERM}, and  the identity
\begin{equation}
\mathcal{S}\,q^{2\mathfrak{n}_{\ell+1}}\mathcal{S}^{-1}\,=\,
q^{2\mathfrak{n}_{\ell+1}}+x\,\tau_q\,\bar{\mathbf{c}}_{\ell+1}\mathbf{c}_{\ell}\,,
\qquad 
\mathcal{S}=1-q^{-1}\,x\,\bar{\mathbf{c}}_{\ell+1}\mathbf{c}_{\ell}\,,
\end{equation}
where $x=\mu\la^{-1}q^{\frac{\mathfrak{n}}{M}}$.
Form  these relations and recalling that  $\mathscr{E}_q(\tau_q\,\mathbf{x})=1+\mathbf{x}$
when $\mathbf{x}^2=0$, it follows that the identity \eqref{idenaux1} 
reduces to 
\begin{equation}
\left(\mathcal{S}\ot 1\right)
\mathcal{E}_{b^2}\left(
\left(q^{2\mathfrak{n}_{\ell+1}}\ot 1\right)\omega_{\ell}\right)
\left(\mathcal{S}\ot 1\right)^{-1}\!=\!
\left(1+
x\,\omega_{\ell}\,
\left(\bar{\mathbf{c}}_{\ell+1}
\mathbf{c}_{\ell}
\ot 1\right)
\right)
\mathcal{E}_{b^2}\left(
\left(q^{2\mathfrak{n}_{\ell+1}}\ot 1\right)\omega_{\ell}\right)\,.
\end{equation}
The only non trivial term in this identity is the one linear in $x$, which can be rewritten as 
\begin{equation}
\mathcal{E}_{b^2}\left(\left(q^{2\mathfrak{n}_{\ell+1}}\ot 1\right)\omega_{\ell}\right)
\left(
\bar{\mathbf{c}}_{\ell+1}
\mathbf{c}_{\ell}
\ot 1
\right)
=
\left(1+q^{+1}\,\omega_{\ell}
\right)
\left(
\bar{\mathbf{c}}_{\ell+1}
\mathbf{c}_{\ell}
\ot 1
\right)\mathcal{E}_{b^2}\left(\left(q^{2\mathfrak{n}_{\ell+1}}\ot 1\right)\omega_{\ell}\right)\,.
\end{equation}
Recalling that $q^{2\mathfrak{n}_{\ell+1}}\bar{\mathbf{c}}_{\ell+1}=q^2\bar{\mathbf{c}}_{\ell+1}$
and  $\bar{\mathbf{c}}_{\ell+1}q^{2 \mathfrak{n}_{\ell+1}}=\bar{\mathbf{c}}_{\ell+1}$ we obtain
\begin{equation}
\mathcal{E}_{b^2}\left(q^2\,\omega_{\ell}\right)
=
\left(1+q^{+1}\,\omega_{\ell}
\right)
\mathcal{E}_{b^2}\left(\omega_{\ell}\right)\,.
\end{equation}
This is the  basic property of $\mathcal{E}_{b^2}(x)$ defined in  \eqref{Eb2-def}.
\paragraph{Verification of $\pi^{\mathcal{F}}\ot \pi^+$ on \eqref{idenaux2}.}
The image of the three operators entering \eqref{idenaux2} is given in 
\eqref{deltaonepiplusNEWferm}, \eqref{1timespiplusXsucc1ferm} and \eqref{1timespiplusXsucc2ferm}.
Their sum is $\Delta(\mathsf{X}_i^{\succ})=\mathsf{X}_i^{\succ}(1)+\mathsf{X}_i^{\succ}(2)+\delta_i$, compare to \eqref{deltaicopr}.
Its image  can be rewritten as
\begin{equation}
\Lambda(\mathsf{y})\,\left[
\left(\pi^{\mathcal{F}}_{\mu}\ot \pi^+_{\la}\right)\Delta(\mathsf{X}_{\ell}^{\succ})
\right]\Lambda^{-1}(\mathsf{y})\,=\,
\left(\mathcal{S}^{}\ot 1\right)
\left(t_{\ell}
\ot \mathsf{m}^{\succ}_{\ell}\right)
\left(\mathcal{S}^{-1}\ot 1\right)\,.
\end{equation}
This equality follows form 
\begin{equation}
\mathcal{S}\,t_{\ell}\,\mathcal{S}^{-1}=
t_{\ell}\left(
1-
q^{-1}\tau_q\,
\bar{\mathbf{C}}_{\ell}\,
\mathbf{c}_M
\right)\,,
\qquad
\mathcal{S}=
1+\left(\sum_{k=1}^{M-1}\,g^{-k}_{\mathfrak{n}}\,\bar{\mathbf{c}}_k\right)
\mathbf{c}_M\,,
\end{equation}
where 
$\bar{\mathbf{C}}_{\ell}:=\sum_{k=1}^{\ell}
g_{\mathfrak{n}}^{-k}\,\,
\bar{\mathbf{c}}_{k}$.
 Following the prescription given in \eqref{realrootDEF} and the relations above,
 the identity \eqref{idenaux2} reduces to 
 \begin{equation}\label{forpoFpiplusAPPENDIX}
\left(\mathcal{S}^{}\ot 1\right)
\mathcal{E}_{b^2}\left(t_{\ell}
\ot \mathsf{m}^{\succ}_{\ell}\right)
\left(\mathcal{S}\ot 1\right)^{-1}\,=\,
\left(1
-q^{-1}\,
t_{\ell}\,
\bar{\mathbf{C}}_{\ell}\,
\mathbf{c}_M
\ot \mathsf{m}^{\succ}_{\ell}
\right)
\mathcal{E}_{b^2}\left(t_{\ell}
\ot \mathsf{m}^{\succ}_{\ell}\right)\,.
\end{equation}
Notice that to simplify the  right hand side we used the following:
for $\mathbf{x}^2=0$  we have $\mathscr{E}_q(\tau_q\,\mathbf{x})=1+\mathbf{x}$.
 The term proportional to 
 $g^{-k}_{\mathfrak{n}}$ in \eqref{forpoFpiplusAPPENDIX} is given by
 \begin{equation}\label{forpoFpiplusAPPENDIXBIS}
\mathcal{E}_{b^2}\left(t_{\ell}
\ot \mathsf{m}^{\succ}_{\ell}\right)
\left(
\bar{\mathbf{c}}_{k}\mathbf{c}_{M}
\ot
1\right)\,=\,
\left(1+q^{-1}
t_{\ell}
\ot \mathsf{m}^{\succ}_{\ell}
\right)
\left(
\bar{\mathbf{c}}_{k}\mathbf{c}_{M}
\ot
1\right)
\mathcal{E}_{b^2}\left(t_{\ell}
\ot \mathsf{m}^{\succ}_{\ell}\right)\,.
\end{equation}
 To derive this relation we also used that $\bar{\mathbf{c}}_s$ with $s=\ell+1,\dots,M-1$ commute with 
 $t_{\ell}$.
 The final observation is that $t_{\ell}\,\bar{\mathbf{c}}_k\mathbf{c}_M=q^{2}t_{\ell,k}\,
 \bar{\mathbf{c}}_k\mathbf{c}_M$ and  $\bar{\mathbf{c}}_k\mathbf{c}_M\,t_{\ell}=t_{\ell,k}\,
 \bar{\mathbf{c}}_k\mathbf{c}_M$ where $t_{\ell,k}$ commutes with $ \bar{\mathbf{c}}_k\mathbf{c}_M$,
  so that 
\eqref{forpoFpiplusAPPENDIXBIS}
 reduces to 
  \begin{equation}
\mathcal{E}_{b^2}\left(q^2\,t_{\ell,k}
\ot \mathsf{m}^{\succ}_{\ell}\right)
\left(
\bar{\mathbf{c}}_{k}\mathbf{c}_{M}
\ot
1\right)\,=\,
\left(1+q^{+1}
t_{\ell,k}
\ot \mathsf{m}^{\succ}_{\ell}
\right)
\left(
\bar{\mathbf{c}}_{k}\mathbf{c}_{M}
\ot
1\right)
\mathcal{E}_{b^2}\left(t_{\ell,k}
\ot \mathsf{m}^{\succ}_{\ell}\right)\,.
\end{equation}
This relation follows 
 the basic property of $\mathcal{E}_{b^2}(x)$, see \eqref{Eb2-def}.
 
 \paragraph{Verification of $\pi^{\mathcal{F}}\ot \pi^+$ on \eqref{idenauxLAST}.}
 Inserting \eqref{deltaonepiplusNEWferm},  \eqref{1timespiplusXprec2FERM} 
and \eqref{piFpipluscommutat} in \eqref{idenauxLAST} and using the prescription \eqref{realrootDEF},
we obtain, after simple manipulations 
\begin{equation}\label{laststepmixedpentagon}
\left(1\ot \mathsf{m}^{\succ}_{\ell}\right)
\mathcal{E}_{b^2}(z_\ell)-
\mathcal{E}_{b^2}(z_\ell)\left(1\ot \mathsf{m}^{\succ}_{\ell}\right)=
\mathcal{E}_{b^2}(z_\ell)\,q\,z_{\ell}\,
\left(1\ot \mathsf{m}^{\succ}_{\ell}\right)\,,
\end{equation}
where
$z_{\ell}:=q^{2\left(\mathfrak{n}_{\ell+1}-\frac{\mathfrak{n}}{M}\right)}
\ot \mathsf{m}^{\prec}_{\ell}$.
To derive this equation we also used $\mathsf{m}^{\succ}_{\ell+1}=q\,\mathsf{m}^{\prec}_{\ell}\,\mathsf{m}^{\succ}_{\ell}$.
Upon observing that
 $(1\ot \mathsf{m}^{\succ}_{\ell})z_{\ell}=q^{2}z_{\ell}(1\ot \mathsf{m}^{\succ}_{\ell})$
the relations \eqref{laststepmixedpentagon} reduces to the basic property of $\mathcal{E}_{b^2}(x)$, see \eqref{Eb2-def}.
  

\subsubsection{Auxiliary for check of $\Delta(\mathscr{M}_{\delta})$.}
\label{sec:forcheckofDeltaMdelta}
The following relation holds
\begin{equation}\label{MdeltaEQ1}
\Lambda(\mathsf{y})
\left[
f_{\delta-(\epsilon_i-\epsilon_M)}\otimes\,
\pi^+_{\lambda}\left(f^{\text{op}}_{\epsilon_i-\epsilon_M}\right)
\right]
\Lambda^{-1}(\mathsf{y})
=
\frac{\la^{M-i}}{q-q^{-1}}\,
\left(
q^{\frac{1}{2}}\,\mathsf{b}_i\,f_{\delta-(\epsilon_i-\epsilon_M)}
\otimes \,1\right)\,.
\end{equation}
\emph{Derivation:} The relations  \eqref{fOPPexplicit} and \eqref{piplusfchecks} imply that
$\pi_+^{\lambda}\left(f^{\text{op}}_{\epsilon_i-\epsilon_M}\right)=\tau_q^{-1}(q^{\frac{1}{M}}\la)^{M-i}\mathsf{y}_M^{}\mathsf{y}_i^{-1}$.
The relation \eqref{MdeltaEQ1} follows upon implementing the action of $\Lambda(\mathsf{y})$ 
as given in \eqref{LambdaonfdeltaminuskM} and \eqref{Lambdaonyy}. \qed

Applying $\pi^{\mathcal{F}}$ to the first tensor factor, \eqref{MdeltaEQ1} reduces to 
\begin{equation}\label{MdeltaEQ1FERM}
\Lambda(\mathsf{y})
\left[\pi^{\mathcal{F}}_{\mu_{\mathfrak{n}}}\left( 
f_{\delta-(\epsilon_i-\epsilon_M)}\right)
\otimes\,
\pi^+_{\lambda_{\mathfrak{n}}}\left(f^{\text{op}}_{\epsilon_i-\epsilon_M}\right)
\right]
\Lambda^{-1}(\mathsf{y})
=
-\frac{(\la_{\mathfrak{n}})^{M}}{q-q^{-1}}\,
\left(
\left(-q^{\frac{\mathfrak{n}-M}{M}}\mu_{\mathfrak{n}}\right)^{i}\,q\,
\bar{\mathbf{c}}_i\mathbf{c}_M
\otimes \,1\right)\,.
\end{equation}


\subsection{The R-matrix in the fundamental representation from the universal R-matrix}
\label{App:Rfund}

Using \eqref{relationfund1}, a simple calculation shows that
\begin{equation}
 \left(\pi^{\rm f}_x\otimes \pi^{\rm f}_y\right)\left(\stackrel{\longrightarrow}{\prod_{\gamma\,\in\, \widehat{\mathbb{A}}_i}}\,
\mathscr{R}^+_{\gamma}\right)\,=\,
1\,+\,\sigma\,\sum_{j=i+1}^M\,\left(\frac{x}{y}\right)^{(i-j)}\,
\SE_{ij}\otimes\SE_{ji}\,,
\qquad
\sigma\,=\,\frac{q^{-1}-q^{+1}}{1-\left(x/y\right)^M}\,.
\label{RAfund}
\end{equation}
Recall that $\mathscr{R}^+_{\gamma}$ is given in \eqref{rgm} wirh $s_{\gamma}=1$ and the ordered set $\widehat{\mathbb{A}}_i$ is defined in \eqref{setA}.
The simple result in \eqref{RAfund} follows from the fact that, for the fundamental representation, the root vectors associated to the set $\widehat{\mathbb{A}}_i$ are nilpotent and commute among themselves. Moreover, the simple dependence on $k$ in \eqref{relationfund1} is responsible for turning infinite products over $k$ into geometric series giving rise to the denominator of $\sigma$. 
Multiplying the factors \eqref{RAfund} according to the order \eqref{rootorder} one finds
\begin{equation}
\left(\pi^{\rm f}_{x}\otimes \pi^{\rm f}_{y}\right)\mathscr{R}^+_{\prec \delta}\,=\,
1\,+\,\sigma\,\sum_{i>j}\,\left(\frac{x}{y}\right)^{(i-j)}\,\SE_{ij}\otimes\SE_{ji}\,.
\label{Rminusev}
\end{equation}
Similarly
\begin{equation}
\left(\pi^{\rm f}_{x}\otimes \pi^{\rm f}_{y}\right)\mathscr{R}^+_{\succ \delta}\,=\,
1\,+\,\sigma\,\sum_{i<j}\,\left(\frac{x}{y}\right)^{(i-j)-M}\,\SE_{ij}\otimes\SE_{ji}\,.
\label{Rplussev}
\end{equation}
The evaluation of $\mathscr{R}^+_{\sim \delta}$ defined in \eqref{rpd} gives
\begin{equation}
 \left(\pi^{\rm f}_{x}\otimes \pi^{\rm f}_{y}\right)\mathscr{R}^+_{\sim \delta}\,=\,\rho(z)\,
\sum_{i,j=1}^M\,\frac{y^M-x^M\,q^{-2\,\delta_{i>j}}}{y^M-x^M\,q^{+2\,\delta_{i<j}}}\, \SE_{ii}\otimes\SE_{jj}\,,
\label{Rdeltafund}
\end{equation}
where
\begin{equation}
\rho(z)\,:=\,
\frac{(q^{-2} z;q^{-2M})_{\infty}\,(q^{2-2M} z;q^{-2M})_{\infty}}{(z;q^{-2M})_{\infty}\,(q^{-2M} z;q^{-2M})_{\infty}}\,
=\,\frac{\varepsilon_{q^M}(-q^{+M}z)\,\varepsilon_{q^M}(-q^{-M}z)}{\varepsilon_{q^M}(-q^{M-2}z)\,\varepsilon_{q^M}(-q^{2-M}z)}\,,
\end{equation}
where $z=(y/x)^M$, $(z;q)_{\infty}:=\prod_{k\geq0}\left(1-z\,q^k\right)$ and $\varepsilon_q(x)$ is defined in \eqref{thetadef}.
To obtain \eqref{Rdeltafund} one uses \eqref{umij}, \eqref{imrootfund} and their Cartan-conjugated analogues. 
Finally the evaluation of \eqref{qtdefSLM} gives
\begin{equation}
 \left(\pi_{x}\otimes \pi_{y}\right)\,q^{-t}\,=\,
q^{\frac{1}{M}}\,\sum_{i,j=1}^M\,q^{-\delta_{ij}}\,\SE_{ij}\otimes\SE_{ji}\,.
\end{equation}
Assembling the pieces together one obtains
\begin{equation}
\left(\pi^{\rm f}_{x}\otimes \pi^{\rm f}_{y}\right)\mathscr{R}^+\,=\,q^{\frac{1-M}{M}}\,\rho(z)\,\mathsf{R}(x,y)\,,
\label{Rfundfundev}
\end{equation}
where
\begin{equation}
 {\mathsf R}(x,y) \,=\,\sum_{i}\,{\mathsf E}_{ii}\otimes{\mathsf E}_{ii}\,+\,
\nu\,\sum_{i\neq j}\,{\mathsf E}_{ii}\otimes{\mathsf E}_{jj}\,+\,
\sum_{i\neq j}\,\kappa_{(i-j)\text{mod}_M}\,{\mathsf E}_{ij}\otimes{\mathsf E}_{ji}\,, 
\end{equation}
\begin{equation}
\nu\,=\,\frac{y^M-x^M}{q^{-1}\,y^M-\,q^{+1}\,x^M}\,,\qquad
\kappa_{\ell}\,=\,\frac{q^{-1}-q^{+1}}{{q^{-1}\,y^M-\,q^{+1}\,x^M}}\,y^{M-\ell}\,x^{\ell}\,,
\end{equation}
One can verify that \eqref{Rfundfundev}  satisfies the intertwining relations \eqref{eqnUnivRAxioms}.
Finally one observes that
\begin{equation}
 {\mathsf R}_{12}(x,y)\,{\mathsf R}_{21}(y,x)\,=\,\mathbf{I}\,.
\label{RR1}
\end{equation}
and (crossing symmetry)
\begin{equation}
\left(\left(\left({\mathsf R}^{-1}(x,y)\right)^{T_1}\right)^{-1}\right)^{T_1}\,=\,\eta(z)\,{\mathsf R}_{12}(q^{-2}x,y)\,,
\qquad 
\eta(z)\,=\,\frac{(1-z\,q^{-2})(1-z\,q^{2M-2})}{(1-z)(1-z\,q^{-2M})}\,.
\label{RRcross}
\end{equation}
where $T_1$ means transposition in the first tensor factor.
Notice that according to the properties of the projection of the  universal R-matrix on evaluation representations,   
see e.g.~chapter 9 of \cite{EFK} one has $\rho(z)=\prod_{k\geq 0} \eta(q^{-2Mk}\,z)$.
For $M=2,3$ the calculation presented in this appendix can be found in \cite{BZG} and \cite{Ex}.

\section{Supplementary material for Section 6}

\subsection{On the cyclicity of $\check{\sr}^{++}$}\label{ciclRpp}
Let
\begin{equation}\label{Xidef}
 \Xi\,:=\,
h(z\,{\mathsf w}_2)\,
h(z\,{\mathsf w}_3)\, 
\dots
h(z\,{\mathsf w}_{M})\,
h(z^{M-1}\,{\mathsf w}_{1\_\,(M-1)})
\,\dots\,
h(z^2\,{\mathsf w}_{1\_\,2})\,
h(z\,{\mathsf w}_1)\,,
\end{equation}
and recall ${\mathsf w}_i\,{\mathsf w}_{i+1}\,=\,q^{-2c}\,{\mathsf w}_{i+1}\,{\mathsf w}_{i}$.
In order to show that $\Xi$ is cyclic we apply the following  procedure
\begin{itemize}
 \item[1.] Apply pentagon ($2\rightarrow 3$) to the last two terms on the left of $\Xi$, i.e.
\begin{equation}
\label{applypent}
 h(z\,{\mathsf w}_2)h(z\,{\mathsf w}_3)\,=\,\,
h(z\,{\mathsf w}_3)\,
 h(z^2\,{\mathsf w}_{2\_\,3})\,
 h(z\,{\mathsf w}_2)\,,
\end{equation}
 \item[2.] Move $h(z\,{\mathsf w}_2)$
 all the way to the right before meeting the last two terms in the product formula for $\Xi$.
This is done without problems since ${\mathsf w}_{k}{\mathsf w}_2={\mathsf w}_2{\mathsf w}_k$ for $4\leq k \leq M$ 
and  ${\mathsf w}_{1\_\,l}\,{\mathsf w}_2={\mathsf w}_2\,{\mathsf w}_{1\_\,l}$
for $3\leq l \leq M-1$.
 \item[3.] Use pentagon again ($3\rightarrow 2$) on the three terms on the right, i.e.
\begin{equation}
h(z\,{\mathsf w}_2)\,h(z^2\,{\mathsf w}_{1\_\,2})\,
h(z\,{\mathsf w}_1)\,=\,h(z\,{\mathsf w}_1)\,h(z\,{\mathsf w}_2)\,,
\end{equation}
 \item[4.] Rewrite
\begin{equation}
\Xi\,=\,
h(z\,{\mathsf w}_3)\,
\tilde{\Xi}\,
h(z\,{\mathsf w}_2)\,,
\end{equation}
and apply the three steps above to $\tilde{\Xi}$ to obtain 
\begin{equation}
\Xi\,=\,
h(z\,{\mathsf w}_3)\,h(z\,{\mathsf w}_4)\,
\tilde{\tilde{\Xi}}\,
h(z^2\,{\mathsf w}_{2\_\,3})\,h(z\,{\mathsf w}_2)\,,
\end{equation}
and so on.
In the last steps one uses
\begin{equation}
h(z^{M-1}\,{\mathsf w}_{2\_\,M})\,h(z\,{\mathsf w}_1)\,=\,
h(z\,{\mathsf w}_1)\,h(z^{M-1}\,{\mathsf w}_{2\_\,M})\,.
\end{equation}
\end{itemize}

\subsection{$\mathsf{r}^{++}$ satisfies the YBE}
\label{app:TBforRpp}

In this appendix we prove that $\rho_z(\mathsf{w})$, related to  $\mathsf{r}^{++}$ via \eqref{frhoform},
 satisfies the relation  \eqref{braidrho}.
The proof we present uses only the identity \eqref{forzintAPP} and 
 is in some respect similar to the proof of the star-star 
relation for elliptic Boltzmann weights given in \cite{BKS13}.

The braid relation \eqref{braidrho} for $\rho_z(\mathsf{w})$ , upon inserting
\begin{equation}
\rho_z(\mathsf{w})\,=\,
\int d\mu(\mathbf{s})\,
\widetilde{\mathcal{K}}_z(\mathbf{s})\,\mathsf{w}(\mathbf{s})\,,
\qquad
 d\mu(\mathbf{s})=\delta(s_{\text{tot}})\,\prod_{i=1}^Mds_i\,,
\end{equation} 
can be rewritten as 
\begin{align}\label{smao2}
 \int\,d\mu(\mathbf{x})\,q^{-2\,(x,2(\Omega-1)t_2+(\Omega^{-1}-\Omega)t_1)}\,
\widetilde{\mathcal{K}}_{z_1}\left(t_1+x\right)\,
\widetilde{\mathcal{K}}_{z_1z_2}\left(2\,t_2\right)\,
\widetilde{\mathcal{K}}_{z_2}\left(t_1-x\right)\,=\,\\
\label{smaoRHS} 
\int\,d\mu(\mathbf{x})\,q^{-2\,(x,2(1-\Omega^{-1})t_1+(\Omega^{-1}-\Omega)t_2)}\,
\widetilde{\mathcal{K}}_{z_2}\left(t_2+x\right)\,
\widetilde{\mathcal{K}}_{z_1z_2}\left(2\,t_1\right)\,
\widetilde{\mathcal{K}}_{z_1}\left(t_2-x\right)\,.
\end{align}
Above we  used the notation $(a,b)=\sum_{i=1}^M\,a_i\,b_i$ and $(\Omega a)_i=a_{i+1}$,
see below for the derivation of \eqref{smao2} from \eqref{braidrho}.
Next, set
\begin{equation}\label{AscrDEF}
 \mathscr{A}_{z_1,z_2}(t_1,t_2):=
 \int d\mu(\mathbf{x})\,q^{-2\,(x,2(\Omega-1)t_2+(\Omega^{-1}-\Omega)t_1)}\,
\frac{\widetilde{\mathcal{K}}_{z_1}\left(t_1+x\right)\,
\widetilde{\mathcal{K}}_{z_2}\left(t_1-x\right)}{\widetilde{\mathcal{K}}_{z_1z_2}(2t_1)}\,.
\end{equation}
It follows from the cyclicity of $\widetilde{\mathcal{K}}_z(\sigma)$, manely
$ \widetilde{\mathcal{K}}_z(\sigma)=\widetilde{\mathcal{K}}_z(\Omega\,\sigma)$,
that
the  identity \eqref{smao2} is equivalent to 
\begin{equation}\label{2Upsilon}
\mathscr{A}_{z_1,z_2}(t_1,t_2)\,=\,\mathscr{A}_{z_2,z_1}(t_2,\Omega^{-1}t_1)\,.
\end{equation}
As explained below one can show that
\begin{equation}\label{AzzII}
\mathscr{A}_{z_1,z_2}(t_1,t_2)\,=\,\int_{\mathbb{R}}d\lambda\,\prod_{k=1}^M\,\frac{\mathbf{s}_b(\alpha_k-\la)}
{\mathbf{s}_b(\beta_k-\la)}\,e^{\pi i \la\,(v_2-v_1)}\,,
\end{equation}
where
\begin{equation}\label{alphabetadefapp}
\alpha=2\Omega(\tau_1-\tau_2)+\tfrac{v_1+v_2}{2}M\,\mathbf{v}_0\,,
\quad
\beta=2(\tau_1-\Omega\tau_2)-\tfrac{v_1+v_2}{2}M\,\mathbf{v}_0\,.
\end{equation}
and $v_i=\frac{1}{2\pi b}\log z_i$, $\tau_a=ib\,t_a$ and $\mathbf{v}_0=\frac{1}{M}(1,1,\dots,1)$.
It is clear from the definition of $\alpha$, $\beta$ that 
\eqref{2Upsilon} is equivalent to the fact that  \eqref{AzzII}  is invariant if  $\alpha_\mapsto -\beta$,
$\beta_\mapsto -\Omega^{-1}\alpha$ and $v_1$ and $v_2$ are exchanged.
This is manifest from recalling that $\mathbf{s}_b(x)\mathbf{s}_b(-x)=1$ and changing integration variable from $\la$ to
$-\la$.
The calculations omitted in the derivation above are given in the following.

\paragraph{From \eqref{braidrho} to  \eqref{smao2}}
We start from the braid relation  \eqref{braidrho} and insert $\rho_z(\mathsf{w})$ as above.
Next,
 reorder the non-commuting exponentials as follows
\begin{equation}
 \mathsf{w}_{1}(s_1)\,\mathsf{w}_{2}(s_2)\,\mathsf{w}_{1}(s_3)\,=\,q^{-\,\alpha(s_1,s_2,s_3)}\,
e^{(\log(\mathsf{w}_{1}),\,s_{1}+s_3)\,+\,(\log(\mathsf{w}_{2}),\,s_{2})}
\end{equation}
\begin{equation}
 \mathsf{w}_{2}(s'_1)\,\mathsf{w}_{1}(s'_2)\,\mathsf{w}_{2}(s'_3)\,=\,q^{-\,\beta(s_1,s_2,s_3)}\,
e^{(\log(\mathsf{w}_{2}),\,s'_{1}+s'_3)\,+\,(\log(\mathsf{w}_{1}),\,s'_{2})}
\end{equation}
where
\begin{equation}
 \alpha(s_1,s_2,s_3)
\,=\,2(s_-,(\Omega-1)s_2+(\Omega^{-1}-\Omega)\,s_+)\,,\qquad s_{\pm}=\frac{s_1\pm s_3}{2}\,.
\end{equation}
\begin{equation}
 \beta(s_1,s_2,s_3)\,=\,2(s'_-,(1-\Omega^{-1})s'_2+(\Omega^{-1}-\Omega)\,s'_+)\,,\qquad s'_{\pm}=\frac{s'_1\pm s'_3}{2}\,.
\end{equation}
These relations follow from  
$\mathsf{w}_{1,i}\, \mathsf{w}_{2,j}\,=\,
q^{2(\delta_{i,j}-\delta_{i+1,j})}
 \mathsf{w}_{2,j}\, \mathsf{w}_{1,i}$, 
which in turns follows from the definitions below \eqref{braodFrho}.
The next step is to take the "coefficient" of 
$e^{2(t_1,\log\mathsf{w}_1)+2(t_2,\log\mathsf{w}_2)}$ so we set
\begin{equation}
2 s_+=2t_1=s_2'\,,\qquad
 s_-=x=s_-'\,,\qquad
 s_2=2t_2=2s_+'\,.
\end{equation}
The rewriting \eqref{smao2} follows.\qed

\paragraph{Simplifying $\mathscr{A}_{z_1,z_2}(t_1,t_2)$.}
Set
$
 y\,=\,ib\,(\Omega-1)x$ and
$ \tau_a\,=\,ib\,t_{a}$. The exponential in the definition of  $\mathscr{A}_{z_1,z_2}(t_1,t_2)$ can be rewritten as 
\begin{equation}\
q^{-2\left(x,2(\Omega-1)t_2+(\Omega^{-1}-\Omega)t_1\right)}\,=\,e^{2\pi i (y,\tilde{\tau})}\,,
\end{equation}
where $\tilde{\tau}=2\Omega\tau_2-(1+\Omega)\tau_1$.
Inserting the delta function in the from
%
%
$\delta(y_{\text{tot}})\,\sim\,\int d\lambda \,e^{2\pi i \lambda\,y_{\text{tot}}}$,
one then finds
\begin{equation}
\mathscr{A}_{z_1,z_2}(t_1,t_2)\,=\,\int_{\mathbb{R}}d\lambda\,\prod_{k=1}^M\,I_k(\lambda)\,,
\end{equation}
where
\begin{equation}
 I_k(\lambda)\,:=\,\int_{\mathbb{R}}dy\,
\frac{\mathbf{s}_b(\hat{\tau}_{k,k+1}-y-v_1+c_b)
\,\mathbf{s}_b(\hat{\tau}_{k,k+1}+y-v_2+c_b)}{\mathbf{s}_b(2\hat{\tau}_{k,k+1}-v_1-v_2+c_b)}\,e^{2\pi i y (\tilde{\tau}_k+\lambda)}\,,
\end{equation}
where $\hat{\tau}=\tau_1$ and $v_a=\frac{1}{2\pi b}\log(z_a)$.
This integration can be done explicitly as
\begin{equation}
\int_{\mathbb{R}}dy\,
\mathbf{s}_b(u-y)\,\mathbf{s}_b(v+y)\,e^{2\pi i y w}=
\frac{\mathbf{s}_b(u+v-c_b)\mathbf{s}_b(-w-\tfrac{u+v}{2}+c_b)}{\mathbf{s}_b(-w+\tfrac{u+v}{2}-c_b)}\,
e^{i\pi w (u-v)}\,,
\end{equation} 
which follows from 
\eqref{forzintAPP}.
We thus conclude that 
\begin{equation}
 I_k(\lambda)\,=\,
\frac{\mathbf{s}_b(\alpha_k-\la)}
{\mathbf{s}_b(\beta_k-\la)}\,e^{\pi i \la (v_2-v_1)}\,e^{\text{linear in $\tau$}}\,,
\end{equation}
where $\alpha$ and $\beta$ are given in \eqref{alphabetadefapp}
and the terms linear in $\tau$ cancel out in the product over $k$.

\section{Comparison with the literature}\label{ShGapp}

In the case $M=2$ closely related models have been studied in the literature by other techniques,
see in particular \cite{ByT,ByT2} and \cite{BMS}.
The purpose of this appendix is to clarify the relation between the representation theoretic
constructions described in this paper and 
the objects constructed in \cite{ByT,ByT2} and \cite{BMS}.

\subsection{Projection to  the lattice-Sinh Gordon model I -- Lax operators}

As a preparation for some of the following discussions let us clarify the
relation between the approach to the lattice Sinh-Gordon model described in \cite{ByT,ByT2}
and the formalism used in this paper in some detail.

Abstractly, one may define the lattice Sinh-Gordon model on the 
kinematical level by defining its $\ast$-algebra
of observables $\CA_{\rm\sst SG}$ in terms of generators $f_k$, $k=1,\dots,2N$ and relations
\begin{equation}
f_{2n}\,f_{2n\pm 1}\,=\, q^2 f_{2n\pm 1}\,f_{2n}\,, \qquad
f_k\,f_{k+l}\,=\,f_{k+l}\,f_k \;\; {\rm for}\;\;|l|>1\,.
\end{equation} 
The time evolution is represented by the automorphism $\tau$ of $\CA_{\rm\sst SG}$,
\begin{equation}
\tau(f_k)\,=\,f_k^{-1}\frac{\ka^2+q f_{k-1}}{1+q\ka^2 f_{k-1}}
\frac{\ka^2+qf_{k+1}}{1+q\ka^2 f_{k+1}}\,.
\end{equation} 
The generators $f_k$ represent initial values for the time-evolution $\tau$ that are 
naturally associated with the vertices of the saw-blade contour $\mathcal{C}$ depicted in 
Figure \ref{lightconelattice}. Equally natural appears to be the contour $\bar{\mathcal{C}}$ 
related to $\mathcal{C}$ 
by means of a spacial translation with length $\frac{1}{2}\Delta$. 
The half-shift $\si^{\frac{1}{2}}$ defined by
$\si^{\frac{1}{2}}(f_k)=f_{k+1}$ alone is {\it not} an automorphism of $\CA_{\rm\sst SG}$.
Let us instead introduce the related automorphism $\tilde{\si}^{\frac{1}{2}}$ by
\begin{equation}
\tilde{\si}^{\frac{1}{2}}(f_{2n-1})\,=\,f_{2n}^{-1}\,,\qquad
\tilde{\si}^{\frac{1}{2}}(f_{2n})\,=\,f_{2n+1}^{}\,.
\end{equation}

The lattice Sinh-Gordon model was defined in \cite{ByT} by means of the Lax matrix
\begin{equation}\label{SGLax}
L^{\rm\sst SG}_n(u)\,=\,\fr{1}{i}e^{-\pi b s}\left(\begin{matrix} i\,\sin\pi b^2\,\SE_{s,n} & e^{\pi b u}\SK_{s,n}^{-1}-e^{-\pi b u}\SK_{s,n}^{}\\
e^{\pi b u}\SK_{s,n}^{}-e^{-\pi b u}\SK_{s,n}^{-1} & i\,\sin\pi b^2\SF_{s,n}
\end{matrix}\right)\,.
\end{equation}
This description is associated to the following representation of the algebra of observables,
\begin{equation}
\pi^{\rm\sst SG}_{}(f_{2n-1})\,=\,e^{-2\pi b\spp_n}\,,\qquad
\pi^{\rm\sst SG}_{}(f_{2n})\,=\,e^{2\pi b(\sx_{n}+\sx_{n+1})}\,,
\end{equation}
where $\sx_n$ and $\spp_n$ generate the usual Schr\"odinger representation of the 
Heisenberg-algebra $[\spp_n,\sx_m]=(2\pi i)^{-1}\de_{n,m}$
on wave-functions $\psi(\mathbf x)=\langle \mathbf x|\psi\rangle$, 
${\mathbf x}=(x_1,\dots,x_N)$.

Another natural representation $\bar\pi^{\rm\sst SG}_{}$
is obtained by composing $\pi^{\rm\sst SG}_{}$ with the automorphism $\tilde{\si}^{\frac{1}{2}}$.
It is naturally associated to the contour $\bar{\mathcal{C}}$.
The operator $\SY_{\infty}$ with kernel 
\begin{equation}\label{SYinfty}
\langle \,{\mathbf x}'\,|\,\SY_{\infty}\,|\,{\mathbf x}\,\rangle\,=\,
\prod_{n=1}^N\,e^{2\pi ix_n'(x_n+x_{n+1})}
\,=\,
\prod_{n=1}^N\,e^{2\pi i(x_{n-1}'+x_n')x_n}\,, 
\end{equation}
is easily seen to satisfy 
\begin{equation}
\spp_n\cdot\SY_{\infty}\,=\,\SY_{\infty}\cdot(\sx_n+\sx_{n+1})\,,\qquad
(\sx_n+\sx_{n+1})\cdot\SY_{\infty}\,=\,-\SY_{\infty}\cdot\spp_{n+1}\,,
\end{equation}
which implies that $\SY_{\infty}$ implements the automorphism $\tilde{\si}^{\frac{1}{2}}$ in the 
representation $\pi^{\rm\sst SG}_{}$.

We are now going to explain how to associate natural representations of the algebra of observables to these 
two contours. To this aim let us note that the monodromy matrix $\SM(\la)$ associated to $\mathcal{C}$ will be 
represented as
\begin{equation}
\SM(\la):=\,L_{2n}^-(\la/\ka)L_{2n-1}^+(\la\ka)\cdots L_{2}^-(\la/\ka)L_{1}^+(\la\ka)\,.
\end{equation}
Considering the contour $\bar{\mathcal{C}}$ leads to the definition of the monodromy matrix
\begin{equation}
\bar\SM(\la):=\,L_{2n}^+(\la\ka)L_{2n-1}^-(\la/\ka)\cdots L_{2}^+(\la\ka)L_{1}^-(\la/\ka)\,.
\end{equation}

In the first case it is natural to regard $\CL_k(\la)=L_{2k}^-(\la/\ka)L_{2k-1}^+(\la\ka)$ as the Lax-matrix 
associated to parallel transport along one physical lattice site, and to compare it with $L_n^{\rm\sst SG}(\la)$.
To simplify notation we will temporarily restrict attention to a specific value of $k$, and drop the subscript $k$ in the
notations. The Lax-matrix $\CL(\la)$ can be represented as 
\begin{align}\notag
\CL_n(\la)\,=\,&\SL^-_{2n}(\la/\ka)\,\SL^+_{2n-1}(\la\ka)\,=\,\left(\begin{matrix}
\su_{2n} & \frac{\ka}{\la}\,\sv^{-1}_{2n}\\
\frac{\ka}{\la}\,\sv_{2n} &\bar\su^{-1}_{2n}
\end{matrix}\right)\left(\begin{matrix}
\su_{2n-1} & {\la}{\ka}\,\sv_{2n-1}\\
{\la}{\ka}\,\sv^{-1}_{2n-1} &\su^{-1}_{2n-1}
\end{matrix}\right)\\
\,=\,&\ka \,
\left(\begin{matrix}   
i\tau_q\,\SE_n &\eta_n^{}\big(\la\SA_n^{}+\la^{-1}\SA^{-1}_n\big)   
 \\
\eta^{-1}_n\big(\la\SA^{-1}_n+\la^{-1}\SA_n^{}\big) &
i\tau_q\,\SF_n \end{matrix}\right)\,,
\label{Laxmult}\end{align}
using the notations  $\eta_n=(\sv^{-1}_{2n}\su_{2n}^{}\su^{-1}_{2n-1}\sv_{2n-1}^{})^{\frac{1}{2}}$, 
$i\tau_q=i(q-q^{-1})=\sin\pi b^2$, and
\begin{equation}
\begin{aligned}
& i\tau_q\,\SF_n=\SB^{-\frac{1}{2}}_n\big(\ka\SA_n^{}
+\ka^{-1}
\SA^{-1}_n\big)\SB^{-\frac{1}{2}}_n\\
& i\tau_q\,\SE_n=\SB^{+\frac{1}{2}}_n\big(\ka^{-1}\SA_n^{}
+\ka\SA^{-1}_n\big)
\SB^{+\frac{1}{2}}_n
\end{aligned}
\qquad
\begin{aligned}
&\SA_n=(\sv_{2n}\su_{2n}\su_{2n-1}\sv_{2n-1})^{\frac{1}{2}}\,,\\[1ex]
&\SB_n=(\sv^{-1}_{2n}\su_{2n}^{}\su_{2n-1}^{}\sv_{2n-1}^{-1})^{\frac{1}{2}}\,.
\end{aligned}
\end{equation}
There is a natural representation of the algebra $\CA_{\rm\sst SG}$ associated to this
set-up, defined by setting
\begin{equation}\label{replc}
\sf_{2n-1}\equiv\pi^{\rm lc}_{}(f_{2n-1}):=\,\SA_n^2\,,\qquad
\sf_{2n}\equiv\pi^{\rm lc}_{}(f_{2n}):=\,\SB_n^{-1}\,\SB_{n+1}^{-1}\,.
\end{equation}
This representation is reducible. One could project onto the eigenspaces of the the central 
elements $\eta_n$. A convenient explicit description of the projection may be given in the representation
where
the operators $(\su_k{})^{\frac{1}{2}}\sv^{-1}_k(\su_k{})^{\frac{1}{2}}$ 
are diagonal with eigenvalues  $e^{\pi b x_k}$.
Let $|\,y_r,y_s\,\rangle$
be a delta-function normalized vector satisfying
\[
\begin{aligned}
&(\su_r{})^{\frac{1}{2}}\sv^{-1}_r(\su_r{})^{\frac{1}{2}}\,|\,y_r,y_s\,\rangle\,=\,
e^{\pi b y_r}\,|\,y_r,y_s\,\rangle\,,\\
&(\su_s{})^{\frac{1}{2}}\sv^{-1}_s(\su_s{})^{\frac{1}{2}}\,|\,y_r,y_s\,\rangle\,=\,e^{\pi bx_s}\,|\,y_r,y_s\,\rangle\,,
\end{aligned}
\qquad
\langle\,y_r',y_s'\,|\,y_r,y_s\,\rangle\,=\,\de(y_r'-y_r^{})\de(y_s'-y_s^{})\,.
\]
Let us furthermore use the shorthand notation
\[
|\,{\mathbf y}\,\rangle:=\bigotimes_{n=1}^{N}|\,y_{2n},y_{2n-1}\,\rangle\,,
\quad{\mathbf y}=(y_1,\dots,y_{2N})\,.
\]
$\eta_n$ is diagonal
in this representation with eigenvalue $e^{\pi b(y_{2n}-y_{2n-1})}$.
The projection $\Pi$ is then defined by simply setting $y_{2n}=y_{2n-1}=x_n$ for $n=1,\dots,N$, which is equivalent to
setting the eigenvalue of $\eta_n$ to one.
It is clear that $\Pi$ maps $\pi^{\rm lc}$ to $\pi^{\rm\sst SG}$.
The projection of $\CL(\la)$ 
will coincide with $\ka^2 L^{\rm\sst SG}(u)$ 
if the parameters are related respectively as
\begin{equation}\label{T-id}
\kappa\,=\,m\Delta\,=\,e^{\pi b s}\,, \qquad
\la\,=\,-ie^{\pi b u}\,.
\end{equation}

It is equally natural to regard $\bar\CL_k(\la)=L_{2k+1}^+(\la/\ka)L_{2k}^-(\la\ka)$ as the Lax-matrix 
associated to parallel transport along one physical lattice site. This Lax-matrix  can be represented 
by a formula similar to \rf{Laxmult}, but with $\SA_n$, $\SB_n$ and $\eta_n$
replaced by $\bar\SA_n$, $\bar\SB_n$ and $\bar\eta_n$, defined respectively as
\begin{equation}\begin{aligned}
&\bar\SA_n=\SC^{-\frac{1}{2}}\cdot\SB_n^{-1}\cdot\SC^{\frac{1}{2}}\,,\\
&\bar\SB_n=\SC^{-\frac{1}{2}}\cdot\SA_n^{}\cdot\SC^{\frac{1}{2}}\,,
\end{aligned}\qquad \bar\eta_n=(\sv^{}_{2n+1}\su_{2n+1}^{}\sv^{-1}_{2n}\su_{2n}^{-1})^{\frac{1}{2}}\,,
\end{equation}
where $\SC^{\frac{1}{2}}$ is the operator representing the translation by one-half of a physical 
lattice site, satisfying $\SC^{-\frac{1}{2}}\cdot\SO_n\cdot\SC^{\frac{1}{2}}=\SO_{n+1}$ for 
each local observable $\SO_n$.
There is another natural representation $\bar{\pi}^{\rm lc}$ 
of the algebra $\CA_{\rm\sst SG}$ associated to this
set-up, defined by replacing   in \rf{replc} the operators  $\SA_n$ and $\SB_n$
by $\bar\SA_n$, $\bar\SB_n$, respectively. The representation $\bar{\pi}^{\rm lc}$ 
is naturally defined in such a way that  the operators $(\su_k{})^{\frac{1}{2}}\sv_k(\su_k{})^{\frac{1}{2}}$ 
are diagonal with eigenvalues  $e^{\pi b {\bar y}_k}$, for $k=1,\dots,2N$, respectively. The natural analog of the
projection $\Pi$ will be denoted $\overline{\Pi}$.

\subsection{Projection to the lattice Sinh-Gordon model II -- Q-operators}

Let us recall that the Q-operators have been defined  as
\begin{align}\label{Q-recall}
\CQ(\la;\bar\mu,\mu):&={\rm Tr}_{\CH_0}^{}\big(\,
\sr^{+-}_{0,{2N}}(\la/\bar{\mu})\,\sr^{++}_{0,2N-1}(\la/{\mu})
\cdots 
\sr^{+-}_{0,{2}}(\la/\bar{\mu})\,\sr^{++}_{0,{1}}(\la/{\mu})\,\big)
\end{align}
Our goal in this subsection is to demonstrate that the projection of $\CQ(\la;\bar\mu,\mu)$ 
to the physical subspace, denoted as  $\SQ(\la;\bar\mu,\mu)$ can be represented in the form
\begin{equation}\label{Q-factor}
\SQ(e^{\pi b w};e^{\pi b\bar{m}},e^{\pi b m})\,=\,
e^{\frac{\pi i}{4}((l-m)^2+(l-\bar{m})^2)}\,
\SY(l;\bar m,m)\cdot\SY_{\infty}\,,
\end{equation}
where the operator
$\SY_\infty$ has been defined above via \rf{SYinfty},
and $\SY(l;\bar m,m)$ is an integral operator with the kernel
\begin{equation}\label{SYmatel}
\langle\,{\mathbf x}'\,|\,\SY(l;\bar m,m)\,|\,{\mathbf x}\,\rangle=
\prod_{n=1}^{N}\,
V_{\bar{m}-l}(x_n'+x_{n+1}^{})
\bar{V}\!{}_{m-l}^{}(x'_{n}-x_{n}^{})\,.
\end{equation}
The special function $V_{u}(x)$ appearing in \rf{SYmatel} is defined 
as 
\begin{equation}
V_{u}(x):=\frac{\mathbf{s}_b(x-\frac{u}{2})}{\mathbf{s}_b(x+\frac{u}{2})}\,.
\end{equation}
We may note that the projection of the Q-operator
onto the physical subspace
is equal to the operator $\SQ_-$ constructed in \cite{ByT}.

In order to derive \rf{Q-factor}, let us start from \rf{Q-recall}, and insert the expressions 
\rf{r+-factor1} for  $\sr^{+-}_{rs}(\la)$ and \rf{R-factor2} for $\sr^{++}_{rs}(\la)$.
It is useful to represent $\sr^{++}_{rs}(\la)$ as
\begin{equation}\label{R-factor3}
\sr^{++}_{rs}(\la)\,=\,\CF_r^{-1}\cdot
\SP_{rs}^{}\,\rho_{\la}^{}(\sg_{{r}s}^{+})\cdot\CF_s\,,
\end{equation}
using the notation $\sg_{rs}^{+}:=\su_r^{}\sv_r^{}\,\su_s^{}\sv_s^{}$. 
By moving all operators $\CF_n$ to the right one may represent
$\CQ(\la;\bar\mu,\mu)$ in the form
\begin{equation}
\CQ(\mu,\bar\mu;\nu)\,=\,\CY(\mu,\bar\mu;\nu)\cdot\SC^{-\frac{1}{2}}\cdot\prod_{n=1}^{2N}
\CF_n\,,
\end{equation}
where 
$\CY_{\infty}^{-1}=\SC^{-\frac{1}{2}}\cdot\prod_{k=1}^{2N}\CF_k$,
\begin{equation}\label{Ydef}
\begin{aligned}
\CY(\la;\bar\mu,\mu):={\rm Tr}_{\CH_0}^{}\Big[& \,
\SP_{0,{2N}}^{} \,\rho_{\la/\bar\mu}^{}\big(\sf_{0,{2N}}^+,\sg_{0,{2N}}^-\big)\,
\SP_{0,2N-1}^{}\,\rho_{\la/\mu}^{}(\sg_{0,{2N-1}}^+)
\cdots \\[-1ex]
&\qquad\qquad\cdots
\SP_{0,{2}}^{} \,\rho_{\la/\bar\mu}^{}\big(\sf_{0,{2}}^+,\sg_{0,{2}}^-\big)\,
\SP_{0,1}^{}\,\rho_{\la/\mu}^{}(\sg_{0,{1}}^+)
\,\Big]\cdot \SC^{\frac{1}{2}}\,.
\end{aligned}
\end{equation}

The strategy will be to evaluate the matrix elements of the
operator $\langle{\mathbf y}'|\CY(\mu,\bar\mu;\nu)|{\mathbf y}\rangle$ 
in the representation introduced in the
previous subsection. We claim that
\begin{align}\label{Ymatel}
& \langle\,{\mathbf y}'\,|\,\CY(e^{\pi b l};e^{\pi b\bar{m}},e^{\pi b m})\,|\,{\mathbf y}\,\rangle=
\\ &\qquad=\,\zeta_b^{2N}\,e^{\frac{\pi i}{4}((l-m)^2+(l-\bar{m})^2)}
\prod_{n=1}^{N}\,
V_{\bar m-l}^{}(y_{2n+1}^{}+y'_{2n})
\bar{V}\!{}_{m-l}^{}(x'_{n}-x_{n}^{})\,e^{\pi i(y_{2n+1}^{}-y'_{2n})^2}\,,
\notag\end{align}
where
$x_{n}=\frac{1}{2}(y_{2n}+y_{2n-1})$.
The function $\bar{V}_w(x)$ is the Fourier-transformation of $V_w(x)$, 
which may be expressed as
\begin{equation}\label{D-FT}
\bar{V}_u(x):=\int dy\; e^{2\pi i xy}\,V_{u}(x)\,=\,\frac{V_{-u-2c_b}(x)}{\mathbf{s}_b(u+c_b)}\,,\qquad
c_b:=\frac{i}{2}(b+b^{-1})\,.
\end{equation}

\begin{figure}[t]
\centerline{\epsfxsize7cm\epsfbox{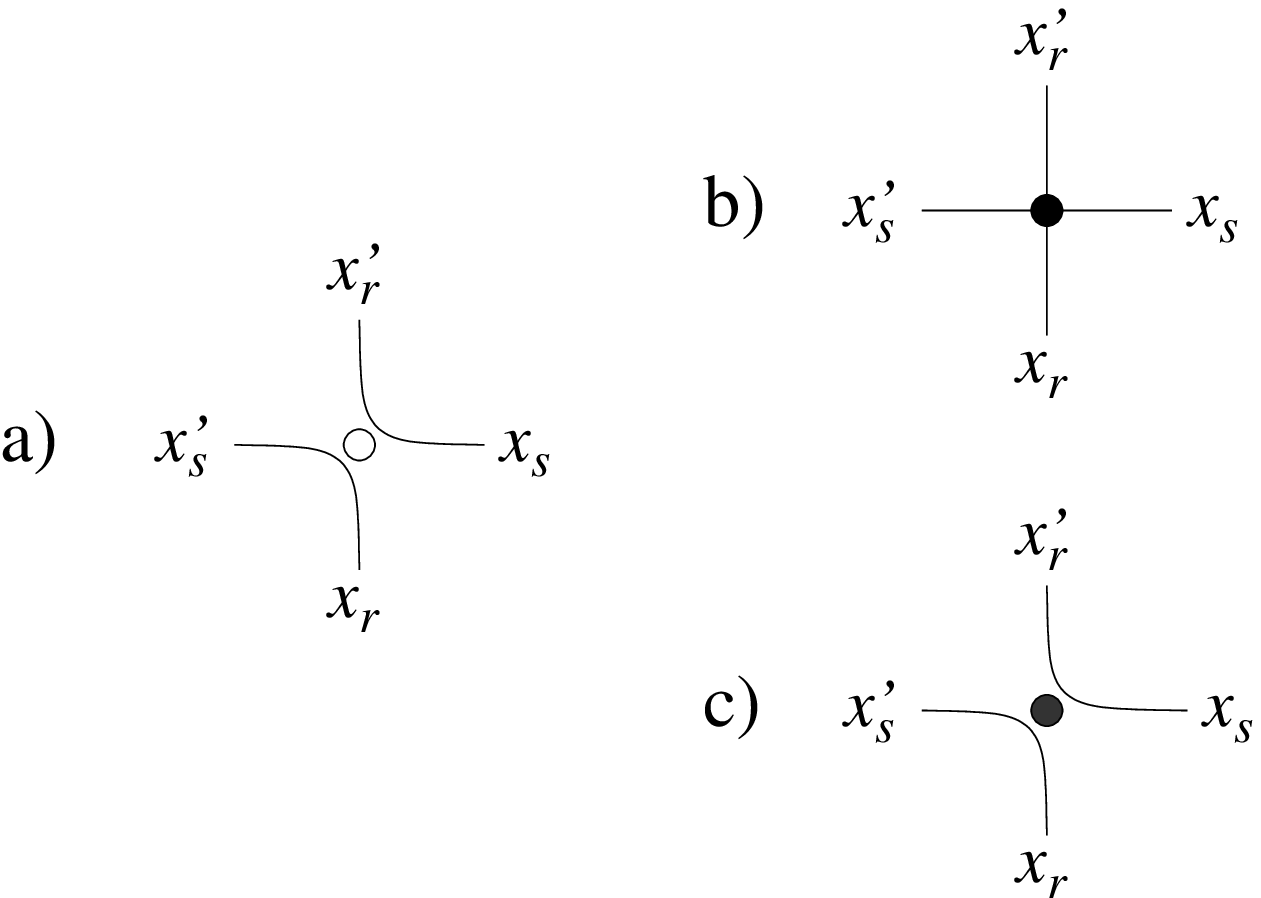}\hspace{.5cm}
\epsfxsize7cm\epsfbox{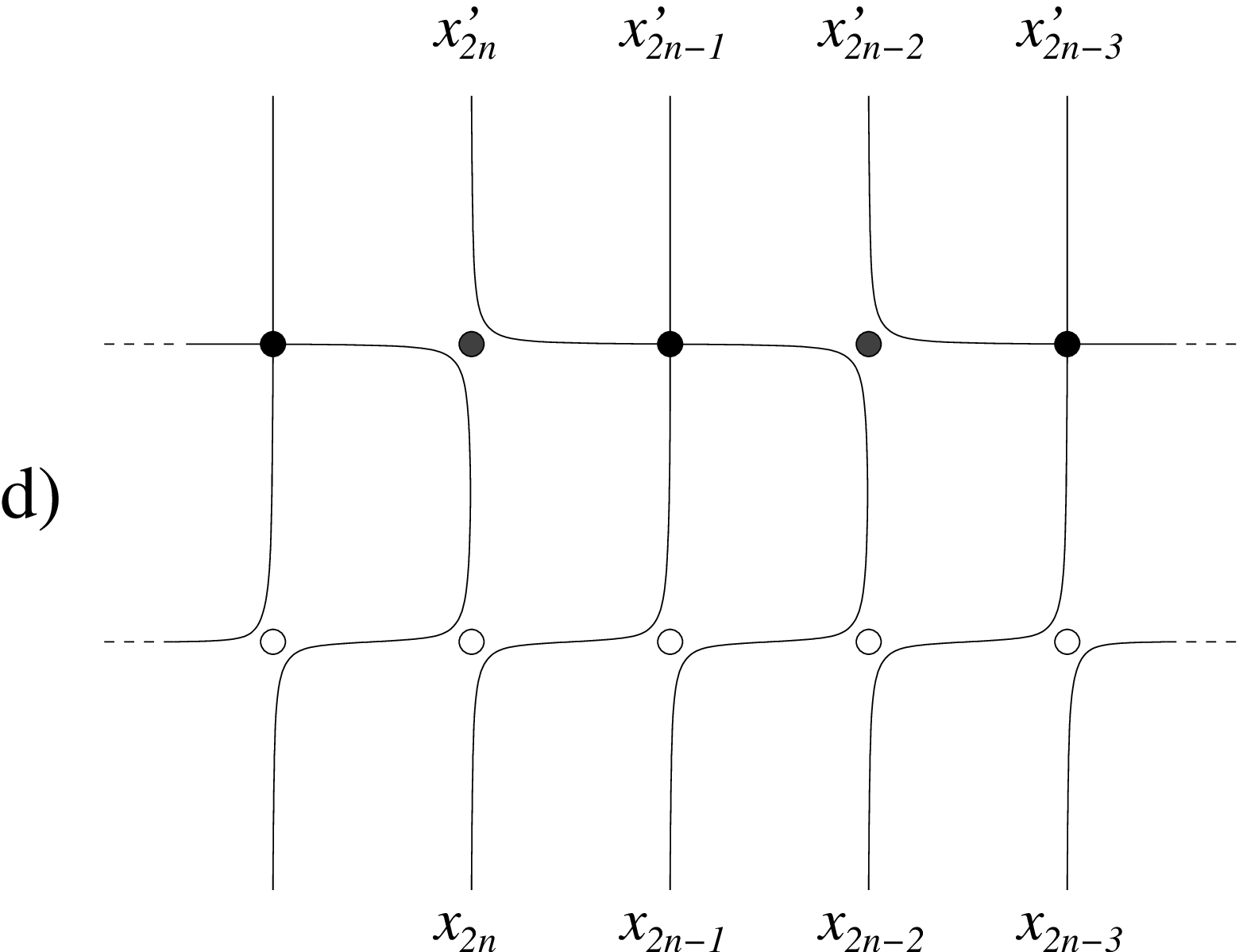}}
\caption{\it Diagrammatic representations for the kernels
defined in equations 
\rf{buildblocks.a}, 
\rf{buildblocks.b} and
\rf{buildblocks.c}, respectively. The labels correspond to the
variables appearing in the formulae \rf{buildblocks}.}
\label{Q-opfig}
\end{figure}

In order to prove \rf{Ymatel}, let us insert the identity operator
in the form $\int dy_{r}\,|\,y_r\,\rangle\langle\,y_r\,|$
in front of each operator $\SP_{r,0}$ in \rf{Ydef},
and let us furthermore insert ${\rm id}=\int \prod_{k=1}^{2n} dy_{k}''\,|\,y_k''\,\rangle\langle\,y_k''\,|$
in front of $\SC^{-\frac{1}{2}}$. This produces
an integral representation for the matrix element on the left hand side of
\rf{Ymatel}.
The building blocks of the integrand are
\begin{subequations}\label{buildblocks}
\begin{align}\label{buildblocks.a}
&\langle\,y_{r}',y_{s}'\,|\,\SP_{rs}\,
|\,y_{r}^{},y_{s}^{}\,\rangle=
\de(y_{r}'-y_{s}^{})\de(y_{s}'-y_{r}^{})\,,\\
&\langle\,y_{r}',y_{s}'\,|\,\SP_{rs}^{}\,\rho_{e^{\pi b w}}^{}(\sg_{{r}s}^{+})
\,
|\,y_{r}^{},y_{s}^{}\,\rangle=\de(z_{rs}'+z_{rs}^{})
\bar{V}_{-w}(y_{rs}'-y_{rs}^{})\,,
\label{buildblocks.b}\\
&\langle\,y_{r}',y_{s}'\,|\, \SP_{rs}^{}\,
{\rho_{e^{\pi b w}}^{}(\sf_{rs}^{+},\sg_{rs}^-)}\,
|\,y_{r}^{},y_{s}^{}\,\rangle=
\de(y_{r}'-y_{s}^{})\de(y_{s}'-y_{r}^{})\,
\,V_{-w}(x_{rs}^{})\,e^{\pi i z_{rs}^2}\,,
\label{buildblocks.c}\end{align}
\end{subequations}
where 
\[
\begin{aligned}
& {x}_{rs}'=\fr{1}{2}(y_{r}'+y_{s}'),\\
& {x}_{rs}^{}=\fr{1}{2}(y_{r}^{}+y_{s}^{}),
\end{aligned}\qquad
\begin{aligned}
& z_{rs}'=y_{r}'-y_{s}',\\
& z_{rs}=y_{r}-y_{s}.
\end{aligned}
\]
Equation \rf{buildblocks.b} follows easily from the identity
\begin{equation}\label{matelF(p)}
\langle\,x'\,|\,F(\spp)\,|\,x\,\rangle\,=\,\bar{F}(x'-x)\,,\qquad
\bar{F}(x):=\int_{\BR}dy\;F(y)\,e^{2\pi ixy}\,,
\end{equation}
where $\sx$, $\spp$ satisfy $[\spp,\sx]=1/2\pi i$, while $|\,x\,\rangle$ and $\langle \,x'\,|$
are eigenvectors of $\sx$ with eigenvalues $x$ and $x'$, respectively.
The delta-distributions allow us to carry out all the appearing
integrations. In order to keep track of the resulting identifications
of variables
it may be helpful to use the diagrammatic representations
of the building blocks \rf{buildblocks} and of the matrix element
\rf{Ymatel} given in Figure \ref{Q-opfig}.

Let $\SY(\la;\bar\mu,\mu)$ be the projection of $\CY(\la,\bar\mu,\mu)$
onto the physical subspace defined by setting all ${z}_n$ to zero. 
It easily follows from \rf{Ymatel} that $\SY(\la;\bar\mu,\mu)$ can
be represented as integral operator with the matrix elements \rf{SYmatel}.

The operator $\CY_{\infty}$ 
satisfies the relations
\begin{equation}
\begin{aligned}
&\CY_{\infty}^{-1}\cdot\sf_{2n-1}^{}\cdot\CY_{\infty}^{}\,=\,
\sf_{2n}^{-1}\,,\\
&\CY_{\infty}^{-1}\cdot\sf_{2n}^{}\cdot\CY_{\infty}^{}\,=\,
\sf_{2n+1}^{}\,,
\end{aligned}\qquad
\CY_{\infty}^{-1}\cdot\eta_{n}^{}\cdot\CY_{\infty}^{}\,=\,\bar{\eta}_n^{}\,.
\end{equation}
This means that $\CY_\infty$ intertwines the representations
${\pi}^{\rm lc}$ and $\bar\pi^{\rm lc}$ respectively.
It follows easily that the projection of $\CY_{\infty}$
onto the physical subspace can  be identified
with the operator denoted $\SY_\infty$, in the sense that
$\overline{\Pi}\cdot \CY_{\infty}=\SY_{\infty}\cdot\Pi$.

\subsection{Comparison with alternative definitions of the Baxter Q-operator}

 A Baxter Q-operator $\SQ^{\rm\sst BT}(u)$
was constructed in \cite{ByT} in such a way that it satisfies
a Baxter-equation of the form
\begin{equation}\label{SG-Baxter}
\ST^{\rm\sst BT}(u)\,\SQ^{\rm\sst BT}(u)\,=\,a^{\rm\sst BT}(u)\SQ^{\rm\sst BT}(u-ib)+d^{\rm\sst BT}(u)\SQ^{\rm\sst BT}(u+ib)\,.
\end{equation}
The coefficient functions $a^{\rm\sst BT}(u)$ and $d^{\rm\sst BT}(u)$ on the right hand sider of \rf{SG-Baxter}
are given explicitly as
\begin{equation}
a^{\rm\sst BT}(u)=d^{\rm\sst BT}(-u)=
e^{-N\pi b s}\,\big[\cosh(\pi b (u-s-\fr{i}{2}b))\big]^N\,.
\end{equation}
The operator $\SQ^{\rm\sst BT}(u)$ constructed in \cite{ByT} can be represented as  the product
$\SQ^{\rm\sst BT}(u)=\SY(u)\cdot \SZ$, with $\SY(u)$ and $\SZ$ being represented by the kernels
\begin{equation}\begin{aligned}
&\langle \,{\mathbf x'}\,|\,\SY(u)\,|\,{\mathbf x}\,\rangle\,=\,
\prod_{r=1}^N\,V_{u-s-c_b}(x_r'+x_{r+1}^{})\,V_{-u-s-c_b}(x_r'-x_{r}^{})\,,\\
&\langle \,{\mathbf x'}\,|\,\SZ\,|\,{\mathbf x}\,\rangle\,=\,\prod_{r=1}^N\,
\bar{V}_{-2s}(x'_r-x_r)\,.
\end{aligned}
\end{equation}

Our aim is to compare $\SQ^{\rm\sst BT}(u)$ with the Q-operators obtained from the
universal R-matrix within the formalism developed in this paper. 
Using formulae \rf{R-factor2} and \rf{r+-factor1}, and following the discussion given in  
Sections \ref{assembly} and \ref{kernels} it
is straightforward to find
\begin{align}
R_{\bar\la,\la; \bar\mu,\mu}(x_K',x_L'|x_{K}^{},x_L^{})\,=\,
& e^{\frac{\pi i}{4}(l-\bar m)^2}V_{\bar{m}-l}(x'_K+x_L')\,
e^{\frac{\pi i}{4}(l-m)^2}\bar{V}_{{m}-l}(x_L'-x_K^{})\\
\times & \,e^{\frac{\pi i}{4}(\bar l-\bar m)^2}
\bar{V}_{\bar m-\bar l}(x_K'-x_L)\,
e^{-\frac{\pi i}{4}(\bar l-m)^2}V_{m-\bar l}(x_{K}+x_L^{})\,,
\notag\end{align}
where $\bar{V}_u(x)$ was defined in \rf{D-FT}.
It follows that the fundamental transfer matrix has the kernel
\begin{align}\label{fundTkernel}
&\langle\,{\mathbf x}'\,|\,\ST(e^{\pi b\bar{l}},e^{\pi b l}; e^{\pi b \bar{m}},e^{\pi b m})\,|\,{\mathbf x}\,\rangle\,=\, 
e^{\frac{\pi i}{4}((l-\bar m)^2+(l-m)^2+(\bar l-\bar m)^2-(\bar l-m)^2)}\times\\
&\quad\times\int dy_1\dots dy_N\,\prod_{r=1}^N
V_{\bar m-l}(y_{r+1}+x_r')\,
\bar{V}_{m-l}(x_r'-y_r^{})\,\bar{V}_{\bar m-\bar l}(y_{r}-x_r)\,
V_{m-\bar l}(y_{r}+x_{r+1}^{})\,.
\notag\end{align}
Setting $\bar l=m$ in \rf{fundTkernel}, for example, one gets
$\SQ(\la;{\bar\mu,\mu}):=\ST(\mu,\la; \bar\mu,\mu)$ with kernel
\begin{align}\label{fundQkernel}
&\langle\,{\mathbf x}'\,|\,\SQ^{\rm\sst SG}(e^{\pi b l};{e^{\pi b \bar{m}},e^{\pi b m}})\,|\,{\mathbf x}\,\rangle\,=\,
e^{\frac{\pi i}{4}((l-\bar m)^2+(l-m)^2+(m-\bar m)^2)}\times\\
&\qquad\times\int dy_1\dots dy_N\,\prod_{r=1}^N
V_{\bar m-l}(y_{r+1}+x_r')\,
\bar{V}_{m-l}(x_r'-y_r^{})\,\bar{V}_{\bar m-m}(y_{r+1}-x_r)
\,.
\notag\end{align}
This  expression can now easily be compared 
with the formulae for the kernel of the lattice-Sinh-Gordon Q-operator 
$\SQ^{\rm \sst BT}(u)$ constructed in 
\cite{ByT}. 
We have
\begin{equation}\SQ^{\rm\sst SG}(q^{\frac{1}{2}}\zeta)\,\equiv\,
\SQ(q^{\frac{1}{2}}\zeta;{\bar\mu,\mu})\,=\,\zeta_b^{4N}\,
e^{\frac{\pi i}{2}((u-c_b)^2+3s^2)}(w_b(u+s))^N\,\SQ_s^{\rm\sst BT}(u)\,,
\end{equation}
if the parameters are related respectively as
\begin{equation}
\begin{aligned} &\mu\,=\,e^{\pi b m}\,=\,\kappa\,=\,m\Delta\,=\,e^{\pi b s}\,,\\
& \bar\mu\,=\,e^{\pi b \bar m}\,=\,\kappa^{-1}\,=\,(m\Delta)^{-1}\,=\,e^{-\pi b s}\,,\end{aligned} \qquad
\zeta\,=\,e^{\pi b l}\,=\,ie^{-\pi b u}\,,
\end{equation}
It follows from \rf{SG-Baxter} that $\SQ(\zeta)$ 
satisfies a Baxter-type equation of the form
\begin{equation}\label{MT-Baxter}
\ST^{\rm\sst SG}(q^{\frac{1}{2}}\zeta)\SQ^{\rm\sst SG}(\zeta)\,=\,
a^{\rm\sst SG}(\zeta)\,\SQ^{\rm\sst SG}(q^{-1}\zeta)+
d^{\rm\sst SG}(\zeta)\,
\SQ^{\rm\sst SG}(q\zeta)\,,
\end{equation}
where 
\begin{equation}
a^{\rm\sst SG}(\zeta)=q^{-\frac{N}{2}}(\zeta/\ka)^{-N}(1-\zeta^2/\ka^2)^N(1-\ka^2\zeta^2)^N\,,\qquad
d^{\rm\sst SG}(\zeta)=q^{-\frac{N}{2}}(\zeta/\ka)^{-N}\,.
\end{equation}
The Baxter equation \rf{MT-Baxter} coincides with the equation derived 
using the representation theory of quantum affine algebras in the main text.

\subsection{Connection with the Faddeev-Volkov model}

We are now going to show how the 1+1-dimensional lattice model studied in this paper is
related to the two-dimensional model of statistical mechanics called Faddeev-Volkov model, 
defined and studied in \cite{BMS}. To this aim it will be useful to introduce the 
Boltzmann weights $W_u(x)$ related to the special function $D_u(x)$ by multiplication with 
a $u$-dependent factor,
\begin{equation}
W_u(x):=\,\Xi(u)\,V_u(x)\,,
\end{equation}
where $\Xi(u):=e^{\frac{\pi i}{4}(2x^2+1+\frac{2}{3b^2}(1+b^4))}\Phi(u)$,
and $\Phi(u)$ is defined as
\begin{equation}
\log\Phi(u):=
\int\limits_{\BR+i0}\frac{dt}{8t}\;\frac{e^{-2itx}}{\sinh (bt)\sinh(b^{-1}t)\cosh((b+b^{-1})t)}.
\end{equation}
The special function $\Phi(u)$ satisfies the functional equations 
\begin{equation}
\Xi(u+c_b)\Xi(u-c_b)\,=\,(w_b(u))^{-1}\,, \qquad \Xi(u)\Xi(-u)\,=\,1\,.
\end{equation}
Together with \rf{D-FT} one finds that $W_u(x)$ is self-dual under Fourier-transformation in the 
sense that
\begin{equation}\label{W-FT} 
\overline{W}_u(x):=\int dy\; e^{2\pi i xy}\,W_{u}(x)\,=\,W_{-u-2c_b}(x)\,.
\end{equation}
Other useful properties noted in \cite{BMS} are
\begin{equation}
W_0(x)=1\,,\qquad \overline{W}_0(x-y)=\de(x-y)\,.
\end{equation}
Let us denote the operator obtained from $\ST$  by the replacement $V_{u}(x)\ra W_u(x)$ and
$\bar{V}_{u}(x)\ra \overline{W}_u(x)$ by $\ST'$. 

It then follows easily from our formula \rf{fundTkernel} 
above that for even number of lattice sites one may identify the kernels 
representing products of fundamental transfer matrices
\begin{equation}\label{prodT}
T_{\bar{\mathbf w},{\mathbf w};{\mathbf s}}^{\rm\sst SG}({\mathbf x}_{N+1}^{},{\mathbf x}_0^{}):=
\langle\,{\mathbf x}_{N+1}^{}\,|\,\Omega_{\rm odd}^{} \ST_{S}'(\bw_M^{},w_M^{})\,\cdots\,
\ST_{S}'(\bw_1^{},w_1^{})\,\Omega_{\rm odd}^{}\,|\,{\mathbf x}_{0}\,\rangle\,,
\end{equation}
where $\Omega_{\rm odd}=\prod_{n=1}^N\Omega_{2n-1}$; we are using the notations
${\mathbf w}=(w_1,\dots,w_M)$, $\bar{\mathbf w}=(\bw_1,\dots,\bw_M)$
and ${\mathbf s}=(s_1,\dots,s_N)$.
Let us temporarily restrict attention to the case that $N$ is even. 
It is easy to see that 
\begin{equation}
T_{\bar{\mathbf w},{\mathbf w};{\mathbf s}}^{\rm\sst SG}({\mathbf x}_{N+1}^{},{\mathbf x}_0^{})=
Z_{\bar{\mathbf w},{\mathbf w};{\mathbf s}}^{\rm\sst FV}({\mathbf x}_{N+1}^{},{\mathbf x}_0^{})
\end{equation}
where $Z_{\bar{\mathbf w},{\mathbf w};{\mathbf s}}^{\rm\sst FV}({\mathbf x}_{N+1},{\mathbf x}_0)$ is the partition 
function of the Faddeev-Volkov model on a rectangular lattice 
which may be explicitly represented as 
\begin{align}\label{ZTkernel}
Z_{\bar{\mathbf w},{\mathbf w};{\mathbf s}}^{\rm\sst FV}({\mathbf x}_{N+1}^{},{\mathbf x}_0^{}):=\!
\int \prod_{n=1}^N\prod_{m=1}^M dy_{n}^{m}\;&\,
W_{w_m-s_n}(y_{n+1}^{m}-y_{n}^{m+1})\,
\bar{W}_{w_m+s_n}(y_{n}^{m+1}-y_{n}^{m})\\[-1ex]
&\times \bar{W}_{{\bar w}_m-s_n}(y_{n}^{m}-y_{n}^{m-1})\,
W_{{\bar w}_m+s_n}(y_{n}^{m}-x_{n+1}^{m-1})\,.
\notag\end{align}
Note that the range of values of the 
parameters considered in \cite{BMS} (motivated by positivity of the Boltzmann weights) corresponds to 
{\it imaginary}  values of $u,u'$ and $s$.


\begin{thebibliography}{GR-AS}

\bibitem[ACDF]{ACDF} 
  D.~Arnaudon, N.~Crampe, A.~Doikou and L.~Frappat,
{\it Spectrum and Bethe ansatz equations for the U(q) (gl(N)) closed and open spin chains in any representation }.
  math-ph/0512037.

\bibitem[AL]{OL}
O. Alekseev, M. Lashkevich, 
{\it 	Form factors of descendant operators: $A_{L-1}^{(1)}$ affine Toda theory}, 
JHEP {\bf 1007} (2010) 095. 

\bibitem[AFZ]{AFZ}
A.E. Arinshtein, V.A. Fateev, A.B. Zamolodchikov,
{\it Quantum s Matrix of the (1+1)-Dimensional Todd Chain}, 
Phys. Lett. {\bf B87} (1979) 389-392. 



\bibitem[AF]{AF} A. Antonov, B. Feigin,
{\it Quantum group representations and the Baxter equation.} 
Phys. Lett. {\bf B 392} (1997), no. 1-2, 115--122.

\bibitem[BaBR]{BBR}
V. Bazhanov, A. Bobenko, N. Reshetikhin,
{\it Quantum discrete sine-Gordon model at roots of $1$:
Integrable quantum system on the integrable classical background}.
Comm. Math. Phys. {\bf 175} (1996) 377--400.

\bibitem[BaFLMS]{BFLMS}
  V.~V.~Bazhanov, R.~Frassek, T.~Lukowski, C.~Meneghelli and M.~Staudacher,
  {\it  Baxter Q-Operators and Representations of Yangians },
  Nucl.\ Phys.\ B {\bf 850}, 148 (2011)
  [arXiv:1010.3699 [math-ph]].
  

\bibitem[BaHK]{BHK} V.V. Bazhanov, A.N. Hibberd, S.M. Khoroshkin,
{\it Integrable structure of $W_3$ Conformal Field Theory, Quantum Boussinesq Theory and Boundary Affine Toda Theory}.
Nucl. Phys. {\bf B622} (2002) 475--547.

\bibitem[BaKMS]{BKMS} 
  V.~V.~Bazhanov, R.~M.~Kashaev, V.~V.~Mangazeev and Y.~.G.~Stroganov,
  {\it $(Z_N\times)^{n-1}$ generalization of the chiral Potts model}.
  Comm. Math. Phys.\ {\bf 138}, 393 (1991).

\bibitem[BaKS13]{BKS13} 
  V.~V.~Bazhanov, A.~P.~Kels and S.~M.~Sergeev,
  {\it Comment on star-star relations in statistical mechanics and elliptic gamma-function identities,}
  J. Phys. A: Math. Theor. 46 (2013) 152001
  [arXiv:1301.5775 [math-ph]].


\bibitem[BaLZ1]{BLZ1}
V.V. Bazhanov, S.L. Lukyanov, A.B. Zamolodchikov, 
{\it Integrable structure of conformal field theory, quantum KdV theory and thermodynamic Bethe ansatz}. 
Comm. Math. Phys. {\bf 177} (1996) 381-398. 


\bibitem[BaLZ3]{BLZ3}
V.V. Bazhanov, S.L. Lukyanov, A.B. Zamolodchikov, 
{\it Integrable structure of conformal field theory. III. 
The Yang-Baxter relation}. 
Comm. Math. Phys. {\bf 200} (1999) 297--324. 

\bibitem[BaLZ4]{BLZ4}
V.V. Bazhanov, S.L. Lukyanov, A.B. Zamolodchikov, 
{\it Integrable quantum field theories in finite volume: Excited state energies}.
Nucl. Phys. {\bf B489} (1997) 487-531. 

\bibitem[BaMS]{BMS} 
  V.~V.~Bazhanov, V.~V.~Mangazeev and S.~.M.~Sergeev,
{\it Faddeev-Volkov solution of the Yang-Baxter equation and discrete conformal symmetry},
Nuclear Physics {\bf B784} (2007) 234-258


\bibitem[BaS05]{BS}
  V.~V.~Bazhanov and S.~M.~Sergeev,
  {\it Zamolodchikov's tetrahedron equation and hidden structure of quantum groups}.
  J.\ Phys.\ A A {\bf 39} (2006) 3295
  [hep-th/0509181].

\bibitem[BaS15]{BS15}
V.V. Bazhanov, S.M. Sergeev,
{\it Yang-Baxter Maps, Discrete Integrable Equations and Quantum Groups},
Preprint  arXiv:1501.06984 [math-ph]









\bibitem[Be1]{Beck1} 
  J.~Beck,
   {\it Braid group action and quantum affine algebras}.
  Commun.\ Math.\ Phys.\  {\bf 165}, 555 (1994)
  [hep-th/9404165].

\bibitem[Be2]{Beck2} 
  J.~Beck,
  { \it Convex bases of PBW type for quantum affine algebras}.
  Commun.\ Math.\ Phys.\  {\bf 165}, 193 (1994).
  [hep-th/9407003].

\bibitem[BoMP]{BMP} P. Bouwknegt, J. Mc Carthy, K. Pilch,
{\it Quantum Group Structure in the Fock Space Resolutions of 
$\widehat{sl}(n)$ Representations}.
Comm. Math. Phys. {\bf 131} (1990) 125--155.


\bibitem[BoGKNR]{Ex}
  H.~Boos, F.~G\"ohmann, A.~Kl\"umper, K.~S.~Nirov and A.~V.~Razumov,
  {\it ``Exercises with the universal R-matrix,''}
  J.\ Phys.\ A A {\bf 43} (2010) 415208.

\bibitem[BCDS]{BCDS}
H.W. Braden, E. Corrigan, P.E. Dorey, R. Sasaki,
{\it 	Affine Toda Field Theory and Exact S Matrices},
Nucl. Phys. {\bf B338} (1990) 689-746. 




\bibitem[BuR]{BR}    D. B\"ucher, I. Runkel, 
{\it Integrable perturbations of conformal field theories 
and Yetter-Drinfeld modules},
J. Math. Phys. {\bf 55} (2014) 111705.



\bibitem[BrZG]{BZG} 
  A.~J.~Bracken, Y.~-Z.~Zhang and M.~D.~Gould,
  {\it ``Infinite families of gauge equivalent R matrices and gradations of quantized affine algebras,''}
  Int.\ J.\ Mod.\ Phys.\ B {\bf 8}, 3679 (1994)
  [hep-th/9310183].



\bibitem[ByT1]{ByT}
A.G. Bytsko, J. Teschner,
{\it Quantization of models with non-compact quantum group symmetry:
modular $XXZ$ magnet and lattice sinh-Gordon model.}  J. Phys. {\bf A39}
(2006), 12927--12981.

\bibitem[ByT3]{ByT2} A. Bytsko, J. Teschner,
{\it The integrable structure of nonrational conformal field theory}.
Adv.Theor.Math.Phys. {\bf 17} (2013) 701-740.

\bibitem[ByT3]{ByT02}	A.G. Bytsko, J. Teschner,
{\it  R-operator, co-product and Haar measure for the modular double of $U_q(sl(2,\BR))$} 
Comm. Math. Phys. {\bf 240} (2003) 171-196.

\bibitem[CP]{CP} 
  V.~Chari and A.~Pressley,
 {\it A guide to quantum groups},
Cambridge University Press, Cambridge (1994).

\bibitem[CM1]{CM1} P. Christe, G. Mussardo,
	{\it Elastic s Matrices in (1+1)-Dimensions and Toda Field Theories}, 
Int.J.Mod.Phys. A5 (1990) 4581-4628.

 \bibitem[CM2]{CM2} P. Christe, G. Mussardo,	
{\it Integrable Systems Away from Criticality: The Toda Field Theory and S Matrix of the Tricritical Ising Model},
 Nucl.Phys. B330 (1990) 465.

\bibitem[Dam1]{Dam1} 
  I.~Damiani,
   {\it A Basis of Type Poincar\'e-Birkhoff-Witt for the Quantum Algebra $\widehat{\mathfrak{sl}}_2$}.
Journal of Algebra  {\bf 161}, 291-310 (1993).
  
\bibitem[Dam2]{Dam2} 
  I.~Damiani,
   {\it La R-matrice pour les alg\'egres quantiques de type affine non tordu}.
Ann.~scient.~\'Ec.~Norm.~sup.~$4^e$ s\'erie, {\bf t.~31}, p. 493-523 (1998).

\bibitem[Dam3]{Dam3} 
  I.~Damiani,
   {\it The R-matrix for (twisted) affine quantum algebras}.
  [1111.4085 math.QA].

\bibitem[DF]{DF} 
  J.~Ding and I.~B.~Frenkel,
    {\it  Isomorphism of two realizations of quantum affine algebra $U_q(\widehat{gl(n)})$}.
  Commun.\ Math.\ Phys.\  {\bf 156}, 277-300 (1993)
 
  \bibitem[DJMM]{DJMM} 
  E.~Date, M.~Jimbo, K.~Miki and T.~Miwa,
{\it Generalized chiral Potts models and minimal cyclic representations of $U_q(gl(n,C))$}.
  Comm. Math. Phys.  {\bf 137}, 133 (1991).


\bibitem[DM]{DM} 
  S.~E.~Derkachov and A.~N.~Manashov,
  {\it  Noncompact sl(N) spin chains: BGG-resolution, Q-operators and alternating sum representation for finite dimensional transfer matrices },
  Lett.\ Math.\ Phys.\  {\bf 97}, 185 (2011)
  [arXiv:1008.4734 [nlin.SI]].
  
\bibitem[Dr1]{Dr1}
 V.G. Drinfel'd,
{\it Hopf algebras and the quantum Yang-Baxter equation}.
Soviet Math. Dokl. {\bf 32} (1985) 254-258.

\bibitem[Dr86a]{D} V.G. Drinfel'd,
{\it Quantum Groups}, in {\em Proceedings of the International Congress of Mathematicians, Vol.~1 (Berkeley, 1986)}, 798--820, Providence, 1987, American Mathematical Society.

\bibitem[Dr86b]{Dr86} 
  V.~G.~Drinfeld,
  {\it ``Quantum groups,''}
  J.\ Sov.\ Math.\  {\bf 41}, 898 (1988)
  [Zap.\ Nauchn.\ Semin.\  {\bf 155}, 18 (1986)].

\bibitem[Dr87]{Dr87} 
  V.~G.~Drinfeld,
  {\it ``A New realization of Yangians and quantized affine algebras,''}
  Sov.\ Math.\ Dokl.\  {\bf 36}, 212 (1988).

\bibitem[EFK]{EFK} 
  P.~I.~Etingof, I.B.~Frenkel and A.~A.~Kirillov,Jr.~,
 {\it Lectures on Representation Theory and Knizhnik-Zamolodchikov Equations},
AMS (1998).

\bibitem[FaR]{FR} 
L.~D.~Faddeev and N.~Y.~Reshetikhin,
 {\it Integrability of the Principal Chiral Field Model in (1+1)-dimension.}
Annals Phys.\  {\bf 167}, 227 (1986).

\bibitem[FaRT]{FaReTa} 
  L.~D.~Faddeev, N.~Y.~.Reshetikhin and L.~A.~Takhtajan,
  {\it Quantization of Lie Groups and Lie Algebras}.
Adv.~Series in Math.~Phys.~{\bf 10} (1989) 299-309 


\bibitem[FaV92]{FV92} L.D.  Faddeev, A. Yu. Volkov,
{\it Quantum inverse scattering method on a space-time lattice}.
Theor. Math. Phys. {\bf  92} (1992) 207-214.

\bibitem[FaV93]{FV93} 
  L.~D.~Faddeev and A.~Y.~.Volkov,
 {\it Abelian current algebra and the Virasoro algebra on the lattice}.
  Phys.\ Lett.\ B {\bf 315}, 311 (1993)
  [hep-th/9307048].

\bibitem[FaV94]{FV94} L.D.  Faddeev, A. Yu. Volkov,
{\it Hirota Equation as an Example of an Integrable Symplectic Map}.
Lett. Math. Phys. {\bf 32} (1994) 125-135.




\bibitem[Fa96]{F}
  L.~D.~Faddeev,
  {\it How algebraic Bethe ansatz works for integrable model.}
  hep-th/9605187.




  
  \bibitem[Fa99]{F99} L.D. Faddeev,	
{\it Modular double of quantum group}, 
Math. Phys. Stud. {\bf 21} (2000) 149-156. 

\bibitem[FaKV]{FKV} L.D. Faddeev, R.M. Kashaev, A.Yu. Volkov,
{\it Strongly coupled quantum discrete Liouville theory. 1. Algebraic approach and duality}, 
Comm. Math. Phys. {\bf 219} (2001) 199-219.

\bibitem[FeF1]{FF1}
B. Feigin, E. Frenkel, {\it Free field resolutions in affine Toda theories}. 
Phys. Lett. {\bf B276} (1992) 79-86.

\bibitem[FeF2]{FF2}
B. Feigin, E. Frenkel, {\it Integrals of motion and quantum groups}, in:  
Proceedings of the C.I.M.E. summer school ``Integrable Models and Quantum Groups'' (Italy, 1993), 
Lecture Notes in Mathematics {\bf 1620}, Springer, 1995.




  
\bibitem[FrH]{FH}
  E.~Frenkel and D.~Hernandez,
  {\it Baxter's Relations and Spectra of Quantum Integrable Models}.
  arXiv:1308.3444 [math.QA].

\bibitem[FrM]{FM}
  E.~Frenkel and E.~Mukhin,
  {\it Combinatorics of q characters of finite dimensional representations of quantum affine algebras}.
  Commun.\ Math.\ Phys.\  {\bf 216}, 23 (2001)
  [math/9911112 [math-qa]].




 \bibitem[FrR]{FrR}
 E.~Frenkel and N.~Reshetikhin,
{\it The q-characters of representations of quantum affine algebras and deformations of W-algebras}.
  [math/9810055 [math-qa]].

\bibitem[FrKS]{FKS}
A. Fring, C. Korff, B.J. Schulz,00
{\it The ultraviolet Behaviour of Integrable Quantum Field Theories, Affine Toda Field Theory},
Nucl. Phys. {\bf B549} (1999) 579-612.

  
  
  \bibitem[HJ]{HJ} D.
Hernandez, M. Jimbo, 
{\it Asymptotic representations and Drinfeld rational fractions.} 
Compos. Math. {\bf 148} (2012) 1593--1623. 

\bibitem[Hi98]{Hi98} 
  K.~Hikami,
  {\it On the fundamental L operator for the quantum lattice W algebra }.
  Chaos, Solitons \& Fractals  {\bf 9}, 853 (1998)
  [hep-th/9404038].

\bibitem[Hi01]{Hi01} 
  K.~Hikami,
  {\it The Baxter equation for quantum discrete Boussinesq equation}.
  Nucl.\ Phys.\ B {\bf 604}, 580 (2001)
  [nlin/0102021 [nlin-si]].
  
  \bibitem[HKKR]{HKKR}	
T. Hoffmann, J. Kellendonk, N. Kutz, N. Reshetikhin,
Factorization dynamics and Coxeter-Toda lattices 
Comm. Math. Phys. {\bf 212} (2000) 297-321. 

\bibitem[Ioh]{Ioh} 
  K.~Iohara,
    {\it  Bosonic representations of Yangian double $\mathcal{D}Y_{\hbar}(\mathfrak{g})$ with $\mathfrak{g}=\mathfrak{gl}_N,\mathfrak{sl}_N$}.
  J.~Phys~A: Math.~Gen. {\bf 29},  15  (1996)
  [q-alg/9603033].

\bibitem[Ip12]{Ip12} 
  I.~C.~Ip,
  {\it Positive Representations of Split Real Quantum Groups: The Universal R Operator}.
 Int.~Math.~Research Notices (2013), 
  [arXiv:1212.5149 [math.QA]].

\bibitem[IpZ]{IZ}
I. C.-H. Ip, A. M. Zeitlin,
{\it Supersymmetry and the Modular Double}.
In: {\it Recent Advances in representation theory, algebraic geometry, and related topics}, 
81-97,  Contemp. Math. {\bf 623}, Amer. Math. Soc., Providence, RI, 2014.



\bibitem[IS03]{IS03}
  A.~P.~Isaev and S.~M.~Sergeev,
   {\it Quantum Lax operators and discrete 2+1-dimensional integrable models}.
  Lett.\ Math.\ Phys.\  {\bf 64}  (2003)  57.

\bibitem[Ito]{Ito} 
  K.~Ito,
  { \it The Classification of Convex Orders on Affine root systems}.
Commun.~in Alg. {\bf 29}:12  (2001) 5605-5630.



\bibitem[J]{J} M. Jimbo,
{\it A $q$-Difference Analogue of $U \left( \mathfrak{g} \right)$ and the
  Yang-Baxter Equation}.
Lett. Math. Phys. {\bf 10} (1985) 63--69.

\bibitem[J85]{J85} 
  M.~Jimbo,
  {\it A q Analog of u (Gl (n+1)), Hecke Algebra and the Yang-Baxter Equation}.
  Lett.\ Math.\ Phys.\  {\bf 11} (1986) 247.

\bibitem[Kac]{Kac} 
  V.~Kac,
 {\it Infinite dimensional Lie Algebras},
3rd ed.~,
 Cambridge University Press, Cambridge (1990).


  \bibitem[KaM]{KM} 
  R.~M.~Kashaev, V.~V.~Mangazeev 
 {\it  Cyclic L operators connected with $U_q(sl(n))$ algebra and related integrable models},
Preprint IFVE-93-20 (1993). 


\bibitem[KaR]{KaR} R.M. Kashaev, N. Reshetikhin, 
{\it Affine Toda field theory as a $3$-dimensional integrable system.} 
Comm. Math. Phys. {\bf 188} (1997) 251--266. 

\bibitem[KiR]{KiR} 
  A.~N.~Kirillov and N.~Reshetikhin,
  {\it q-Weyl Group and a Multiplicative Formula for Universal R-Matrices}.
  Commun.\ Math.\ Phys.\  {\bf 134}  (1990) 421-431.



\bibitem[KhST94]{KST94} 
  S.~M.~Khoroshkin, A.~A.~Stolin and V.~N.~Tolstoi,
  {\it Gauss decomposition of trigonometric R matrices }.
  Mod.\ Phys.\ Lett.\ A {\bf 10}  (1995) 1375.
  [hep-th/9404038].


\bibitem[KhT91]{KT1}
S.~M. Khoroshkin, V.~N. Tolstoy,
\emph{The universal {$R$}-matrix for
  quantum untwisted affine {L}ie algebras},
{Funct. Anal. Appl.} \textbf{26}
  (1992) 69--71.
  
 \bibitem[KhT91a]{KT91a}
S.~M. Khoroshkin, V.~N. Tolstoy,
{\it The universal {$R$}-matrix for
  quantum untwisted affine {L}ie algebras},
{Funct. Anal. Appl.} \textbf{26}
  (1992) 69--71.
 
 \bibitem[KhT91b]{KT91b}
S.~M. Khoroshkin, V.~N. Tolstoy,
{\it Universal {$R$}-matrix for
  quantized (super)algebras},
{Commun. Math. Phys. } \textbf{141}
  (1991) 599--617.
 
\bibitem[KhT92]{KT} S.M. Khoroshkin, V.N. Tolstoy,
{\it The Uniqueness Theorem for the Universal $R$-matrix}.
Lett. Math. Phys. {\bf 24} (1992) 231--244.

\bibitem[KhT92]{KT92}
S.~M. Khoroshkin, V.~N. Tolstoy,
{\it The Uniqueness Theorem for the Universal {$R$}-matrix},
{Letters in Math. Phys.} \textbf{24}
  (1992) 231--244.

\bibitem[KhT93a]{KT93a}
  S.~M.~Khoroshkin and V.~N.~Tolstoy,
  {\it ``On Drinfeld's realization of quantum affine algebras,''}
  J.\ Geom.\ and Phys.\  {\bf 11} (1993) 445-452.


\bibitem[KhT93b]{KT93b}
  S.~M.~Khoroshkin and V.~N.~Tolstoy,
  {\it ``The Cartan-Weyl basis and the universal R-matrix for quantum Kac-Moody algebras and super algebras,''}
  Proc.~of the Int.~Workshop on Math.~Physics. ``Quantum symmetries'' (1993) pp.~336-251.

\bibitem[KhT94]{KT94} S.M. Khoroshkin, V.N. Tolstoy,
{\it Twisting of quantum (super)algebras. Connection of Drinfeld's and Cartan-Weyl
realizations for quantum affine algebras}.
arXiv:hep-th/9404036.


\bibitem[KhT2]{KT2}
  S.~M.~Khoroshkin and V.~N.~Tolstoy,
  {\it ``Twisting of quantum (super)algebras. Connection of Drinfeld's and Cartan-Weyl realizations for quantum affine algebras,''}
  [hep-th/9404036].

\bibitem[Ko]{Ko}
  T.~Kojima,
  {\it Baxter's Q-operator for the W-algebra WN}.
  J.\ Phys.\ A A {\bf 41} (2008) 355206.

\bibitem[KuSk81]{KuSk81}
  P.~P.~Kulish and E.~K.~Sklyanin,
  {\it Quantum Spectral Transform Method. Recent Developments }.
  Lect.\ Notes Phys.\  {\bf 151} (1982) 61.


\bibitem[LS]{LS} 
  S.~Levendorskii and Y.~Soibelman,
   {\it  Quantum Weyl Group and Multiplicative Formula for the R-Matrix of a Simple Lie Algebra}.
Funct.\ Analysis and its Appl.\  {\bf 25}  (1991) 143-145.

\bibitem[LSS]{LSS} 
  S.~Levendorskii, Y.~Soibelman and V.~Stukopin,
   {\it  Quantum Weyl Group and Universal Quantum R-matrix for Affine Lie Algebra $A^{(1)}_1$}.
Lett.\ in Math.\ Phys.\  {\bf 27}  (1993) 253-264.

\bibitem[Lu97]{L97} S.L. Lukyanov,
{\it Form-factors of exponential fields in the affine $A^{(1)}_{N-1}$ Toda model},
Phys. Lett. {\bf B408} (1997) 192-200.

\bibitem[Lu00]{L00} S.L. Lukyanov,
{\it Finite temperature expectation values of local fields in the sinh-Gordon model},
Nucl. Phys. {\bf B612} (2001) 391-412. 

\bibitem[LuZ]{LZ}
	S.L. Lukyanov, A.B. Zamolodchikov,
{\it Quantum Sine(h)-Gordon Model and Classical Integrable Equations}, 
JHEP {\bf 1007} (2010) 008. 

\bibitem[Lus]{Lu} 
  G.~Lusztig,
 {\it Introduction to quantum groups},
Progress in Mathematics {\bf 110},
Birkh\"auser (1993).

\bibitem[Mo]{Mol}
  A.~Molev, M.~Nazarov and G.~Olshansky,
  {\it Yangians and classical Lie algebras}.
  Russ.\ Math.\ Surveys {\bf 51} (1996) 205. 


\bibitem[NaTa]{NaTa} 
  M.~Nazarov and V.~Tarasov,
 {\it Yangians and Gelfand-Zetlin bases}.
  Publ.\ Res.\ Inst.\ Math.\ Sci.\ Kyoto {\bf 30} (1994) 459. 


\bibitem[PT99]{PT99}
B. Ponsot, J. Teschner,
{\it Liouville bootstrap via harmonic analysis on a noncompact quantum group},
Preprint arXiv:hep-th/9911110.

\bibitem[Ra13]{Raz13} 
  A.~V.~Razumov,
  {\it Monodromy operators for higher rank}.
  J.~Phys.~A:\ Math.\ Theor. \  {\bf 46}, 385201 (2013).

\bibitem[ReSe]{ReSe} 
  N.~Y.~.Reshetikhin and M.~A.~Semenov-Tian-Shansky,
  {\it Central extensions of quantum current groups}.
  Lett.\ Math.\ Phys.\  {\bf 19}  (1990) 133.


\bibitem[RiT]{RT}
D. Ridout, J. Teschner
{\it Integrability of a family of quantum field
theories related to sigma models.}
Nuclear Phys. {\bf B853} (2011) 327--378. 
Preprint [arXiv:1102.5716]

%
\bibitem[Ro]{Ro} 
  M.~Rosso,
  {\it An Analogue of PBW Theorem and the Universal R-Matrix for $U_{\hbar}sl(N+1)$}.
  Commun.\ Math.\ Phys.\  {\bf 124}  (1989) 307-318.

\bibitem[Stu]{Stu}
  V. Stukopin,
  { \it Quantum Double of Yangian of Lie Superalgebra $A(m,n)$ and computation of Universal R-matrix}.
  [math/0504302].



\bibitem[Tar92]{Tar92} 
  V.~Tarasov,
  {\it Cyclic monodromy matrices for sl(n) trigonometric R matrices}.
  Commun.\ Math.\ Phys.\  {\bf 158}  (1993) 459.

\bibitem[Tol]{Tol} 
  V.~N.~Tolstoy,
  { \it Extremal projections for quantized Kac-Moody  superalgebras and some of their applications}.
 Springer Lecture Notes in Physics \textbf{370}  (1990) 118-125.

\bibitem[Tol2]{Tol2} 
  V.~N.~Tolstoy,
  { \it Extremal projectors for contragradient Lie (super)symmetries (short review)}.
  [1010.4054].


\bibitem[Vo]{Vo} A.Yu. Volkov, 
{\it Noncommutative hypergeometry}
Comm. Math. Phys. {\bf 219} (2001) 199-219. 

\bibitem[WQ]{RW}
  M.~Rossi and R.~Weston,
 {\it A Generalized Q operator for U(q)(affine sl(2)) vertex models}.
  J.\ Phys.\ A A {\bf 35} (2002) 10015
  [math-ph/0207004].


\bibitem[Ya]{Ya}  Hiroyuki
Yamane, 
{\it Quantized enveloping algebras associated with simple Lie superalgebras 
and their universal $R$-matrices.}
Publ. Res. Inst. Math. Sci. {\bf 30} (1994), no. 1, 15--87. 

\bibitem[Y]{Y}
H. Yamane,,
{\it Quantized enveloping algebras associated to simple Lie superalgebras and their universal R-matrix},
{Publ.~Res.~Inst.~Math.~Sci~} \textbf{30}
  (1994), 15--87.


\bibitem[Z00]{Z00} Al. B. Zamolodchikov, 
{\it On the thermodynamic Bethe ansatz equation in sinh-Gordon model.} 
J.Phys. {\bf A39} (2006) 12863-12887. 

















\end{thebibliography}
\end{document}